UNIVERSITÀ DEGLI STUDI DI TERAMO
DIPARTIMENTO DI SCIENZE DELLA COMUNICAZIONE

# ASPECTS OF MECHANICAL DESIGN FOR AN INFRARED ROBOTIC TELESCOPE IN ANTARCTICA: IRAIT

TESI DI DOTTORATO DI RICERCA IN
ASTROFISICA COMPUTAZIONALE
CICLO XVIII

CANDIDATO:
Igor Di Varano

TUTORS:
Prof. Gino Tosti,
Prof. Oscar Straniero

COORDINATORE:
Ch.mo Prof. Franco Eugeni

# CONTENTS

**ACKNOWLEDGMENTS**

I would like to express my deepest gratitude to Professor Gino Tosti, who has so kindly been the supervisor of this thesis. Its dedication for the IRAIT project and his great knowledge on the matter have both been a constant target for my work.
Furthermore, I would extend my gratitude to Professor Oscar Straniero for having strongly stimulated all my enthusiasm for this research.
I must thank the technical staff and, particularly, Giuliano Nucciarelli, for his precious advice that made it possible, for me, to find all I needed for the actual work.
I am especially grateful to Professor Maurizio Busso, from Perugia University, for having encouraged me and for having had faith in my project.
Finally, many thanks to my colleagues from Osservatorio Astronomico di Teramo, for their cordiality and attention showed about my work.

# INTRODUCTION

The purpose of this thesis is to focus attention on the mechanical aspects in designing an infrared telescope, IRAIT (International Robotic Antarctic Infrared Telescope), entirely robotic and remote controlled, which must operate at Dome C, on Antarctic Plateau. Before illustrating in detail the choice criteria for the various mechanical components in order to satisfy stress requirements and structural verification, adopted test issues, and other technical solutions, firstly a few questions need to be answered. They mainly concern the preference for Dome C as probably the best observing site in the world, the scientific targets, instruments and tools necessary to reach such goals.

Of course a mechanical project, to be worth, needs to match also other abilities. Material requirements, truss configuration, or joint elements must be selected in agreement with the optical restraints, electronics, and, above all, astronomical requirements. We have to consider that the system must be unattended as well, so that everything must be compatible with software and hardware tasks. Of course the selection of mechanical components must be done according to the available budget asset.

- In the first chapter of this work a general overview of exceptional site characteristics and scientific experiments conducted at Dome C is presented, with data collected from the last campaign.
- The second chapter contains the state of art of IRAIT project, with a brief description of camera, optical layout, mechanisms of the secondary and tertiary mirrors, hardware and software control, and it also introduces the discussion about mechanics.
- Aspects of mechanical structure of the telescope are analyzed more widely in the third chapter. Here the results of the structural analysis that I conducted through a finite element method software are presented. They concern the behavior of single parts, subassemblies and overall structure to active loads applied. It is shown that, as a matter of fact, thermal stress can be reckoned as the most influent of all static loads. I've also made an estimation of eigenfrequencies of some critical subassemblies, to study the dynamic response of the system aiming at the best insulation from vibrations, for a good performance of the telescope.

- The argument of the fourth chapter regards the criteria adopted to select some machine elements, such as the bearings, gears, drives and joints.
- On the basis of displacements and stress checks, I determined the systematic mechanical errors, that together with astronomical ones, affect each observation. Such errors are described in chapter five. They are mainly due to flexures, gear mating, axes misalignments, and other causes. Their presence must be taken into account and corrected through appropriate algorithms in the telescope control software.
- The sixth chapter contains results retrieved from my thermal analysis, made under a CFD (Computational Fluo-Dynamic) software. These data include thermal response and heat fluxes in unsteady state inside the electronic boxes, on the basis of data collected by sensors on the last campaign at Dome C.
- In the last chapter my conceptual design of the mechanical interface and the layout of the rack with the AMICA camera and ancillary components is reported. Considerations about mounting operations and camera mechanical alignment are also exposed.
- Finally, the conclusions illustrate the future integration between simulated data and analysis expected from tests, which are planned to be done in a climatic chamber; suggestions about future developments of the project are briefly discussed .

# CHAPTER 1 Characteristics of the Antarctic Site Dome C

## 1.1 General characteristics of Antarctic Plateau

Antarctica, even if well known as the less welcoming land for a human being, has always fascinated his mind since the end of the XVIII century, when James Cook first circumnavigated polar regions. Soon many expeditions followed. One of the missions most frequently mentioned is the one led by Ross, who sailed a bay, named after him, moving further to a distance of 1100 km from the Pole (1839-43). The first to reach South Pole was the Norwegian Roald Amundsen on 16th December 1911, followed by Scott, only 3 weeks later. Interest of scientific projects in Antarctica arose among several nations starting from the year 1882, when the first IGY (International Geophysical Year) was inaugurated. The main attention focused on purposes such as Earth magnetic field, or auroras australis. Anyway Italy was involved with exploration activities in the IGY only in 1957, when an Italian ambassador in Wellington, Macchi di Cellere, asked the Italian Government to let an assistant join in scientific activities at Base Scott, on Ross Island. In spite of Italian adhesion to the IGY, there were no independent missions, but data and information collected by other countries were only processed. In this occasion the world-wide scientific community created an organization in order to promote and coordinate scientific research in Antarctica, the SCAR (Scientific Committee on Antarctic Research).

In 1959 an agreement among 12 major countries was stipulated, which led to the Washington Treaty, in 1961. It ratified the total demilitarization and denuclearization of all countries under 60° latitude South. Due to this delay, Italy had to demonstrate a real interest in research activities as well as in exploitation of mining resources.

Italy subscribed the Treaty only after 20 years, and in 1985 the PNRA (Programma Nazionale di Ricerche in Antartide) was established. At the end of 1991 a quinquennial planning for research activities of Italy in Antarctica was approved.

It intended to operate within a program, according to the SCAR targets. In this contest technological and scientific research should be carried out in a full international visibility .

Main areas of interest are geophysics, oceanology, biology, glaciology, astronomical observatories and geodetics, cosmology and astroparticles. Among all, priorities are provided for international cooperation in the analysis of global climatic processes, as well as the development and testing of new technologies. Some of these missions have already achieved successful results, like BOOMERANG experiment, which has revealed by a balloon-borne instrument and from the measurement of CMB angular power spectrum, the flatness of the universe. Therefore astrophysical ctivities extend from cosmological studies to Sun-Earth interactions, passing through the detection of high energy particles related to auroras phenomena, or by means of installation of visual, IR and submillimetric telescopes. In fact the exceptional transmittance and transparency of the site permit observations in a broad range of the spectrum, especially at those wavelengths not yet totally explored . Two sites have been selected as permanent bases by PNRA. The first one is Mario Zucchelli Station, in Terra Nova Bay, along the coast of the Ross Sea; the other is Dome Concordia, on the Antarctica Plateau.

### 1.1.1 Mario Zucchelli Station

Baia Terranova Station (BTS), now called Mario Zucchelli Station (MZS) in memory of ENEA Antarctica Project Leader, who prematurely died in November 2003, was built on a little peninsula, on the Northern coast of Victoria Land, at 74°41’ latitude South, and 164°03’ longitude East. All facilities, including the main building and hangars, spread out over a region of 50000 $m^2$, which is reachable by plane in October-November, and only by ship from December to February.
Because of very strong katabatic winds, blowing during Winter months, all the campaigns in MZS are held on the long Austral day. As a rule they are distributed in 2 or 3 periods, and the maximum number of participants is limited to seventy. For external activities helicopters, twin otters, and mobile facilities are available.
The OASIS laboratory (an Astrophysical Observatory for investigations in sub-millimetric field), installed in Terra Nova Bay during the Second Italian Antarctic Expedition, which was held between December 1986 and February 1987, carried out a set of data in order to investigate the local atmospheric transmission at millimetric wavelengths. It demonstrated the advantages which would come out from the realization of a far infrared telescope, due to the very low vapour content, estimated to be 1.5 mm in absence of continental wind (see Dall’Oglio G., DeBernardis P., 1987) .
ARENA is a consortium among various research laboratories, universities, funding agencies, and industrial companies, involving seven European countries, and Australia. It is the first European initiative to coordinate astrophysical programmes, with related infrastructure implementation in Antarctica. It is aimed at fostering cooperation between small groups, which are already joining in current Antarctic astronomical projects, or planning to start new ones[1].

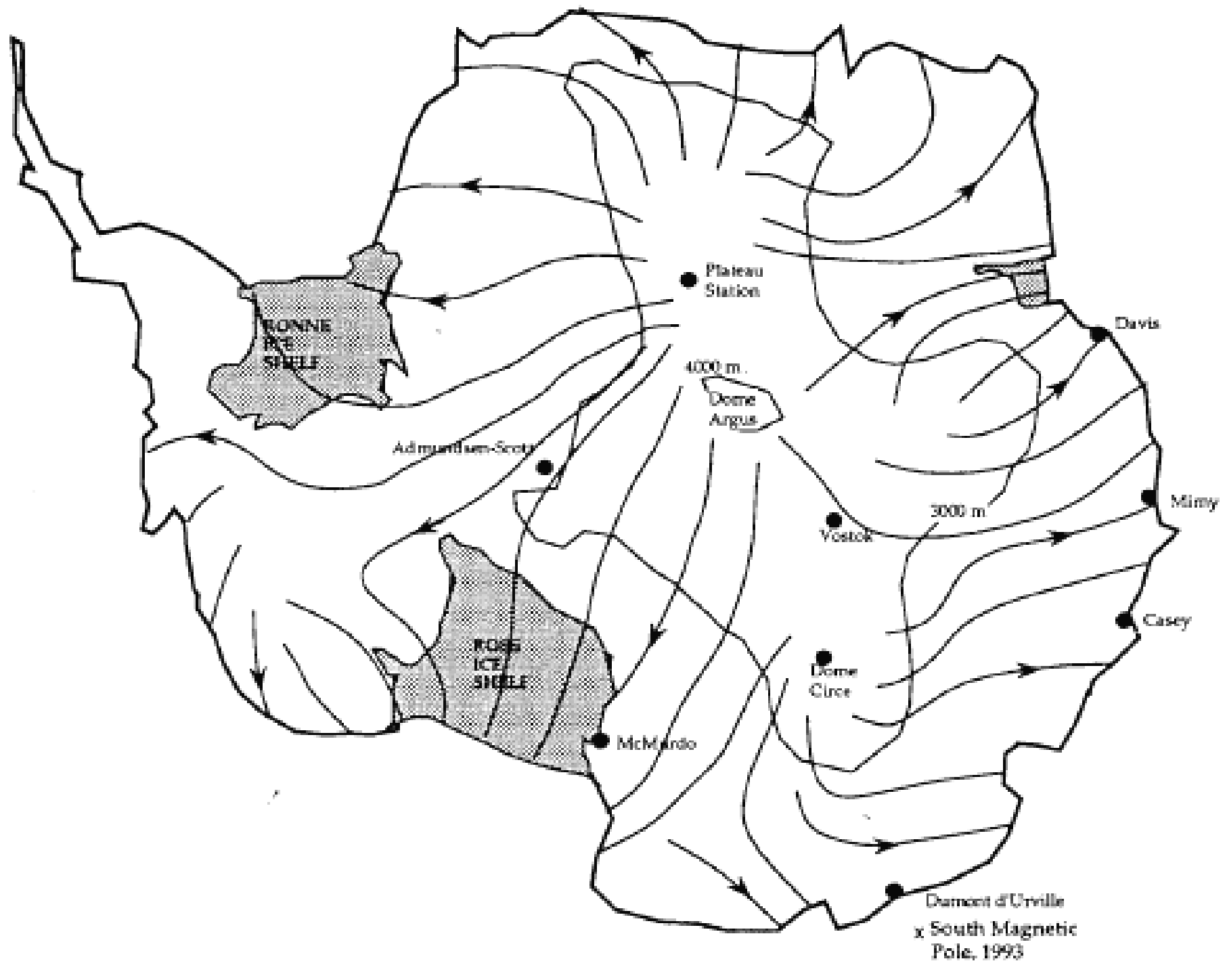

**Figure 1.1** A map of Antarctica continent with katabatic wind fluxes represented.

### 1.1.2 Dome C Concordia Station

Dome Concordia Station (75°06' South and 123°24' East), is 3233 m above sea level, on the Antarctic Plateau. It is 1200 km far from Terra Nova Bay and 1100 km from Dumont d'Urville Base. In 1993 ENEA and IPEV (French Polar Institute) signed an agreement , in which they decided to build up a scientific base at Dome C (Concordia). The major "peaks" of the continent are called "Domes": they're quite different from Transantarctic Mountains, because they are flat, just like plateaus. Dome A (Argus, 4100 m) is probably the best observing site on the planet, but it is very difficult to be reached; the other is Dome F (Fuji, 3810 m) (M. Candidi, A. Lori, 2001).
The construction of the permanent station began in summer 1999-2000 and has been completed this year. It consists of 2 polygonal towers, each of 18 sides, on 3 floors, connected by a hanging

1 *Antarctic Research: a European Network in Astronomy,* Call identifier FP6-2004-Infrastructures-5, November 2004.

tunnel. One building is quiet, the other noisy and hosts workshops and a power station. Each of them has a weight of 150 t, and is supported by six hydraulic feet with a maximum elongation of 40 cm, so that they prevent sinking into snow and possible misalignments.
There is also a summer camp, composed of 7 containers for the sleeping area, 12 for the living area, a power-station with other tents and containers for storage, and several services. It can accommodate 13 persons during winter, and 32 on summer.

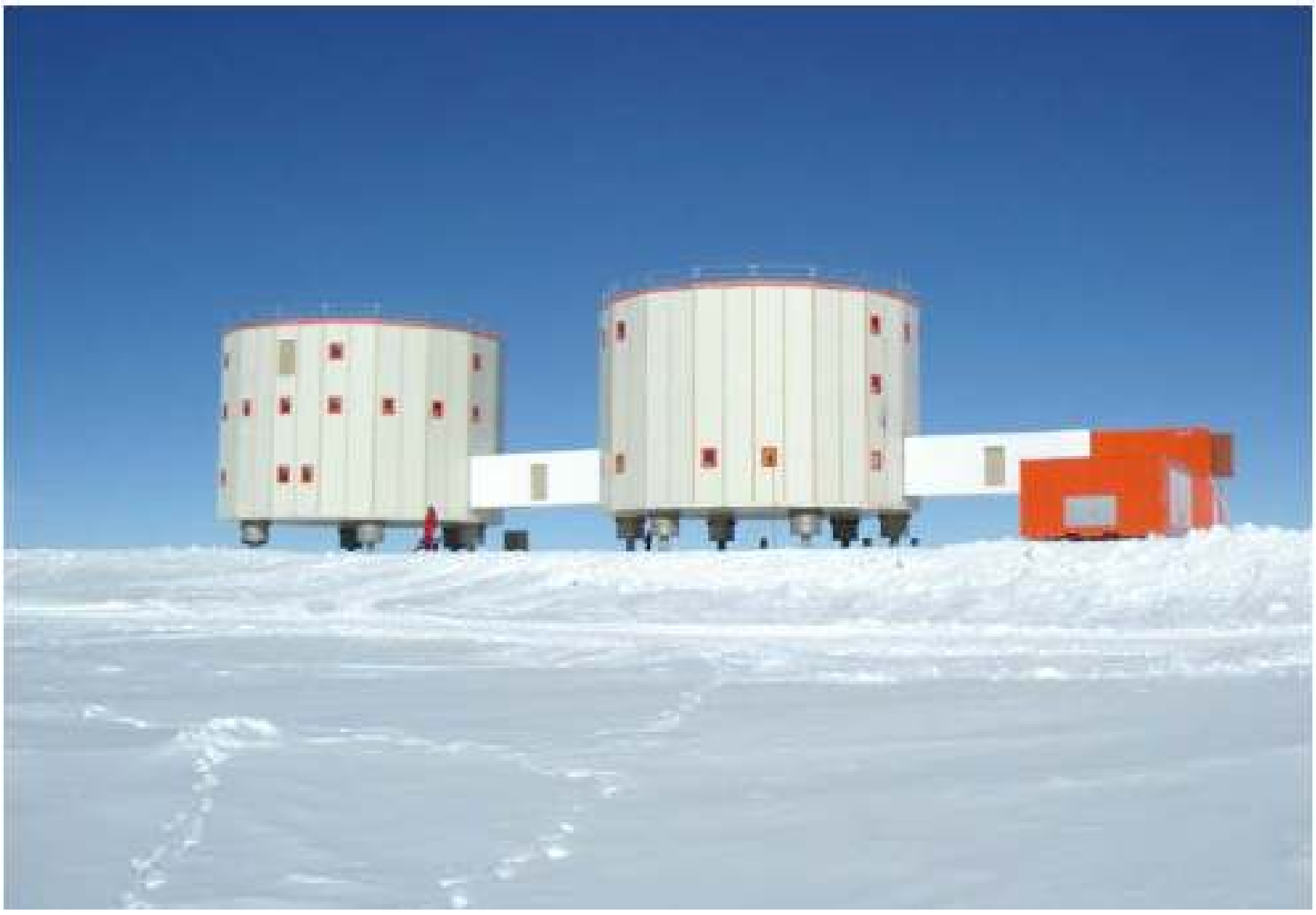

**Figure 1.2 A picture of the three main buildings at Concordia base.**

## 1.2 Antarctica as an astronomical site

The principal characteristic of Antarctica Plateau is the very low temperature, both on minimum and average values. It is like a huge slab of ice 2.5 km thick, with an elevated albedo, which remains at a constant temperature. Average summer temperature is -25 °C, and in winter it falls down to -70 °C. The remarkable transparency in MIR, for the very low vapour content (about 0.25 mm), permits an efficient heat loss through the atmosphere. Temperature trend is very singular in this place: it goes down very rapidly to a minimum equilibrium point and it remains steady all the season. As a consequence there's an inversion layer a few hundreds meters above the ground; it was estimated to be 300 m at South Pole, where the temperature is 22° warmer (J. S. Lawrence, 2004) than on the ground, whereas at Vostok a gradient of 25 °C was measured along a 500 m layer. This uniformity of territory and atmosphere implies a very slow response time to environment changes such as the *global warming*. All experiments conducted on the Plateau were done assuming that the dynamical atmospheric turbulence follows Kolmogorov behaviour, that is a steady turbulence. On the basis of this model, structure constants can be used to determine fluctuation intensity of a physical quantity dependent on time. For example velocity profile is given by the formula: $D = <[V(x) - V(x+r)]^2> = C_V^2 r^{2/3}$ for $l<<r<<L$.

where $l$ and $L$ are respectively the inner and outer scale (M.Azouit, J.Vernin, 2005) of the turbulent motion. The relationship between inner and outer scale is approximately given by: $l_0 \cong L_0 / \mathrm{Re}^{3/4}$ .

The same can be repeated for temperature. Another structure constant of refraction index, $C_N$ ,can be found, knowing that it is proportional to P and inversely proportional to T:

$$C_N^2(h) = \left( 80 \cdot 10^{-6} \frac{P(h)}{T(h)^2} \right)^2 C_T^2(h).$$

Hence, it is possible to determine the Fried parameter by the formula:

$r_0 = \lambda^{6/5} [\sec(z)]^{-3/5} \left[ \int_0^H C_N(z) dz \right]^{-3/5}$ where $z$ represents the zenith distance.

The seeing $\varepsilon$ is in inverse proportion to $r_0$, and is given by:

$\varepsilon \approx \lambda / r_0$

Other basic parameters are the coherence time $\tau_0$ and isoplanatic angle $\theta_0$, which determines the range over which atmospheric phase fluctuations are coherent. They can be obtained respectively by the relations:

$\tau = 0.3 \frac{r_0}{\langle V \rangle}$, where $V$ stands for the wind velocity and $\theta_0 = 0.314 \times \left( \frac{r_0}{\langle h \rangle} \right) \cos\zeta$ .

With $\zeta$ the zenith distance and $h$ the characteristic height of turbulent layer .

In the transparent near and mid-infrared regions (up to 20μ), there's an improvement of 20%–80% in sensitivity at Dome C, consistent with the decrease in sky spectral brightness. In less-transparent regions at the edges of absorption bands (between 2.7 and 7 μ), the decrease in emission combines with the increase in transmission to generate a substantial benefit for the high-plateau sites (a factor of 2–3 for Dome C), which should allow these wavelengths to be examined considerably better than at temperate latitudes. In the mid to far-infrared, sensitivity increases with respect to South Pole up to a factor of 100 are predicted for Dome A and 10 for Dome C.

In the spectral region between 20 and 40 μ, and in the range 5-25 μ, windows may be broader, more stable and less affected by absorption and emission than in any other place on Earth. Subarcsec seeing conditions allow diffraction limited imaging (at least at near and mid-IR wavelengths) without complex optics [5].

A moderate size telescope on the Antarctic Plateau can be as powerful as an instrument of far aperture, operating in temperate sites.

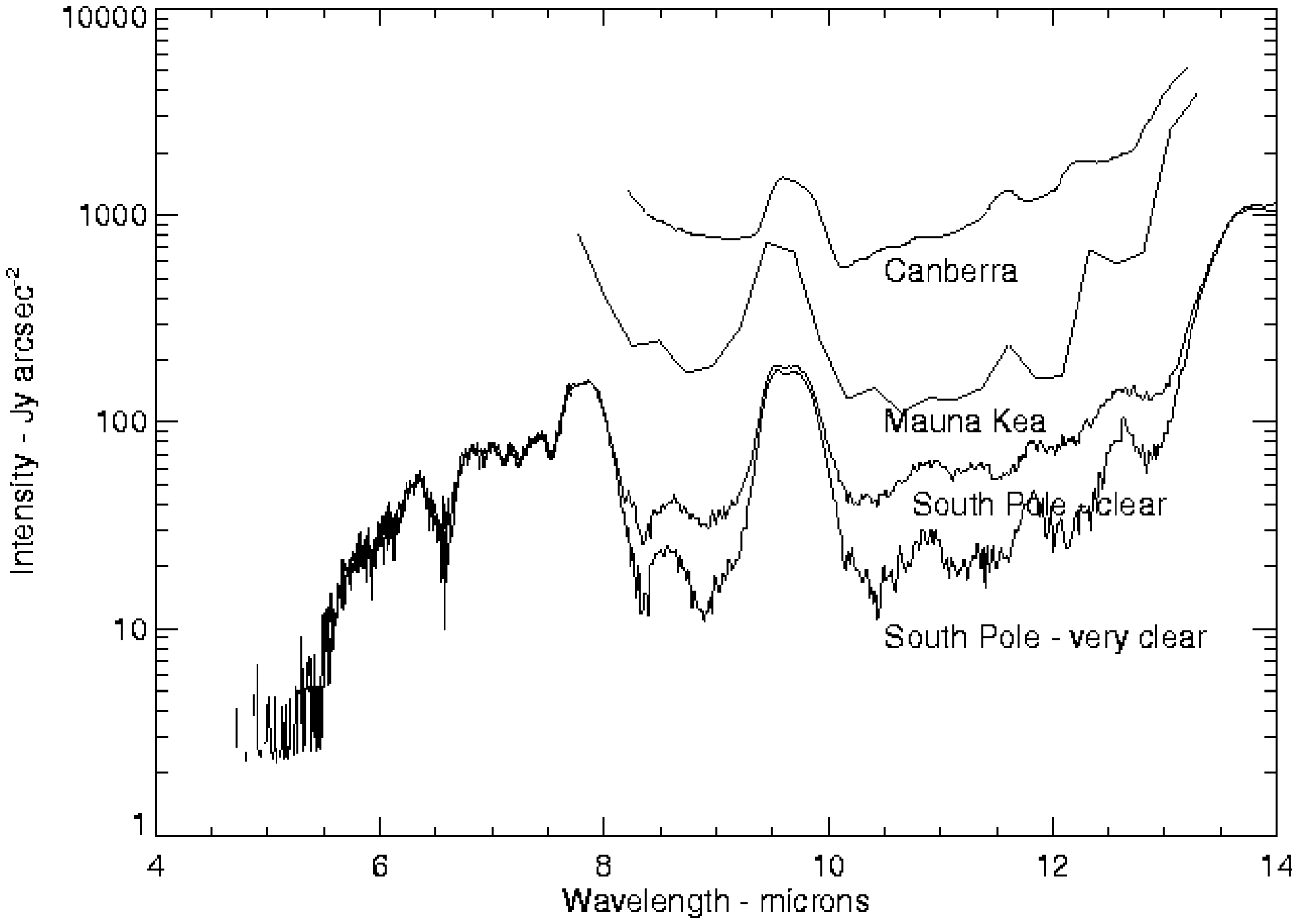

**Figure 1.3** This plot shows the exceptionally low sky brightness of South Pole compared to other Best Sites on Earth [by Chamberlain et al. 1999].

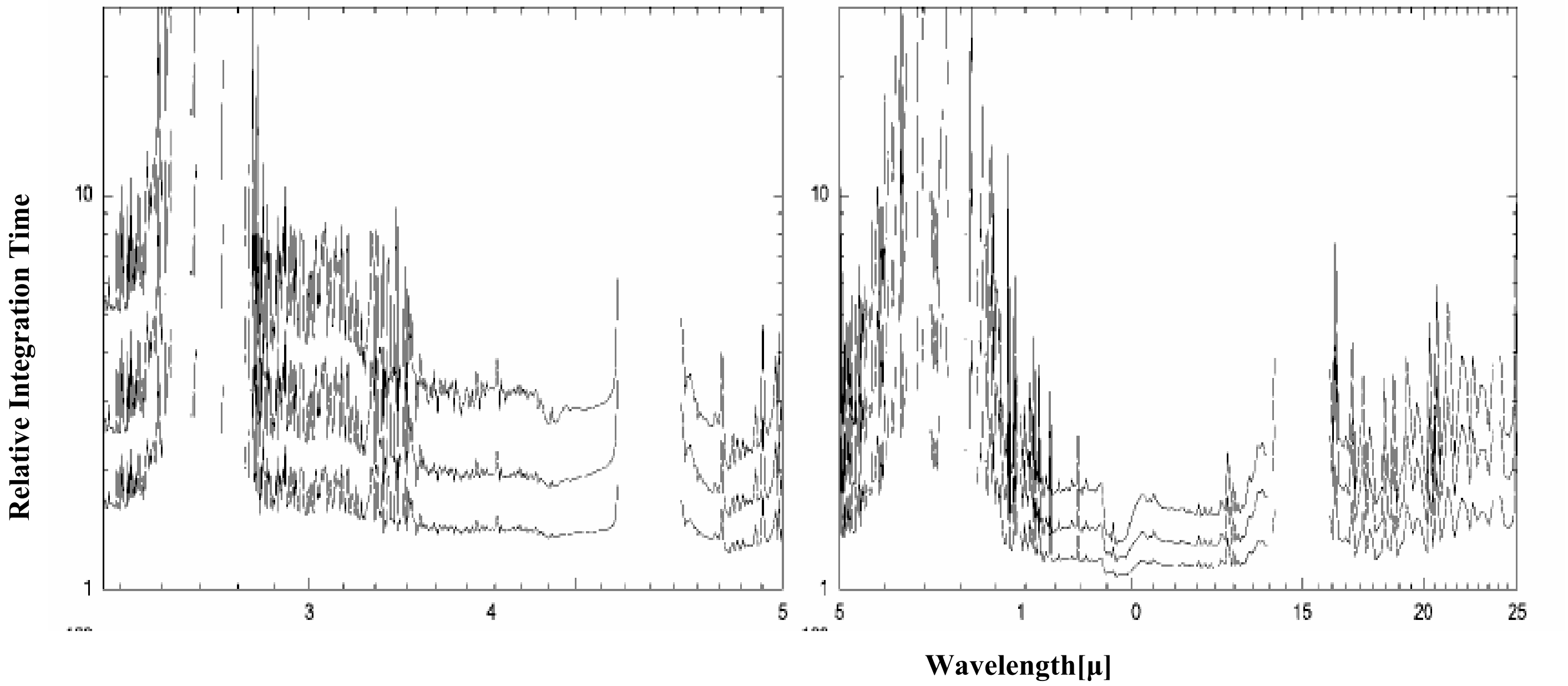

**Figure 1.4** Ratios of integration time in NIR & MIR at Dome C (bottom curve), Dome F(middle curve) and Dome A (top curve).

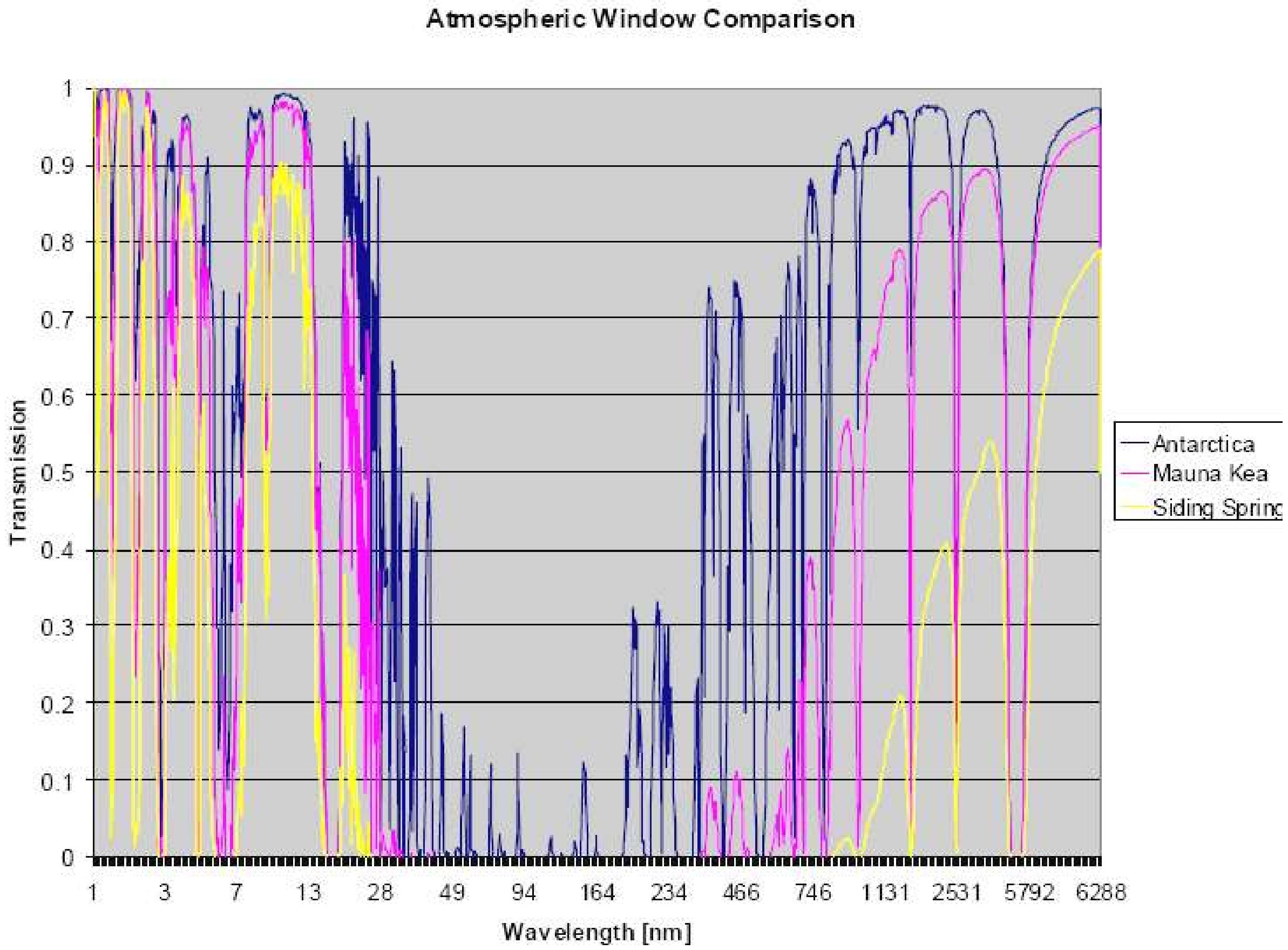

**Figure 1.5** Plot of atmospheric window measured in three different sites, taken by *The history of astrophysics in Antarctica,* [B.T. Indermuehle et al.]

### 1.2.1 Dome A

HEAT project, (the High Elevation Antarctic Terahertz Telescope), is a project of a 0.5-meter far-infrared telescope, equipped with heterodyne receiver/spectrometer system. It includes a consortium formed by Caltech, University of Arizona and other European Research Centres.
It is supposed to be completely unattended for remote operation and it is planned to be installed at the summit of Dome Argus, the highest peak on the Antarctic plateau. It will operate in the atmospheric windows between 150 and 400 μ, the most important spectral range for the study of formation of galaxies, stars, planets, and life. It will have high aperture efficiency and excellent atmospheric transmission (Walker, Christopher et al.).

### 1.2.2 South Pole

Seeing measurements in the optical bandwidth were taken, following a site-testing program led by the University of South Wales (Australia) and University of Nice (France), at the Amundsen-Scott South Pole Station. A set of microthermal sensors, placed at 3 levels on a 27 m-high mast, covering a period of 4 months observations, revealed a mean value over 0.64 arcsec. The significant decrease of the optical turbulence over the height of the mast, measured in the upper (17−27 m) and lower (7−17 m) sections, with mean values of 0.37 arcsec and 0.46 arcsec respectively, confirmed the presence of the inversion temperature layer ( R.D. Marks et al).
At South Pole, these factors combine together to generate an average wind speed of 6.3 m/s.
The atmosphere can be divided in two regions: a highly turbulent boundary layer from 0 up to 220m, with a strong temperature inversion and wind shear, and a very stable free atmosphere.
Near the top of the boundary layer, the inversion begins to flatten out at around 200-250m, where microthermal turbulence becomes quiet, even though the inversion continues weakly 100 m further or more. The temperature profile generally levels off weakly and smoothly (R.D. Marks, 2002).

## 1.3 Environmental conditions and astronomical site testing at Dome C

One of the advantages of this site is the very low wind speed. In fact on the Antarctic Plateau katabatic winds are completely mitigated. This is a favourable aspect also for the possibility in the future to install large telescopes. Data provided by AASTINO revealed average wind speed to be the best at Dome C, and it is expected to be constant in winter. The highest value measured is 16 m/s and, in any case, half than at South Pole.

Another basic factor is the diamond dust phenomenon, or rather microscopic ice crystals getting in everything, which cause serious problems to motor boxes and to the balls and rollers of bearings, forming a sort of lubricant coating.

The very low humidity and pressure level make the contribution of convection effects almost negligible, so that instead of freezing even overheating problems may occur. Therefore, besides the insulation thickness of boxes for the parts to warm up, it is necessary a preliminary estimation of both thermal powers and heating cycles.

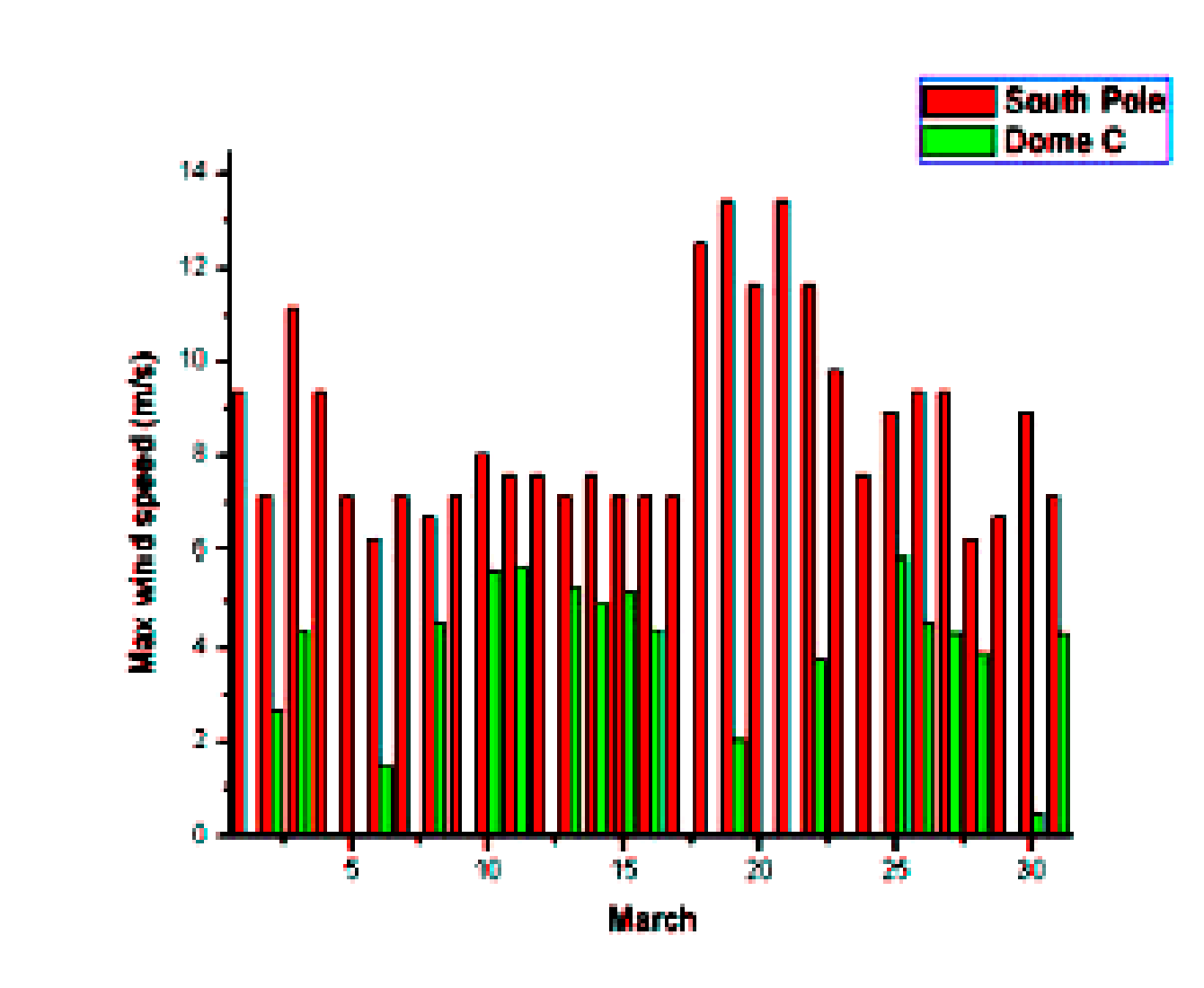

**Figure 1.6** Average wind speeds collected simultaneously at Dome C and at South Pole.

| | |
|---|---|
| **Average air temperature** | -50.8 °C |
| **Minimum temperature** | -84.6 °C |
| **Summer average temperature** | -30 °C |
| **Winter average temperature** | -60 °C |
| **Average wind speed** | 2.8 m/s |
| **Maximum wind speed** | 16 m/s |
| **Mean air pressure** | 644 mbar |
| **Yearly snowfall** | 2-10 cm 35 days/year |
| **Absolute humidity** | 2 g/m$^3$ |
| **Snow pressure** | 0.1 kg/cm$^2$ |

**Table 1-1 Climatic conditions at Dome C.**

### 1.3.1 AASTINO

AASTINO (Automated Astrophysical Site Testing International Observatory) was commissioned to evaluate atmospheric window in the atmospheric window from UV to submillimetric.
It is a remote autonomous laboratory, built at Dome C in January 2003, which comprises instrumentations for a set of experiments. The first one, installed in January 2004,is Multi-Aperture Scintillation Sensor (MASS), with the attempt to measure wintertime seeing. It uses a low cost small telescope and is able to measure seeing contributions from six layers at different altitudes (0.5, 1, 2, 4, 8 and 16 km); anyway it is insensitive to the layer below 500 m. For this reason SODAR (Sonic Detection And Ranging) instrument has been employed to determine the contribution from the layer between 30 and 500m. It is an acoustic radar capable of detecting the temperature fluctuation constant $C_T^2$ by the intensity of the echo reflected off turbulent cells. It can also reconstruct a three dimensional profile of the wind velocity using two beams oriented at 45° from zenith.
The integration time to obtain an acceptable signal to noise ratio is about 30 minutes. The SODAR has been working since summer 2002/2003 (T. Travouillon et al., 2005).
The results gave a superb value for the median seeing of 0.27 arcsec, and below 0.15 arcsec 25 % of the time. This means that a telescope placed at Dome C would compete with one other that is 2 to 3 times larger at the best mid-latitude observatories. Furthermore, an interferometer placed there would give performances similar to those required for a space mission.
The best seeing ever recorded at Dome C was 0.07 arcsec, which represents the lowest value reported anywhere. Even if it is expected that the turbulence conditions at Dome A will be superior even to Dome C, the complete lack of infrastructure at this site (it has never been visited) means that Dome C may be a preferable location (J. S. Lawrence, Michael C. B. Ashley, 2004).

### 1.3.2 CONCORDIASTRO

Concordiastro is a collaboration between Nice and Naples. It aims to check out the solar seeing quality at Dome C site, installing a 40 cm telescope designed to acquire both high-spatial resolution and full-disk images. CONCORDIASTRO/Italy is the solar physics part of this project, and it intends to probe the astronomical quality of Dome C during the long Antarctic day. These are ideal conditions for fields of research such as astroseismology, which requires high spatial and temporal resolution.

Winter observations have been done by means of the DIMM (Differential Image Motion Monitor), that measures turbulence parameters as scintillation, isoplanatic angle, outer scale and coherence time. This has been achieved by GSM experiment (Generalized Seeing Monitor) expressly designed for Antarctica. The profile of the turbulence was obtained by balloon borne micro thermal sensors.
The principle on which DIMM experiment is based is to derive seeing from the differential motion of two images of the same star. Forty frames per second have been sampled. Therefore seeing was calculated by the integration on $\tau$, $2\tau$, $3\tau$ and $4\tau$, and extrapolated to 0.
Anyway seeing was not appreciable, around 1 arcsec, because of weather conditions of the moment. Canopus (*α* Eri, *V=-0.9*) was selected for seeing measurements during daytime.
The main program wants to perform turbulence analysis on the solar and lunar limbs and correlate them to seeing estimations with the DIMM, and, among all, to test the mechanical resistance to the polar temperatures of an optical stellar coronagraph.

The experiment consists of a set of four telescopes, operating during last summer and winter. Three of them are identical and will be used for site testing. They have been specially customized for Antarctic conditions. At their focus thermostated and insulated boxes were placed, hosting a digital CCD camera.

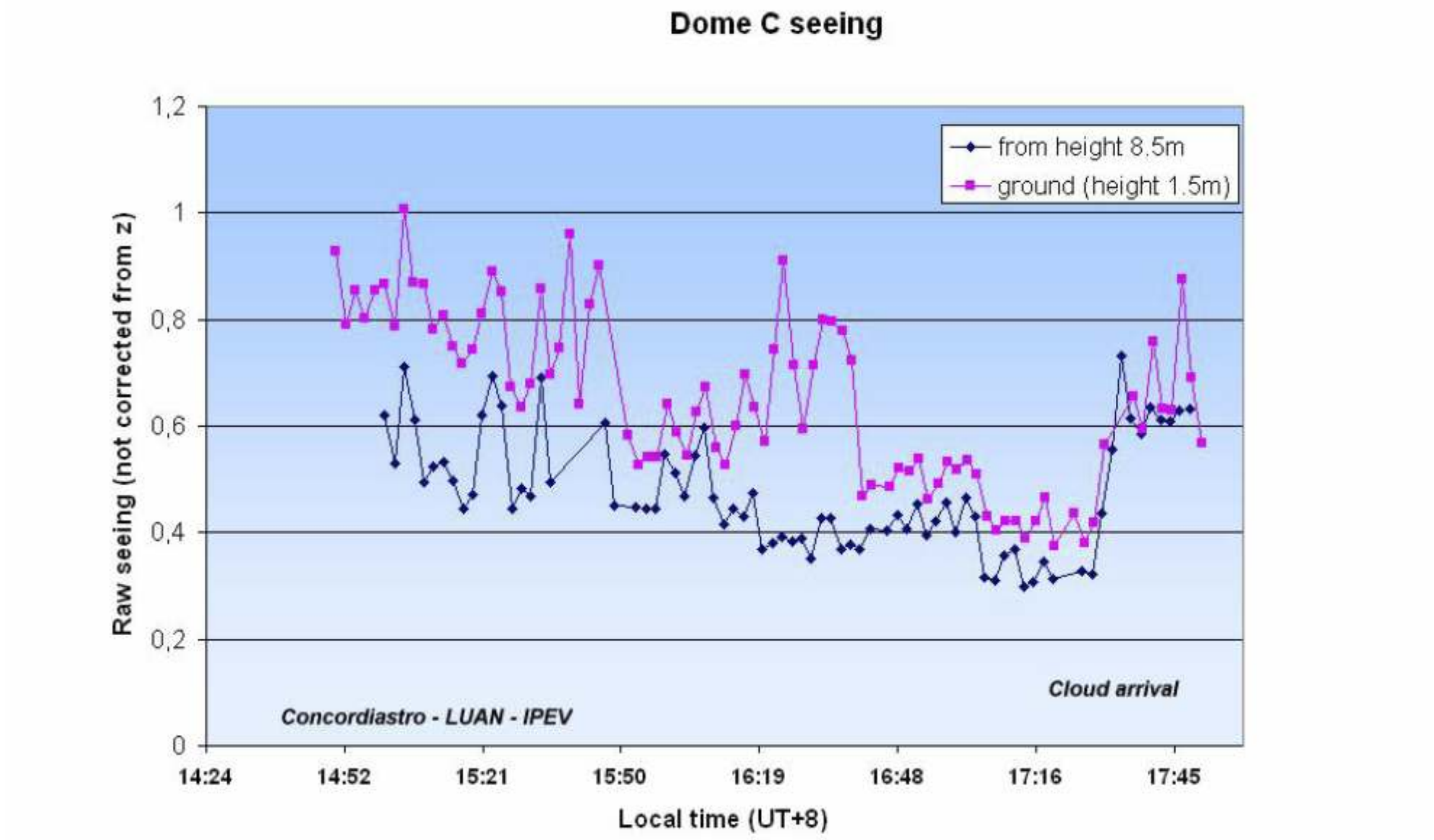

**Figure 1.7** Seeing measurement collected by CONCORDIASTRO observatory during the day.

Therefore, on the basis of data collected by astronomical tests, we can assert that high altitude, sky transparency, the very low water vapour content make Dome C as one of the best places on Earth for observations in the near, mid infrared, and submillimetric windows.

# CHAPTER 2 THE IRAIT PROJECT

## 2.1 Project overview

In this chapter the attention is focused on the opportunities given by IRAIT project, and general aspects concerning, focal instrument, optics, control operations, and the phases of transport to the final destination are also listed. Mechanical design, instead, is explained in the next chapters.

The history of IRAIT project, born from an original idea of prof. Maffei and dr. G. Tosti in 1997, aimed to place a telescope in Antarctica to exploit the near and mid infrared windows. A prototype was made by MARCON company, and was installed in Coloti, near Perugia, with the purpose of making some tests on the control electronics and on the software (TCS), as well as on motors and drive train. A Cassegrain telescope with a 80 cm aperture and alt-azimuthal mount was chosen, because it permits to reduce the overall size and have a lower center of gravity with optimized layout of devices, and, beside that, needs a smaller enclosure. At first it was thought to refurbish the same prototype to face extreme environment conditions, but soon it was found out that materials and mechanics were not suitable for them.

Since 2001 the project has become more ambitious when from local it turned into international, including famous European research groups, as the University of Granada (UGR) and the Institut of Estudis Espacials de Catalunya (IEEC), as well as University of Nice for near infrared instruments (DENIS) for the second Nasmyth focus. In this way the IRAIT project has turned into the acronym of International Robotic Antarctic Infrared Telescope, receiving financial supports for the following years 2004-2006. Meanwhile Teramo Observatory (OACT) has joined the team as supervisor of the project of an infrared camera AMICA (Antarctic Mid- Infrared CAmera) , to be mounted on the first Nasmyth focus. This project is financed by INAF and involves also Padua, Brera (Milan) and Turin Observatories.

Considering optical requirements and potentiality of the Antarctic site which allows to exploit infrared windows “hidden” in temperate sites, it has been decided to use two separate arrays, one for NIR (2-5 μm), the other for MIR (8-25 μm), through two different channels, working in alternation. The first one is supplied by Raytheon: it is a InSb 256 x 256 type, with a 30 μm pixel size, corresponding to a scale of 0.538 arcsec/pixel and a FOV of $2.30 \times 2.30$ arcmin$^2$.

The effective quantum efficiency is approximately 80% over almost all the spectral range and the integration capacity is of $2{\cdot}10^5$ e- . A DRS Si:As IBC 128 x 128 model, with a spectral response of 2-40 μ bandwidth, Moderate-Flux type, has been chosen as MIR array. It has a pixel size of 50 μ, which gives a pixel scale of 1.345 arcsec/pixel. This resolution is expected to provide an adequate sampling of the PSF at 10 and 20 μm. For both arrays read out electronics comprising 4 output channels is provided.Main operating parameters are shown in the table below.

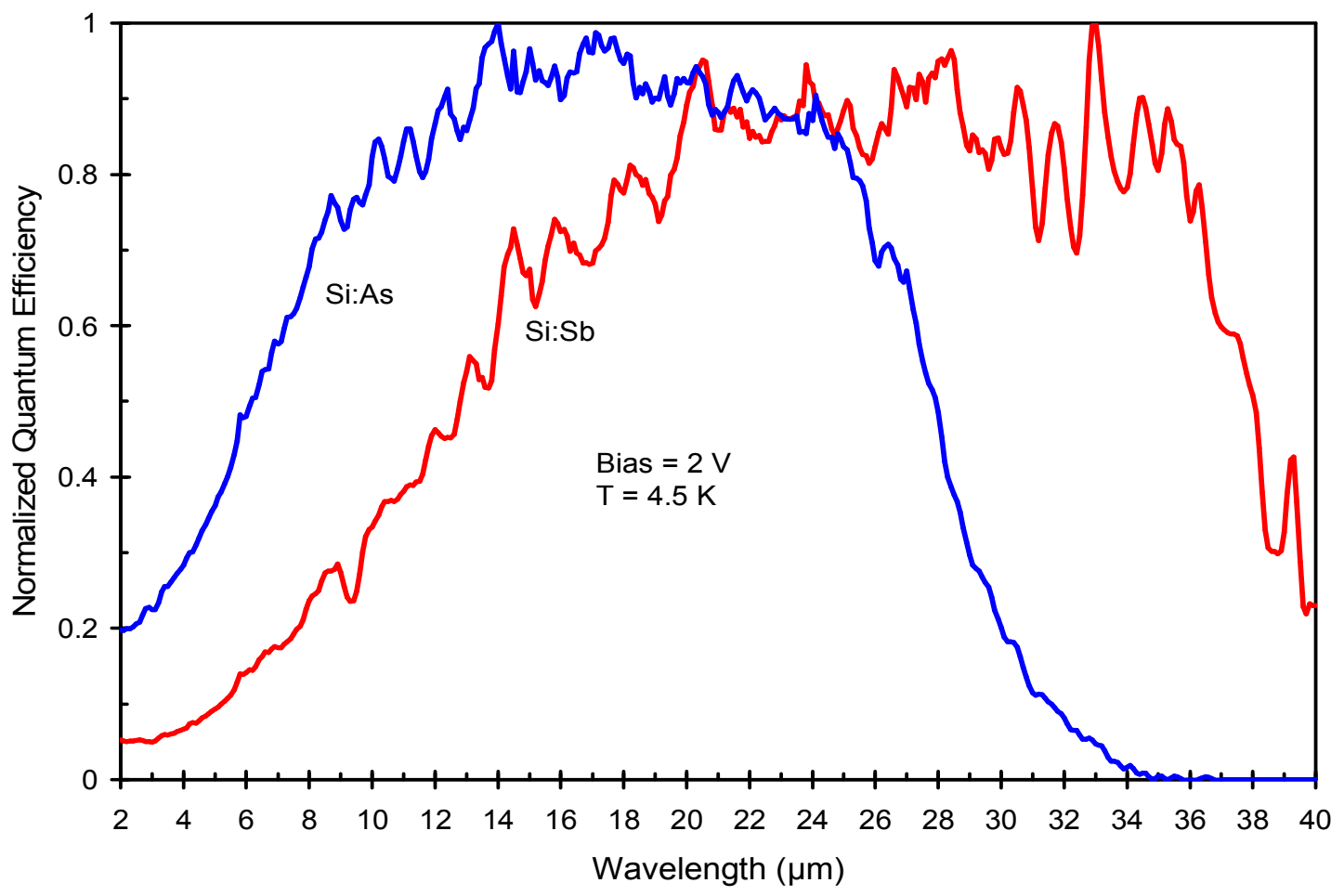

**Figure 2.1** Diagram of QE vs wavelength [by DRS Technologies].

The optical system will comprise only mirrors, in order to avoid focusing and aberration problems. The total demagnification factor is 1.47 : 1. The sole transmission elements will be entrance window made of KRS-5 and filters (Dolci, M.). A closed-loop cryocooler guarantees to maintain the temperature permanently at 5 K, so that maintenance requirements will be reduced to the minimum. The two astronomical standard broad band filters N (10 μm) and Q (20 μm) will be also available (Tosti, G., et al., 2004).

| | NIR (256x256) InSb | MIR (128x128) Si:As |
|---|---|---|
| **Pixel size [μ]** | 30 | 50 |
| **Pixel scale [arcsec/mm]** | 0.538 | 1.345 |
| **FOV** | 2.30’× 2.30’ | 2.87’x2.87’ |
| **Quantum efficiency** | >80% | 55% |
| **Integration capacity** | $2*10^5$ e | $10^7$ e |
| **Charge capacity** | 0.06 pF | 1.3 pF |
| **Dark current** | < 30 to $3x10^9$ $e^-$/s | < 60 to $3x10^8$ $e^-$/s |
| **Operating temperature** | 15-30 K | 2 ÷ 14 K |
| **Frame rate** | 18 Hz | 500 Hz |
| **Max output change** | ~1.5 V | ~1.4 V |
| **Number of outputs** | 4 | 4 |
| **Input-referred Read noise (e-)** | 2300/400 rms | 500 rms |
| **Power dissipation [mW]** | <40 | <10 |

**Table 2-1 Operating data of the two arrays.**

## 2.2 Scientific purposes

There are several scientific scopes that can be carried out, taking advantages of high sky transmission, very stable atmospheric conditions and continuity of observations, especially in the six months of winter night.
Surveys are planned to be made in selected regions of Magellanic Clouds and of galactic molecular clouds, in order to study star formation mechanisms at different metallicities and masses. We could reach good statistics on young stellar objects, brown dwarfs and circumstellar phenomena.
A significant improvement in the C-star/M-star statistics is expected, thanks to the observation of features present in their mid-IR energy distribution, due to silicates and to SiC. Simple surveys of ionic line emission in the infrared can be made to understand the ionization conditions and the kinematics of Planetary Nebulae.
Many bright infrared galaxies, as a serendipitous outcome of our IR survey, can be observed. We plan to collect data over several seasons, so that we reach sufficient statistics on extragalactic objects and provide a catalogue of these objects in the southern sky. On individual objects, we shall be able to provide information on the infrared flux of many Active Galactic Nuclei (whose IR emission is often unknown).
Among the many scientific goals, we recall the study of the final stages of stellar evolution, characterized by a strong mass loss (in particular extreme AGB and post-AGB stars, supernovae and Low mass X-ray binaries), the study of stellar formation processes, the search of cool stars (brown dwarfs and L-type) to determine the mass function of low and very low mass stars. Another challenging purpose is the research of extra solar planets and solar system minor bodies, with the help of suitable tracking techniques for fast moving objects. We plan to use information provided by our surveys mainly to study the thermal emission of Near Earth Asteroids.

## 2.3 IRAIT operational modes

IRAIT must be completely robotic and remote controlled, so that software and hardware must satisfy severe requirements of fail safe and reliability. Moreover it must operate in winter, when Dome C base is not accessible by vehicles and maintenance is substantially reduced to minimum operations. For this reason it was decided to use the container on the wood platform as a control station: it will be heated and emergency maintenance operations can be accomplished. Controllers and electronic proximity devices for both telescope and camera are located inside modular insulated boxes, that can be easily detached and transported to the workshop.
Telescope movement control is achievable through a set of levels: an electric level provided by the driver, an electro-mechanical one provided by the motor, a general system control via PLC, and human and manual one provided by PCs. The real core of the system is PLC, as it confers reliability to the system: in fact, in case of complete power interruption, it can receive and process data retrieved by several detectors (thermocouples, heaters, encoders, limit switches, and so on) regardless of the operating system. In this case the only drawback is that the user cannot input new data. It is a model by ABB 90 Series CPU 07 KT 97 Ethernet. The main units (controller, PLC, PC) are connected via Ethernet. Furthermore a wireless point - to - point connection between the station and the container is expected for redundancy.
The difficult operating condition at Dome C have required the development of a control software capable of performing all unattended operations as for a manually operated telescope/instrumentation.
The IRAIT control software, mostly written in C++, follows an object oriented approach. From the analysis of some scientific use cases as issues, a model of the system based on classes was created. The graphic user interface (GUI, WxWindows package) consists of external packages; the archive queries are forwarded through a MySQL application program interface, FITS image are stored by means of CFITSIO package. Custom software drivers provided by the companies were also used to program hardware devices.
A multiplatform application running under both Windows XP and Linux is used. The Doxygen software tool has been employed to create the documentation.
The IRAIT control software comprises three main packages: the Observatory Control Software (OCS) running on a PC resident in the Base Concordia Control Room; the Telescope Control Software (TCS) running on a dedicated PC1, and the Infrared Camera Control Software (ICCS) running on another computer.
The OCS, structured in different processes, was designed to support remote access to the telescope from Concordia station and, occasionally, from Europe. In a remote way, it is also possible to prepare, modify and upgrade the scheduling, to make use of the data retrieved by AMICA; control the progress of the schedule; recover system alarms; interrupt and reboot the system; backup the system parameters and scientific data and make some tests on different parts of the system.

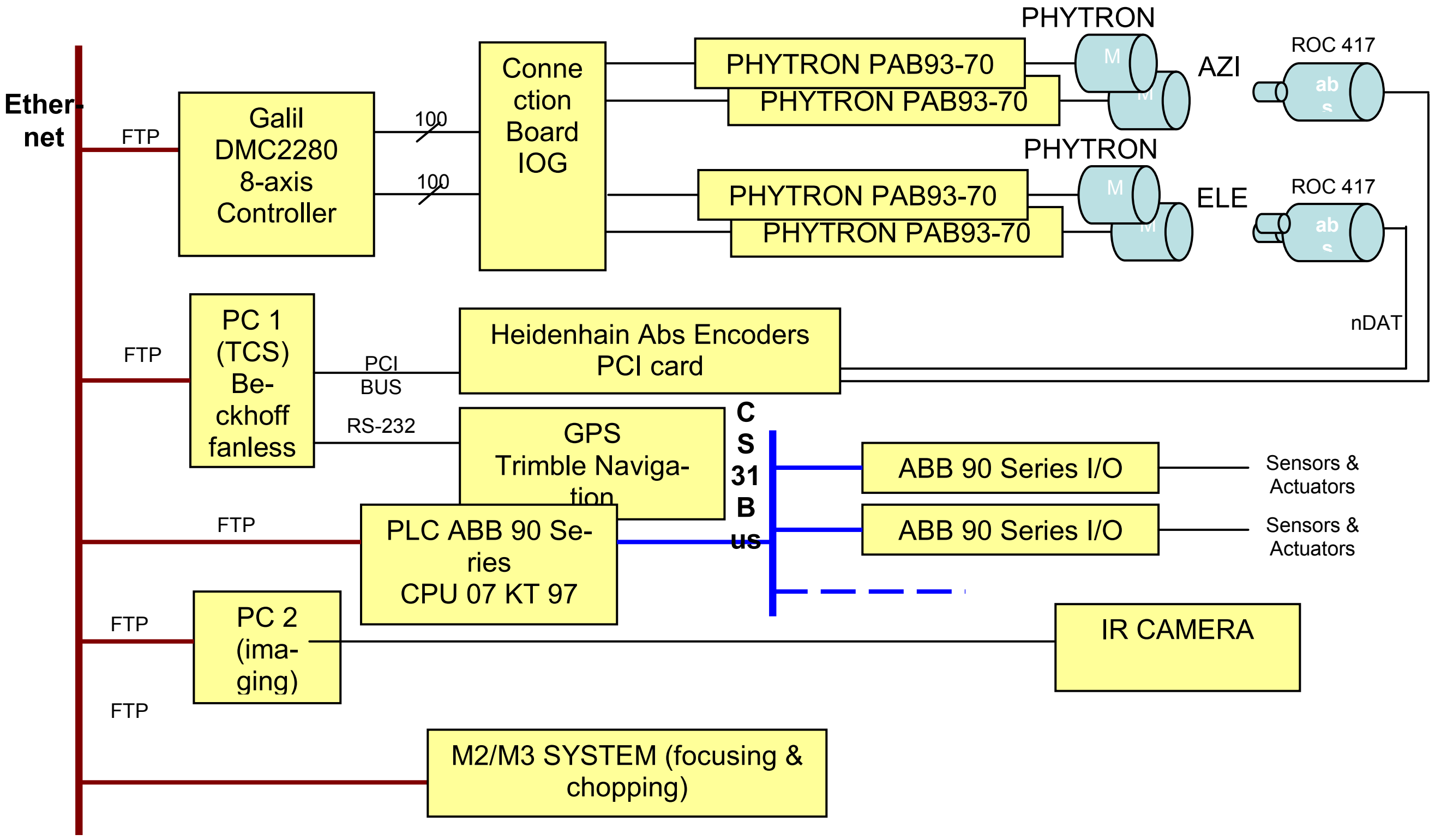

**Figure 2.2 A layout of IRAIT electronic connections.**

## 2.4 IRAIT Optical layout

The telescope has a Nasmyth optical configuration with two foci, which gives the advantage of having more space to mount the focal plane instrumentation on the altitude axis, rather than under the primary cell. The primary mirror is bored at the center, to host the support for a tertiary mirror tiltable of 180°, and to give the possibility of switching between the two foci. Since it is an infrared telescope, it is necessary to have a high f # to reduce the sky background and to assure an elevated signal to noise ratio. It is a Cassegrain - like reflector, with a 80 cm primary mirror and a focal length of 16.932m.

Secondary mirror, with a 130 mm diameter, is under-dimensioned to a 750 mm size mirror, in order to avoid seeing primary edges. Tertiary mirror has an inclination of 45°, to switch between two positions of Nasmyth foci: it is seen from the focal plane as a circular shape mirror of 120 mm diameter. All mirrors are gold coated, in order to reach a high reflectivity on the NIR &MIR spectral range (0.5-30 μ). The resulting f number is f/# =21.1651.

The optical features of the three mirrors are summarized in the table below.

| | **M1** | **M2** | **M3** |
|---|---|---|---|
| **Radius [mm]** | 4800.00 | -920.470 | FLAT |
| **Aperture [mm]** | 800.000 | 130 | |
| **Conic constant** | -1 | -1.83000 | - |
| **Material** | SITALL | SITALL | SITALL |
| **Coating** | PROTECTED GOLD | PROTECTED GOLD | PROTECTED GOLD |
| **Ellipse major /minor axis (mm)** | - | 130.45709/130.43674 | 122.76 / 86.80 |
| **X/Y/Z offset [mm]** | | 0.176 / 0 / -0.00307 | -0.555 / 0 / -0.555 |
| **Thickness [mm]** | 133 | 21.74 | 20.46 |
| **Surface roughness** | λ/8 @ 633 nm minimum | λ/8 @ 633 nm minimum | λ/8 @ 633 nm minimum |
| **Central hole diameter [mm]** | 120 | - | - |
| **Mirror distances [mm]** | **M1-M2**: 2005.00<br>**M2- M3**: 1705.00<br>**M2- Array**: 2786.74<br>**M3- Array**: 1081.74 | | |
| **Focal length [mm]** | **16932** | | |
| **Resulting f/#** | 21.1651 | | |
| **Focal plane scale [arcsec/mm]** | 12.181894 | | |
| **Beam diameter on TFP [mm]** | 59.104109 | | |

**Table 2-2 IRAIT optical features.**

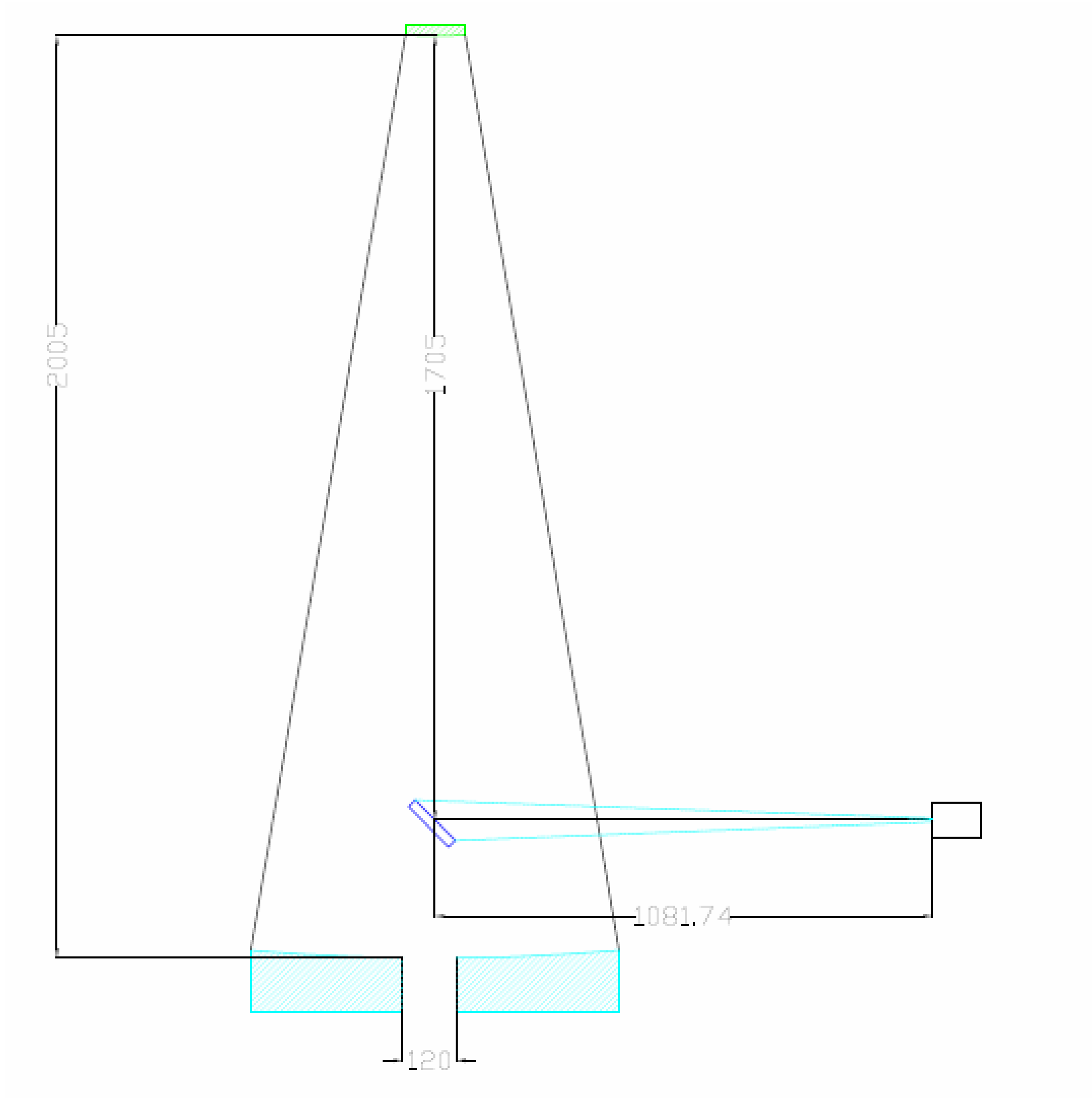

**Figure 2.3 Optical scheme of IRAIT.**

## 2.5 Secondary and tertiary mirror drivers

The NTE company is in charge of the project and building of the moving mechanism for the secondary and tertiary mirror, and the relative subsystem. Required drives for secondary mirror motion are focusing and chopping.

The former is realized by means of a linear actuator, which includes a motor and a reducer by Pythron (VSS32.200.1.2.UHVC and VPGL 32 i-50 UHVC units). The stroke of the actuator is 100 mm, and the screw pitch is 2 mm per revolution. Two limit switches prevent the actuator from getting over the limits.
The second is provided by a custom piezoelectric system, capable of working at -80° C, manufactured by PiezoMechanik technology with a tilt range of ± 4.5 mrad (equivalent to 5x5 arcmin in the sky) (Catalan, A., 2004). It uses eddy current sensors with high resolution, resistance in harsh environments (Bru, R., 2005). Maximum chopping frequency is 25 Hz, so compatible with the lower frequency of top ring (of the optical tube), which was estimated to be about 80 Hz. It was also checked that reaction forces are not amplified by the dynamic loads of the optical tube.
The MARCON company *Costruzioni Ottico Meccaniche* is in charge of building a set of the three mirrors. Sitall was chosen as material, with a gold coating and a SiO protection layer, machined at λ/8 rms.

| | |
|---|---|
| Amplitude | 0.00225 rad |
| Nominal frequency | 5 Hz |
| Maximum frequency | 25 Hz |
| Linear actuators radius | 0.075 |
| Pivot Point | 0.040m |
| M2 thickness | 0.0225m |

**Table 2-3** M2 main characteristics.

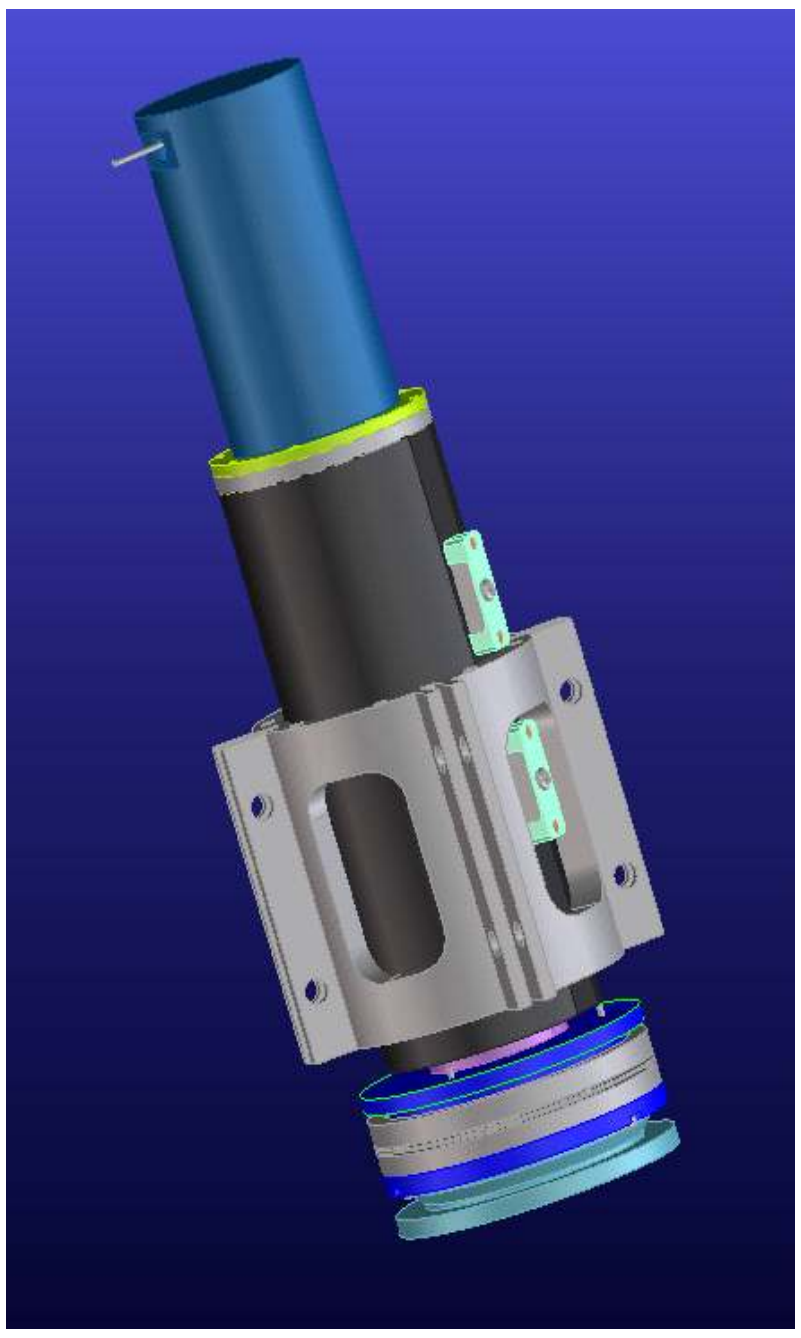

**M2 Drive general view**

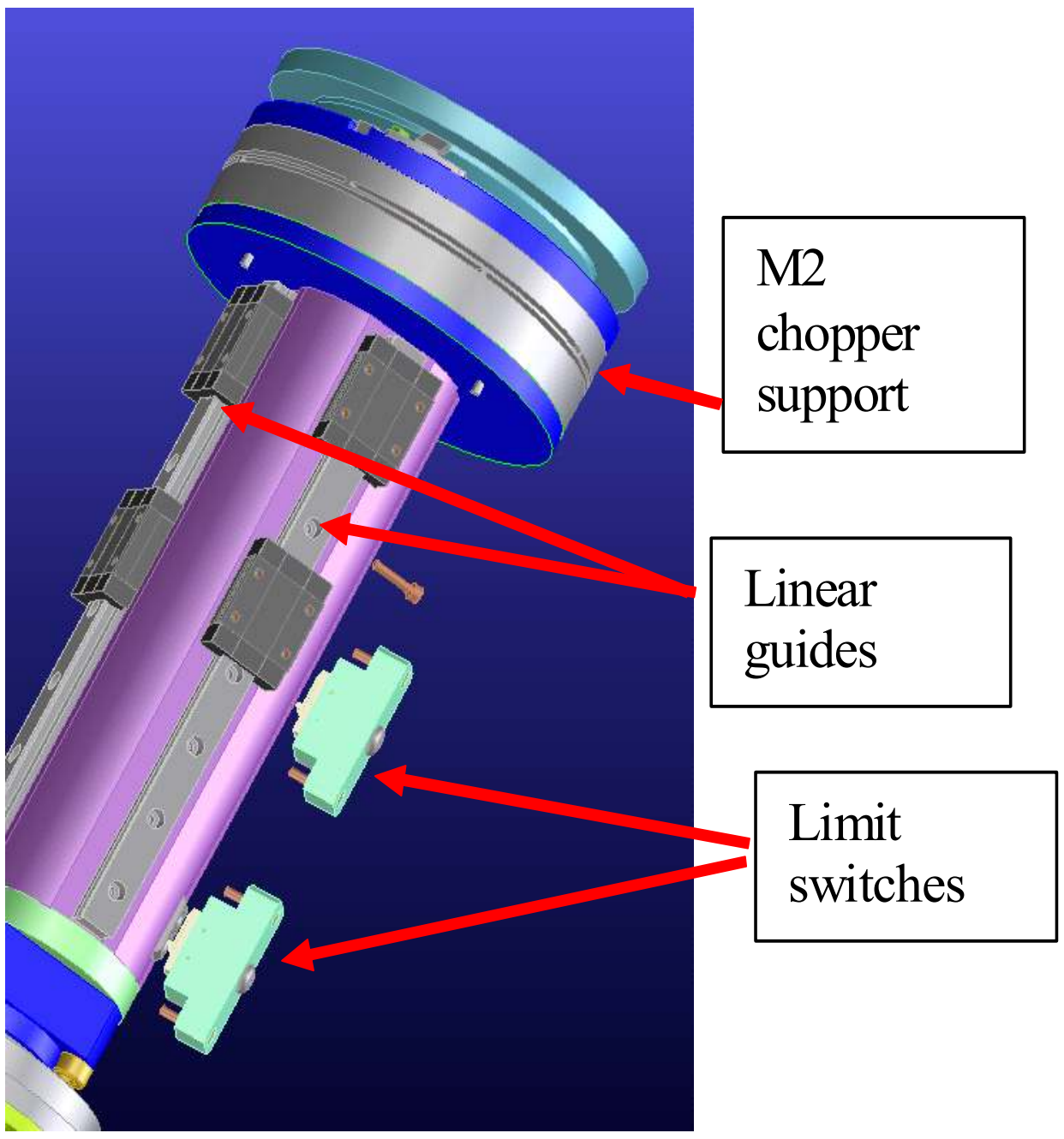

**M2 Focuser without external cover**

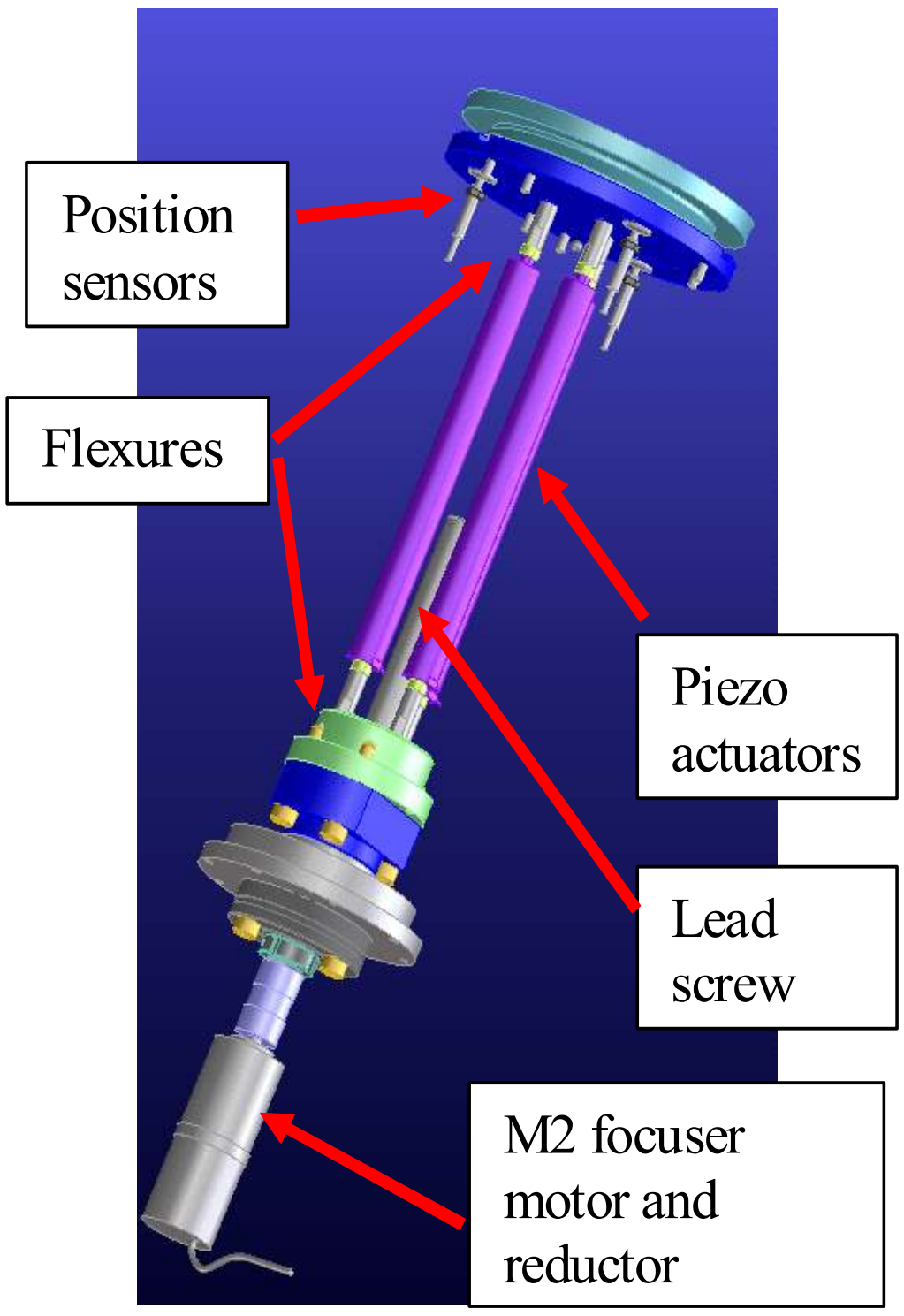

**M2 Focuser without external cover and linear guides**

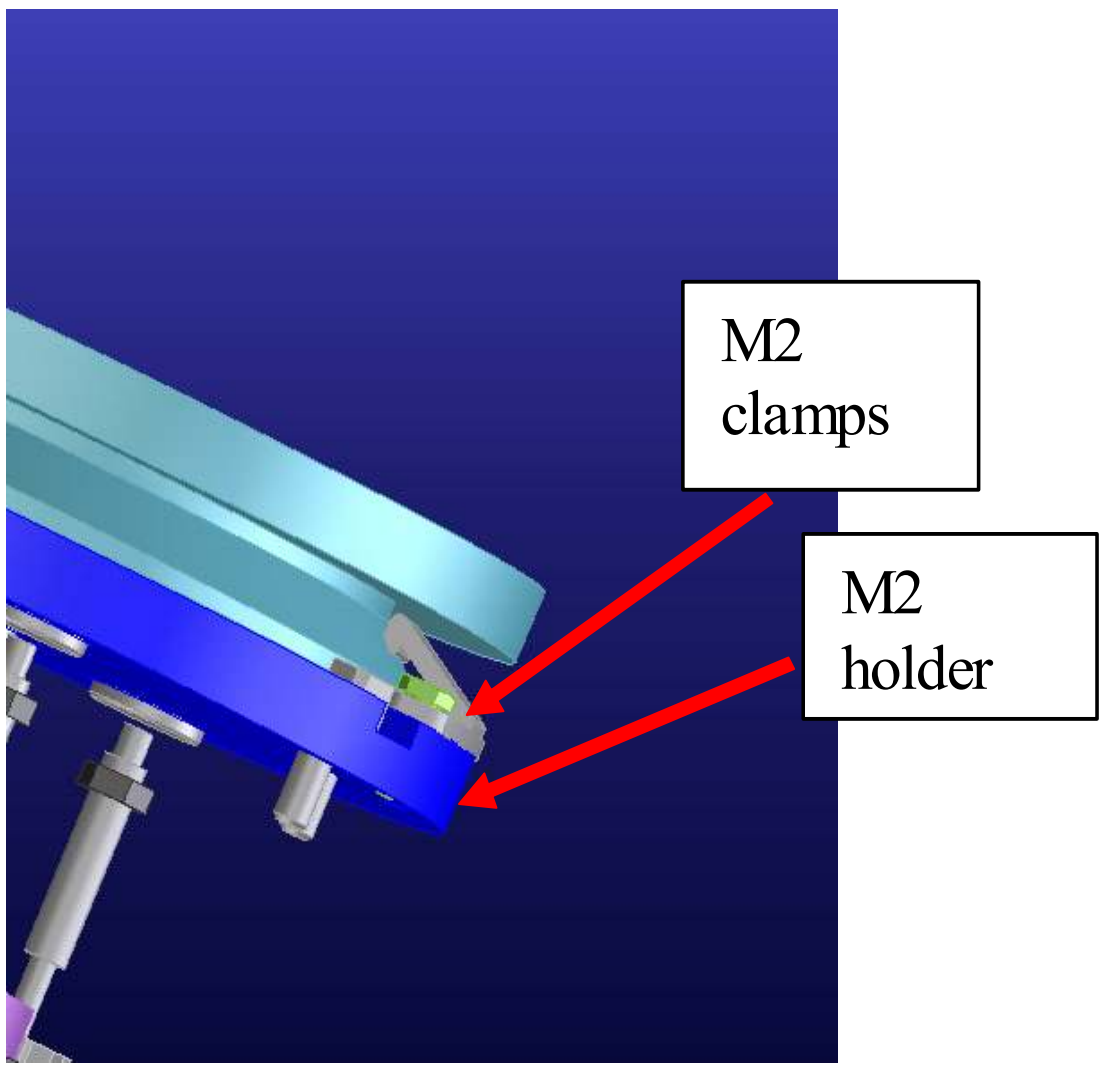

**M2 mirror holder**

**Figure 2.4 Some pictures of M2 drive system [by courtesy of NTE].**

## 2.6 Transportation and logistics at Dome C

After having discarded a few hypotheses about the transport, installation of the telescope and its enclosure at Dome C, as the one formulated by Tecnomare ( Gasperoni,F., et al., 2003) concerning the mounting of a big dome, we came to the final idea of using a tent as enclosure, which is far lighter, less expensive, and easier to manage than a traditional dome. The telescope will travel alone, assembled in a modified ISO 20 container, in the position shown in figure 2.5. It will be bolted on the base chassis, in two distinct parts: the fork mount with optical tube and altitude drive boxes in horizontal position will be fixed on the front end; the azimuth bearing and the plate supporting drive system in azimuth will be located at the center, in order to fit the internal size of the container. The base chassis is anchored to a flat ISO 20 of 40 cm height through a special mechanical interface of M30 thread holes.
On the free side of the chassis other utilities will be installed to balance the overall weight. The part that needs to be moved for assembling *in situ* is estimated to be about 2000 kg.

Another open top ISO 20 container will host the rotating platform and the chassis with legs that bear the upper structure with the tent. Lighter and more delicate parts, that are more sensible to shocks and vibrations, as mirrors, will fly to the destination by Twin Otter, a light plane with a load capacity up to 1000 kg, with a journey time of about 5 h from MZS . The same transport modality is expected for the camera with the rack.

The container will be taken on board of a commercial Italian ship in agreement with PNRA, sailing from Salerno or La Spezia harbour to Hobart in Tasmania. Once there, all the material will be moved to French ship Astrolabe and proceed towards Dumont D'Urville. For the 2006-2007 campaign departure dates on schedule are:

- R2 29/30 December 2006;
- R3 28/29 January 2007;
- R4 on 17/18 February 2007

The best date is probably R3, so the parts must be delivered at least 50 days before, within the second week of December 2006.

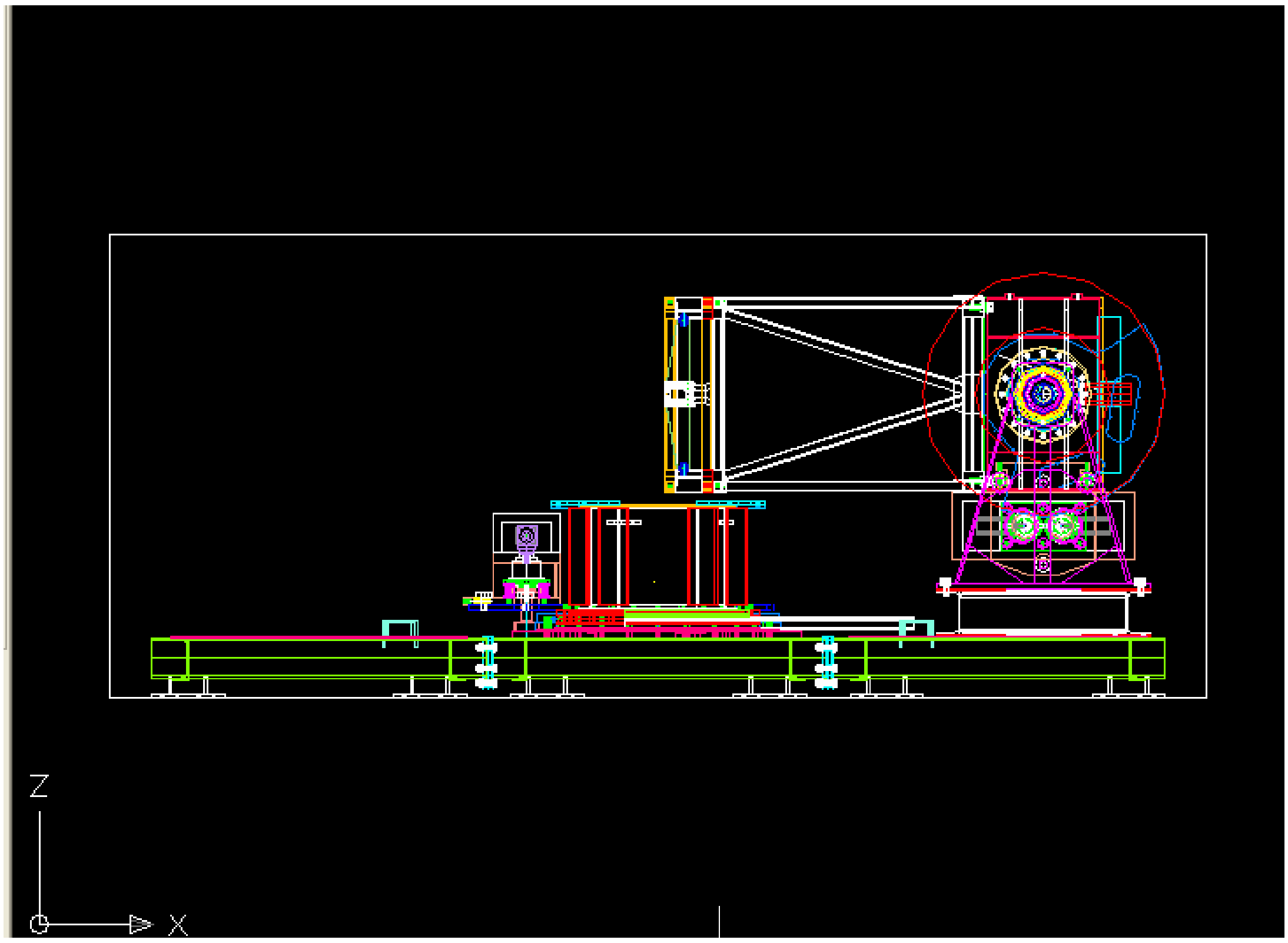

**Figure 2.5 Telescope layout during transport.**

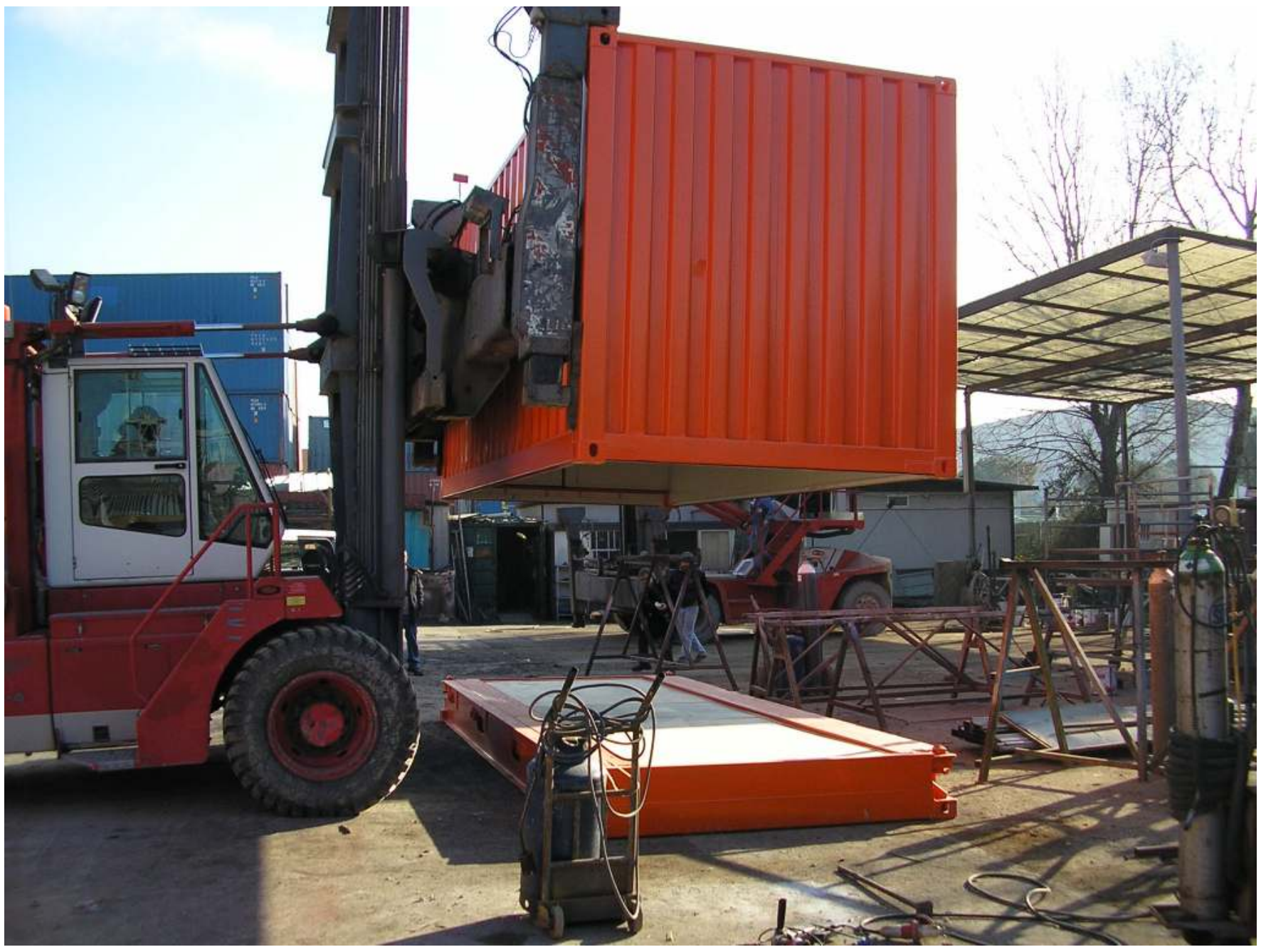

**Figure 2.6 A photograph of modified ISO 20 with separate custom floor. The walls will be removed and the base, with the telescope anchored on it, will lay on the wood platform.**

The containers will be transported by means of ski-equipped aircrafts from Cape Prud 'Homme, to Dome C by tractor train in ten or twelve days, depending on weather conditions and on the amount of snow. Anyway, even if by the end of January next year the containers arrive at Cape Prud 'Homme, in any case, they must be unused for a year, for PNRA disposition, so that it will be fully operative for the summer campaign in 2007.

What mostly affects load conditions, besides thermal variations, is undoubtedly the transport, since there are inertial accelerations of 4g, in all directions, with an esteemed frequency of 15 Hz. Therefore a proper definition of the interface to the container basis is necessary to prevent shocks and vibrations propagation.

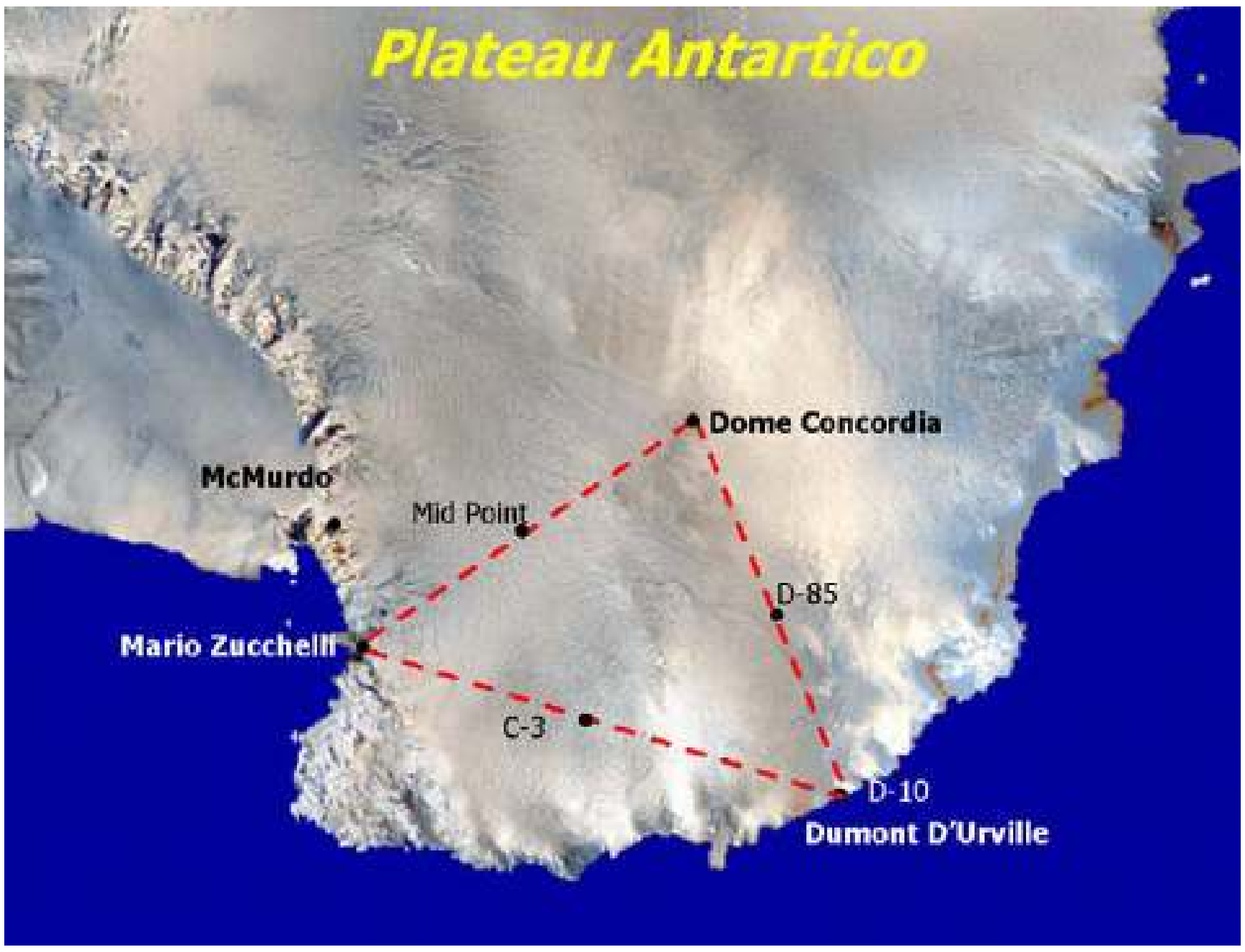

**Figure 2.7 Map of the transport routes for Dome C.**

The telescope will be placed at 500 m from Concordia Station, in a straight line with the AASTINO experiment, in a windward area, sufficiently distant from the Towers whose emissions would consistently perturb observing conditions. It will be also reachable on foot, considering that in winter-over activities it is not possible to use any vehicle.

The platform where the container will be placed, now under construction, is raised up of 4 meters respect to the ice level in order to prevent the diamond dust. The necessary power supply for the whole system is 20 kW, and PNRA has assured that, at this distance, there are no problems of voltage drop.

The container walls will completely open, once placed in the assigned area. The telescope structure will be covered by a tent of Boeing tissue, with an aluminium coating. It will be mounted on the trampling level of the telescope, with an external diameter of 5.2 m, a medium thickness of 0.318 mm, capable of resisting to temperatures down to -68 ° C. The bearing frame is made of box tubes of 80x40x3 mm dimensions.

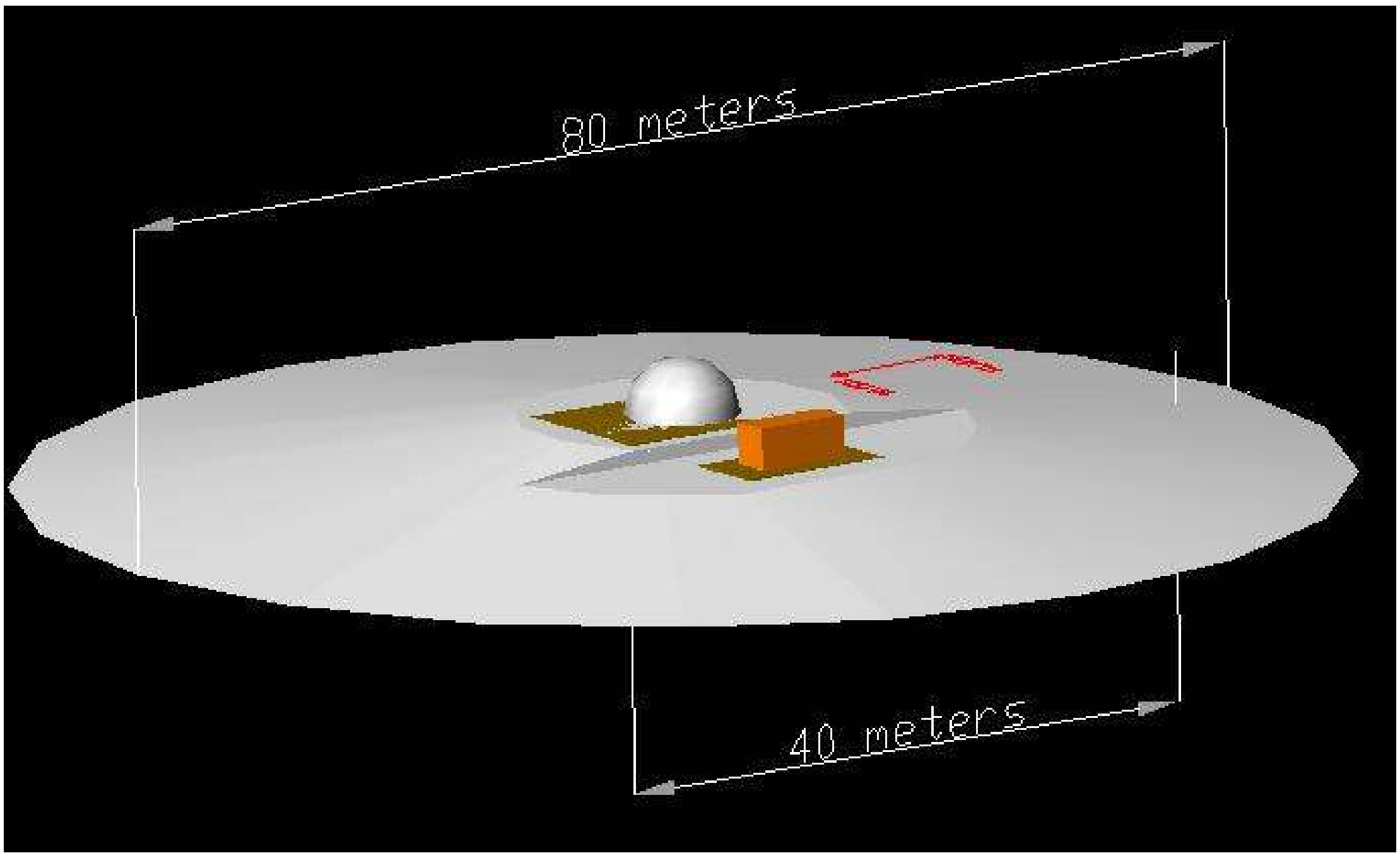

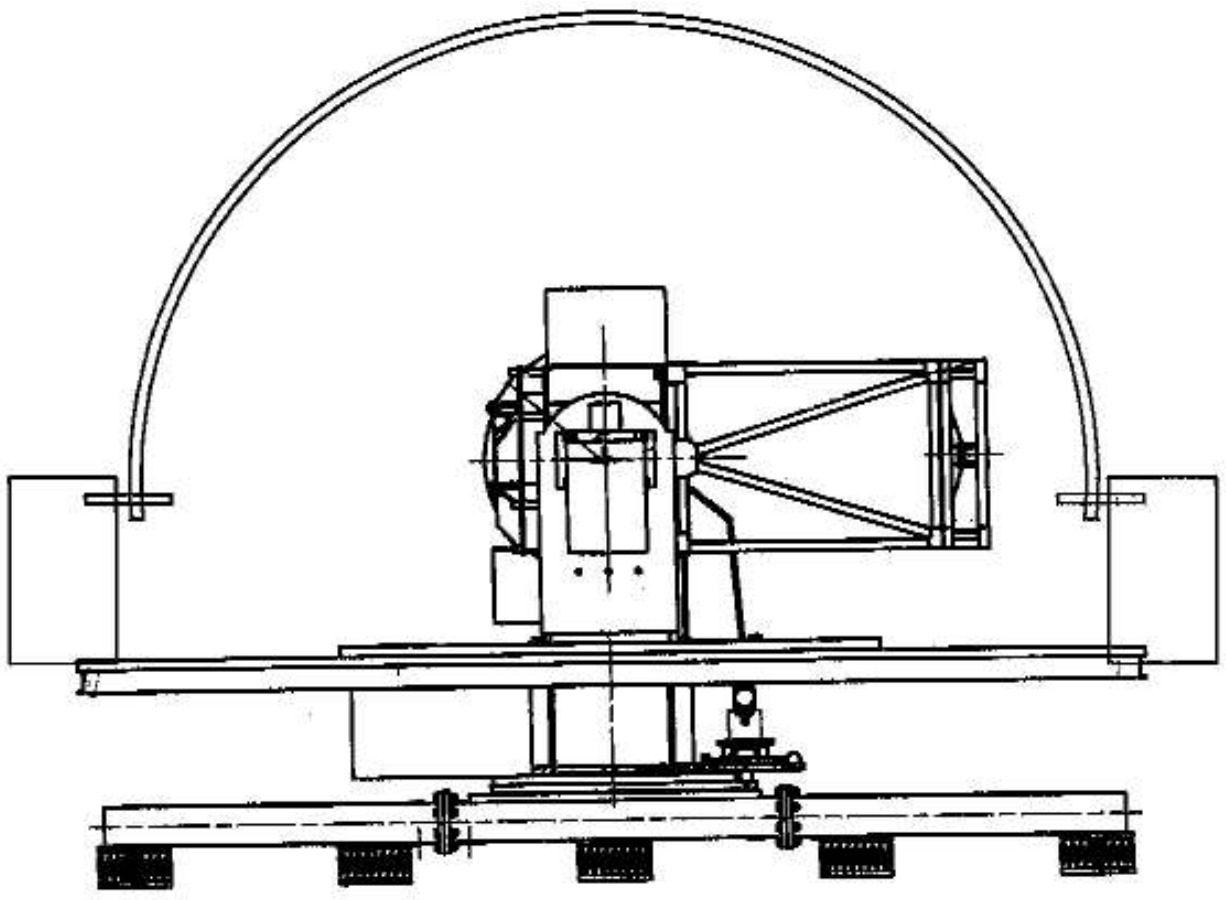

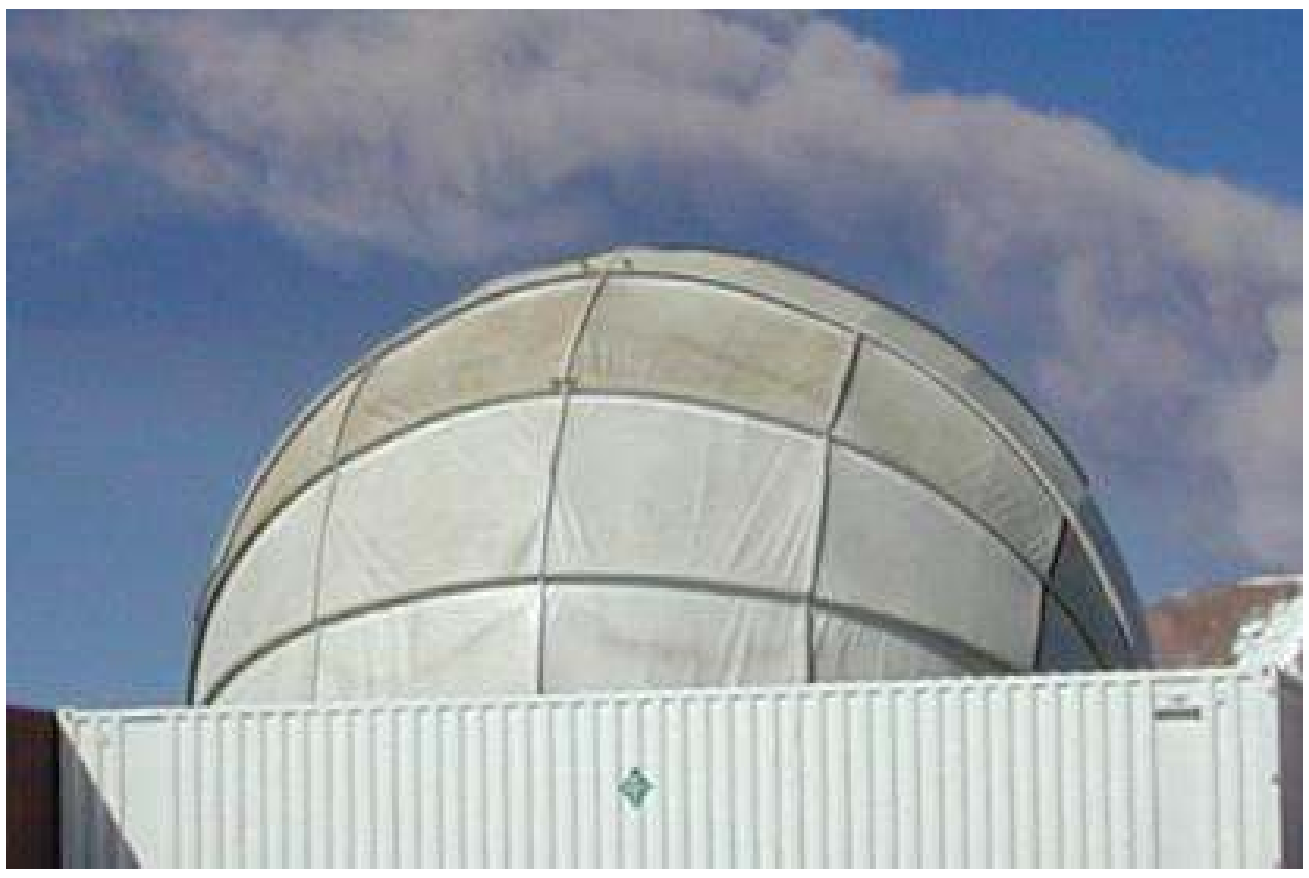

**Figure 2.8 An overall view of lay-by with the tent configuration.**

# CHAPTER 3 IRAIT Mechanical Structure

## 3.1 Main characteristics of an alt–az mount telescope

Once we had determined the optical parameters and the scientific targets that we intended to pursue, we passed to the selection of an alt-az mount. In fact this configuration permits to have a more compact structure with a lower center of mass, and as a consequence it is less expensive than an equatorial mount . The design of primary mirror cell is also simpler, because it tips in only one direction with respect to the gravity, so it reduces the dimensions of additional radial or axial supports (fig. 3.1) .

Another advantage is that in our case there's no necessity to mount a field derotator, as parallactic angle was found to be very small and at latitudes very close to South Pole. Anyway, one of the main drawback is the blind spot due to a singularity region, encountered by azimuth velocity in the neighborhood of zenith point. The dimension of this spot strictly depends on the dynamic response and servo-system of the telescope.

On the other hand, in an equatorial fork mount there' s no need of field derotator because, once a celestial object is pointed, tracking is provided by the only rotation in right ascension at a constant rate. Access to camera and handling of equipment in this case is the same as in an equatorial mount, for the fact it has a Nasmyth configuration. In the next paragraphs the entities of tracking velocity and acceleration limits are discussed. On the basis of mount selection we produced an initial sketch of the telescope (see fig. 3.2). Taking as main constraint mirror distances, we have started to develop the optical tube design at first, then fork arms and, thus, all the rest of it.

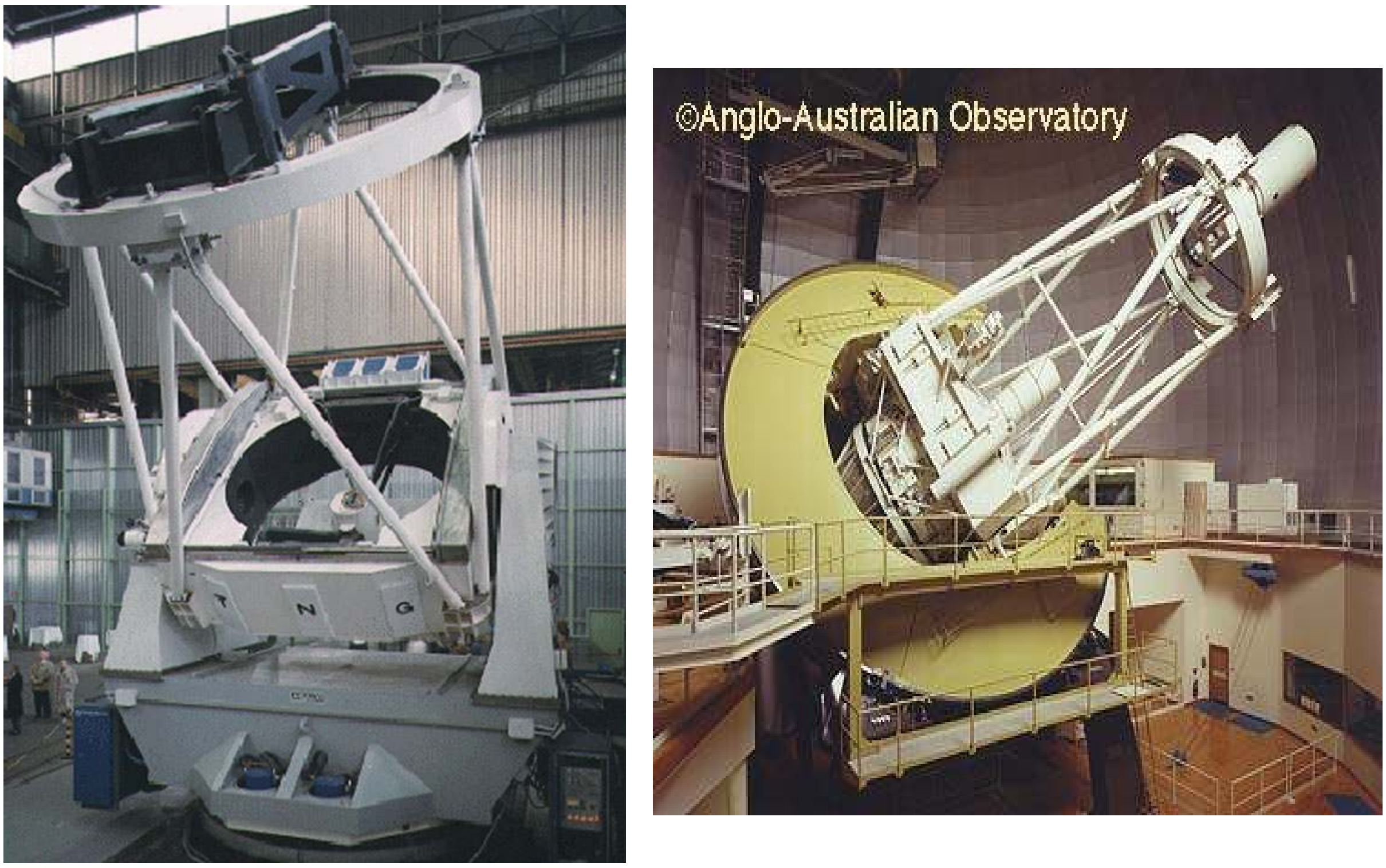

**Figure 3.1** A comparison between an alt-az and an equatorial mount. The first is a picture of TNG telescope (Cassegrain like, with an aperture of 3.6 m). The second is the Anglo-Australian 3.9 m telescope with equatorial mount .

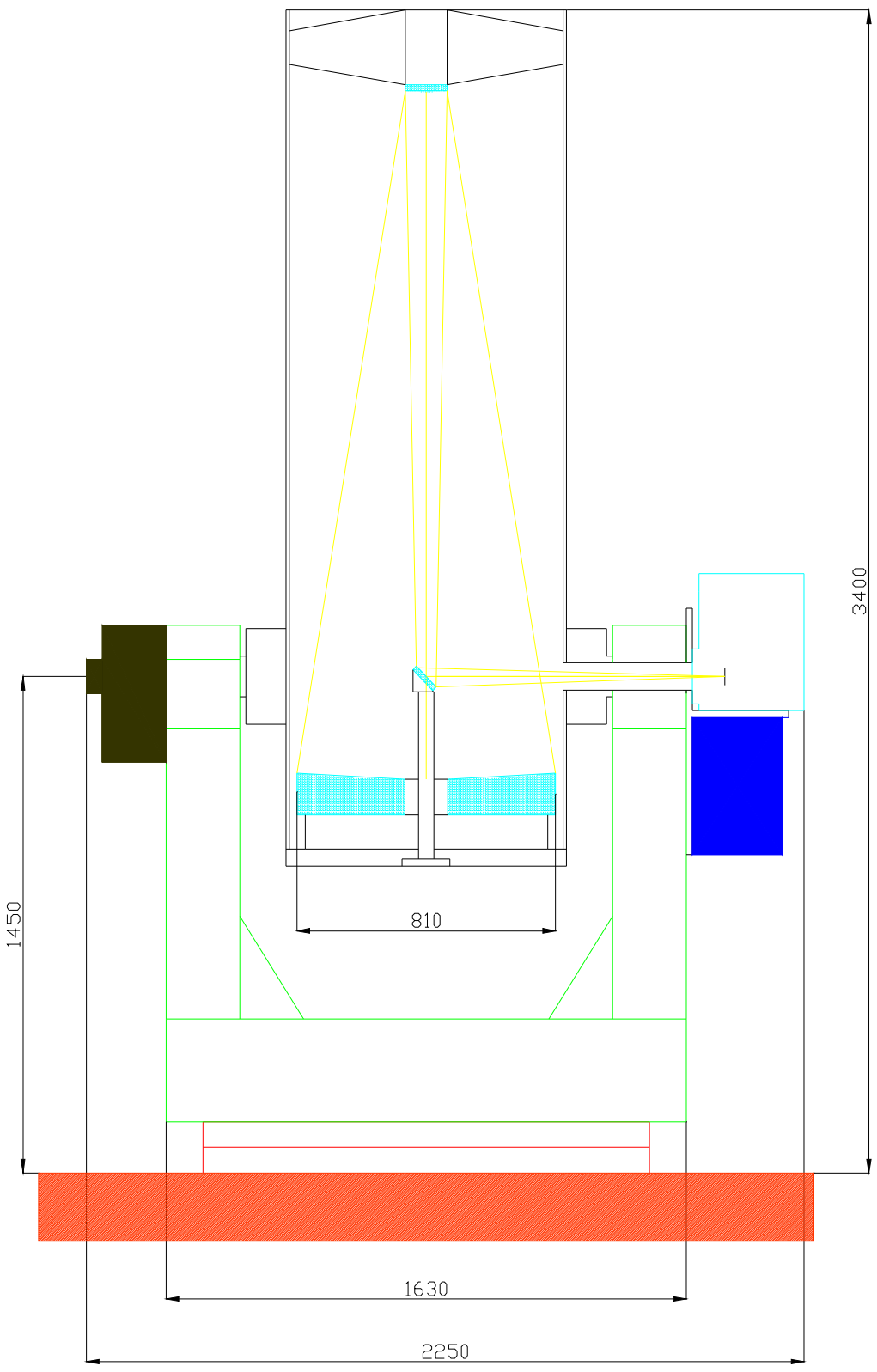

**Figure 3.2** A sketch of the initial concept of IRAIT, on the basis of the optical scheme.

## 3.2 Drive rates and tracking limits

### 3.2.1 Field-rotation correction

In every star catalogue celestial objects are identified by astronomical coordinates α and δ. In order to pass from the equatorial system to the horizontal one, we have to convert declination δ and hour angle H, to zenith distance *z* and azimuth *A*. The following formulae, given by most of spherical astronomy textbooks, are used:

$$z(H) = \arccos(\sin\varphi \sin\delta + \cos\varphi \cos\delta \cos H) \qquad (1)$$
$$A(H) = \mathrm{arctg}[\sin H/(\sin\varphi \cos H - \cos\varphi\, \mathrm{tg}\delta)] \qquad (2)$$
$$p(H) = \mathrm{arctg}[-\sin H/(\sin\delta \cos H - \cos\delta\, \mathrm{tg}\varphi)] \qquad (3)$$

where *p* is the parallactic angle, defined as the difference between hour angle of an object and its vertical circle. They are expressed as function of the hour angle *H*. Both *p* and *A* are defined in the range between -180° and 180° (westward positive), centred at the observer's meridian (Schmidt, G.,2004).

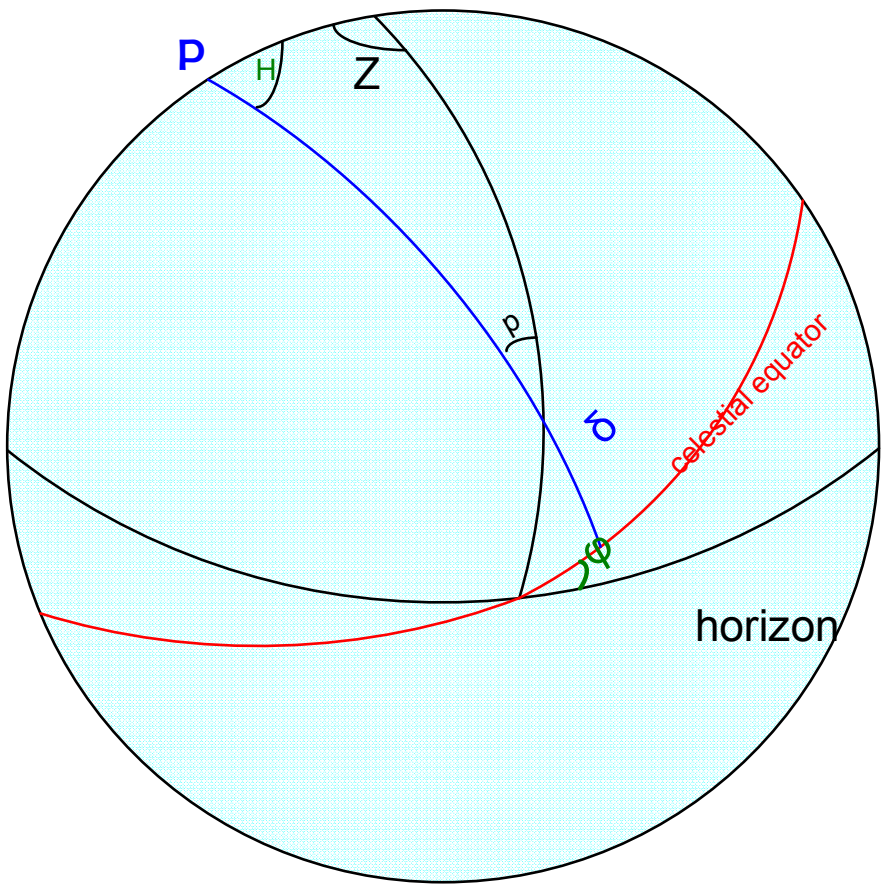

**Figure 3.3** A representation of parallactic angle.

By differentiating the three variables with respect to *t* we can determine the velocity rates of drives and derotation system:

$$z' = \cos\phi \sin A \qquad (5)$$

$$A' = \frac{\sin\phi - \sin\delta \cos Z}{\sin^2 Z} \qquad (6)$$

$$p' = \cos\ \phi \cos\ A / \sin\ Z \qquad (7)$$

It is clear from (6) and (7) (Smart, 1962) that critical velocities are reached close to zenith point, where the denominator is zero. Of course velocity is limited by the drive choice on the basis of the inertia of the moving system and it cannot arise arbitrarily. We have decided to use motor drives for both azimuth and altitude rotations with a maximum velocity of 1.5 degrees per second, as the best compromise between tracking requirements and blind spot reduction (as shown in the next paragraph).

We can determine the parallactic angular velocity (field rotation rate) for different altitude circles by equation (8), passing to rad/hr, considering that *dH/dt = 15°/hr = 0.262* rad/hr:

$$dp/dt = -0.262 \cos\varphi \cos A / \sin Z \quad \text{rad/hr} \qquad (8)$$

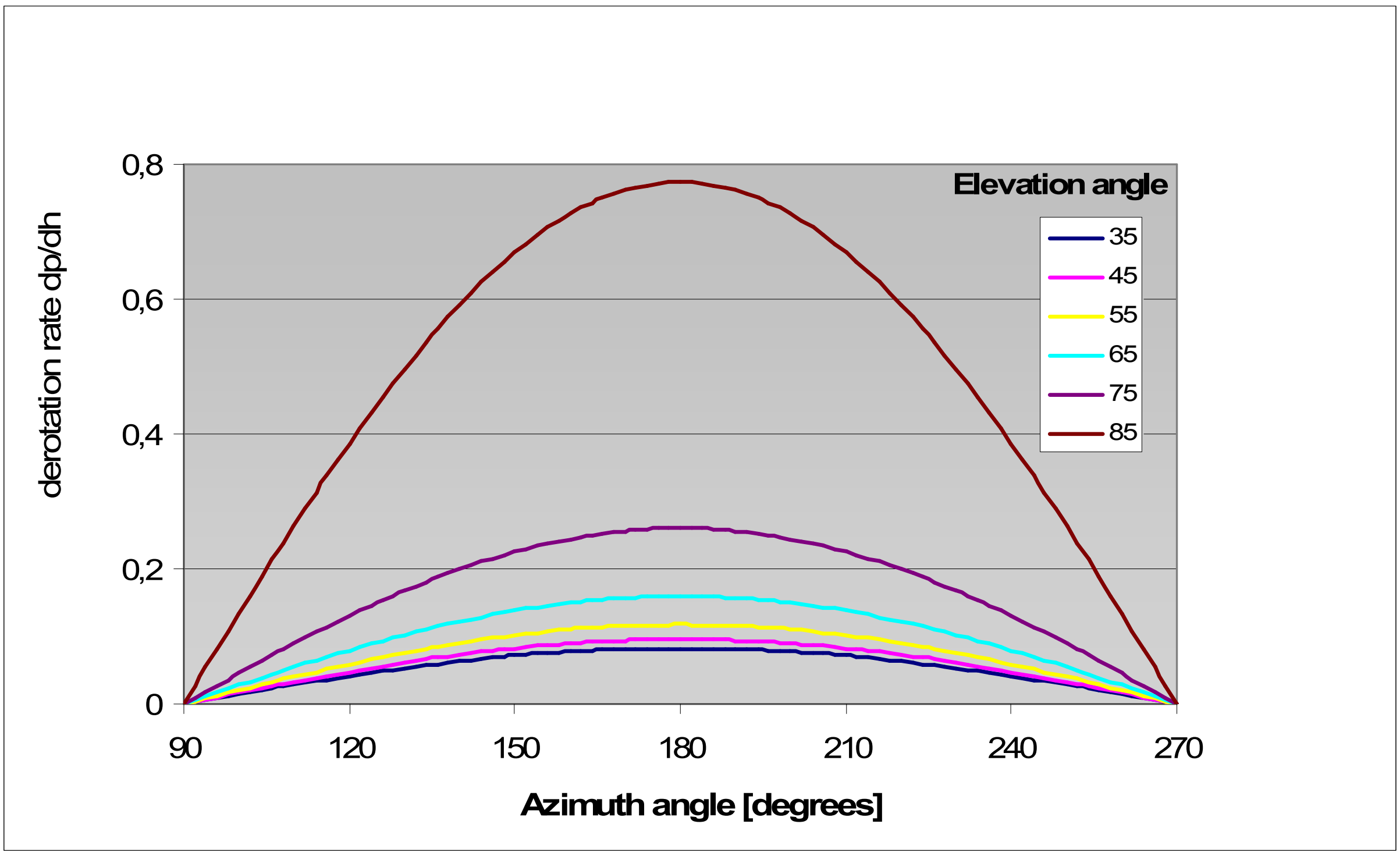

**Figure 3.4** Field rotation rate versus azimuth for different elevation angles.

Maximum value of *dp/dh* (velocity rate of parallactic angle with hour angle), relative to an elevation angle of 85 degrees, is 0.772 rad/hr =0.737 deg/min= 0.012 deg/sec, a rather low velocity that for short exposures does not necessarily require a field derotator.

### 3.2.2 Tracking limits of velocity in azimuth near zenith

By substituting the equation (1) in (6) we obtain the rate of change of A’ as a function of the hour angle. The maximum rate is reached at the transit, when it assumes a value given by the relationship:

$$A' = \cos\delta / \sin(\phi - \delta) \qquad (9)$$

So we have plotted the maximum velocity of the sky near zenith at Dome C latitude (φ=-75°6’25’’), compared to that of a temperate site (φ= 45°), at three different declinations (fig. 3.5). The lower the latitude, the larger is A'. The choice of a maximum driving velocity of 1.5 degree/sec was suggested by the possibility of making some preliminary tests in Italy. In fact this value, equivalent to 90 degrees/min is compatible with the maximum velocity required of 101 degrees/min. Anyway, luckily at Dome C, it lowers to about 37 degrees/min, and we are sure that, with such motors, there will be no tracking problems near zenith .

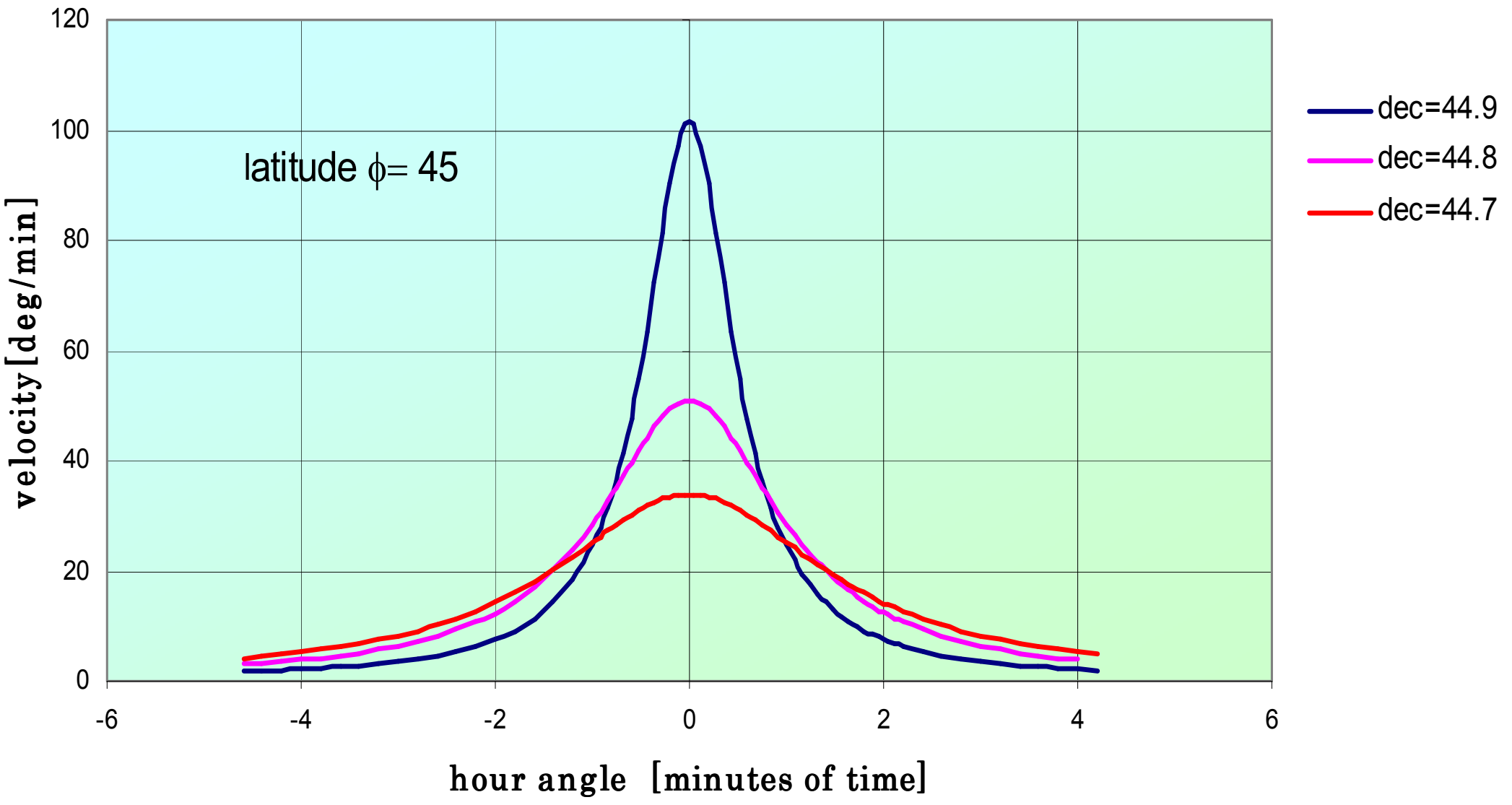

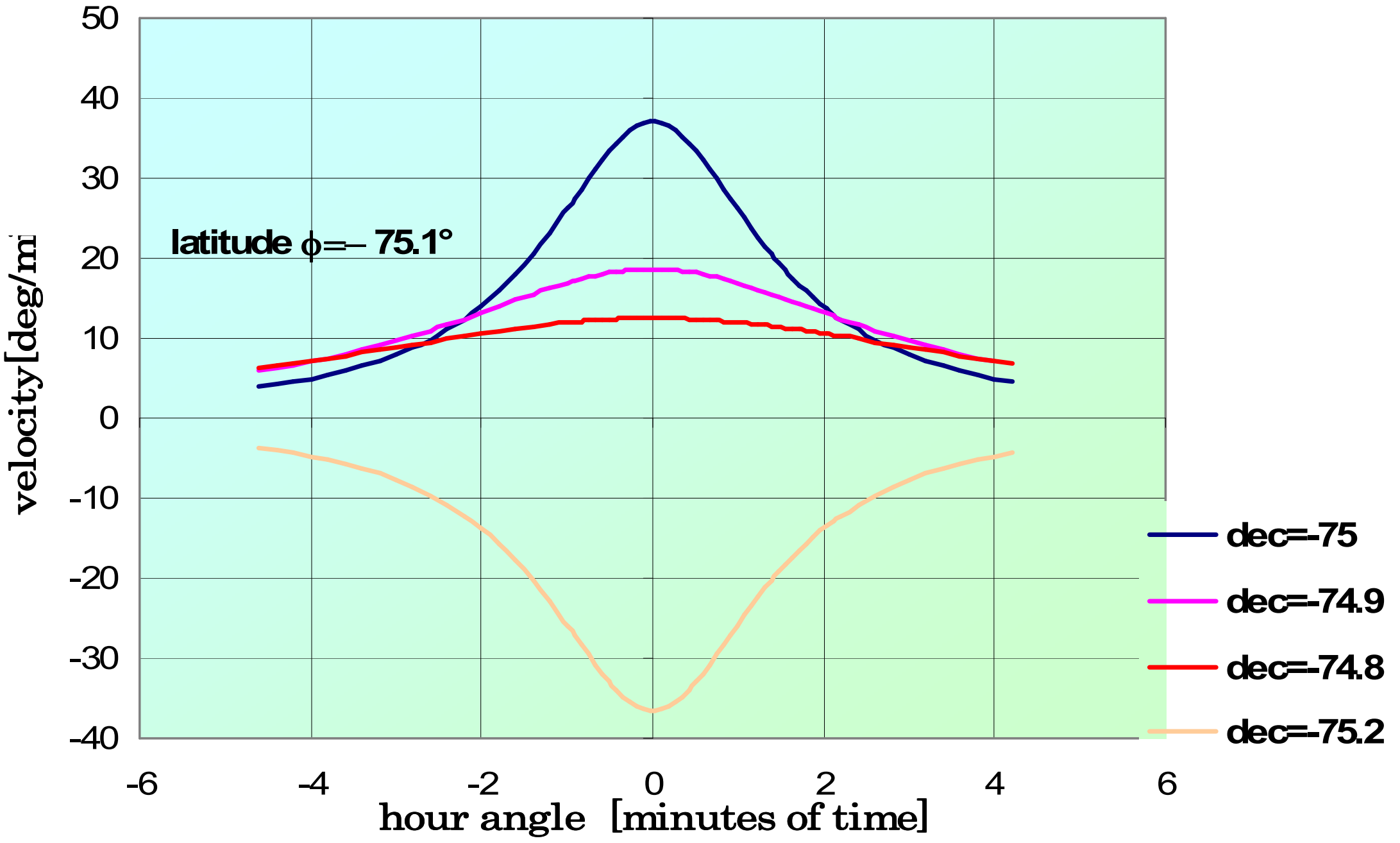

**Figure 3.5** Two plots of azimuth velocity as a function of hour angle, at different declinations. You can note he symmetry around the meridian and the consequent deceleration of motor at the transit, with a change of sign when (φ-δ) is lower than zero.

### 3.2.3 Blind spot

Here is presented an algorithm by Borkowski to determine the size of blind spot. First of all we must determine the range of declinations where it is not possible to track an object. This is provided by the equation:

$$\phi - arctg \frac{cos\ \phi}{|V| - sin\ \phi} < \delta < \phi + arctg \frac{cos\ \phi}{|V| + sin\ \phi} \tag{10}$$

Substituting the known quantities, knowing that V=1.5°/s =360 times diurnal sky rotation (it is a scalar), we obtain: -75.13° < δ <- 75.05°. Objects with declination comprised in this range begin to move faster than the maximum velocity of the telescope, with a corresponding hour angle:

$$H_0 = -ar\cos \frac{\sin\delta\left(\frac{1}{2V} - \sin\phi\right) + \sqrt{1 + \left(\frac{\sin\delta}{2V}\right)^2 - \frac{\sin\phi}{V}}}{\cos\delta\cos\phi} \tag{11}$$

obtained by solving equation (1) under condition A'=V, for H=$H_0$. To avoid discontinuity problems when the quantity $\varphi$- $\delta<0$ we insert the condition:

$$A_0 = A(H_0) + \pi[1 - sign(\varphi - \delta)] \quad (12)$$

In order to determine the western boundary of the spot it is necessary to solve the following transcendent equation (valid for positive hour angles):

$$H_+ = H_0 + [A(H_+) - A_0]/V \quad (13)$$

To do this we can utilize Newton's method, with the first iteration step given by:

$$H_+ = H_1 - \frac{A(H_1) - A_0 - (H_1 - H_0)V}{A'(H_1) - V} \quad (14)$$

and choosing as guess point:

$$H_1 = H_0 - \pi \frac{\sin A_0}{|V|} \quad (15)$$

Starting from declination δ=-75.059°, and with an hour angle of $H_0$=0 h 0.48 min, we have determined the size of blind spot.

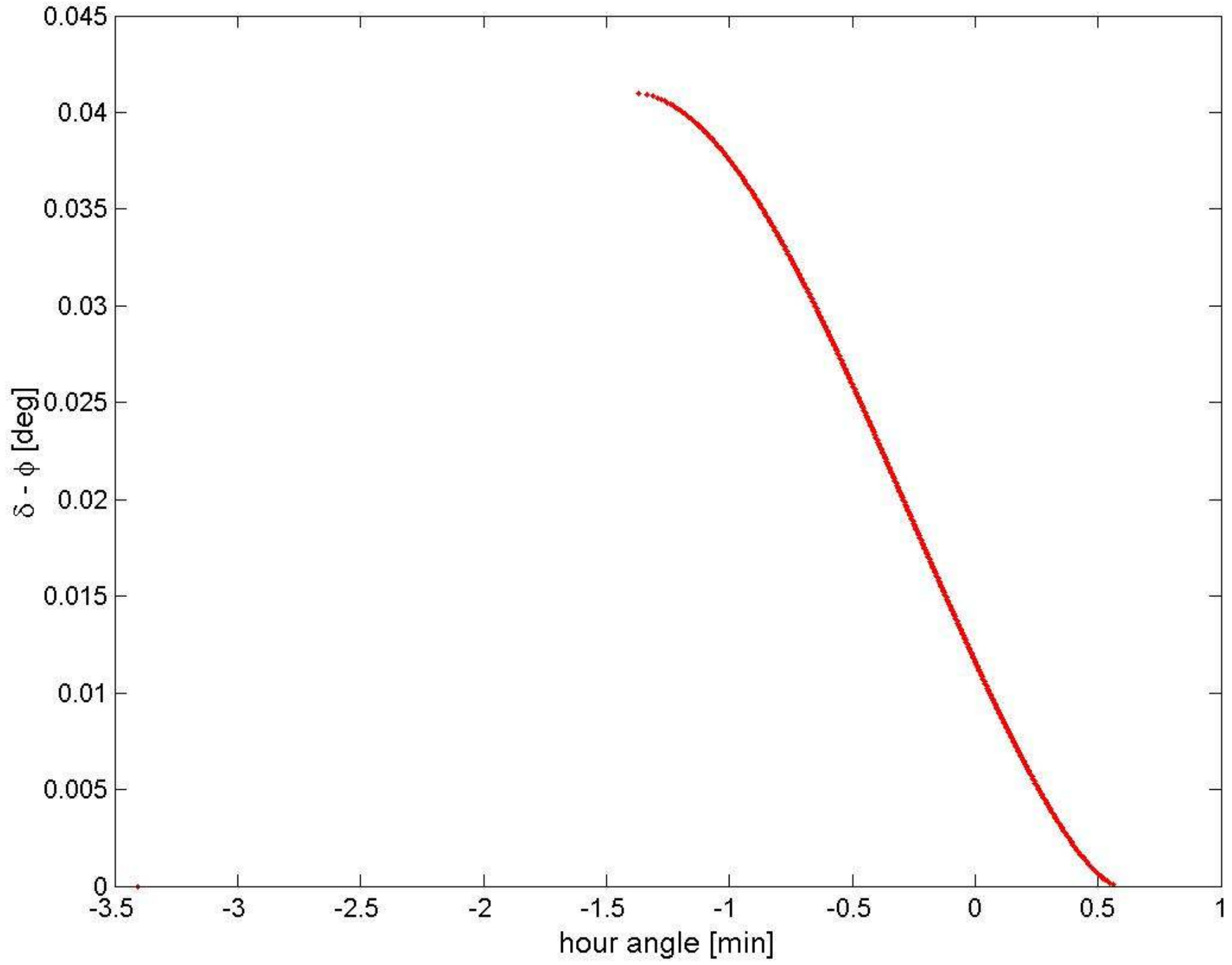

**Figure 3.6** Blind spot size near zenith position with a velocity of 360 times diurnal motion.

## 3.3 Mechanical design criteria

Here, a description of the method we have developed to accomplish the mechanical realization of IRAIT telescope, is summarized.

The flow chart picks out the various phases of the design plan. First of all standard parts commercially available have been pursued, based on design specifications. We have been in contact with such companies as Marcon for optics and telescope mechanics, RKS-SKF for bearings, REDEX, Alphariduttori for gear boxes, Fomblin and Solvay for greases, etc. For each component we have gathered primary data as material, weight, size and, also, their main technical features.

Then we have passed through the CAD preprocessing phase. The geometry of the different parts has been created in Mechanical Desktop, a software by Autodesk. A database of weights, centers of mass and moments of inertia has been generated, and different assemblies relative to several work hypotheses have been drawn. Having determined the moment of inertia of the rotating subassemblies along alt and azimuth axes, we have selected appropriate motors.

In the next step, regarding FEM (Finite Element Method) processing, the model is examined under such packages as ANSYS, SAP2000, to make stress and strain tests, applying known loads and using the best choice for external restraints.

This step is anything but simple, as the drawing cannot be input directly into analysis from the CAD program; instead, the model needs to be simplified for an easier meshing procedure, and some disturbing shapes as holes or sharp edges, for example, need to be cleaned up. Indeed, convergence criteria may fail if the choice is too coarse or even too fine, as the time spent in analysis noticeably increases. A proper selection of solid or shell elements (typically tetrahedral or hexahedral) depends also on the type of analysis: degrees of freedom for element nodes, thermal and structural constraints.

At the end, after the results were found to be suitable with optical and mechanical restraints, we have proceeded to purchase the items in the market, and built the custom parts.

Throughout the project, some practical issues have been taken into account.

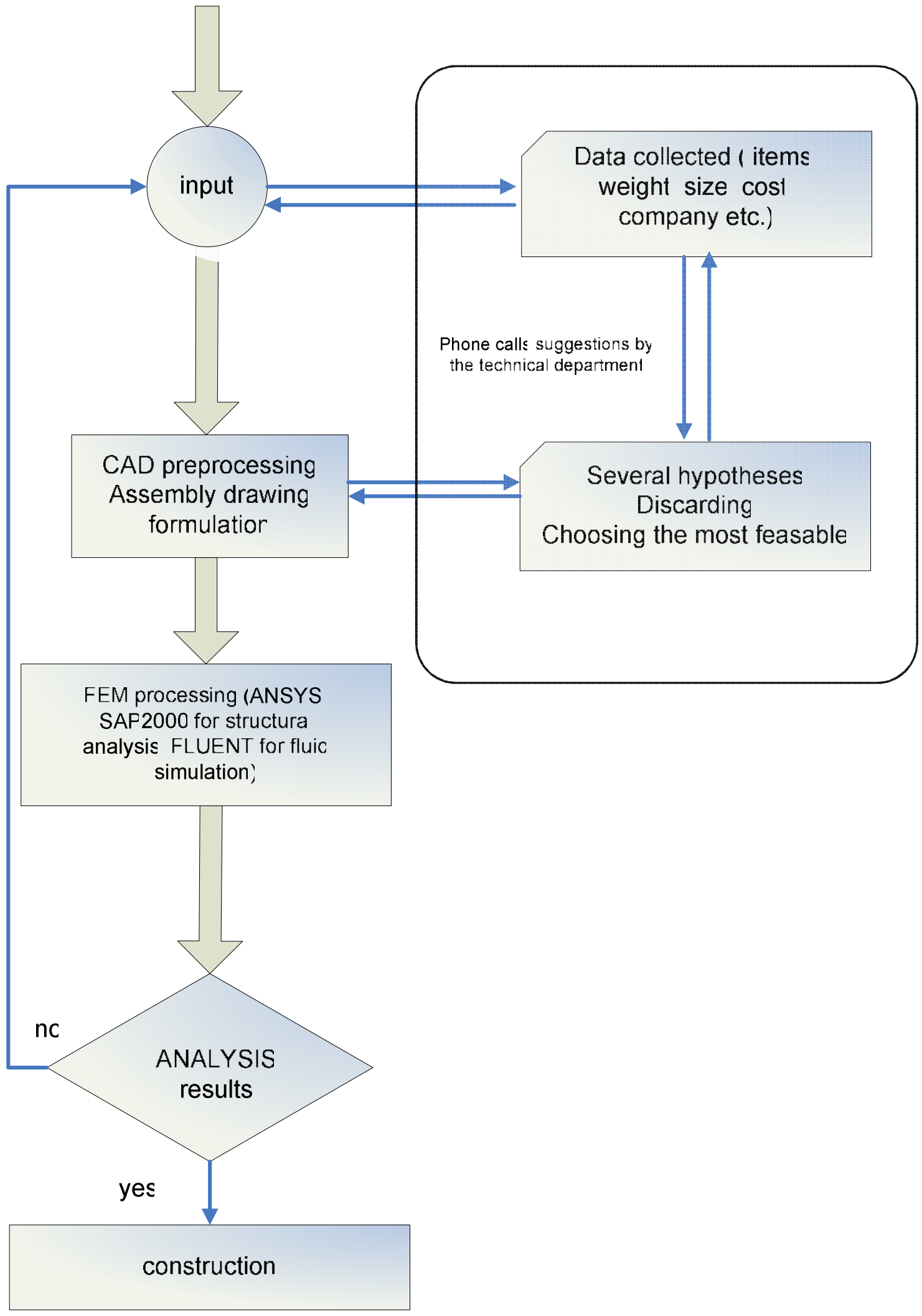

**Figure 3.7** Flowing chart of the IRAIT mechanical design procedure.

Possibly equal and uniform parts have to be used, in order to reach uniformity, symmetry and to reduce costs and delivering times as well.

We have pursued an easy dismounting for critical parts, more subject to maintenance, as the elevation bearings, drives or electronic boxes. Useful space for screwing needed to be increased, as the bolts must be at least M10, for the fact that all objects must be handled with gloves.

During the project, different systems of backlash reduction have been examined. First we thought of a classical system worm wheel- toothed gear with preloaded spring, as in case of Coloti telescope. We thought also of a system including double pinion with spiral torsion spring. However, since it was necessary a further toothing on the outer ring of cross bearing, which was too expensive, we abandoned this solution. Finally, a system provided by Dual Drive has been opted for, including a torsion bar with two pinions in counter-rotation, as it avoids any backlash error.

For precise rotations, radial and axial runouts of bearings must be limited: for this reason a significant preload has been added to the bearings, specified by the manufacturer. There is a non linear relationship between preload tension and deformation: it would be quite essential to study in what way the preload changes in a climatic room, at various operating conditions and with different types and quantities of greases.

Connection between elements of two different materials is particularly critical, as different coefficients of thermal expansion are associated. Bolts and joints connecting the parts assembled in Italy must not be fastened with the preload indicated by the ISO standard, as it increases tensions at lower temperatures. The value of this preload must be agreed together with the manufacturer.

Welded parts must be treated with annealing technique, in order to release residual stresses left by manufacturing process. Safety factor is typically increased of a factor 5, considering the whole structure is subject to shocks of 4g during transport. It is also important to implement a dynamic analysis of the optical tube, besides a static one.

Surfaces exposed to atmospheric agents, electronic boxes, optical tube and fork must all be carefully coated. The other parts, which cannot bear the environmental conditions, must be heated and protected by thick insulator layers.

Among all the factors, of course, temperature remains the main restraint. The telescope is planned to work during summer. Anyway, for the last dispositions of PNRA, it will have to stay parked at the reserved lay-by at Dome C for a year. In this context the container can be warmed up by a heater of 3 kW, controlled by a timer every 10 minutes, so that temperature does not fall below -40 ° C.

In any case, for reliability and a fail safe system, each mechanical part have to resist to a minimum temperature of -80° C.

Electronic devices must be preserved from cooling below -20 °C, and protected by insulation boxes. It has been pointed out, by the last campaigns at Dome C, that even the smallest slits and apertures are exposed to the diamond dust drift. For this reason good sealing system must be provided. Labyrinth sealings would be the best ones, even if it is too expensive for the overall budget.

Peculiar transportation conditions have required the development of a stiff base chassis, with pipe section bars, capable of damping shocks and fluctuations. The different modules are fixed upon the chassis, and must be assembled after the arrival at Dome C. Therefore, there's a list of operations planned to be made *in situ*: mirrors mounting and their calibration, alignment of optical and mechanical axis for focal plane instruments, fastening of the group including fork mount and optical tube on the azimuth: special cranes will be at our disposal for this type of operations.
In the next paragraphs the duties of main mechanical elements are explained with more detail.

## 3.4 Choice of materials

For the choice of materials we must take into account thermal expansions and related stresses, both *in situ*, due to solar radiation, and the ones due to thermal gradient between manufacturing phase in Italy and the operative one in Dome C. A good rule is to select, as far as possible, the same materials for the various components and to keep a uniform thickness. The most recommended material for such applications is austenitic steel with low percentage of C and high of Ni (>3.5%), like for example FeE355Ni6, with a breaking stress of $\sigma_m = 490 \div 640$ N/mm$^2$.
For casting parts, a steel with high tenacity is indicated as the best one, like G12Ni14 (with 3% Ni and 0,15% C) . The estimated tenacity for a stainless steel, as, for instance, for AISI 304 is about 125 J KCU[2] , higher than that of common carbon steels (C40) which is 25 J.
An accurate selection of lubricants and greases is fundamental too. Fomblin company has been contacted for this purpose, and they suggested us a grease with a PFPE oil base and a working temperature down to -80 °C.

## 3.5 Telescope design

The telescope structure comprises essentially three subsystems: the optical tube, the fork mount and the base chassis. These parts were designed by taking care of the internal container dimension; in particular, the base chassis is bolted through appropriate joints to the flat of a modified ISO20 container. It primarily needs to be stiff enough to bear the shocks during transportation. We can distinguish (as in fig. 3.5) other three subsystem interfacing with the telescope, to be mounted *in situ*:

1. Fixed platform, which is leaned on the wooden platform through six legs, and sustains the upper tent;

[2] KCU refers to a probe with engraved U, which is subject to a Charpy test with a machine of useful energy of 300 J (see UNI EN 10045).

2. Azimuth co-rotating platform, consisting of a truss structure with an internal ribbed rail, guaranteeing a safe protection of boxes, cables and slip ring from ice storage. It will be covered by a wood layer;
3. The tent with its motors and additional components.

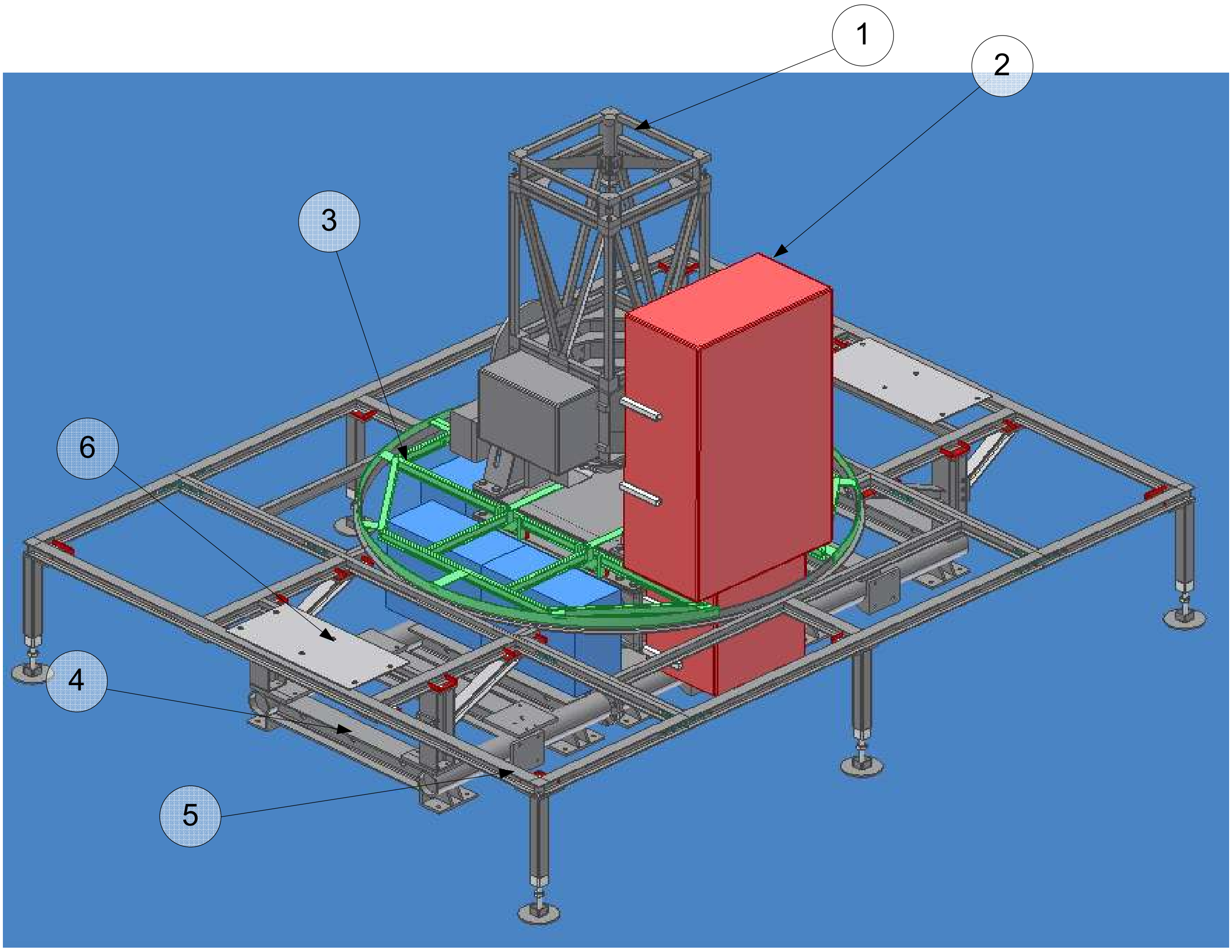

**Figure 3.5** IRAIT assembly and AMICA rack perspective view. 1) telescope; 2) AMICA rack with upper and lower cabinets; 3) azimuth co-rotating platform; 4) base chassis interfacing with container floor; 5) fixed platform; 6) tent interface joint plates.

The structure is substantially made of steel Fe 460 B, welded, subsequently cooked with surface treatment, in order to prevent thermal shocks at environmental conditions, especially during winter months. It has been manufactured with machine tools; mechanical interfaces between all the parts need to be properly worked with a finishing process. MARCON company has been in charge of building and assembling the whole carpentry.

The estimated centre of mass of the system is at an height of 1500 millimeters from the floor of the container on the vertical axis.

The overall system must be stiff enough to have eigenfrequencies greater than 5-10 Hz. Rotation in alt-az is provided by brushless motors and coaxial reducers. Absolute encoders, RON model 727, provided by Heidenhain with a resolution of 18000 pulses/ rev for a closed loop control on both axes, will be used. A slip-ring mounted inside the azimuth bearing, centered on the axis, will avoid any cable problem, removing limitations to angular stroke in both senses of rotations .

## 3.6 *The optical tube*

The optical tube must be light, above all, in order to reduce the overall weight, and, therefore, the moment of inertia, with a better selection of drive system in altitude: as a consequence, it must satisfy less severe requirements. While developing technical drawings we have tried to keep the moment of inertia around horizontal axis as low as possible. The optical tube subassembly can be intended, on its turn, composed by the following parts:

1. Serrurier truss;
2. Supporting structure of optical tube;
3. Secondary mirror mount and supporting structure with spiders;
4. spur gear sector;
5. primary mirror cell with mechanical interface for M3 mount.

### 3.6.1 Serrurier truss design

#### 3.6.1.1 **Basic requirements**

It is well known that the largest flexure occurs when the optical tube is horizontal. Misalignment between M1 and M2 resulting from the displacement of M2 vertex must be compatible with optical requirements. Therefore, the main function of this component is that of preserving a constant distance between M1 and M2. Serrurier strut of the optical tube can be compared (in first approximation) to a cantilever beam with a load applied at one end. The nature of applied loads is essentially given by: self weight, wind load, active load of M2 motorized focus.

In the first stage of our project we have considered a static analysis including contributions of self weight and two genres of thermal stresses: the first, due to the gradient to which the whole structure is subject during transport from Italy to Antarctica, has been assumed of $\Delta T = -70$ °C; the second, due to the different state of a part of the strut exposed to sun, and the other one remaining in the shadow.

Assuming the coordinate system in figure 3.8 and 3.9, it must be checked that $\Delta x/l<min(FOV1,FOV2)$, where $\Delta x$ is the maximum displacement relative to both ends of upper beam, and $l$ the length of the beam, considering the smaller NIR array of 2.29×2.29 arcmin$^2$.

The second restraint is $\Delta z/d<min(FOV1,FOV2)$, where $d$ represents the distance between M1 and M2 (2005 mm).

Firstly we have studied a simple model in 2D with two diagonal beams that converge to the top ring, and the other with beams that converge towards M1 cell. Then the analysis was extended to a 3D case, discovering that the original configuration with convergence node to M1 was better.

#### 3.6.1.2 Nodes displacements: 2D and 3D comparison (static analysis)

Outcomes are those retrieved by a simple model made in SAP 2000: it is a strut of 7 steel beams, with 3 external restraints ( a pin and a roll) . A span load, distributed along the upper beam (from node 2 to 7), equal to the maximum weight of optical tube plus accessories, was taken into account.

The different geometries are called *OThp1* and *OThp2*. Thus, in *OThp1* we're interested in calculating the difference of the two displacements, relative to nodes 2 and 3: $|x_2-x_3| = 0.0721$mm; and to nodes 20 and 5 $|z_{20}-z_5|=0.233$mm.

Instead, in *OThp2* both values are lower: $|x_2-x_4|=9.47*10^{-3}$ and $|z_{17}-z_3|=0.016$ mm.

For this reason, it appears that the second configuration is better: in fact, supposing all nodes to be hinges, so that they are only free of rotation, the layout of central elements (three hinges arch) reduces the flexure of vertical load. In fact, most of Serrurier struts have two by two convergent elements toward the M2 mount, in small telescopes as well as in large telescopes like TNG, for example: in general this configuration confers enough stiffness to damp vibrations from the M2 chopping and focusing.

Anyway, after a more careful analysis of the particular load case the structure must bear, as local temperature gradient is the most relevant of all, actual geometry revealed to respond in a better way. The 3D model was put into ANSYS, and maximum displacements and stresses were calculated.

Actual load conditions have been simulated supposing that upper beams are subject to a span load equal to the weight. Restraints were put on the for nodes of the basis, in order to allow thermal expansion in two directions coplanar with basis (Y and Z).

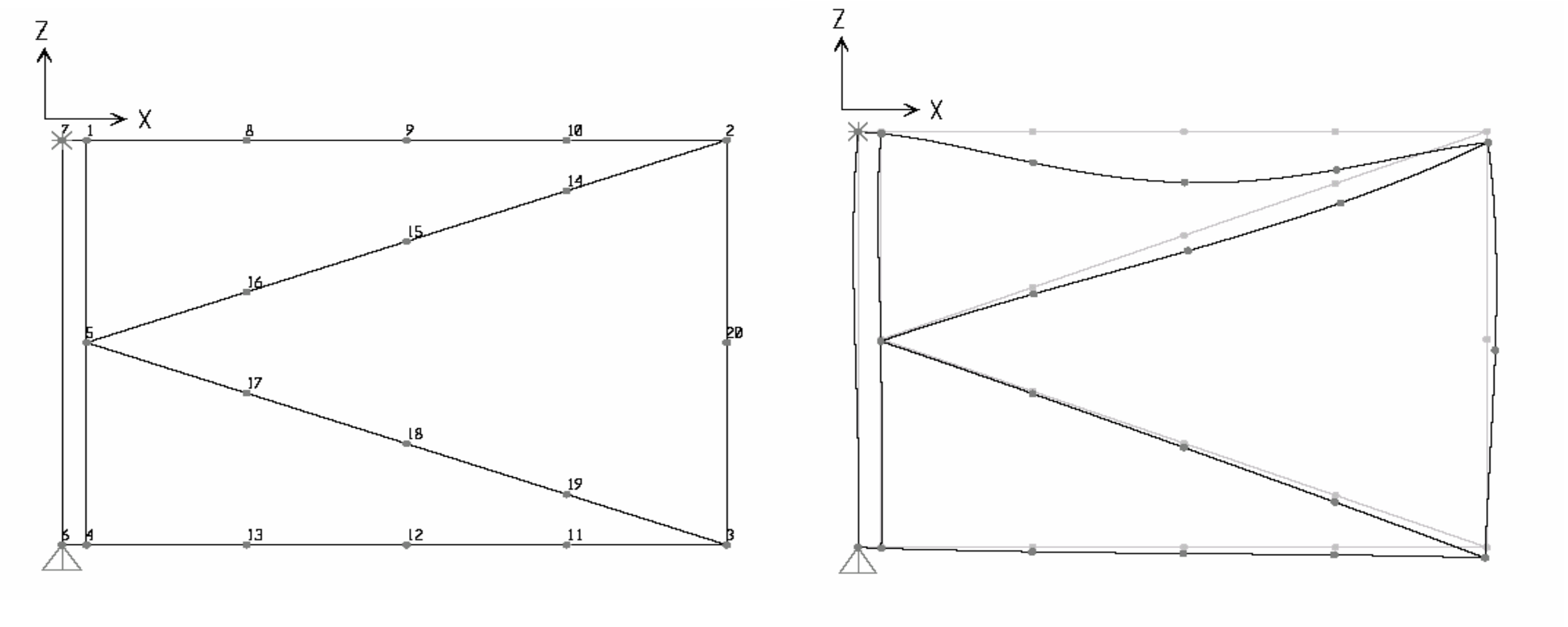

**Figure 3.8** A plot of the strut with nodes in undeformed and deformed configuration, in OThp1 model, obtained by SAP2000.

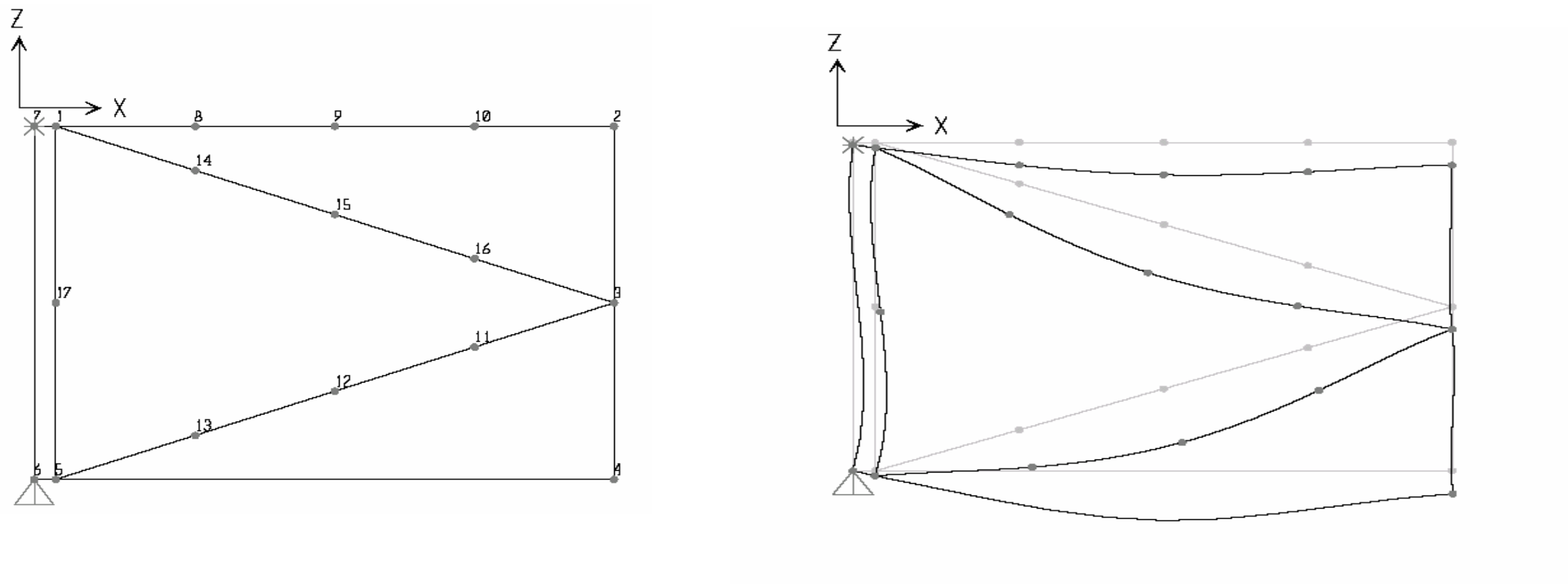

**Figure 3.9** A schematic of strut with nodes in undeformed and deformed configuration, in OThp2 model.

### 3.6.1.3 **Flexure diagrams**

Box tube beams (of dimensions 50x50x5 mm) have been taken into consideration. The material adopted for simulation is elastic, isotropic steel, and, also, beams of constant section with tension, compression, torsion, and bending capabilities (i.e. BEAM4 in ANSYS definition) were chosen for elements. The figures 3.11,3.12, 3.13 show displacements in the global coordinates UX,UY,UZ as well as Von Mises stress along the strut. With a thermal gradient of -30 ° applied on half of the tube, and a span load on upper beams of -3300 N/m, a maximum stress of 56 N/mm$^2$ was obtained, whereas the maximum displacement is UZ (0.913 mm). Flexure errors fulfill the optical features, as they are comprised in the allowed range.

| NODE | FX | FY | FZ | MX | MY | MZ |
|---|---|---|---|---|---|---|
| 1 | 6607.7 | 6634.8 | -5.3754 | 0 | 0 | 0 |
| 26 | -6663.9 | 0 | -467.85 | 0 | -23.410 | 480.37 |
| 53 | 6876.3 | 6603.7 | | 0 | -30.040 | 239.00 |
| 78 | -6820.0 | 41.850 | 473.23 | 0 | 0 | 0 |

**Table 3-1** Joint reactions at four nodes on the basis.

| DISPLACEMENTS | NODE NUMBER | VALUE |
|---|---|---|
| UX MAX | 70 | 0.1208 mm |
| UY MAX | 112 | 0.9108 mm |
| UZ MAX | 70 | 0.1904 mm |
| flexure error 1 (Δy/d) | - | 1.56’ |
| flexure error 2 (Δx/l) | - | 26.4” |

**Table 3-2** Largest flexure errors.

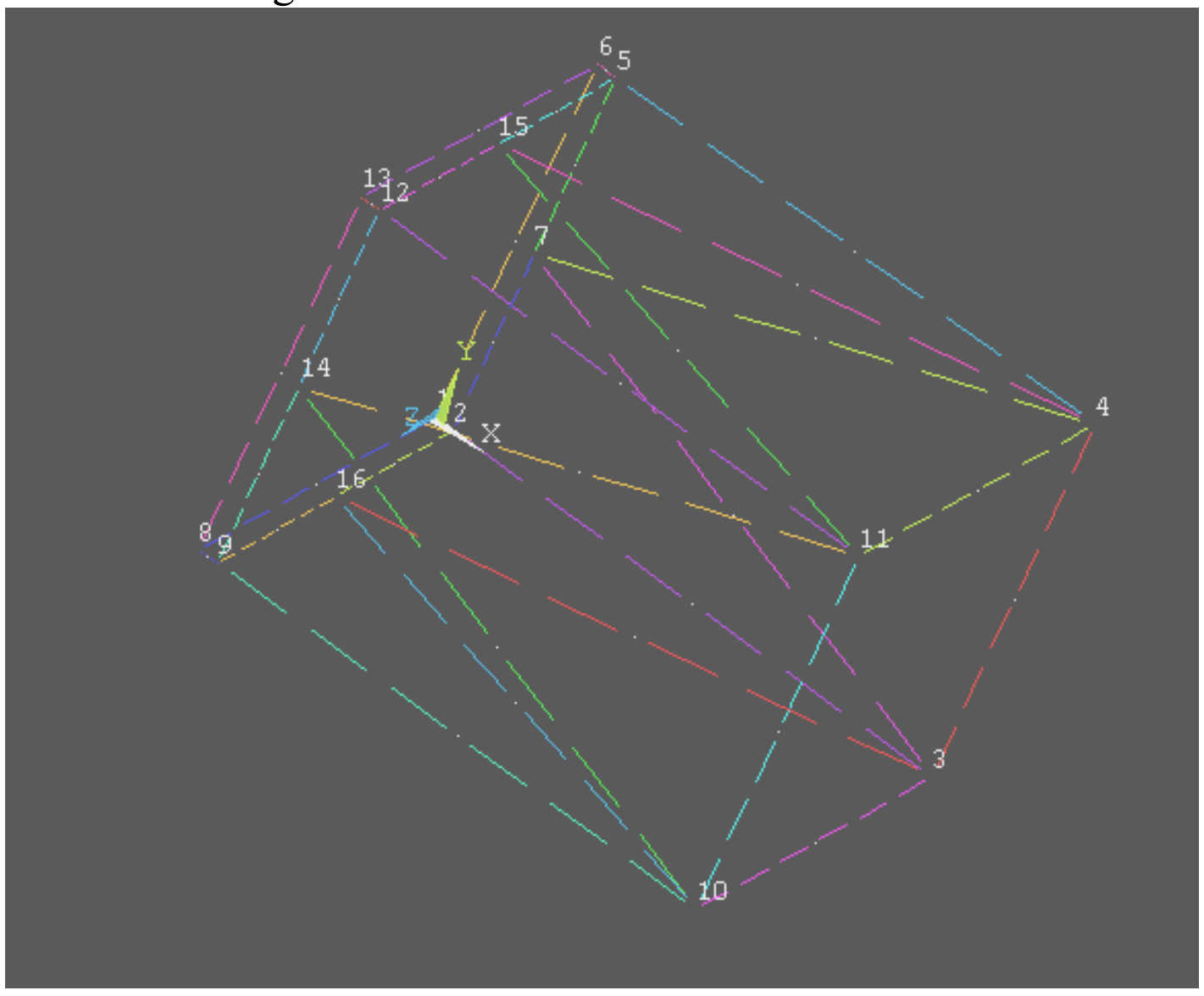

**Figure 3.10** OT picture with numbered keypoints.

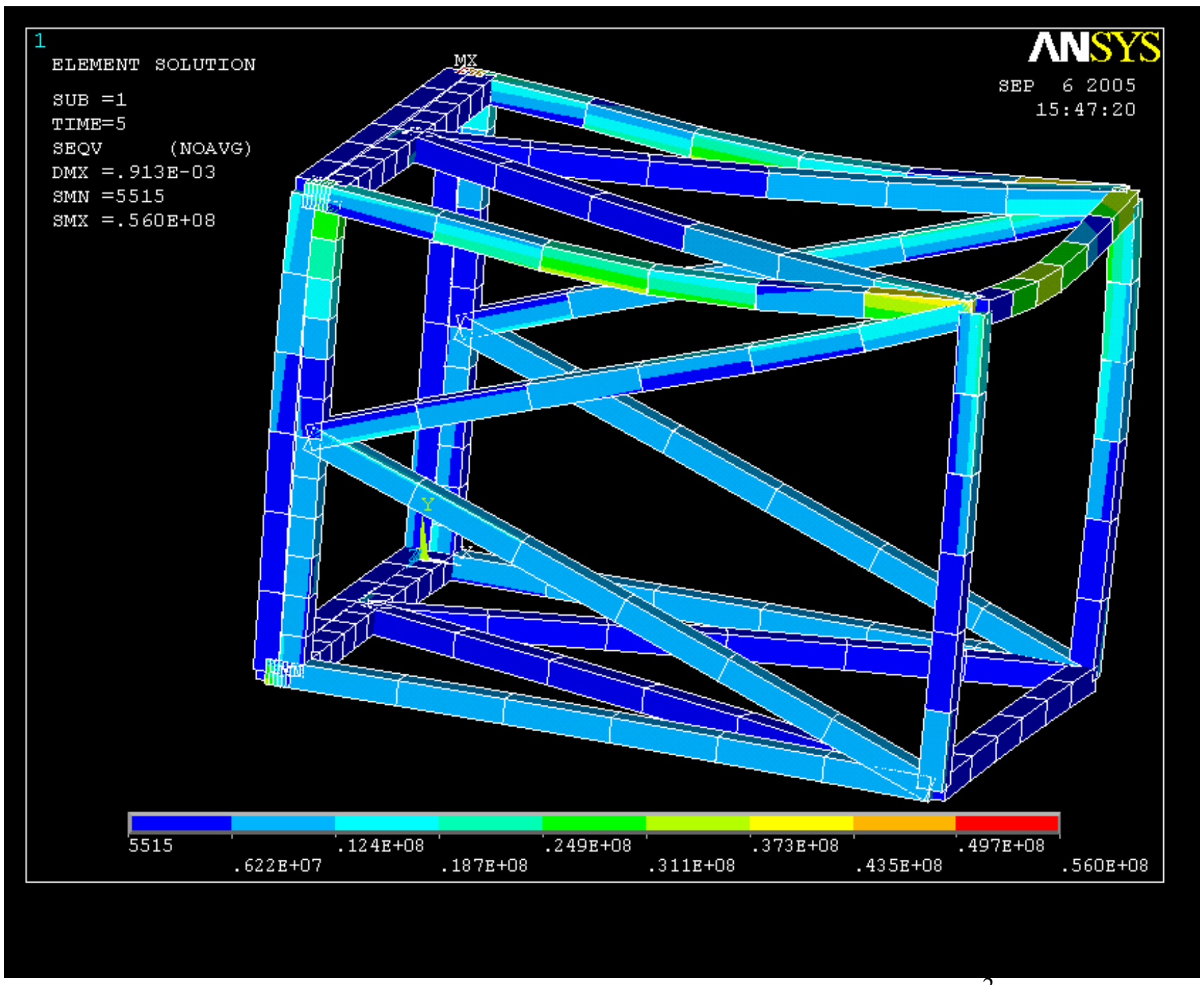

**Figure 3.11** Von Mises stresses plot. Values are indicated in $N/m^2$.

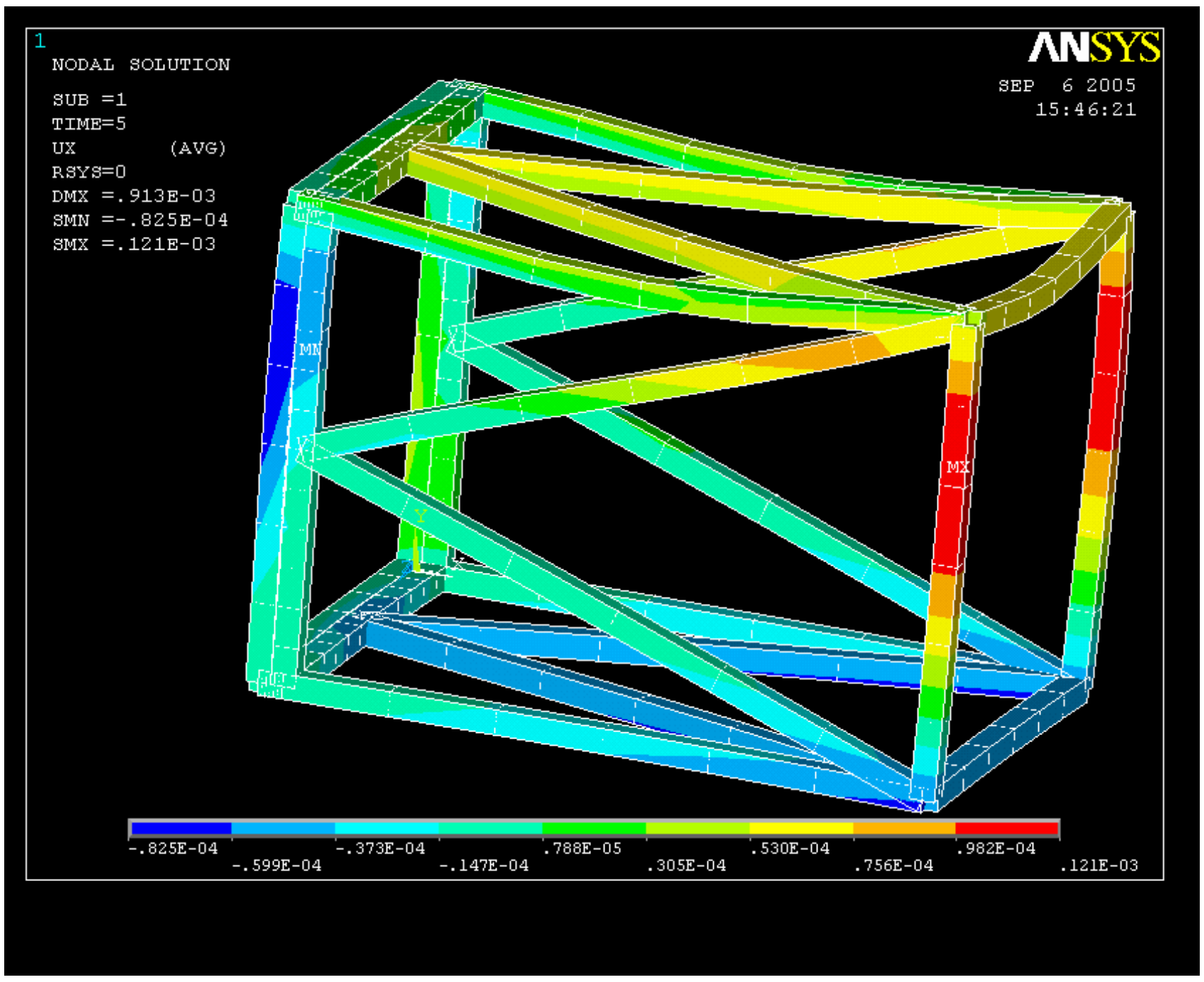

**Figure 3.12** Plot of displacements along X direction.

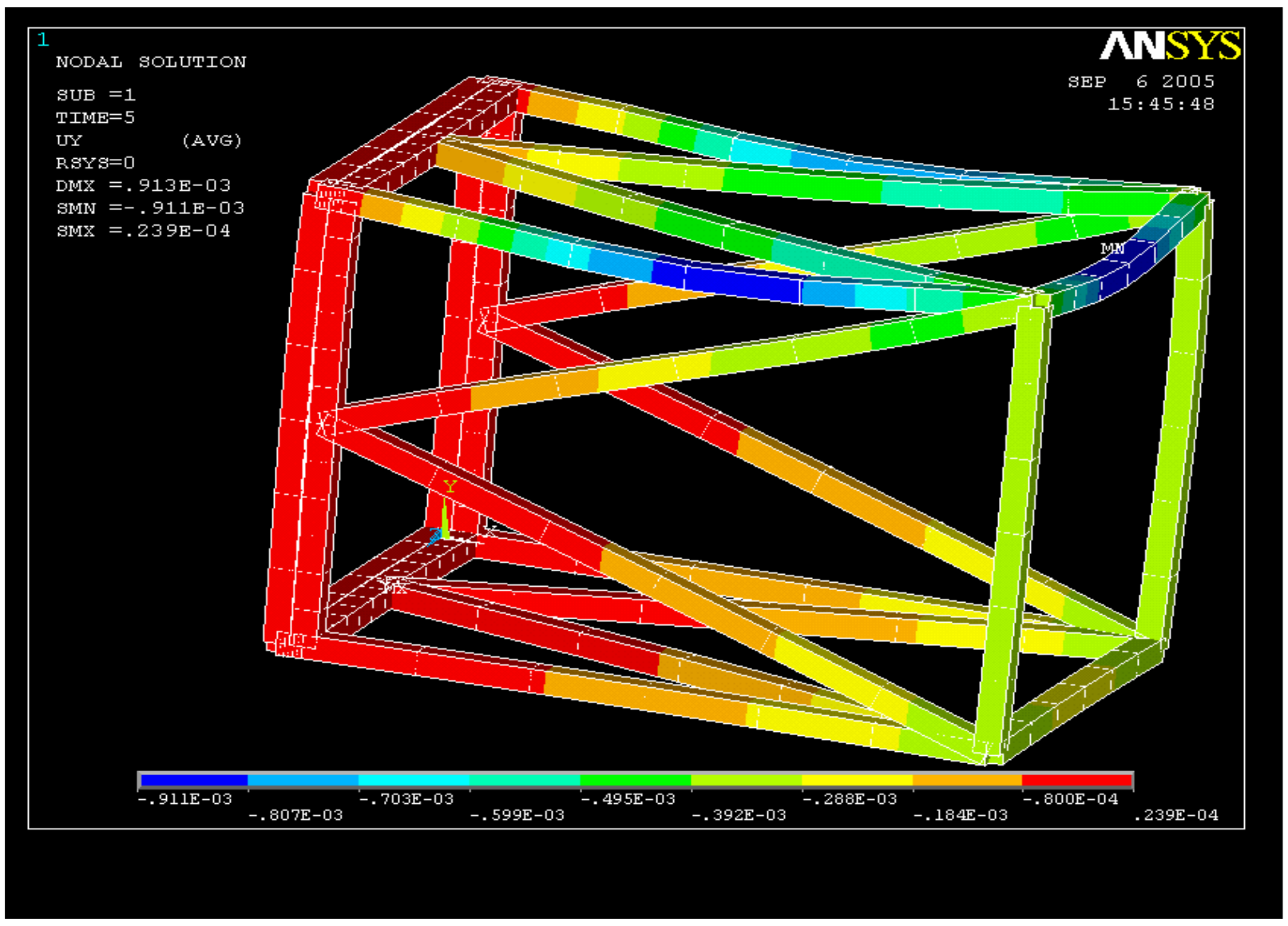

**Figure 3.13** Plot of displacements along Y direction.

In the last phase of the project a more detailed analysis has been carried out, integrating the initial truss model in the complete optical tube, and in the whole structure of the telescope at last, including the fork mount.

Moreover, for maintenance and easy dismounting requirements, the truss comprises other functional elements:

- part of the rapid assembling system of secondary mirror mounting;
- a centering system for the optical tube supports, by means of corner blocks with M14 holes. These blocks provide more stiffness at the nodes and also allow a precise alignment at interfaces, with an easy construction.

The same type of connection is also intended for primary cell bearing structure.

We have chosen box tube beams 50x50x5 mm, welded together, interposing shaped joints to strengthen the most critical points of the truss.

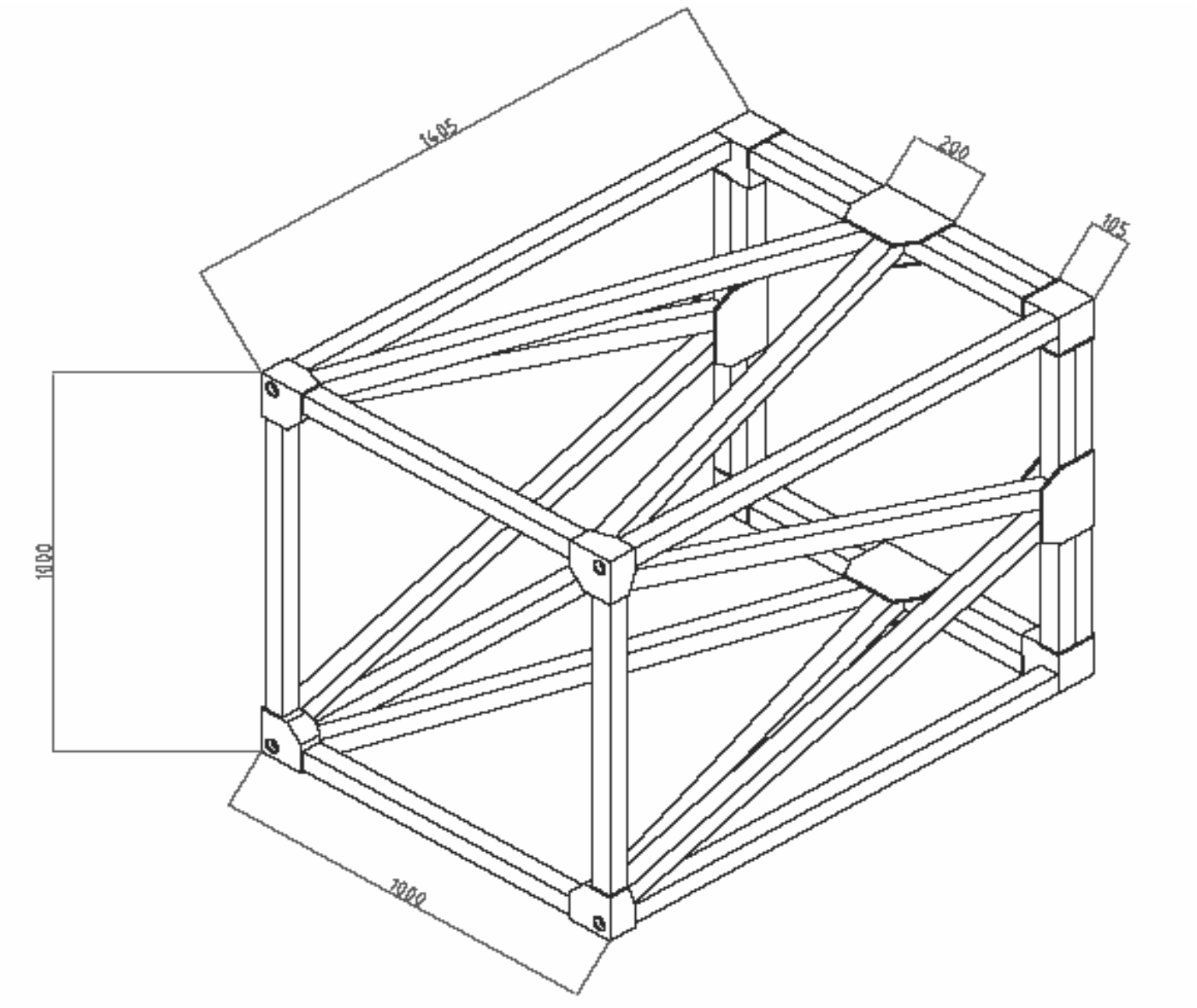

**Figure 3.14** Axonometric view of Serrurier truss, with main dimensions.

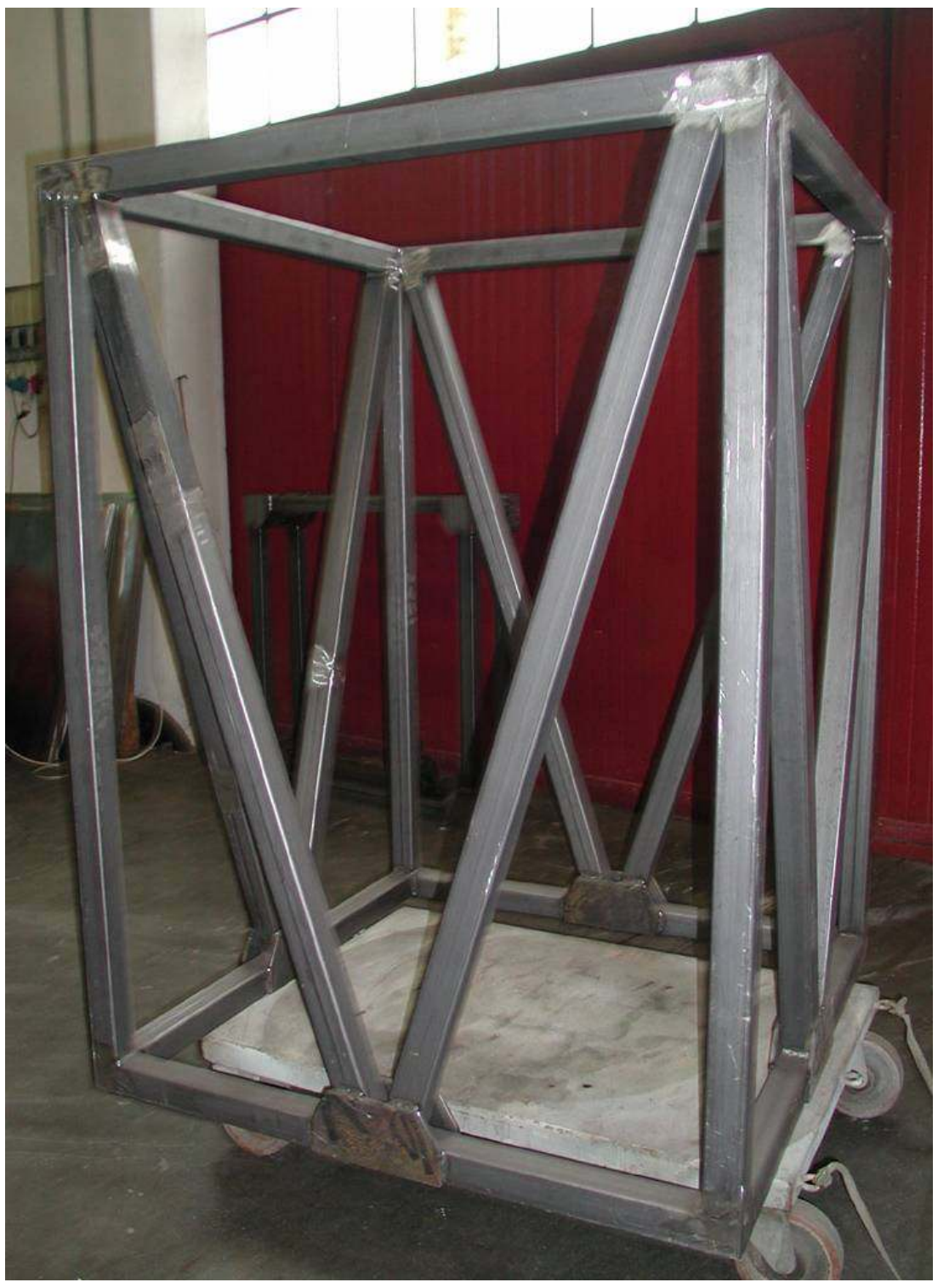

**Figure 3.15** A picture of the truss just after the manufacturing process (by MARCON courtesy).

### 3.6.2 The supporting structure of the optical tube

It is made of a unique welded block with a central aperture of 870 mm diameter in the top face, and two hollow axles with internal diameter of 100 mm, along altitude axis, for optical beam passage towards the foci. The following parts can be distinguished:

1) Upper plate of 10 mm thickness equipped with a fastening system to mount Serrurier truss;
2) a mechanical interface for one electronic box;
3) a mechanical interface for primary mirror cell and part of the rapid assembly system of it, designed by Marcon;
4) a flange to bolt the tooth sector housing;
5) two axles with cylindrical centers and seatings to mount the inner rings of taper roller bearings;
6) Withdraw sleeves and metal rings providing the correct preload to the bearings;
7) housing system for lubricant seals;
8) a mechanical interface for the altitude encoder, located on the axle relative to the second Nasmyth focus.

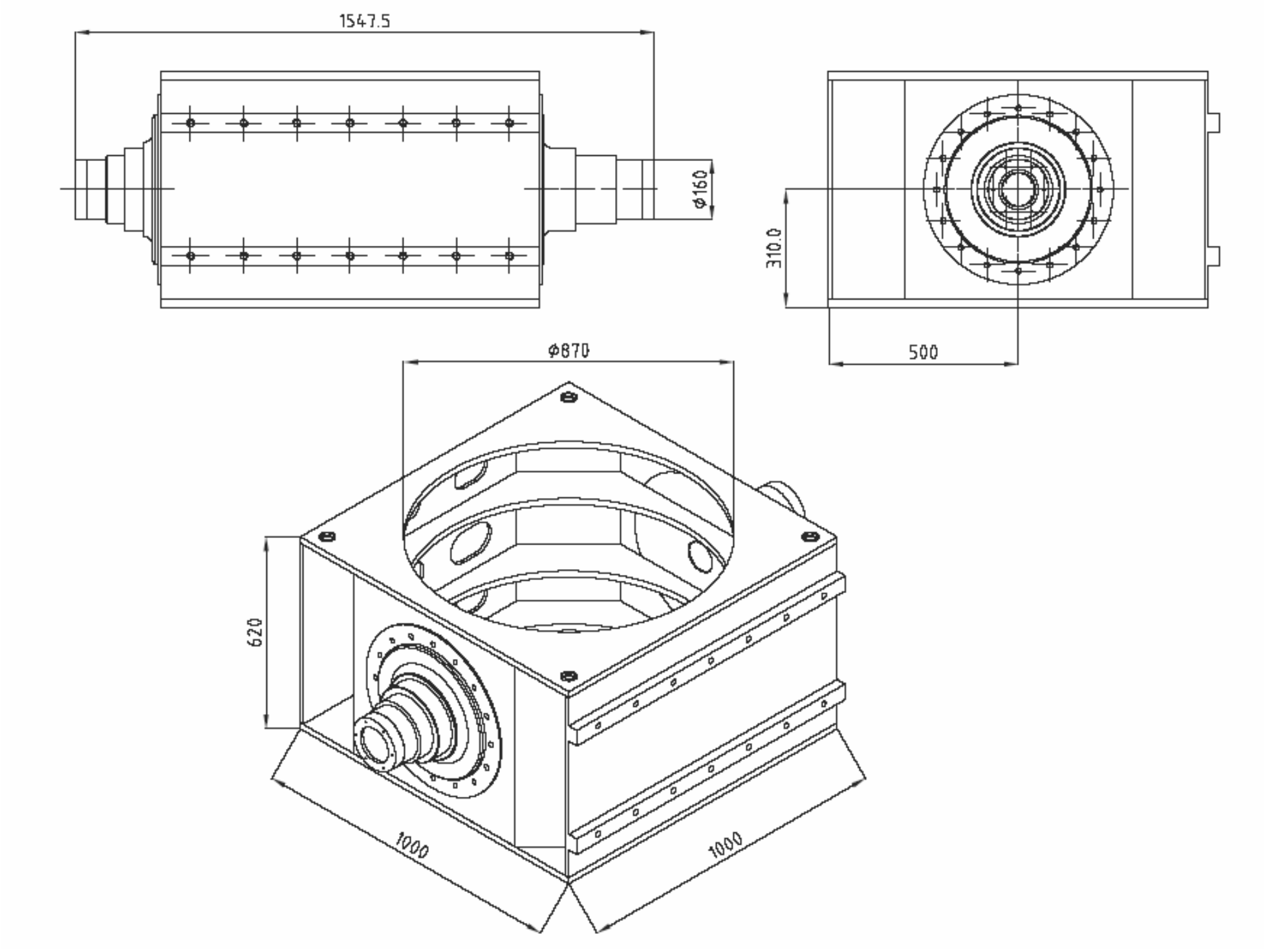

**Figure 3.16** A drawing of the structure that hosts the M1 cell.

These parts are shaped metal sheets of 10 and 20 millimeter thickness. The machining, achieved by means of numerically controlled machine tools, mainly consists of: squaring, leveling with positioning and execution of drillings and threading. For tolerances of bearing seatings, instructions by SKF have been followed.

### 3.6.3 Top ring

The telescope top ring has an upper structure supporting the secondary mirror and focus assembly. Likewise the S. strut, the top ring is conceived as made of 50x50x5 mm hollow box tubes. Four corner blocks allow interfacing and accurate centring to the truss by means of 4 M20 x 1.5 nuts. It provides also a rapid dismounting from the Serrurier strut.

There are four triangular spiders, at 90° degrees : each of them is bolted with 2 point contact to the M2 mount and one point contact hinge to the external truss, in order to provide an easier preload to the bolts. Spiders must not be bolted with standard preload, but instead the right play must be conferred, considering the different temperatures between Italy and Antarctica.

Tolerances between the geometrical centers of M2 and M2 mount must be in the range specified by NTE (Bru R.,A.Catalan et al., 2005).

| XY plane | 0.7 MM RADIUS |
|---|---|
| Z | ±0.5 mm |
| Rx | ±0.125 mrad |
| Ry | ±0.125 mrad |
| Rz | ±0.125 mrad |

**Table 3-3** M2 integration errors.

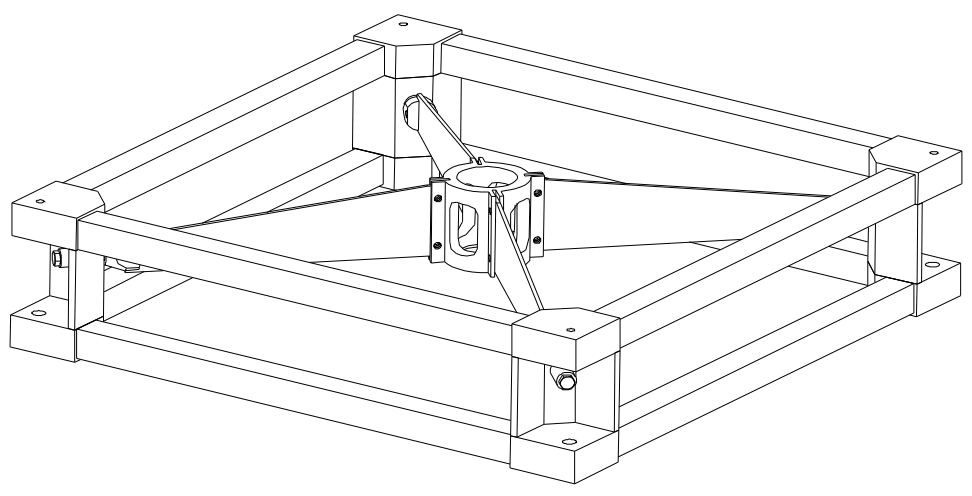

**Figure 3.17** Axonometric view of secondary mirror strut.

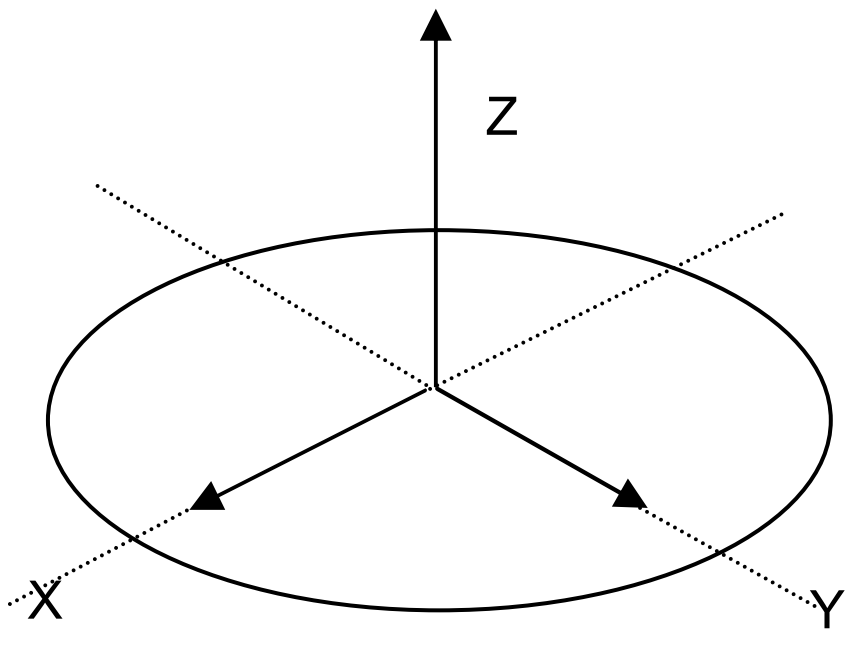

**Figure 3.18** M2 coordinates.

### 3.6.3.1 Estimation of vibration modes of the top ring

An important issue was to check the eigenfrequencies of the top ring in its final configuration. In fact the parts composing the top ring has undergone some changes during the project; as the spiders in the first configuration, it had two contact points on the truss. There was also a reduction of thickness in the flange in order to allocate the new linear actuator provided by Pythron. Therefore the last dimensions of internal and external diameter are respectively 104 and 130 mm. The flange supporting the M2 subsystem has screws of the type M8x1.25.

Two different load cases were imposed: one considering the proper weight, a torque around Z axis due to the motor, a thermal gradient ΔT=20 ° assigned to a pair of spiders; the other equal to the first apart from a constant mean environment temperature of -40°C instead of the gradient, uniformly applied to all surfaces. Fixed constraints were set on the four holes at the mechanical interface with the upper truss. In these points all degrees of freedom are restrained, except the rotation along the transverse direction of each spider.

In the first case, as load conditions we have taken into account: a proper weight, including focuser and chopper subsystem, and secondary mirror, estimated to be about 371.48 N; a moment around vertical axis, whose value indicated by NTE is 500 Nmm. It has been determined on the basis of the following assumptions: the moment of inertia of the mass to move is $9 \cdot 10^{-3}$ kgm$^2$ ; knowing the maximum chopper frequency being 25 Hz, and maximum angular acceleration of the mirror being $a_{max}=A\omega^2 cos\theta t =A\ (2\pi f)^2 =55.5$ rad/s$^2$. The torque due to angular acceleration is: $M=I \cdot a$=500Nmm. Then a uniform thermal load of - 20 °C has been assigned to the spiders, indicated in fig. 3.19 as 1 and 4; and of -40 °C to 2 and 3. A static analysis has been performed with the two load conditions as indicated, in a meshed model with 7919 solid elements. The maximum stress reached under thermal gradient is almost the allowable one, being $\sigma_{VM} = 237 < 250$ N/mm$^2$. Instead, in the second case Von Mises stress is drastically reduced to 5.166 N/mm$^2$, so that it is far more preferable to shield the structure with panels.

A modal analysis has also been conducted to study the modal response of the top ring to the vibrations induced. The first 8 eigenfrequencies have been calculated:

$f_1$= 82.66 Hz ; $f_2$=150.109; $f_3$=151.001; $f_4$=151.022; $f_5$=224.375; $f_6$=233.84; $f_7$=415.655; $f_8$=417.483

As the lower mode is fundamental to check the damping of the oscillations, it is confirmed that 82 Hz is enough, so that chopping dynamic forces will not be amplified by the optical tube behavior. In the fig. 3.20 - 3.22 mean displacements are plotted for some of them, and the frequency response in the most significant range of the first two modes (50-200 Hz) is presented. Where not specified, mean amplitudes (USUM) are in microns. Plots in fig. 3.24 show that peak value for amplitude equal to $4.8 \cdot 10^{-2}$ mm is reached in the first mode, and it reduces significantly to $2.5 \cdot 10^{-3}$ for the second mode.

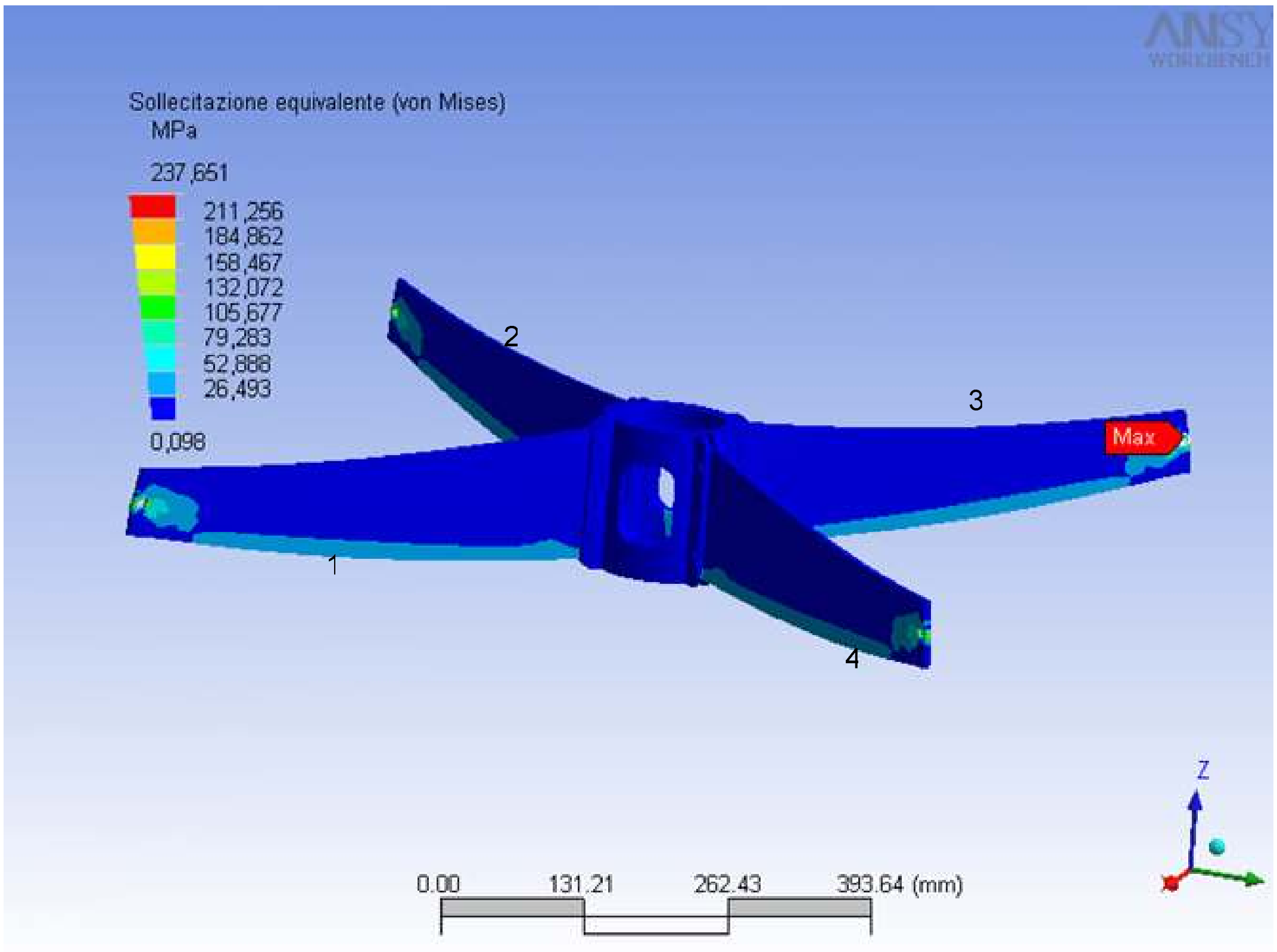

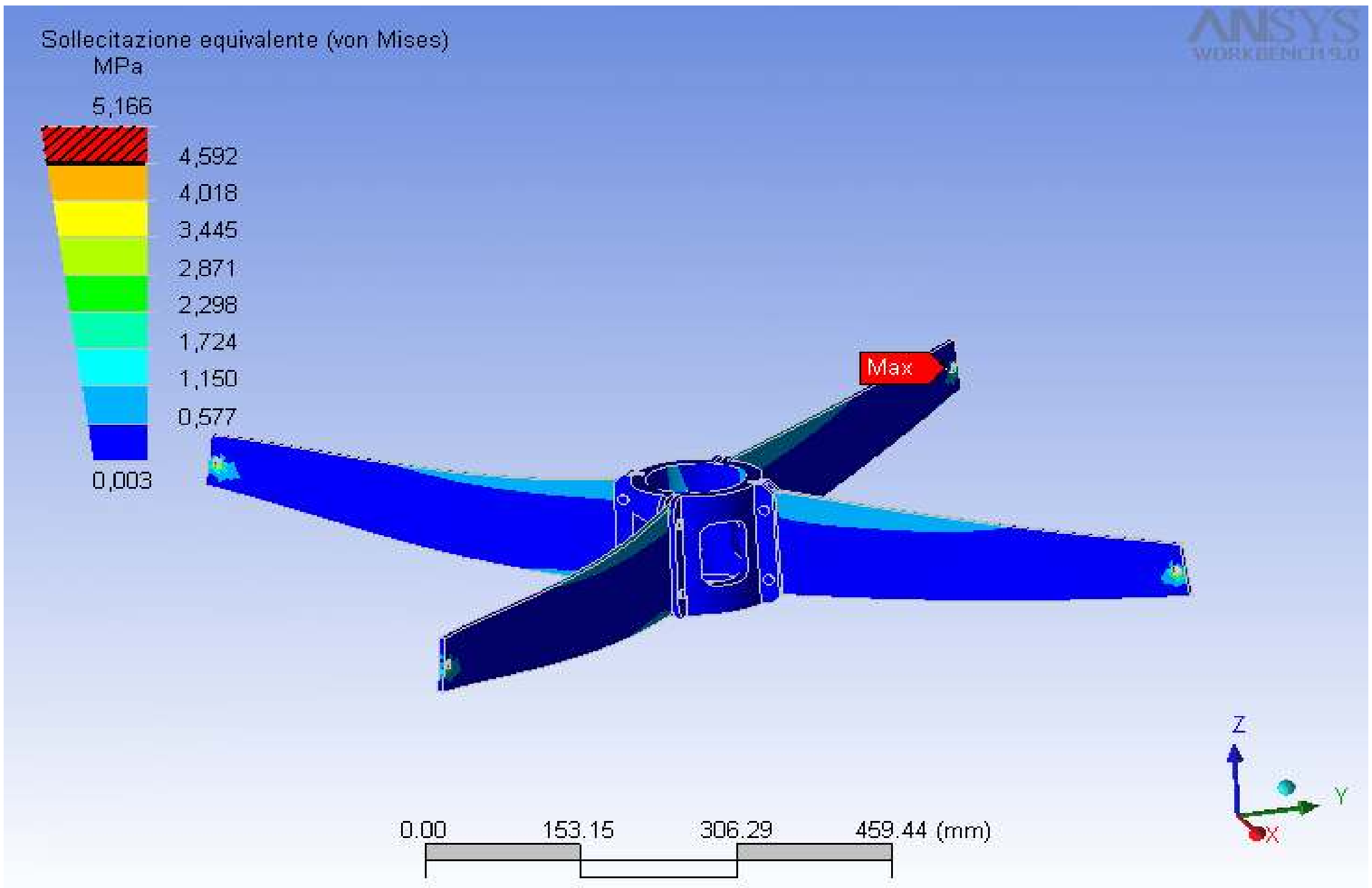

**Figure 3.19** An overview of Von Mises stress distributed on the surfaces under two load cases. The amount of stress in case of constant temperature applied (with no gradient) is noticeably lower.

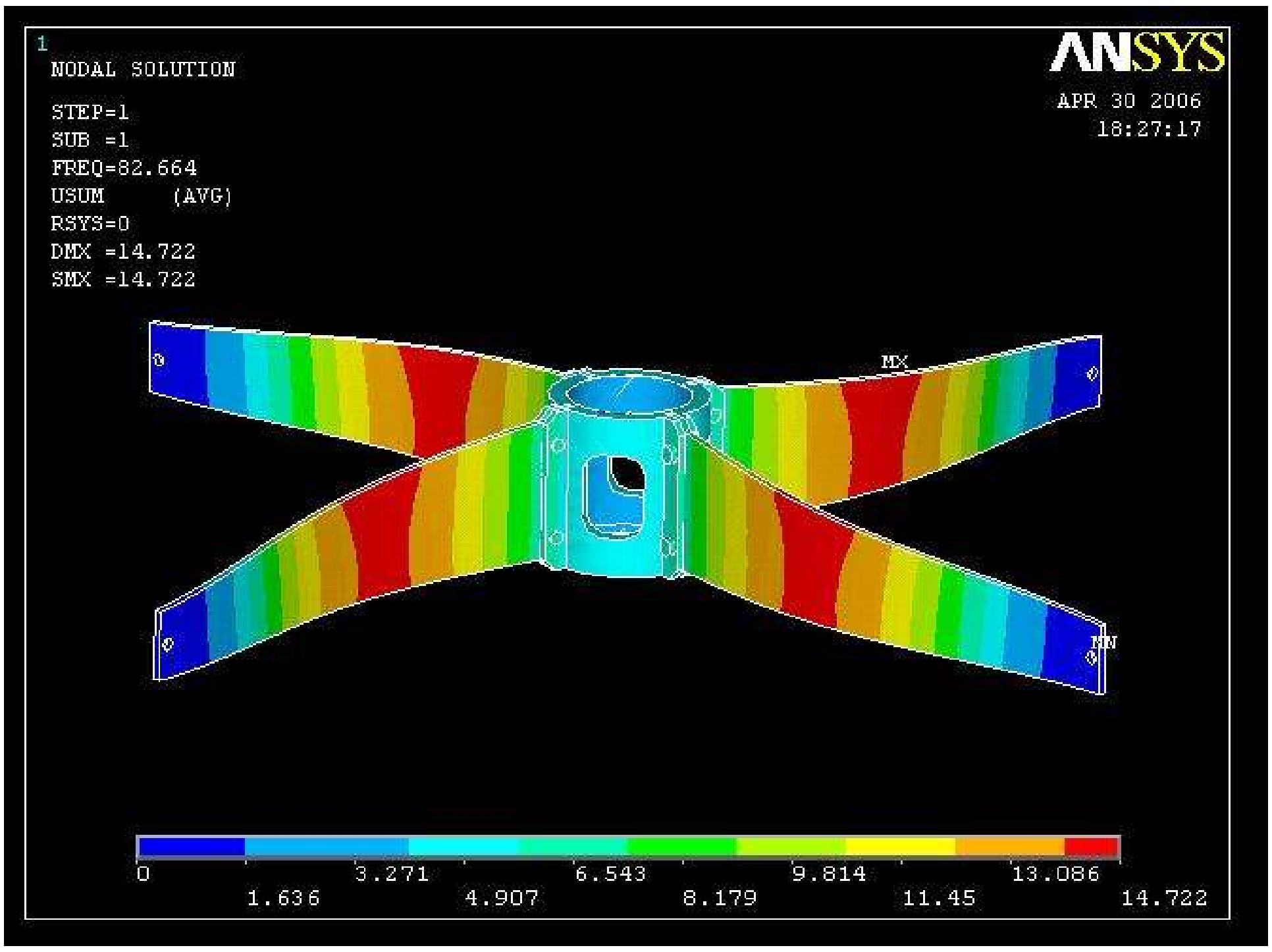

**Figure 3.20** Average amplitude values for the first vibration mode.

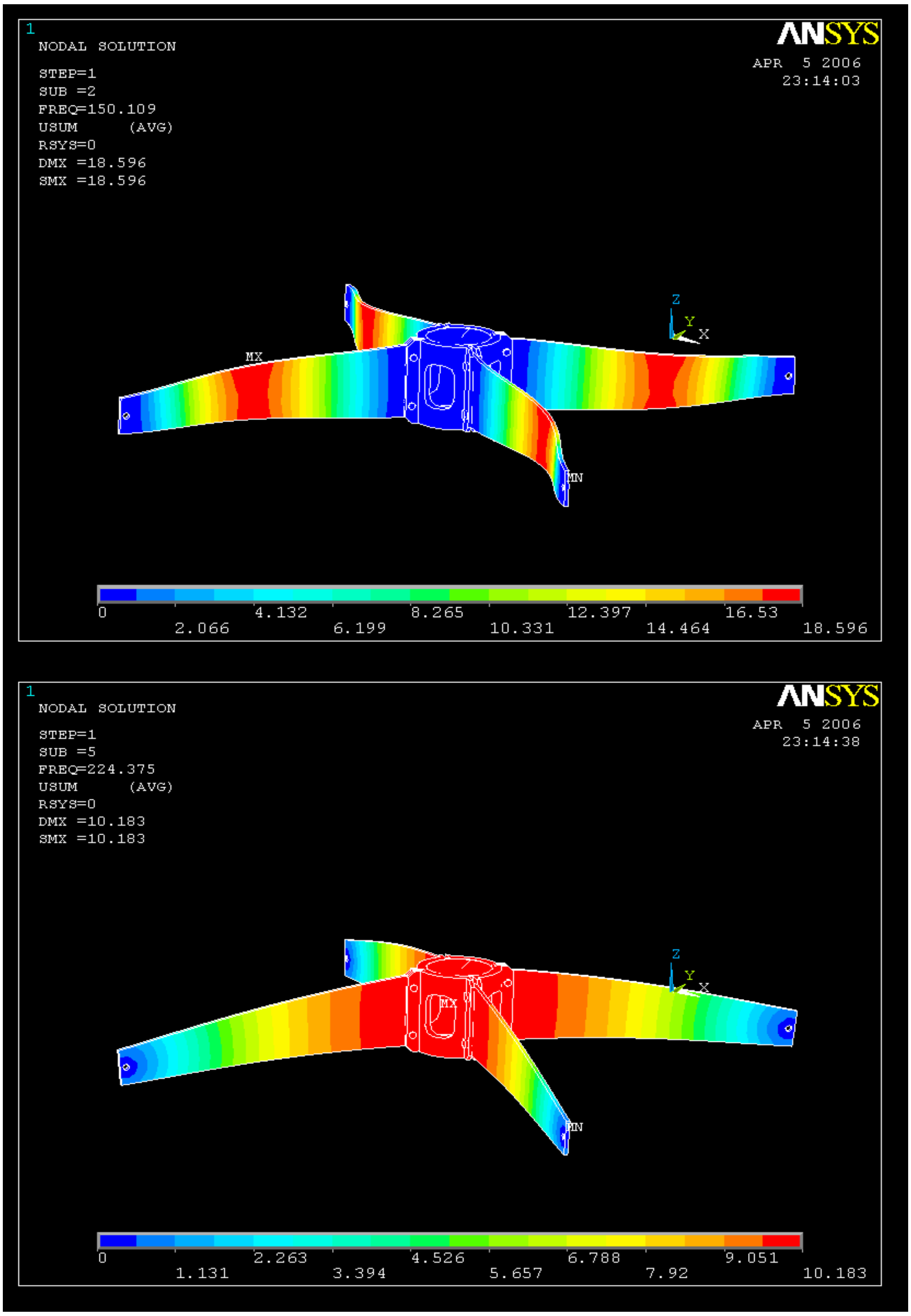

**Figure 3.21** Average amplitude values for the second and fifth mode.

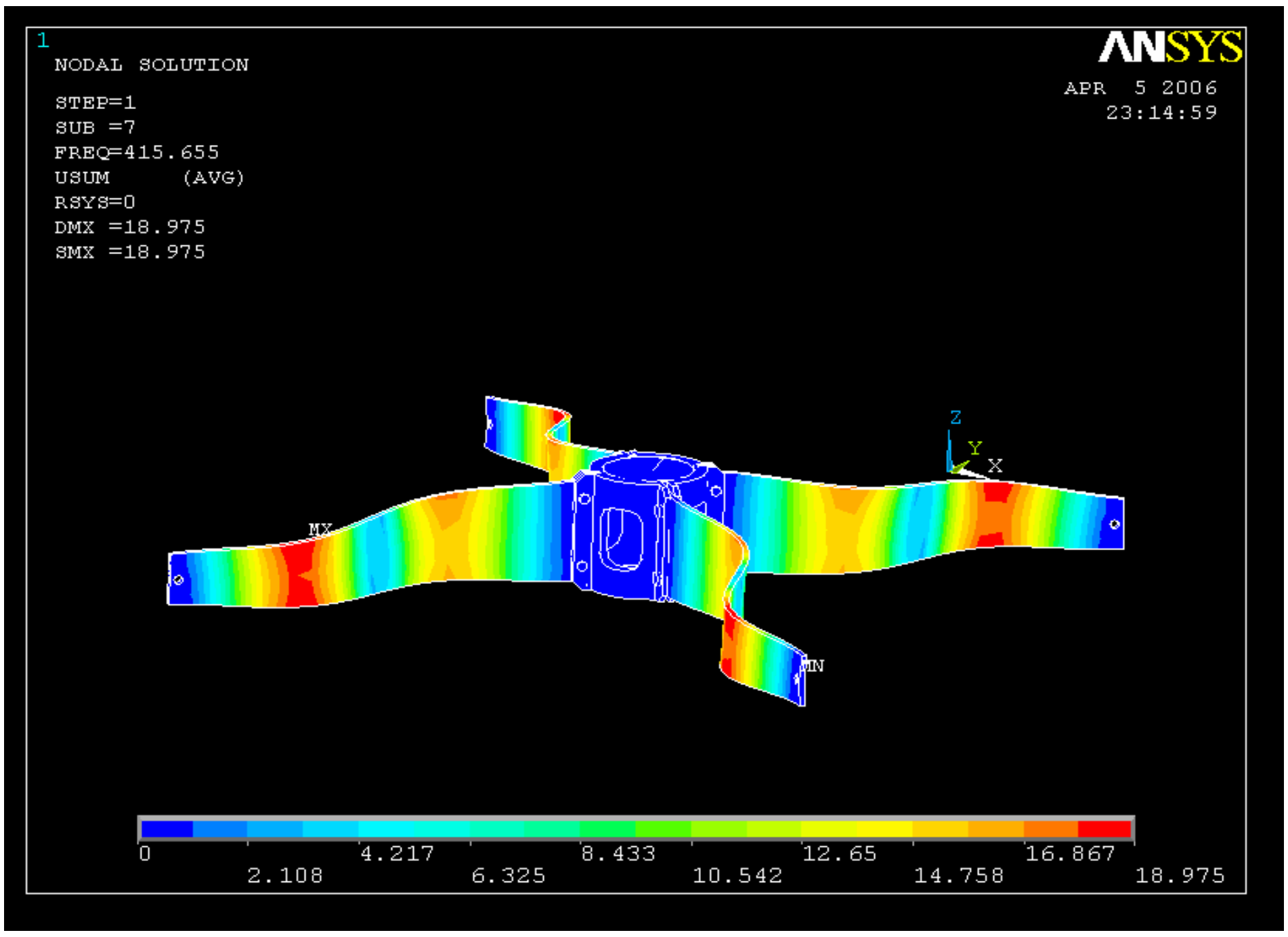

**Figure 3.22** Displacement for the 7th vibration mode.

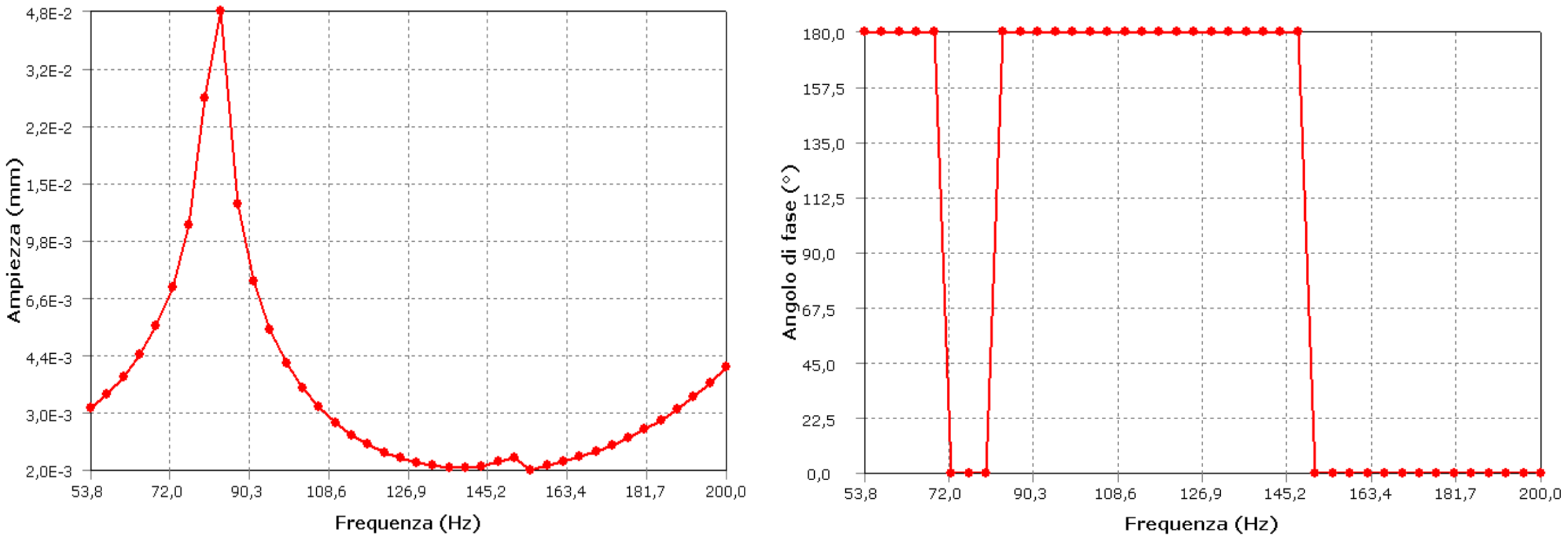

**Figure 3.23 Diagrams of amplitude and phase in the direction of maximum displacements.**

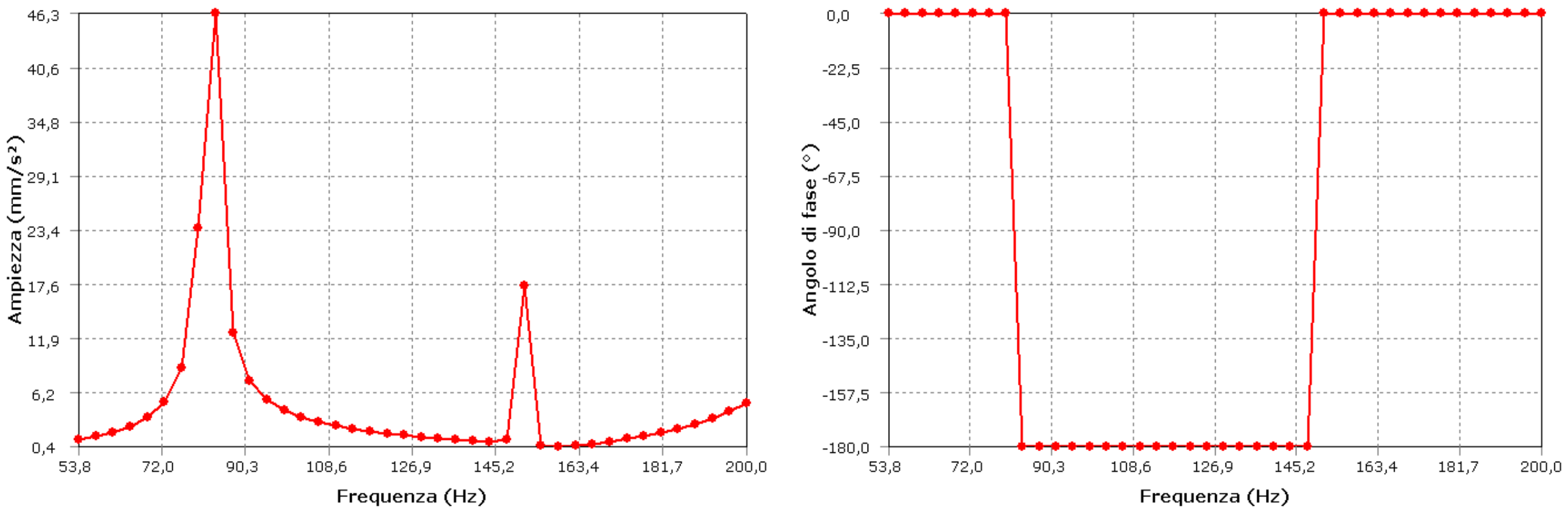

**Figure 3.24** Diagram of amplitude and phase in the direction of maximum acceleration (forces).

### 3.6.4 Optical tube analysis: second formulation

A static analysis of the real optical tube geometry has been lead under ANSYS, in order to determine the stress state and the reaction forces on the altitude axles, or in other terms the axial and radial loads which are effectively transmitted to the roller bearings. We have focused attention on a single case including two thermal loads of constant temperature through the beam sections, gravity, wind static action, a pressure load due to M1 weight, and a distant force representing the weight of optical tube electronic box. It was assigned a load of -40 ° C to one side, and another of -20 ° C to the opposite.

An external restraint to all 6 degrees of freedom was applied on semi-axle of the second Nasmyth focus, and a restrain to 5 DOF, leaving free only sliding along altitude axis, was applied to the other axle. Load conditions and restraints applied are summarized in table 3-3.

| **Load** | **Type** | **Magnitude** | **Direction** |
|---|---|---|---|
| Gravity | Force | 8501.2 N | -Z |
| wind | Force | 58.5 N<br>$1.6289 \cdot 10^{-5}$ MPa | Y |
| M1 weight | pressure | 2000 $N/m^2$ | -Z |
| Box weight | Moment | Mx=800Nm,<br>My=25Nm | Point of application: face 1 |
| Temperature | -20 °C | -- | Applied to 30 faces on AMICA side |
| Temperature | -40 °C | -- | Applied to 30 faces on 2nd Nasmyth f. side |

**Table 3-4** Loads classification. X is directed along elevation axis, Z is vertical positive upwards, Y the remaining axis.

### 3.6.5 Comparison between analytic approach and numerical results

The first Nasmyth focus semi-shaft can be modelled with a cantilever beam with a lumped force at one end. It is half the proper weight of OT; within the hypothesis it is equally distributed between the two arms. It is also considered an horizontal contribution of tension due to thermal load. In fact, leaving a degree of freedom along X on the opposite shaft, the effect of the gradient is that of pulling the shaft. The component of this type of load is predominant, as discussed before. The flexure moment $M_y$ is due to the vertical force. The shaft consists of three section beams, which are hollow tubes with an internal diameter of 86 mm, and the outer one respectively of 160, 180 and 220 mm. The reason of this variation is to create appropriate housings to fix the internal ring of the bearings. Joint reactions are determined by solving the equilibrium equations in the XZ plane:

$$\begin{cases} M_y = Fl = 1147662 \ Nmm \\ X = EA\alpha\Delta T = 720588.36 \ N \\ Z = F = 4250.6 \ N \end{cases}$$

| **Sections** | **A1** | **A2** | **A3** |
|---|---|---|---|
| $D_i$ [mm] | 86 | 86 | 86 |
| $D_o$ [mm] | 160 | 180 | 220 |
| Area [$mm^2$] | 14297.38 | 19638.1 | 32204.46 |
| Mom of Inertia Ix=Iy [$cm^4$] | 2948.4788 | 4884.4853 | 11230.5025 |

**Table 3-5** Section properties.

Von Mises stress for the most critical section, A1, is the following:

$$\sigma_{eq} = \sqrt{\left(\frac{X}{A_1}\right)^2 + \left(\frac{M}{W_1}\right)^2 + 3 * \frac{4}{3}\left(\frac{Z}{A_1}\right)} = 50.404 \text{ N/mm}^2.$$

Where $W$ stands for resistance module: $\dfrac{\pi(D_0^4 - D_i^4)}{32 D_0}$.

In reality, joint reactions retrieved by the finite elements model are quite different, because of the contact surfaces among the beams and between the shafts and the plates of the primary mirror (M1)cell.

Here are presented reaction forces, a plot of Von Mises stress and the deformations. With the same procedure, noticed in the previous paragraph, the flexure error was also determined.

Maximum stress exerted on the peripheral surface is 52.517 MPa (see fig.4.9), almost the same as in the analytical model; anyway it is not constant through the section since there’s a buckling effect on the hollow shaft: it behaves like a thick tube subject to external pressure.

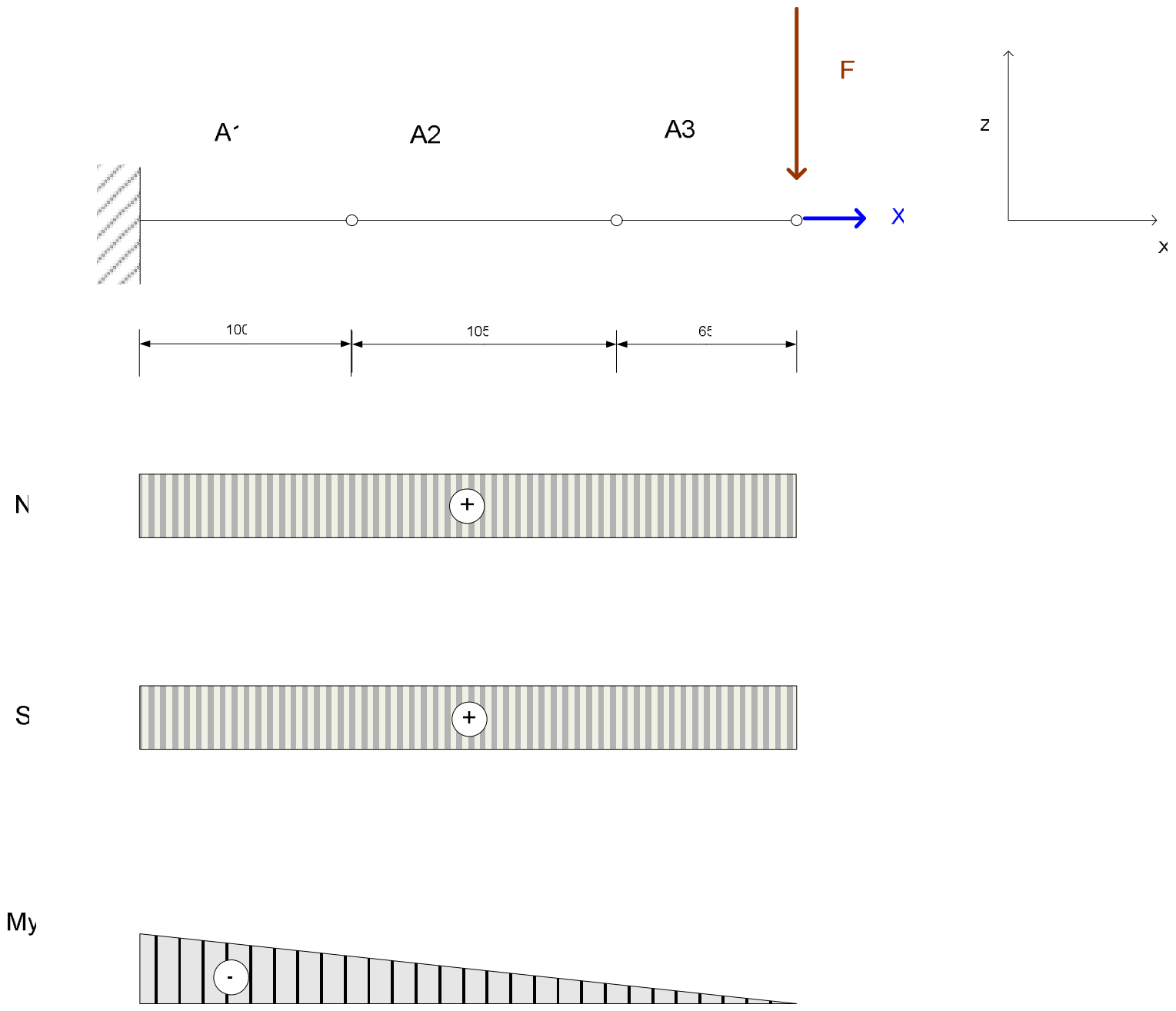

**Figure 3.25** Cantilever beam model for the 1st Nasmyth focus axle.

| axle | $F_x$ [N] | $F_y$ [N] | $F_z$ [N] | $M_x$ [Nmm] | $M_y$ [Nmm] | $M_z$ [Nmm] |
|---|---|---|---|---|---|---|
| 1 | 0 | 19,609 | 4505,63 | -1574,4 | -576590 | -26774 |
| 2 | -0,22652 | -37,807 | 3995,56 | $-7{,}9504*10^5$ | 134670 | -11083 |

**Table 3-6 Joint reaction forces and moments.**

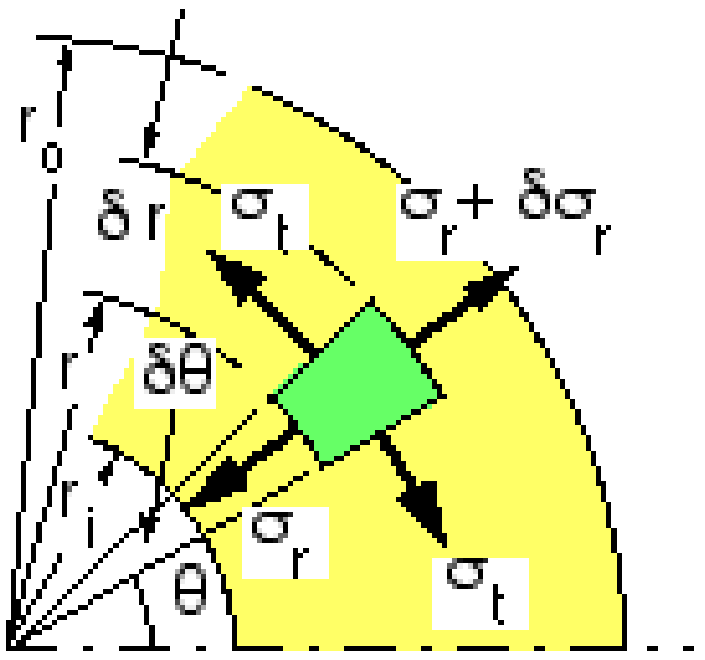

**Figure 3.26** A representation of tangential and radial stresses acting on an element belonging to the section of a thick cylinder.

Two equilibrium equations can be written for an infinitesimal element in the radial and tangential directions:

$$(\sigma_r + \delta\sigma_r)(r + \delta r)\delta\theta - \sigma_r r\delta\theta - 2\sigma_\theta \delta r \tan\left(\frac{\delta\theta}{2}\right) = 0$$

Neglecting the terms of greater order, we obtain the Clapeyron (1850) equation:

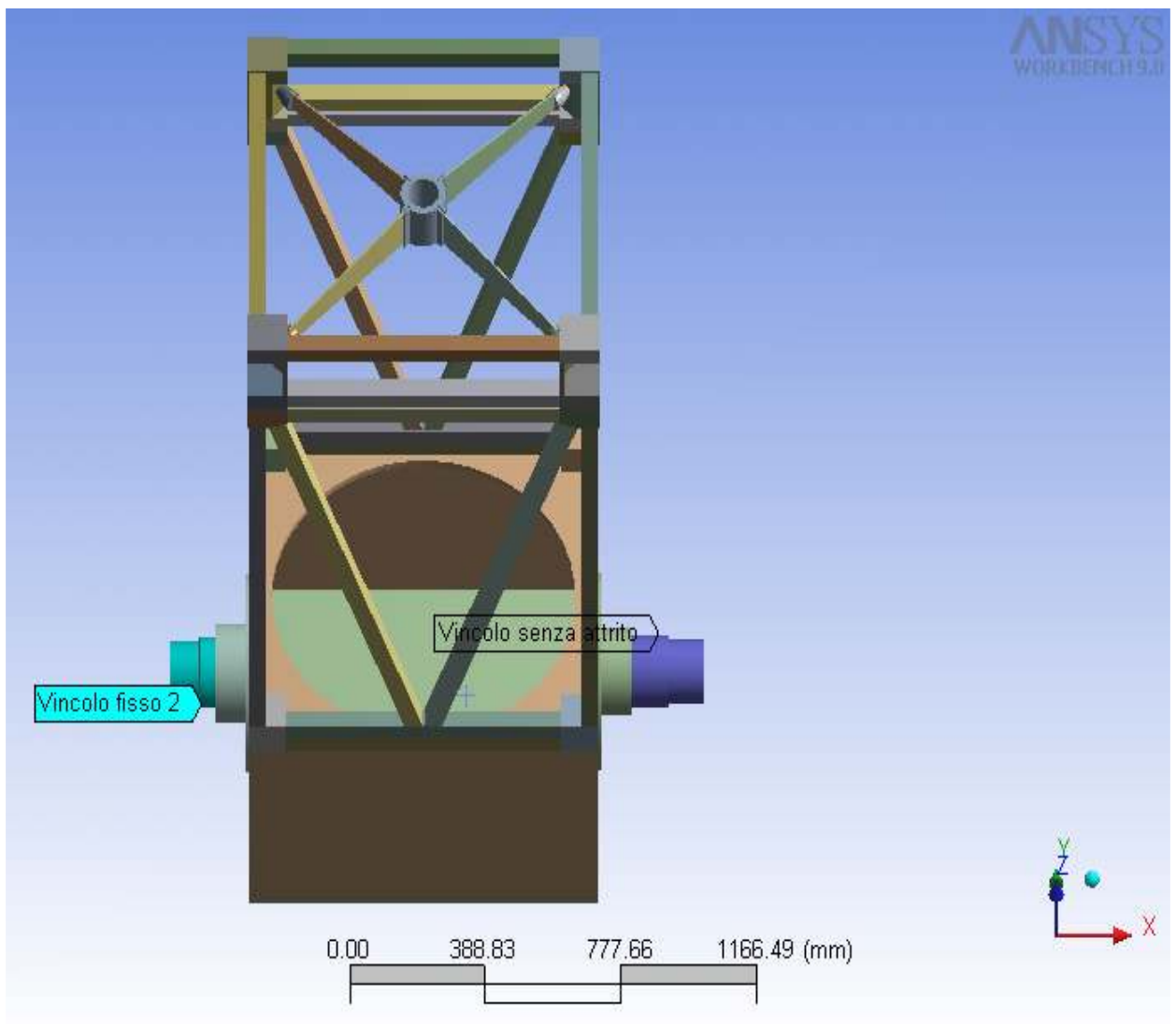

**Figure 3.27** A picture with joint constraints highlighted.

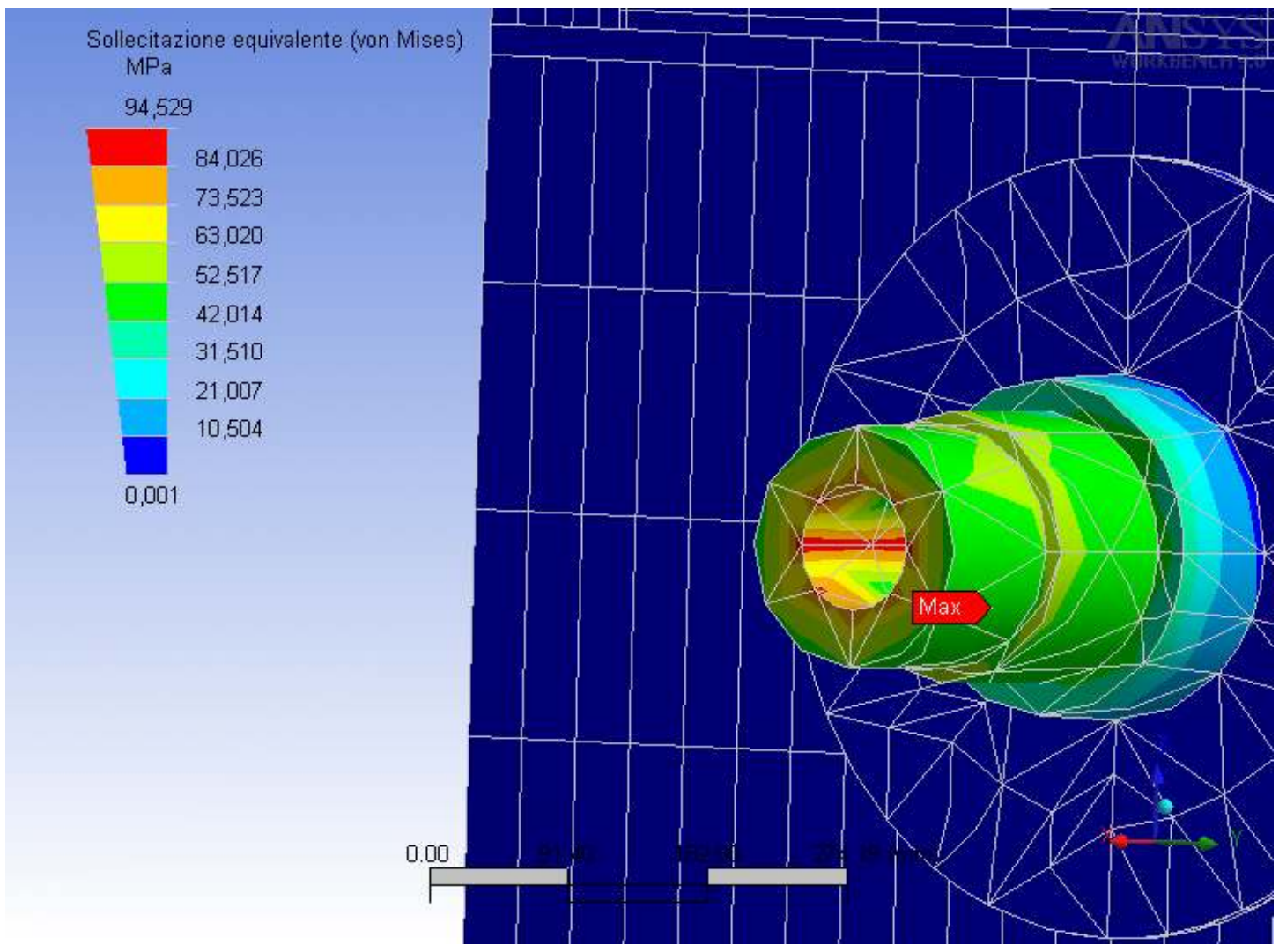

Figure 3.28 Von Mises stress plot.

$$r\frac{d\sigma_r}{dr} = \sigma_\theta - \sigma_r$$

Within the hypothesis of plane strain ($\varepsilon_z = 0$), stress-strain relationships are valid:

$$\begin{cases} \sigma_r = \dfrac{E}{1-\nu^2}(\varepsilon_r + \nu\varepsilon_\theta) \\ \sigma_\theta = \dfrac{E}{1-\nu^2}(\varepsilon_\theta + \nu\varepsilon_r) \end{cases}$$

Expressing all as a function of displacement *u*, provided that $\varepsilon_r = \dfrac{du}{dr}$, $\varepsilon_\theta = \dfrac{u}{r}$, we finally have:

$$\begin{cases} \sigma_r = -A + \dfrac{B}{r^2} \\ \sigma_\theta = -A - \dfrac{B}{r^2} \end{cases} \quad \text{with} \quad A = \frac{r_0^2}{r_0^2 - r_i^2} p_e, \quad B = \frac{r_0^2 r_i^2}{r_0^2 - r_i^2} p_e,$$

Where the initial conditions applied are: $p_e \neq 0$, $p_i = 0$.

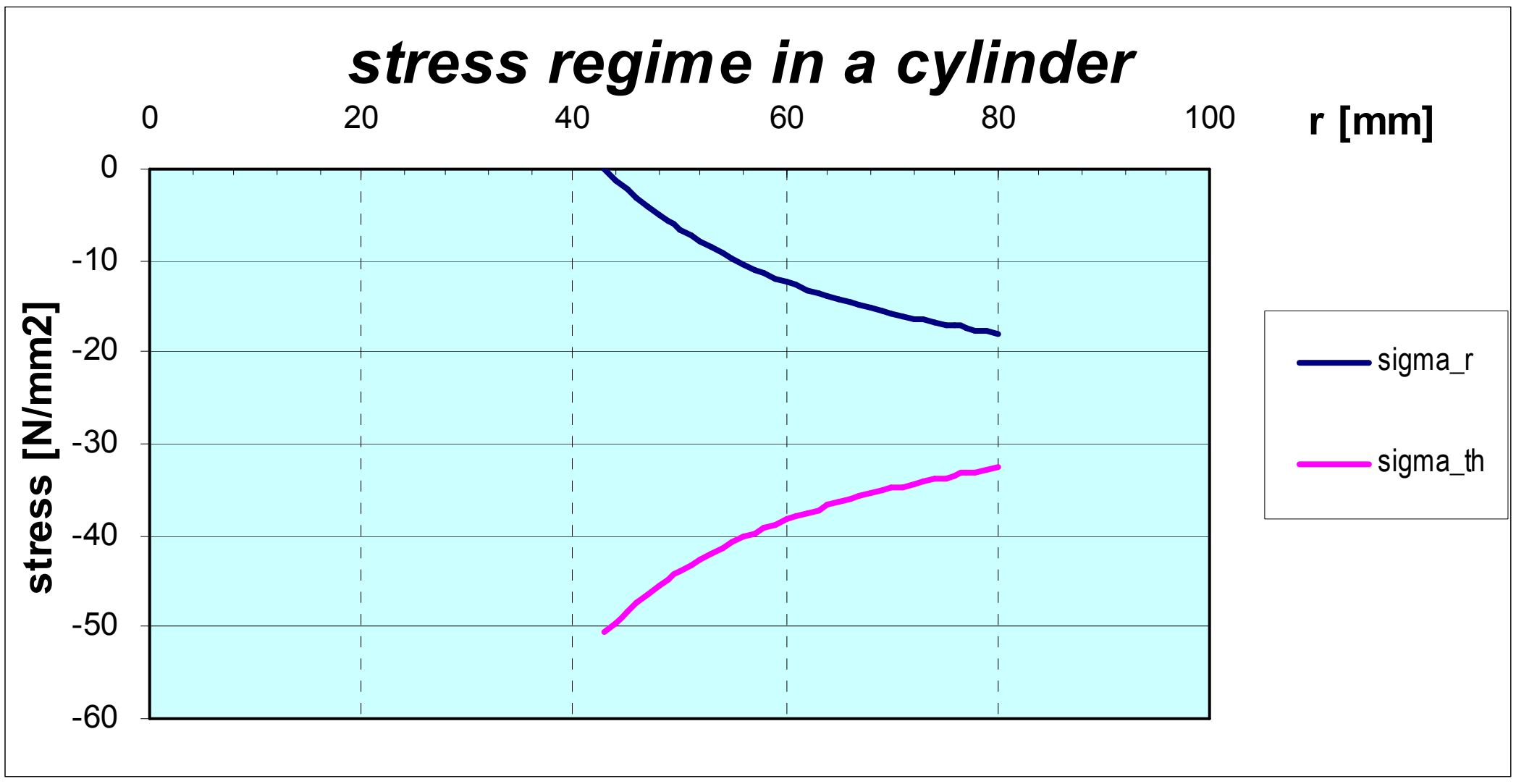

**Figure 3.29** Plot of the radial and tangential stress components vs radius. Negative values indicate compression status.

The compression load in this case is acting as an external pressure of $p_e$=18 MPa.

As far as the displacements and the flexure errors are concerned, the most critical is the Z component, yet in the vertical position, as it is shown on table 3-7. For this reason it will be a primary task to thermally regulate the telescope coating it with a special reflective varnish, or applying Mylar shields, for example.

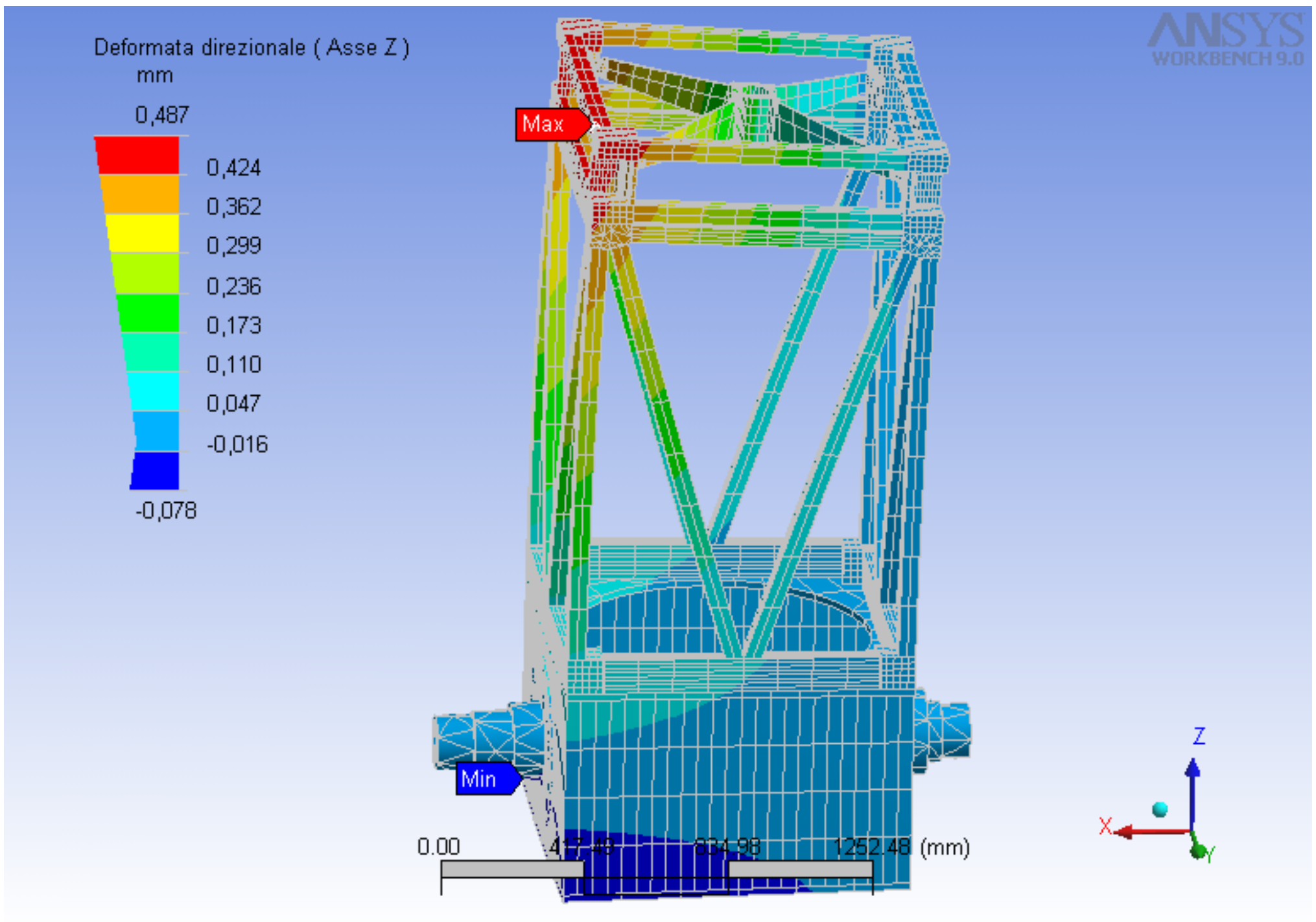

**Figure 3.30** Contour plot of deformation along Z axis.

| DISPLACEMENTS | VALUE | Node number |
|---|---|---|
| Δx MAX | -0.439 | 15077 for M2, 49190 for M1 |
| Δy MAX | -0.0204 | “ “ |
| Δz MAX | 0.487 | 22734, 24092 for the ends of upper truss |
| flexure error 1 (Δx/d) | 45.16 arcsec | |
| flexure error 2 (Δy/d) | 2.1 arcsec | |

**Table 3-7 Estimated flexure errors.**

### 3.6.6 Spur gear sector for altitude motion

The sector of a spur gear whose diametral pitch is 1260 mm with two counter- rotating pinions, made of steel 18 Cr Ni 8, provide the motion on elevation axis. It has milled holes to reduce the weight, and an aperture angle of 145°. It has 145 teeth and drive train is the same as on azimuth. The main characteristics of the gear are indicated in table 3.1. The sector is rigidly bolted on the semiaxle flange on second Nasmyth by means of 16 screws M14x2.

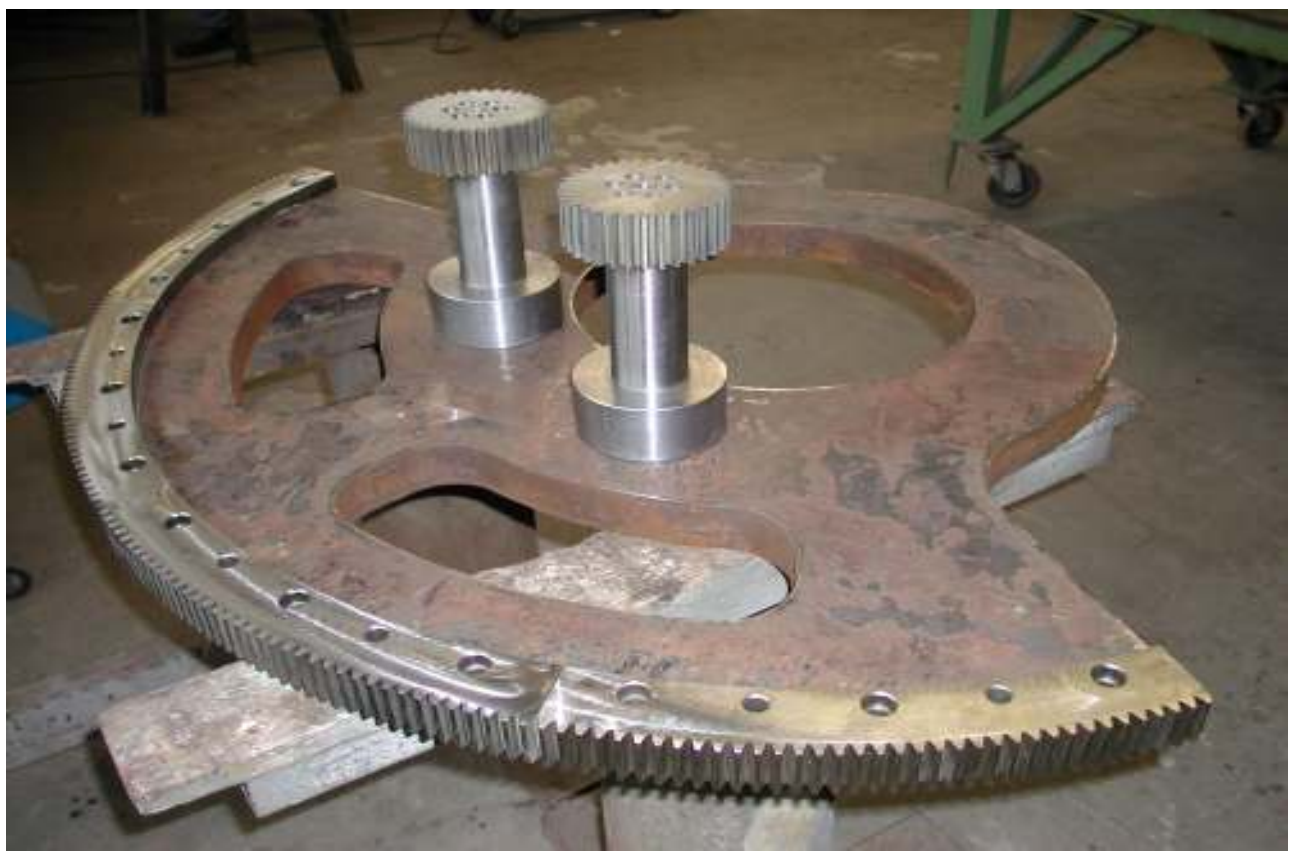

**Figure 3.31** A front view with principal dimensions of the spur gear sector.

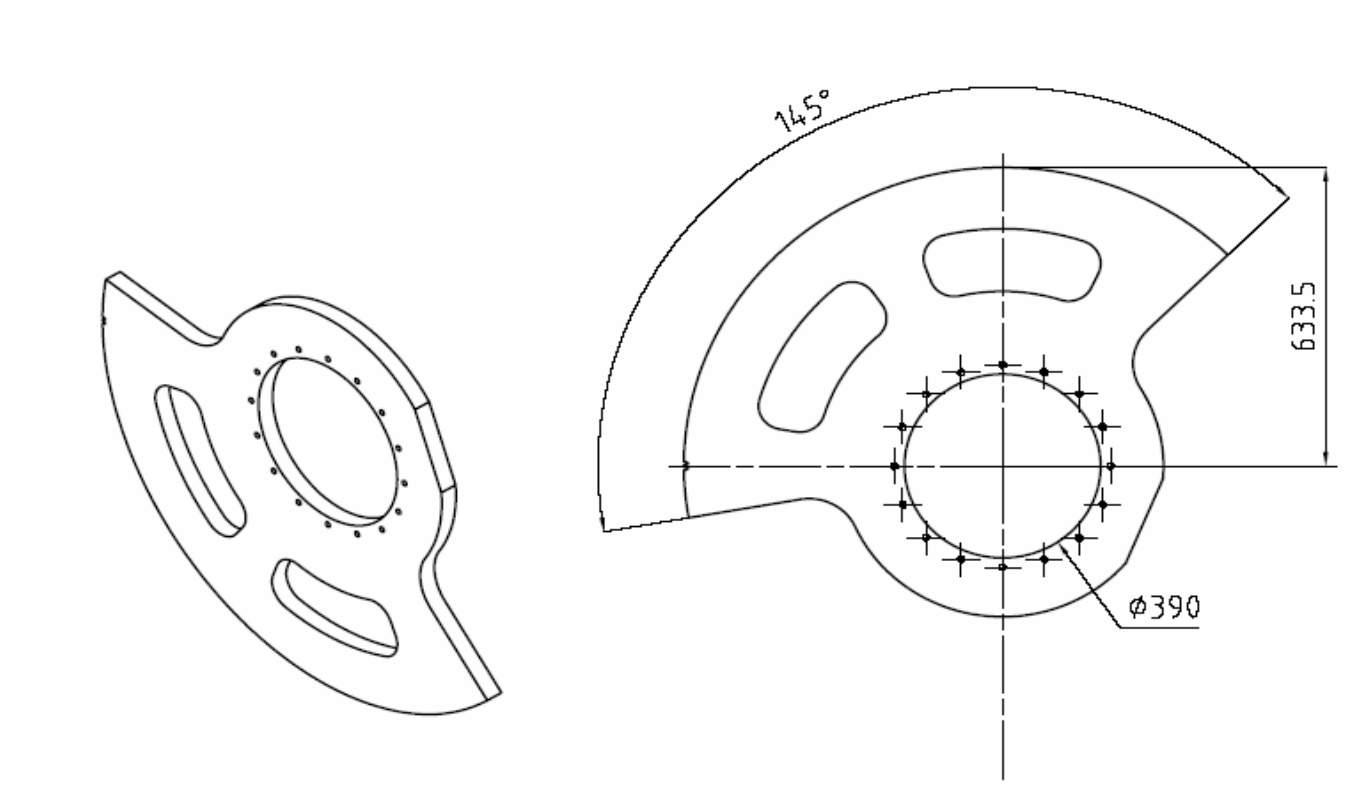

| | |
|---|---|
| Weight [kg] | 154.7 |
| Pressure angle | 20° |
| Total teeth number | 145 |
| Module | 3.5 |
| Width [mm] | 35 |
| Pinion Teeth number | 32 |
| Diametral pitch [mm] | 1260 |
| Pinion Diametral pitch [mm] | 112 |

**Table 3-8** Design parameters for toothed sector.

## 3.7 Fork structure

The fork subsystem can be considered made of four main parts: the two arms of the two foci, a platform with H profilate bars 230x240 on the perimeter where the arms are mounted, a basic frame with an internal hole of 800 mm, which is bolted on the internal ring of azimuth bearing. Small plates 10 mm thick are welded at the corners and have the function to align, in a correct way, the arms, once installed at Dome C. In fact, provided that the telescope assembled such as operating do not fit the internal size of the modified ISO 20, it was thought to assemble part of it after transportation. In other words, an appropriate housing on the base chassis has been designed, dedicated to the two arms together with the optical tube. Separately, at the centre of the chassis, the basic frame is bolted on the azimuth bearing.
The basic frame of the arms in turn comprises:

- a flange for fastening and calibrating the azimuth motion assembly;
- connecting elements for the support of azimuth motion assembly;
- a protection box for azimuth motion assembly;
- a mechanical interface for the inner ring of the cross-roller bearing, with eight welded stiffening ribs;
- a mechanical interface for the electrical box of the fork;
- a mechanical interface for counterweights.

The above mentioned corner plates are welded on the platform and have three holes M30, arranged at right angle, to fasten the bolts supporting the upper structure.

These holes must be executed after welding, taking into account residual stresses after manufacturing process. The basic frames represent a mechanical linkage among the arms and cross roller bearing. We needed to introduce the large pipe of the external diameter equal to internal ring of the bearing, in order to create a compartment to mount the slip ring and the absolute azimuth encoder. The larger size of the slip ring is the length, equal to 559.30 mm. Contact surfaces between external ring of azimuth bearing and fork base plate have to be manufactured with tolerances specified by RKS.

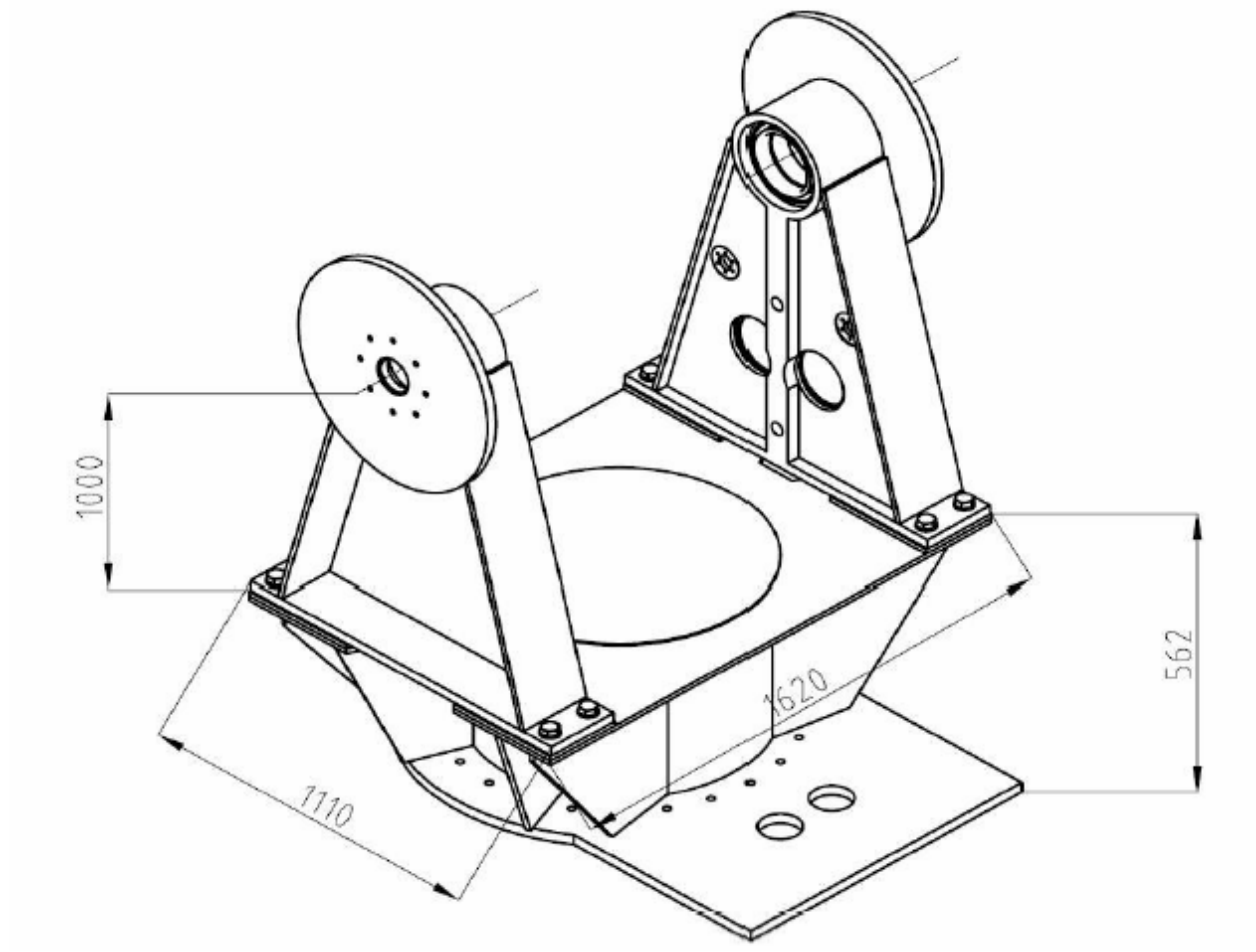

**Figure 3.32** An axonometric view of the fork.

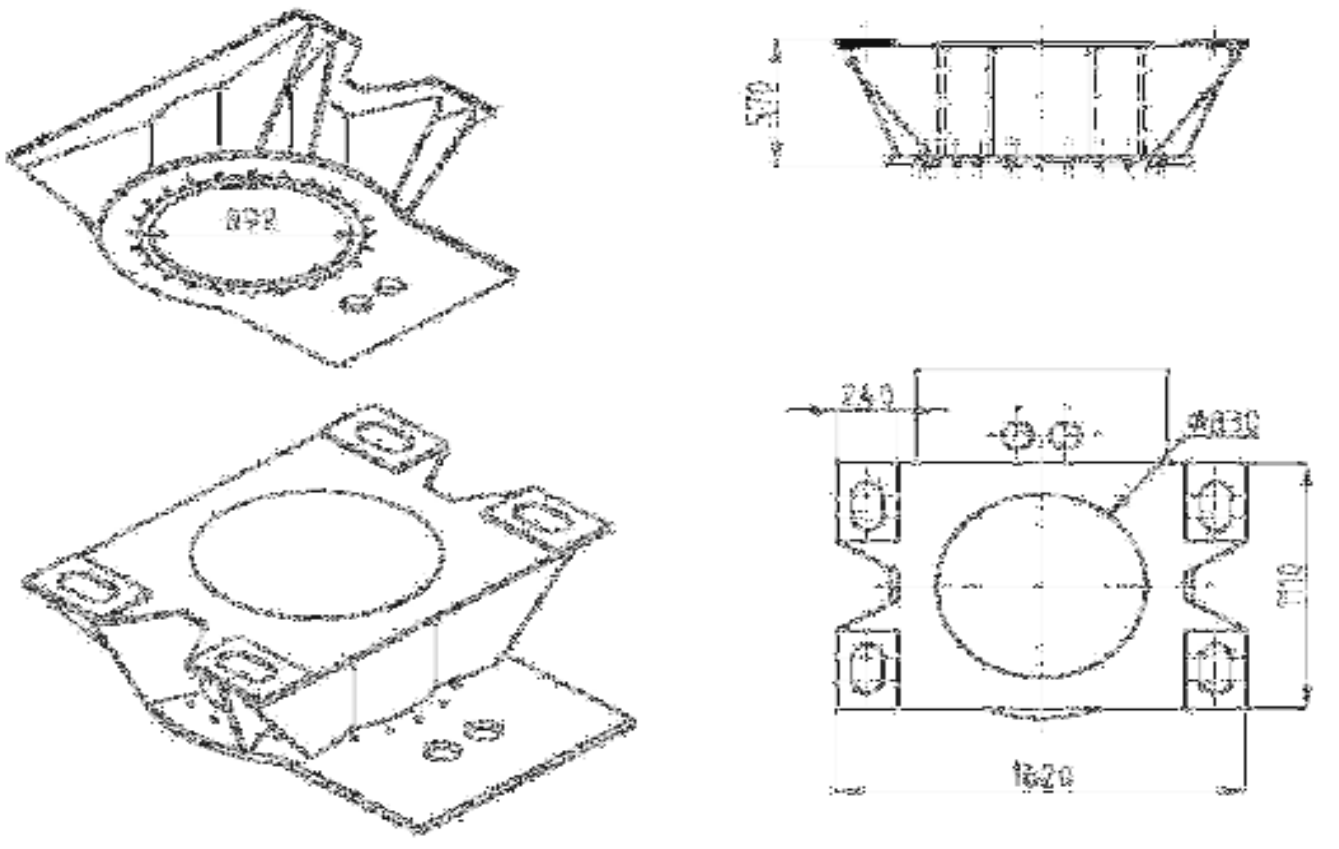

**Figure 3.33** Axonometric, front and top view of the mechanical interface between the fork and the azimuth bearing with overall dimensions indicated.

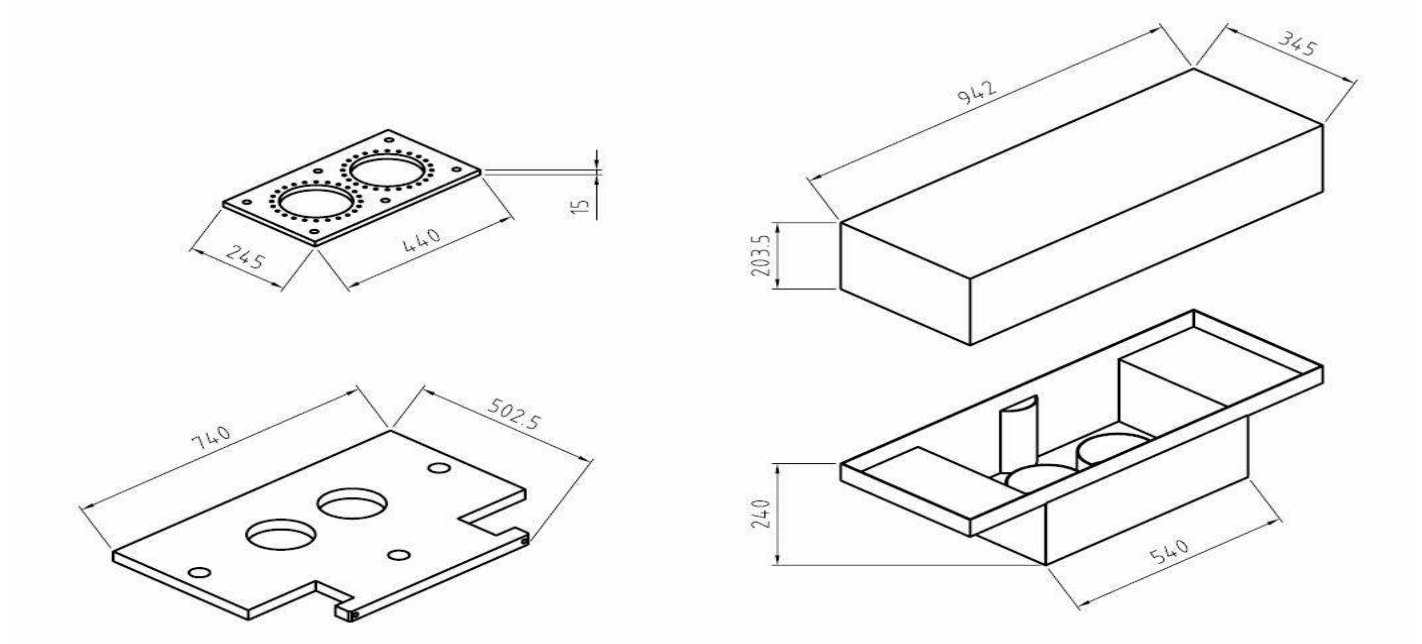

**Figure 3.34** **I**nsulating motor boxes and views of connection plates between pinions and boxes for both azimuth and altitude axes.

### 3.7.1 Fork arm of the first Nasmyth focus

It includes:

- an external ring seat for bearings ;
- races for seals;
- interface flange with AMICA system;
- a drilled base plate used to fasten and set the correct alignment;

It is realized with metal sheets respectively of 15, 20 and 30 mm thickness, with angle weldings. Milled bores on side and bottom plates are present to allow the cables passage. There's a carter plate to reinforce the frame moreover. Four M16 holes on the external face of upper housing are provided for the interface flange with the camera.

The bores on the top of the housing have the function to fasten a cable carrier for the connections relative to the optical tube.

The mounting seats for altitude bearings are finished with tolerances depending on what specified by SKF catalogue for types 32032X ed un 32936, which have to be mounted in the "O" configuration, as indicated by SKF itself. The vertical column is the frame supporting bending moments and bulk loads, and has a rectangular section of 3200 $mm^2$ (40x80mm).

### 3.7.2 Interface flange with AMICA

#### 3.7.2.1 Flange analysis

The main scope of the flange (component n° 44 of the IRAIT drawing) is that of assuring correct mounting of AMICA camera and to provide the alignment of optical and mechanical axes within admissible tolerance.

The state of stresses and strains distributed on the flange[3] has been examined. The convention used for reference system is that indicated in figure 3.35: X is vertical axis, Z direction is parallel to optical axis and Y the remaining axis.

A vertical force of -5000 N is applied to the point of coordinates (0, 0, 1000), assuming that the origin coincides with the centre of the hole corresponding to the entrance window. A set of 8 holes M14 at a radius of 120 mm from the centre are employed to bolt the cryostat. The thickness of the flange is 26.6 mm. This static load simulates the overall weight of AMICA rack with upper and lower cabinet and the mechanical interface necessary to mount it.

[3] See Appendix for techinical drawing of the flange.

Four holes M16 used to fix the flange on the bearing housing are chosen as constraints. The only degree of freedom left for each of them is rotation around Z.

A set of six M14 holes was initially provided to mount the rack, with a wheelbase of 200 mm in X. In this case fastening can be done from behind the flange. Applied strength gives as contributions a torque and bending moment, whose resultants are shown in figure 3.35.

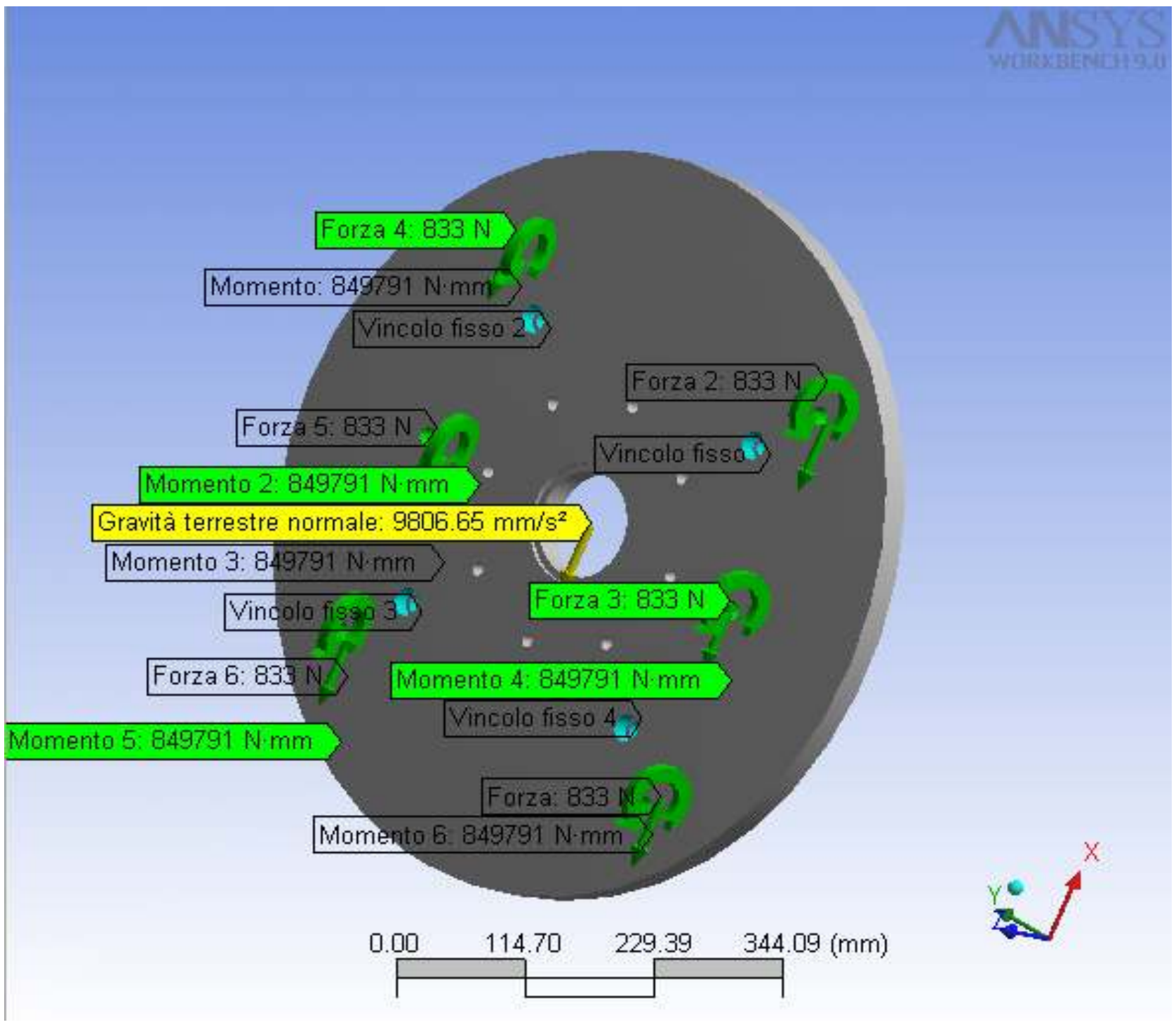

**Figure 3.35** Overview of acting forces and moments.

The maximum stress criterion is satisfied as it is inferior to that admissible ($\sigma_{max}$ = 143 MPa <$\sigma_{adm}$ =250 MPa) (see fig. 3.36). The larger strain is reached on the outer diameter, superior and inferior, and the values in the three directions are respectively in the ranges:

$-0.0141 \leq U_x \leq 0.0143$ *mm*

$-5\cdot10^{-3} \leq U_y \leq 5\cdot10^{-3}$ *mm*

$-0.087 \leq U_z \leq 0.087$*mm*

Maximum mean values of force and moment occur in the hole indicated as 3.

| Constraints | Fx [N] | Fy [N] | Fz [N] | Mx [Nmm] | My [Nmm] | Mz [Nmm] |
|---|---|---|---|---|---|---|
| 1 | 1049.9 | -121.32 | -6166.7 | 59176 | $3.572*10^{-5}$ | -504.59 |
| 2 | 1056 | 105.64 | -6165.4 | -58516 | $3.5698*10^{-5}$ | -310.13 |
| 3 | 1028.3 | -108.22 | 6165.8 | 58735 | $3.5864*10^{-5}$ | -602.64 |
| 4 | 1075.4 | 123.91 | 6166.4 | $3.6122*10^{-5}$ | $3.5631*10^{-5}$ | -20.756 |

**Table 3-9** Joint reactions forces and moments.

Within the above mentioned conditions, there's no sensible deformation on the mounting holes of the camera. Anyway, supposing that it even assumes maximum displacement in Z, the maximum tilt angle in Y is 0.087/200 rad =1.49 arcmin (see fig. 3.38). Therefore it is far lower than 0.5°, which is the largest acceptable optical tolerance.

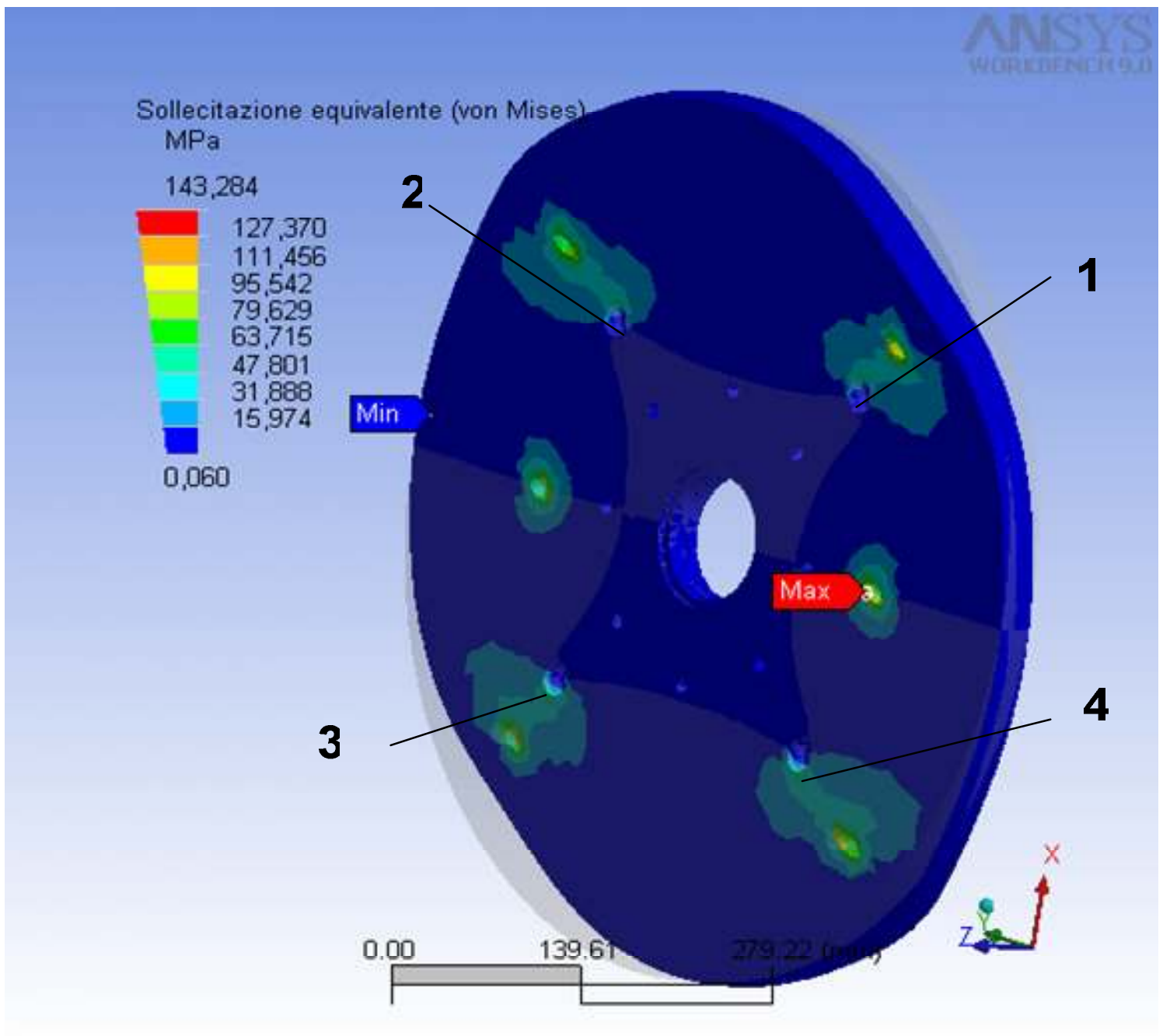

v

**Figure 3.36** Von Mises stress distribution along the flange.

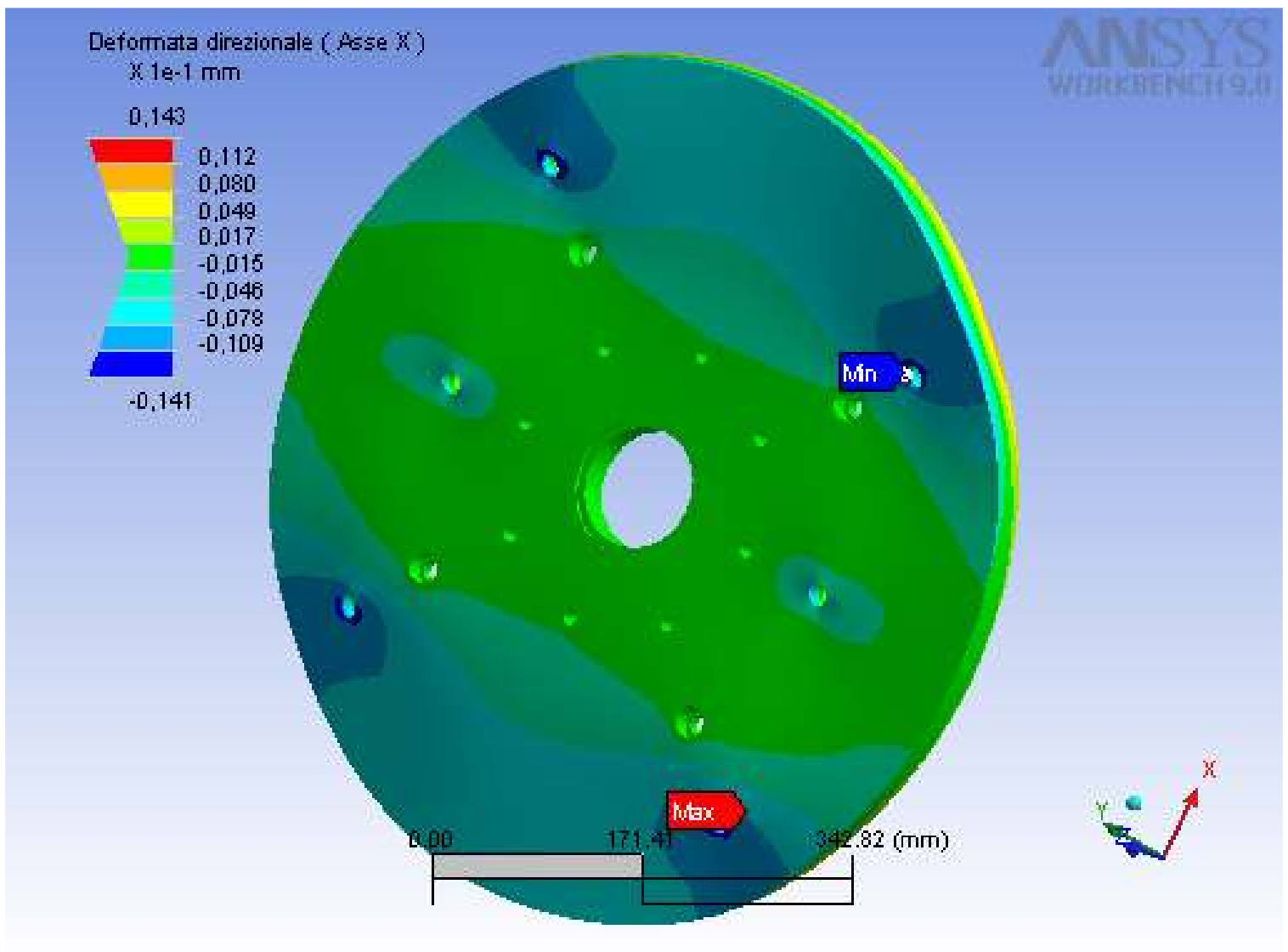

**Figure 3.37** Deformation plot along X.

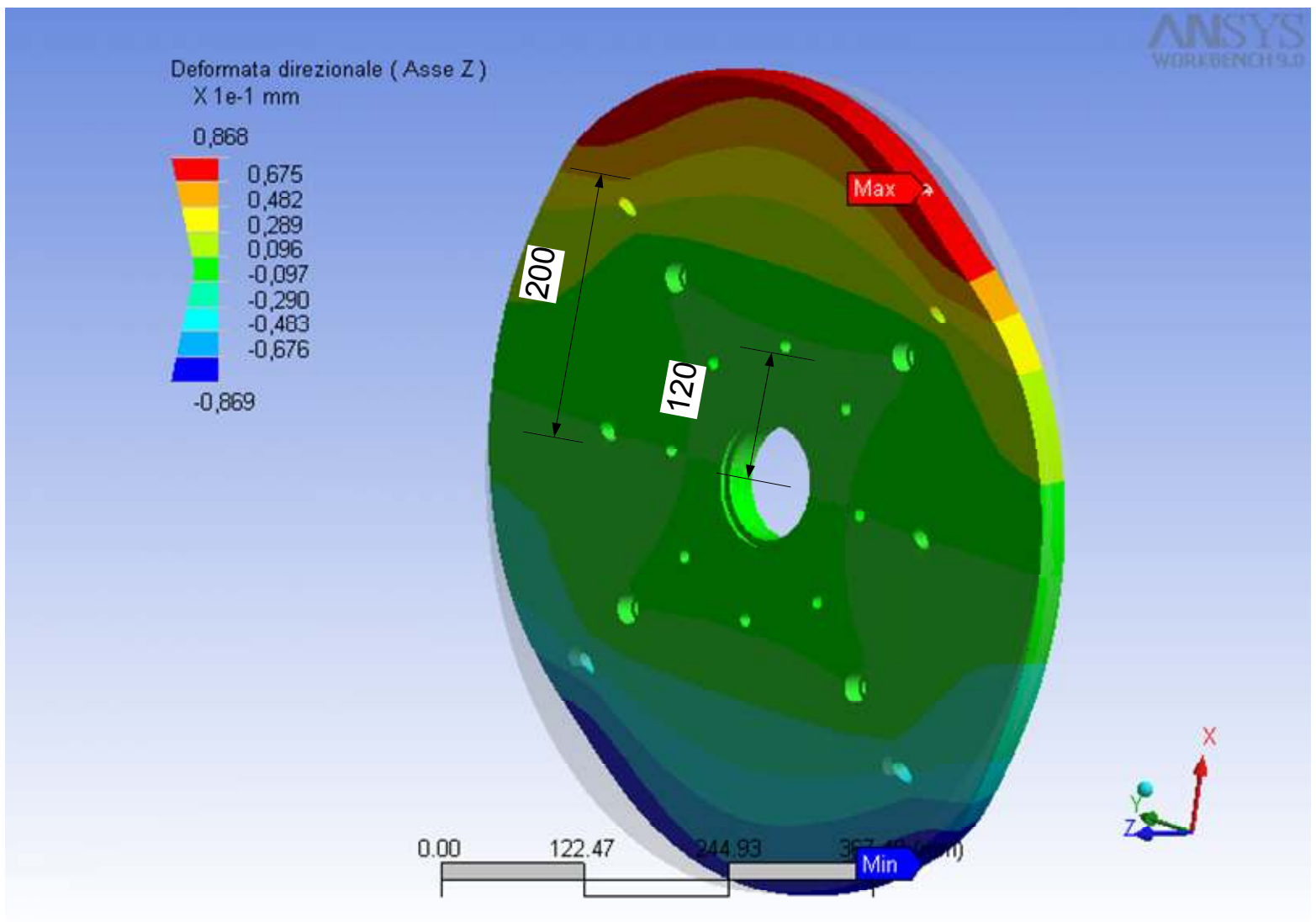

**Figure 3.38** Deformation plot along Z.

### 3.7.2.2 Bolt selection

Here is described the method of dimensioning the M16 as supporting bolts and it's intended to show the possibility of applying the load previously described. Let's assume that the flange is stiffer than the bolts, with negligible strains, so that it rigidly rotates around the centre in Z: the two upper bolts are subject to tension, while the others are subject to compression, and the bending moment $M_y$ gives the contribution of a triangular emisymmetric load.

On the basis of active loads a verification of the largest allowable tensions on employed bolts was made.

The condition to be checked is that global tension force , given by the sum of preload ($N_0$) necessary to fasten the bolt, and the external force, due to bending moment, are lower than 80% of the yield stress of resistance section.

The distance along X between the bolts is d=260 mm, and by the moment equilibrium around Y, we obtain the tension force:

$$F_t = \frac{Pz}{d} = \quad 19.23 \quad kN$$

$$N_0 + F_D \frac{K_b}{K_b + K_p} < 0.8\sigma_R A_R$$ where:

$N_0 = 0.7\sigma_R A_R$ is the preload ;

$K_b = \frac{EA_R}{L}$ and $K_P = \frac{EA_{eq}}{L}$ are respectively the stiffness coefficient of bolt and flange; and $A_R$ ,$A_{eq}$ are resistance and equivalent bolt areas, and they are scheduled in handbooks; L is the length of the threaded rod (81 mm). The condition is fulfilled as it is: 72.22<81.64 kN.

Therefore the bolts selected for the application are four M16x2, of class of resist5ance 8.8, whose feature are illustrated in the next table.

In fig. 3.40 the values of bolt tensions are plotted with the variation of application point of the force on X.

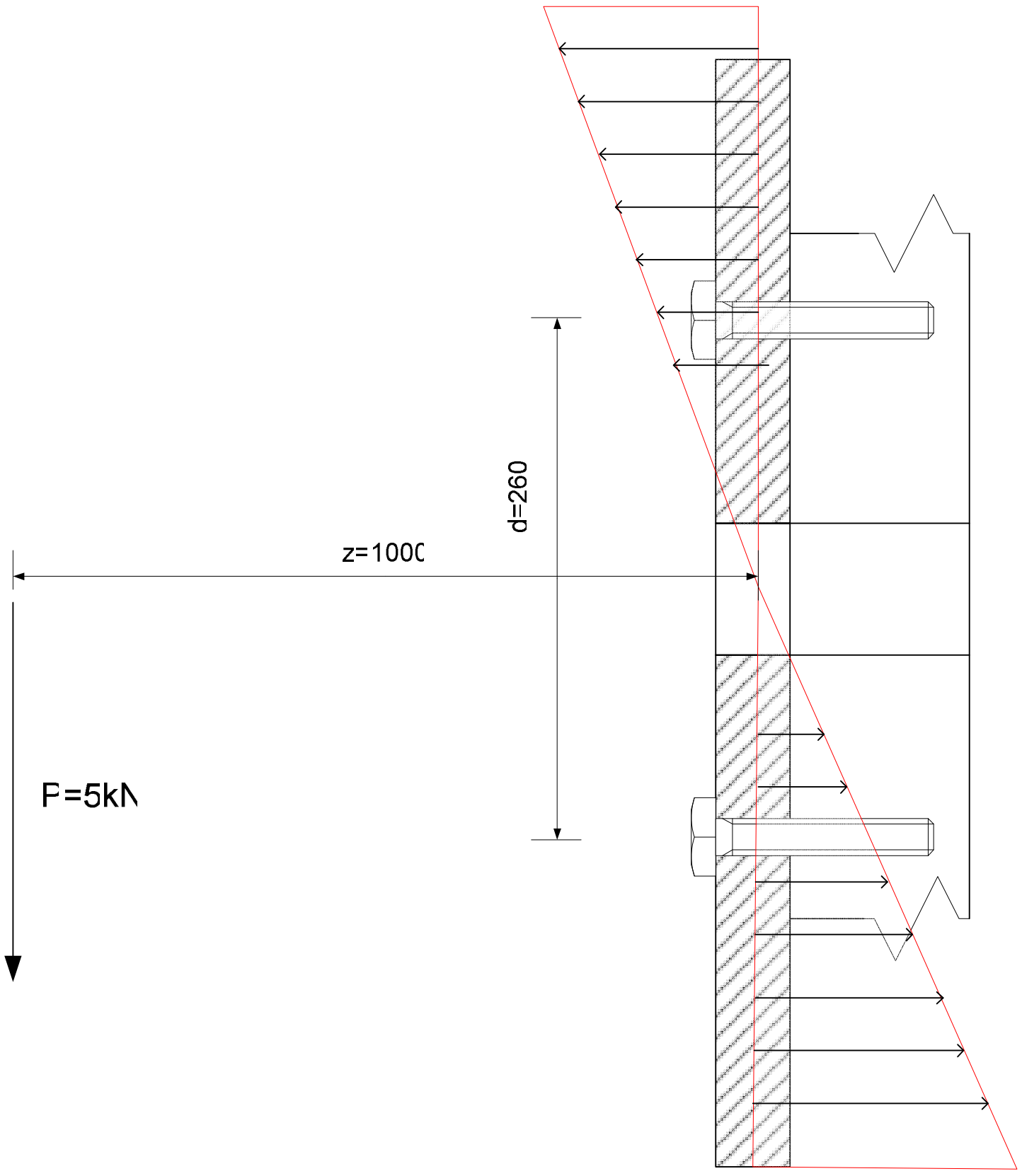

**Figure 3.39** A section view of the state of tension and compression forces on a pair of bolts.

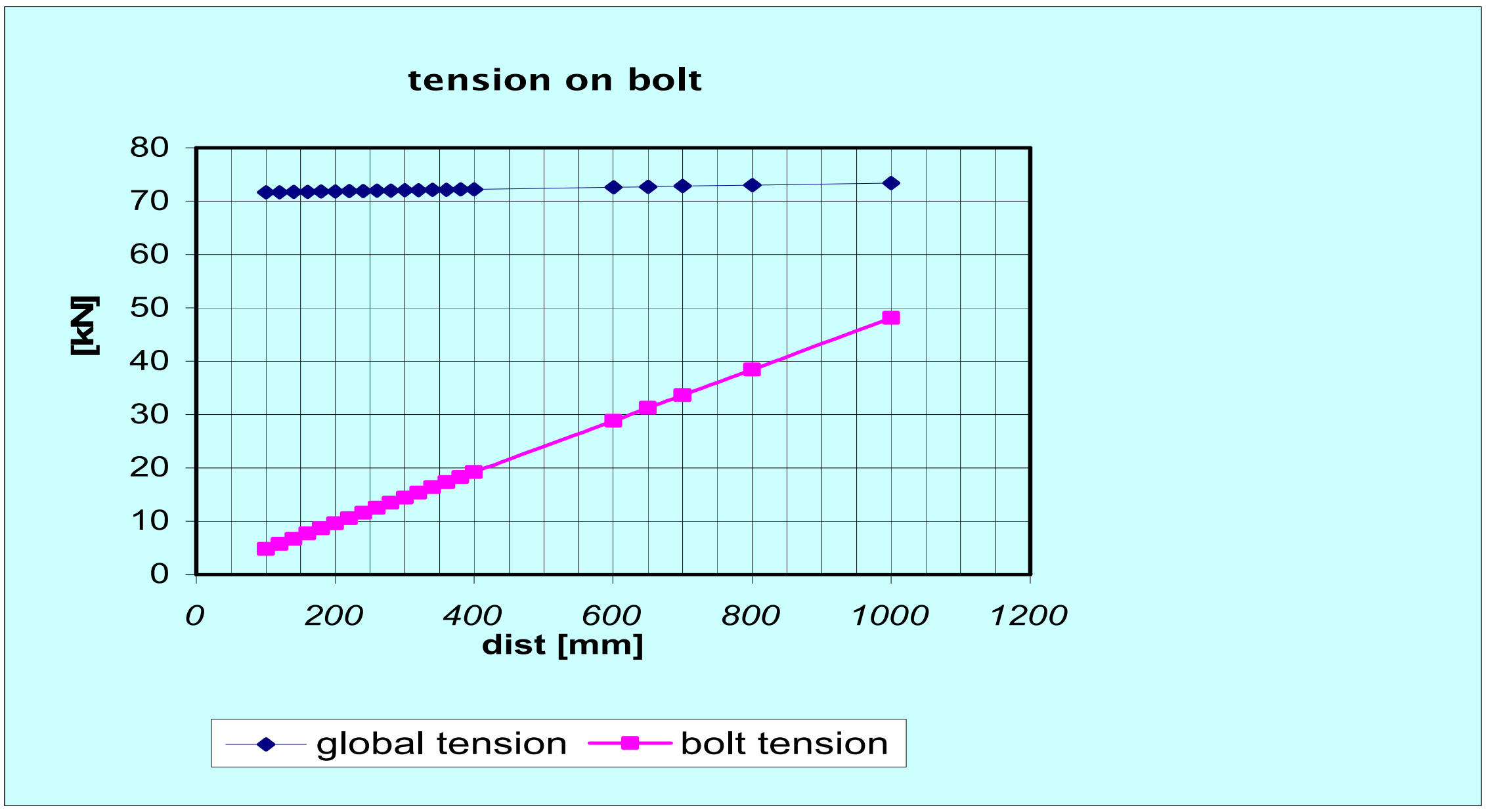

**Figure 3.40** Plot of tensions with variation of force application point along Z.

| Screw type | Resi-stance class | Limit brea-king stress [MPa] | Yield stress [MPa] | Young modu-lus [MPa] | Pitch [mm] | Prelo-ad Force [kN] | Prelo-ad Torque [Nm] | $A_r$ | $A_{eq}$ (on the flange) | Safety factor (80% yield stress) |
|---|---|---|---|---|---|---|---|---|---|---|
| **M16** | 8.8 | 800 | 640 | 210000 | 2 | 73790 | 235.88 | 157 | 1124.87 | 1.25 |

**Table 3-10** Main characteristics of M16 bolt.

### 3.7.2.3 Structural analysis for fork arm at first Nasmyth focus

The model of the fork arm was submitted to a static analysis in order to determine the points with maximum stress and strains. Loads applied are the same as in the flange case, i.e. reaction forces and moments calculated on the 4 M16 holes have been substituted: this is the most critical case as these forces are concentrated rather than distributed. Although the plotted Von Mises stresses are very high in the holes, most acceptable values are those spanned along the lateral plates, in the neighborhood of the weld joints with the housing bearing. In fact the taper extreme of the hole in the drawing is seen as a singularity by the meshing process of the program, so it does not correspond always to actual conditions. The larger value on them is about 63.753 MPa, and so it's lower than $\sigma_{adm}$= 250 MPa. We're interested in determining the tilt angle in Y and Z to match optical tolerance and also that of the axle bearings. In fact, even if taper roller bearing with single row has the ability to accommodate larger misalignments than other types of bearing, this limit is within 4 arcmin. So the following conditions must be checked, taking into account to opposite points on the internal faces of the housing bearing (see fig. 3.42):

$$\frac{\Delta U_z}{l} = \frac{-3.166 * 10^{-2} + 0.1027}{230} = 1.06 arc\min, \quad \frac{\Delta U_x}{d} = \frac{-0.19726 + 0.15469}{250} = \text{-0.58 arcmin}$$

Indicating with l the depth (X direction) of the arm= 230 mm and d the inner diameter where external ring of the bearing is fixed. These values are in the admissible range, respecting the tolerances of taper bearings.

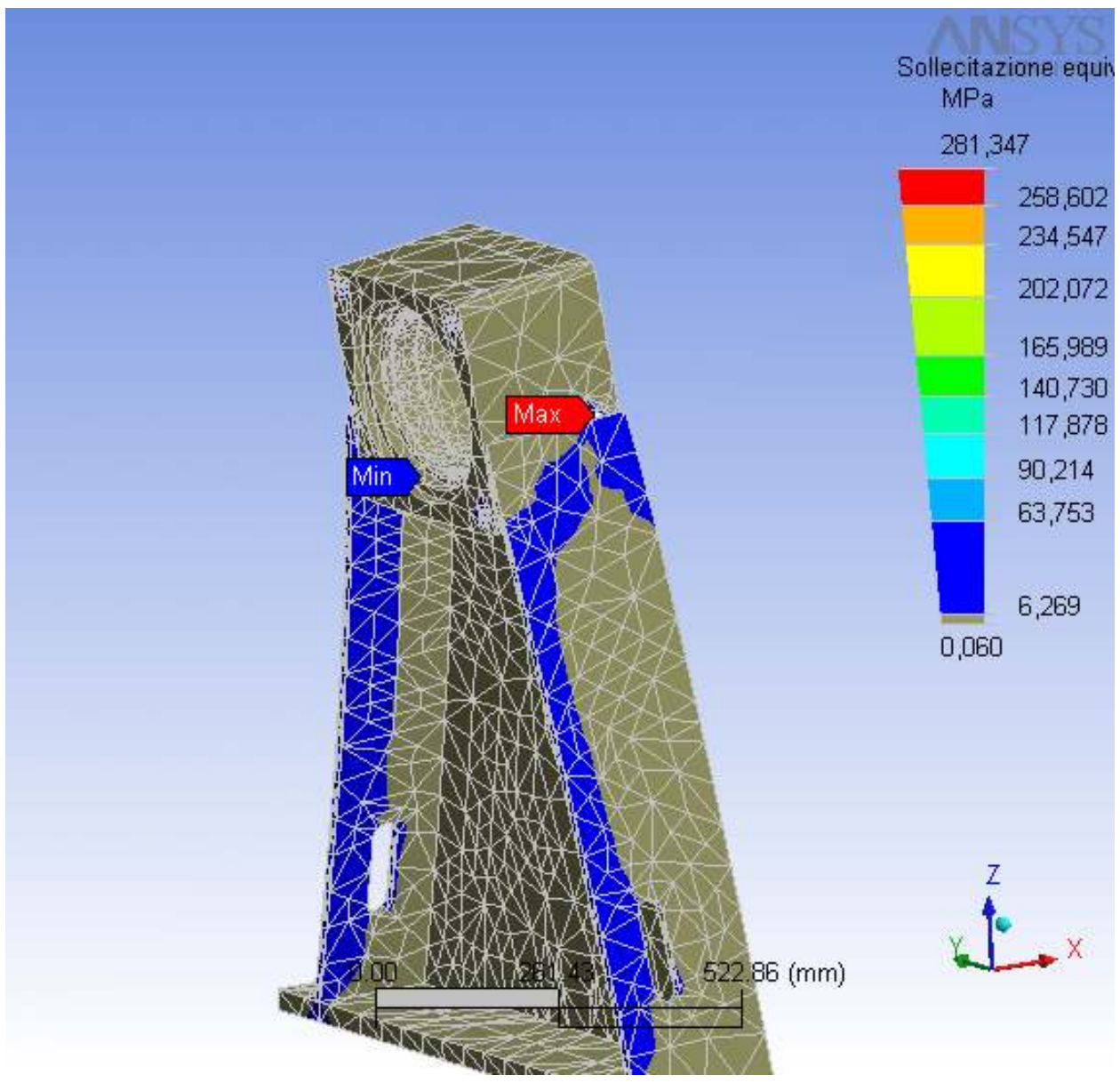

**Figure 3.41** Von Mises stresses diagram.

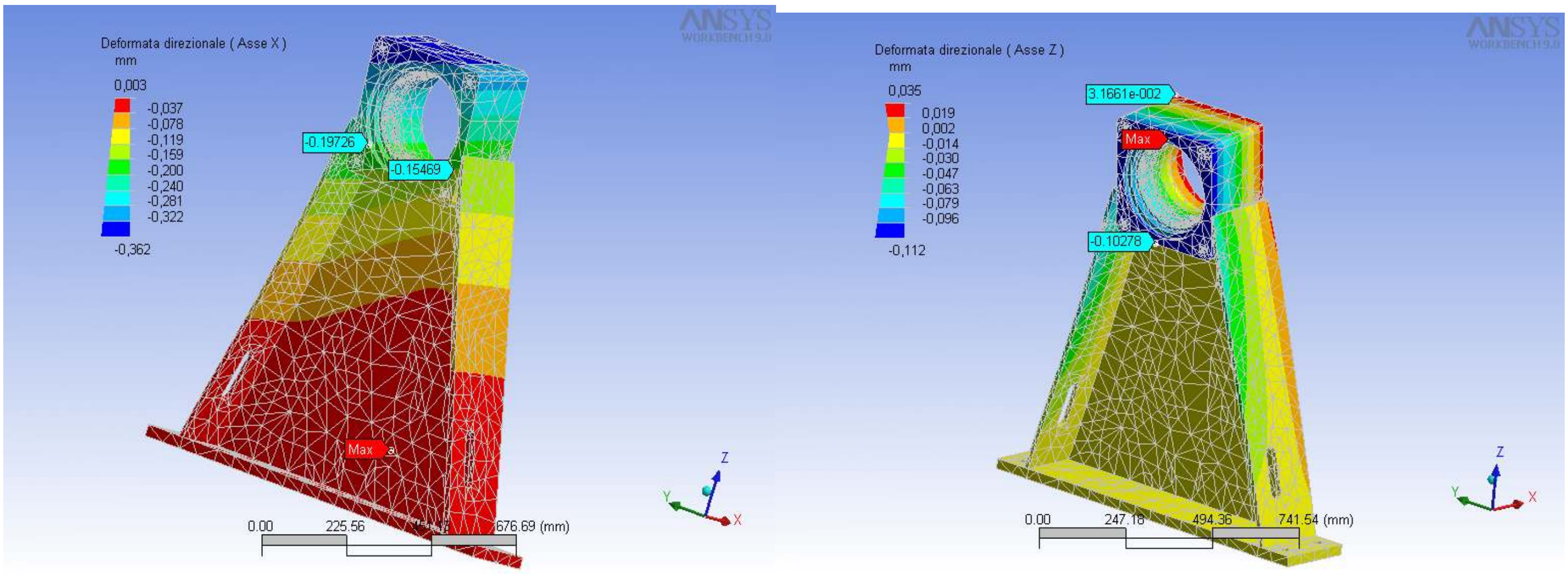

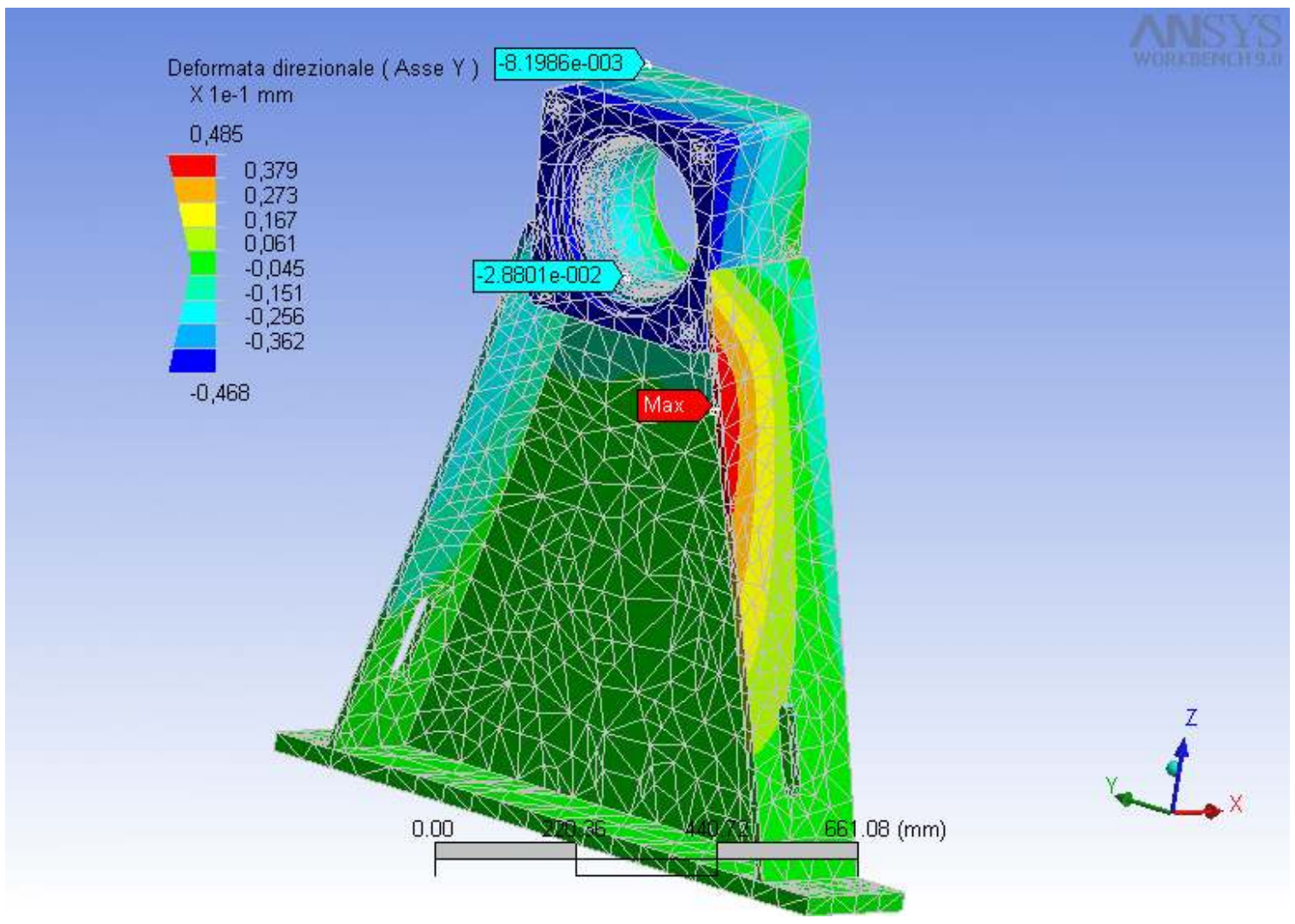

**Figure 3.42** Plots of deformations along X,Y,Z to which the structure of fork arm is subject.

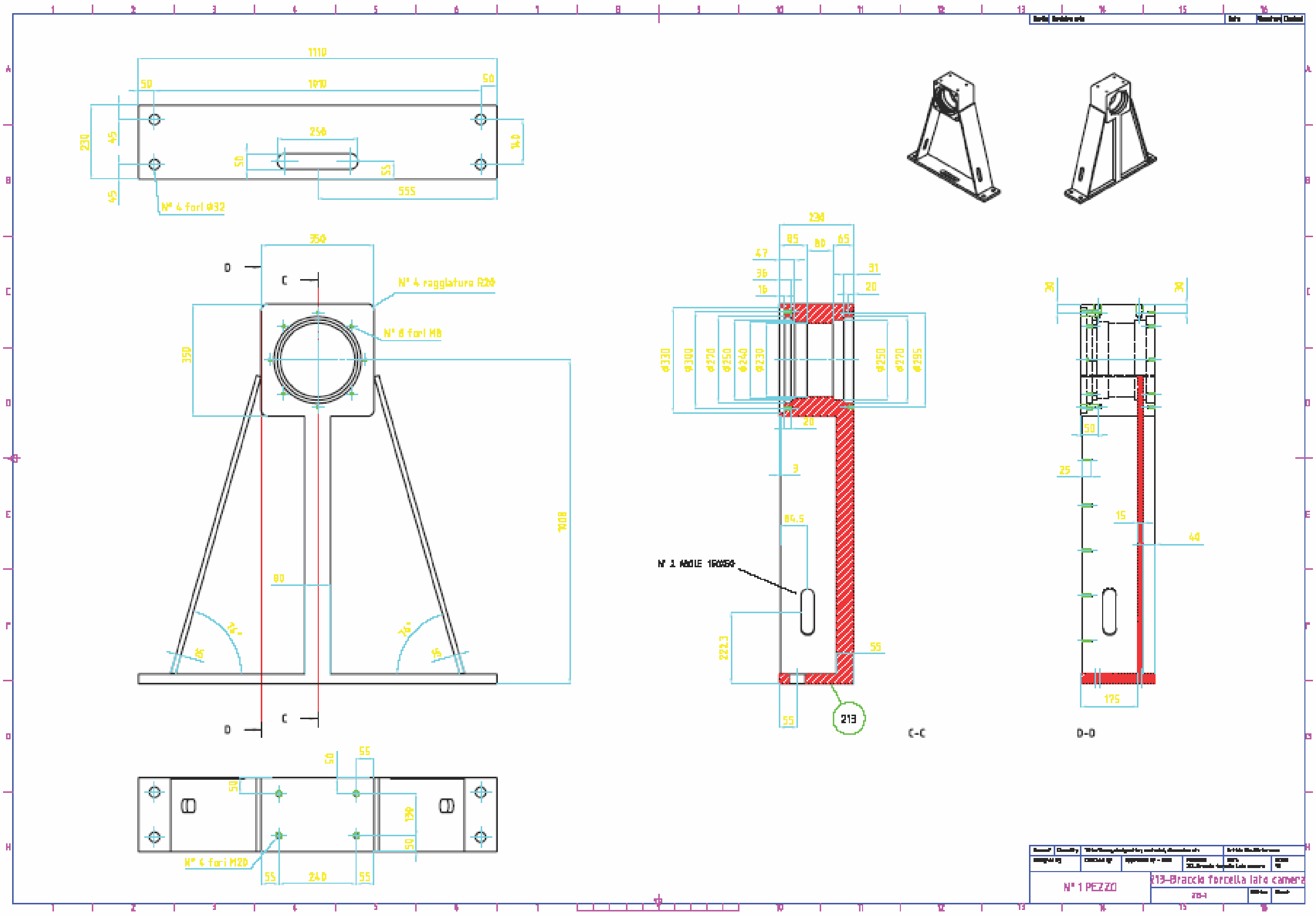

**Figure 3.43** Technical drawing with dimensions of the fork arm for the first Nasmyth focus.

### 3.7.3 Fork arm on the second Nasmyth focus

This arm has a more complex geometry if compared to the other, because it must support the motor box and have enough space to allocate the absolute encoder. The main difference from the first focus arm is in the central plate and the adjusting screws necessary to mount the altitude drive box. There are seatings for insertion of two taped roller bearings, in order to be adjusted in contrast. The right preload to be assigned is a serious duty: in fact, as bearings of these dimensions in general are designed for capacity far superior to our conditions, they must be preloaded for a precise motion. Two holes of 270 mm diameter have been provided for the pinions insertion at a distance between axes of 230 mm .
At the moment focal plane instrument to put on has not yet been decided, and it is under discussion.

#### 3.7.3.1 Loads analysis

The model in figure 3.4a has been submitted to a structural analysis under ANSYS, using solid elements (4-node quadrilateral shell) and with the following loads: gravity force; forces and flexure moments due to the bolts of the flange, supposing a weight of 1000 N at a distance of 1m along the elevation axis; torque due to forces exchanged between the pinion and the spur gear sector depending on the equivalent inertia of the optical tube and, mostly, on the bearings preload; reaction forces of the axle retrieved from previous analysis (see paragraph 3.6.5) on the internal surface of center hole.
All degrees of freedom on the base perimeter are constrained.
The estimated proper weight of the part is 252.84 kg. The moment of inertia around elevation axis, calculated by Mechanical Desktop, assigning a standard mild steel on all the components of the optical tube assembly, is 1283.62 kgm$^2$, and the preload indicated by SKF catalogue is 130 Nm, so that total torque is:
$M_t = J_x \dot{\omega} + M_{prel} = 163.3\, Nm$ , where $\dot{\omega}$ =1.5 °/s$^2$= 0.026 rad/s$^2$ represents the angular acceleration of the system. Maximum stress values are reached in the regions around the bolts, neglecting also, in this case, singularity points internal to the holes.

With the same procedure described above tilt angles were calculated. For the tilt in Y: $\frac{\Delta U_z}{l} = \frac{-0.179 - 1.468 * 10^{-2}}{230} = 2.88 arc\min$;

$\frac{\Delta U_x}{d} = \frac{-0.036 + 0.182}{250} = 2 arc\min$.

These values are in agreement with optical and mechanical tolerances (see fig. 3.46 and 3.47). A modal analysis with the static force extended to three eigenfrequencies, has been led. The result is that lower frequency is 89.384 Hz which is compatible with motor standard working frequency of 50Hz, as the damping effect is good outside the range $\frac{\omega_{mot}}{\sqrt{2}} < \omega < \sqrt{2}\omega_{mot} = 70.72\ \text{Hz}$.

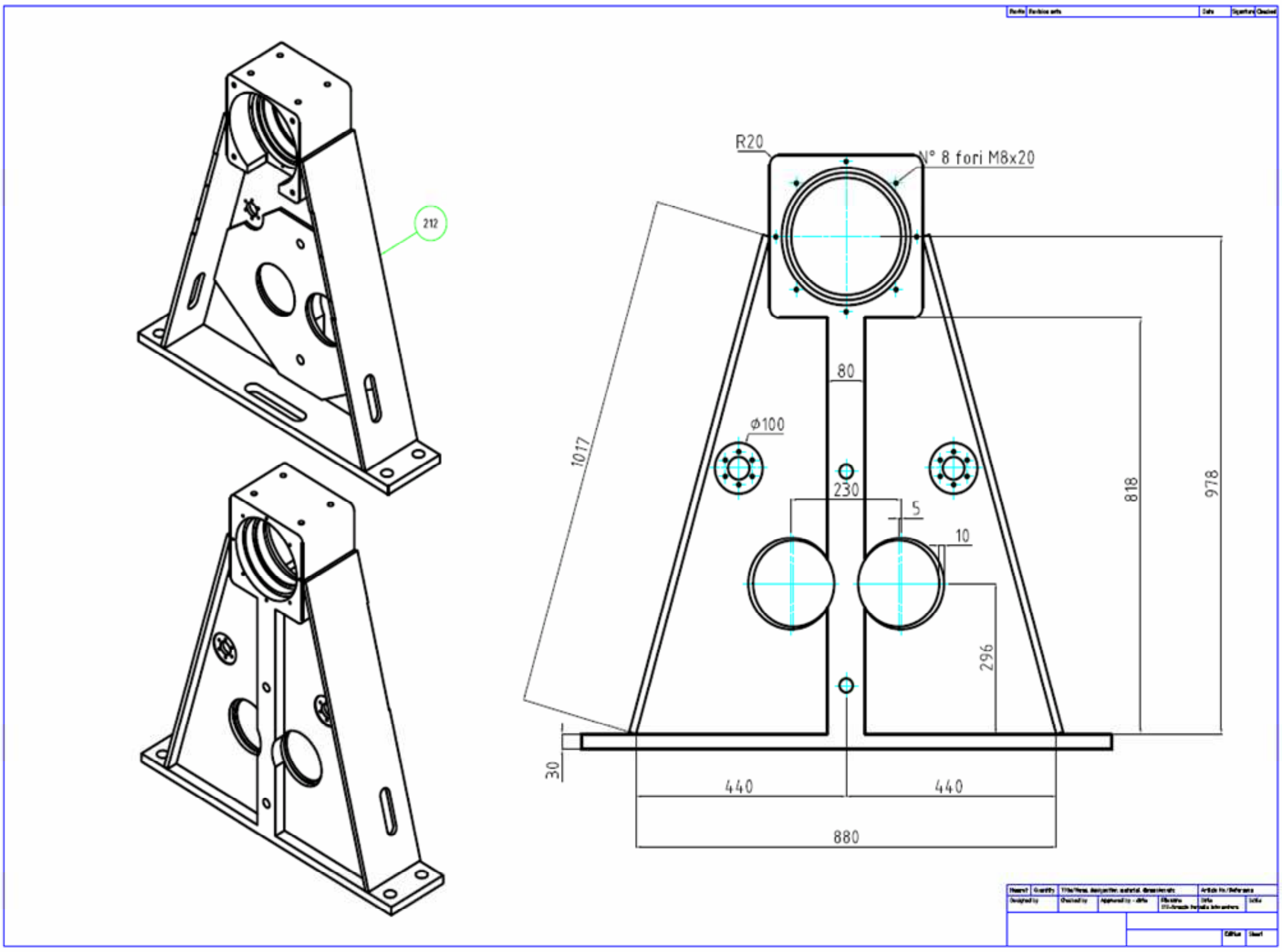

**Figure 3.44** Drawing of the fork arm for the second Nasmyth focus with main dimensions indicated.

.

| Loads | Fx [N] | Fy [N] | Fz [N] | Mx [Nm] | My [Nm] | Mz [Nm] |
|---|---|---|---|---|---|---|
| Weight | - | - | -2523.8 | - | - | - |
| Bolt holes 1 Ø16 mm | - | - | -250 | -36250 | $-2.5*10^5$ | 0 |
| 2 Ø16 mm | - | - | -250 | 36250 | $-2.5*10^5$ | 0 |
| 3 Ø16 mm | - | - | -250 | 36250 | $-2.5*10^5$ | 0 |
| 4 Ø16 mm | - | - | -250 | -36250 | $-2.5*10^5$ | 0 |
| Central hole | 0.22652 | 37.807 | -3996 | $7.9504*10^5$ | $-1.3467*10^5$ | 11083 |
| Motor torque | - | - | - | $1.633*10^5$ | - | - |

**Table 3-11** Overview of loads acting on fork arm 2.

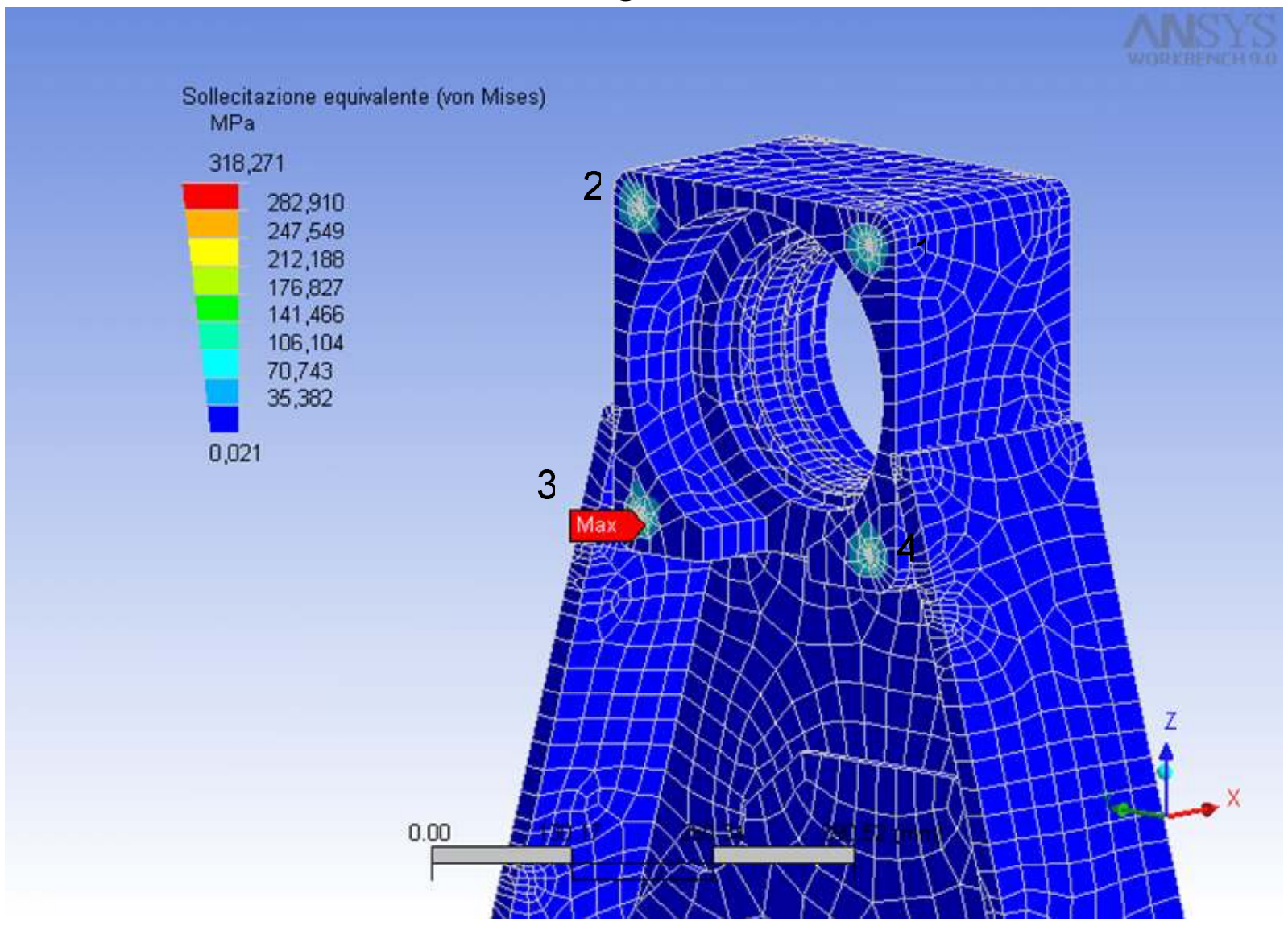

**Figure 3.45** Von Mises stresses plot on the arm.

| | Fx[N] | Fy[N] | Fz[N] | Mx [Nm] | My[Nm] | Mz[Nm] |
|---|---|---|---|---|---|---|
| Reaction forces on the base | -0.2643 | -43.658 | 3139.1 | -1072.6 | 1203.5 | -11.768 |

**Table 3-12** Reaction forces and moments at constraint nodes applied on the base perimeter.

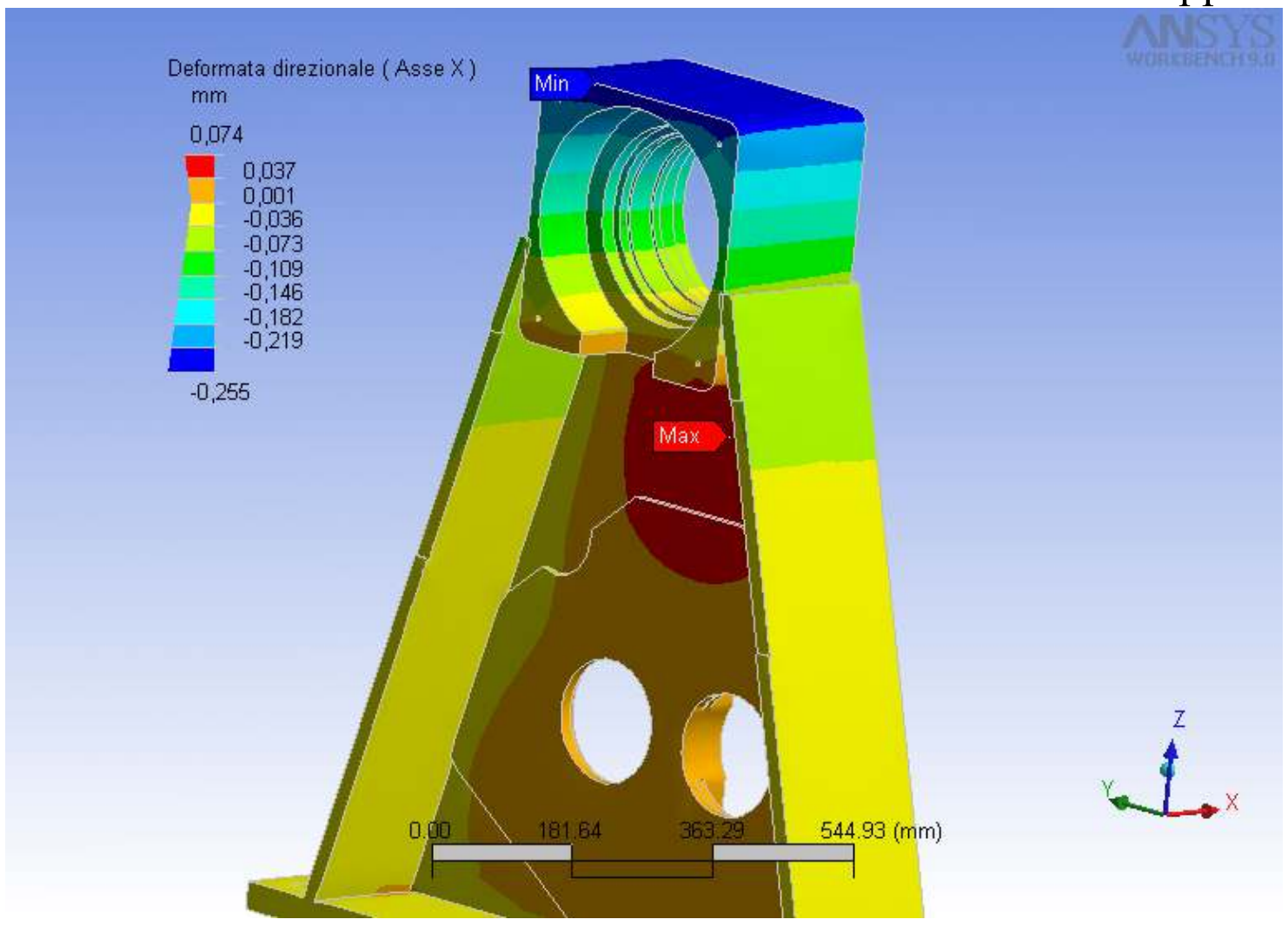

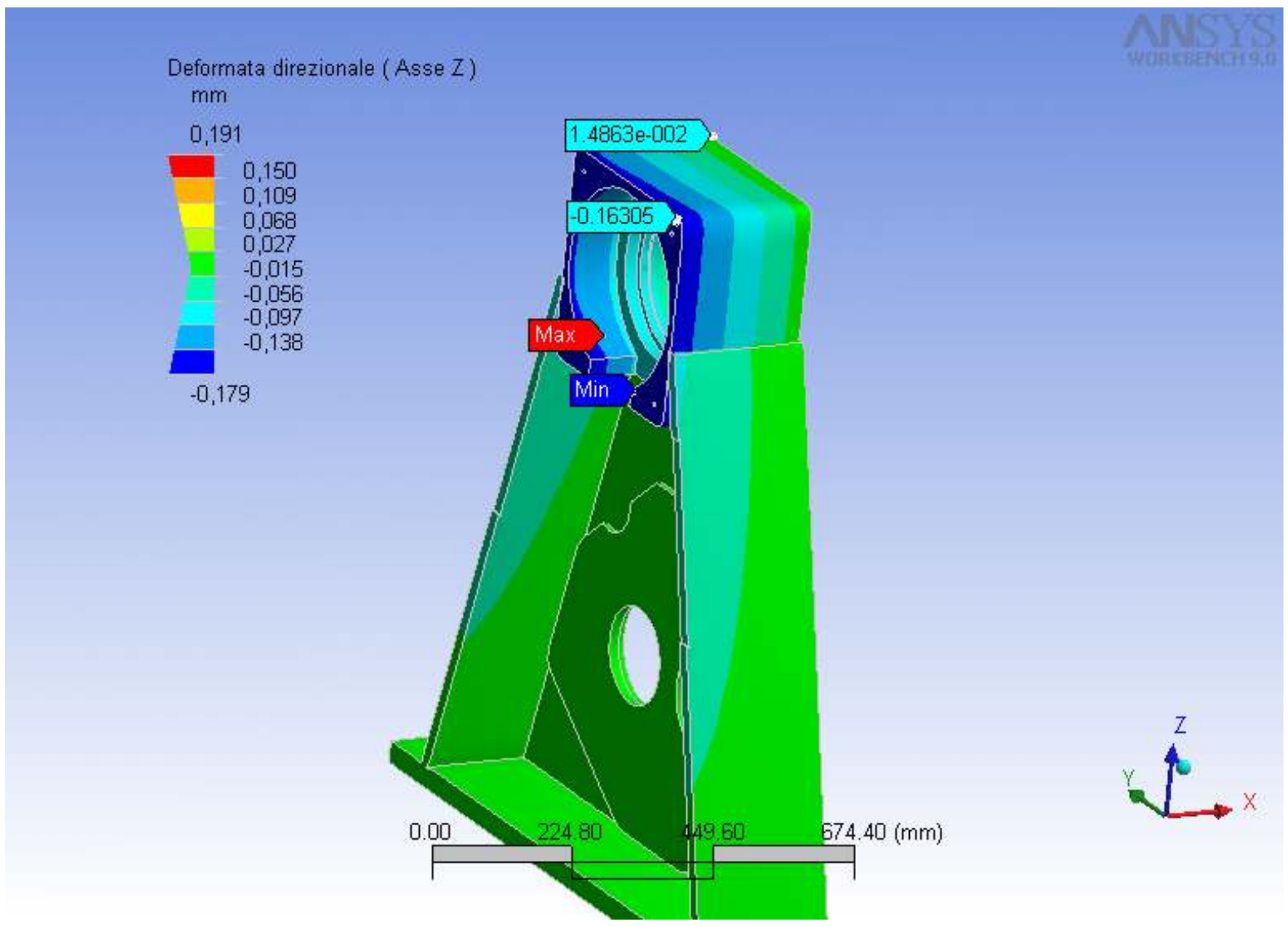

**Figure 3.46** Strains plots along two direction of interest.

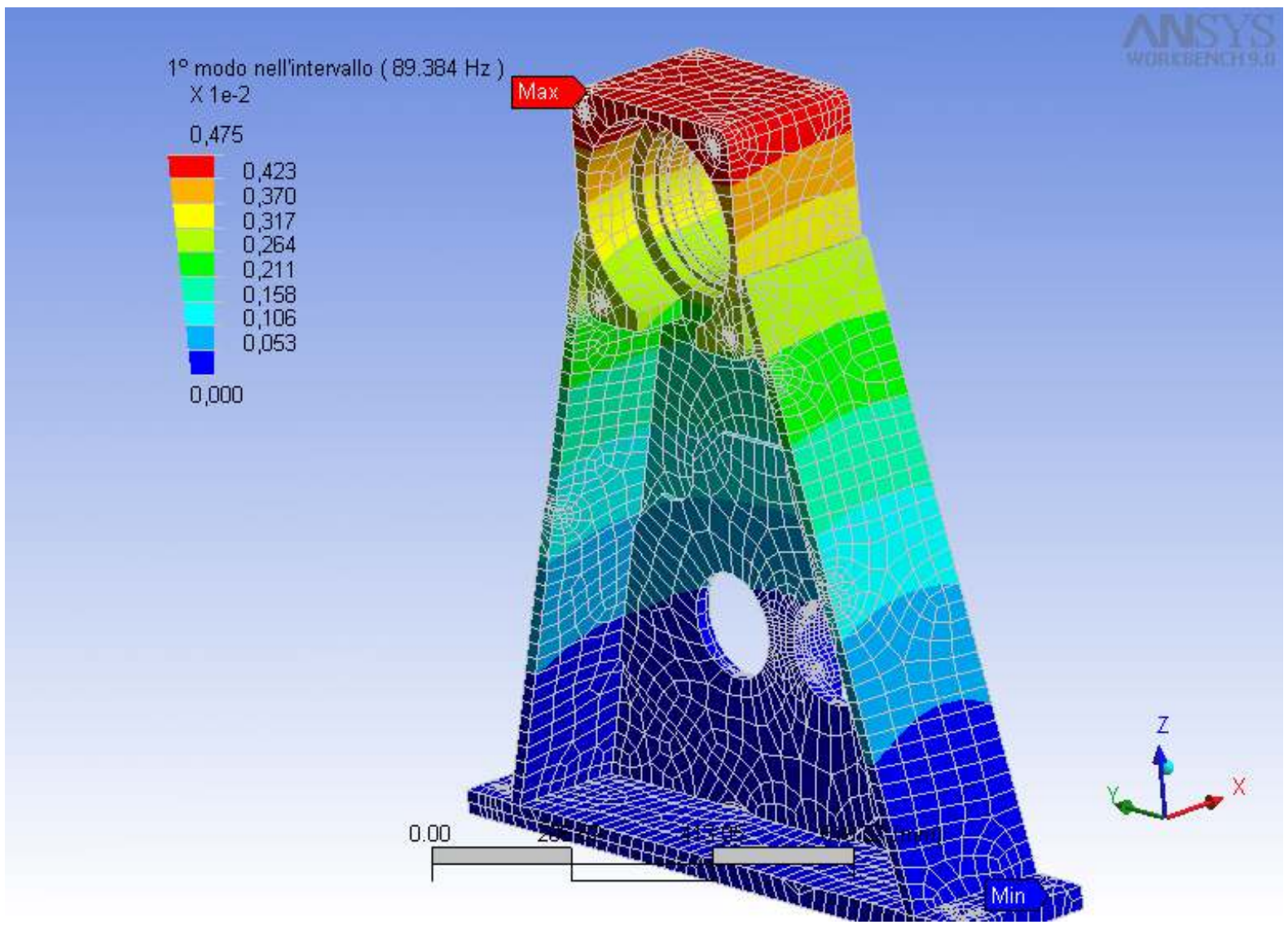

**Figure 3.47** Plot of the first vibration mode of the fork arm 2.

## 3.8 Base chassis

The base chassis constitutes a mechanical interface which connects the outer, fixed ring of azimuth bearing to the container floor; besides that, it is a system that must provide enough robustness and stiffness, needed to damp vibrations and shocks due to transportation. Dimensioning the beams and linkage joints of the chassis is, substantially, a problem of fatigue analysis. The load conditions can be split in two, distinguishing navigation and traverse phases. In fact they are a bit different cases, as, even if the forces module is the same, load frequencies change. It can be seen as a cyclic load oscillating in the range [-4Mg , 2Mg], and in case of navigation with a mean frequency of 0.5 Hz, while in case of traverse a maximum value of 20 Hz can be considered. With a travel of duration respectively of 50 days and 10 days, we have $2 \cdot 10^6$ millions of cycles that are not negligible for a stress reducing factor. For this reasons (stiffness and fatigue limits) the section of the profilate bars we chose have been oversized. Overall dimensions, shown in fig. 3.48, are those compatible with the modified ISO20 container (see next paragraph): it interfaces with the floor through 24 M30 holes. Two arrays of longitudinal segmented pipe bars, bolted by means of twelve stiffened brackets to the basement, 20mm thick, are used to resist to torsion. Pipe bar selected has an outer diameter of 230 and inner one of 180 mm. Cross double T beams and C channels are employed to reduce vertical displacements, limit bending effects and to support the plate where the outer ring of azimuth bearing is mounted. More details are contained in the Appendix B.3.

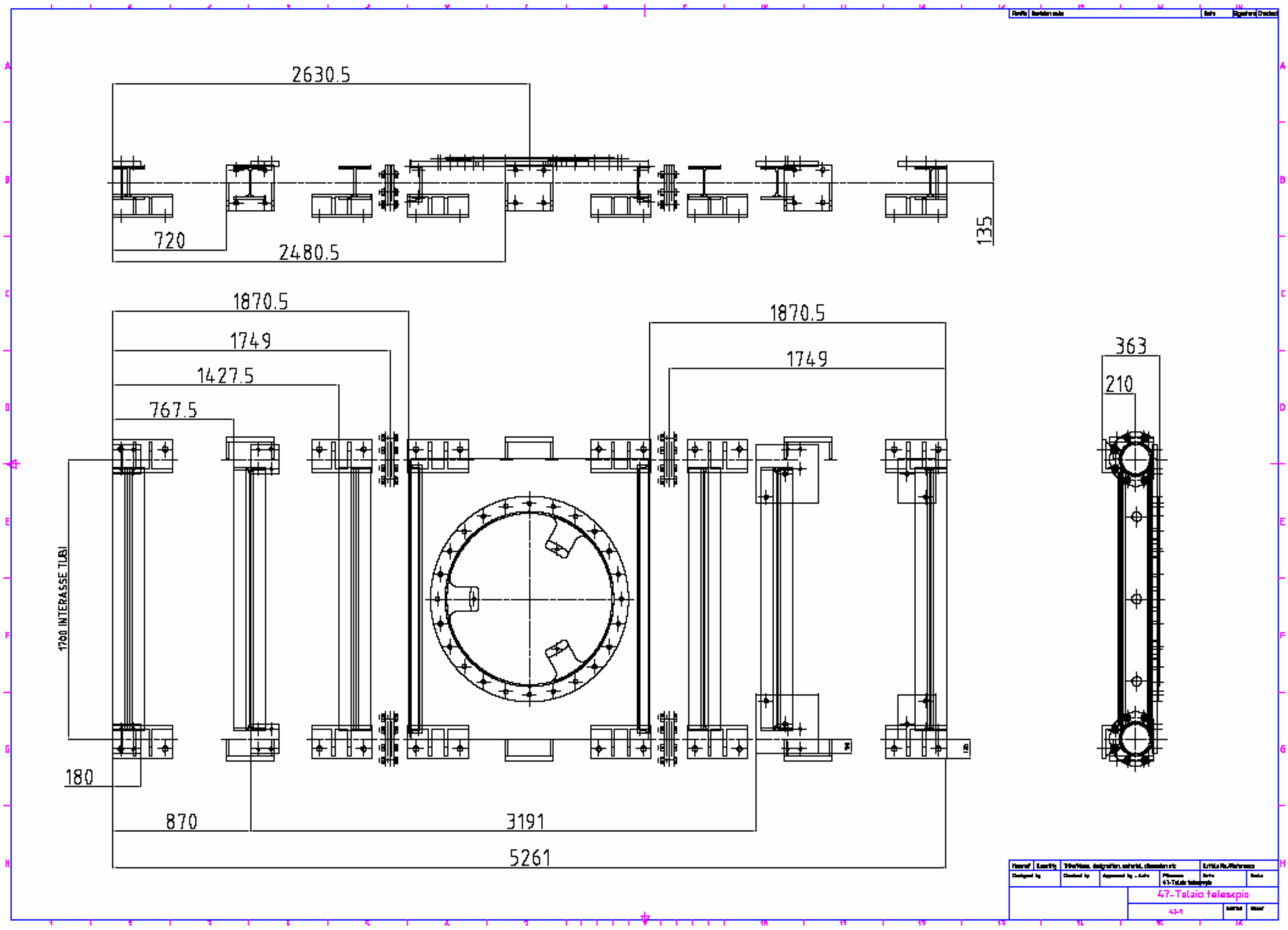

**Figure 3.48** Technical drawing of base chassis with main dimensions indicated.

In the beginning of the project, we contacted Mascherpa, a company that distributes antishock isolators. Looking over the problem, they suggested us to use a set of twelve wire rope isolators, oriented along the perimeter of the chassis, working under compression. As they have a resonance frequency of 3.6 Hz, they can damp requested 3g inertia loads, at a frequency of 15 Hz. Anyway this solution revealed to be too expensive for the available budget, and it has been discarded.

## 3.9 Shipping container

The compatible dimensions, for the container transporting the telescope structure, are those of a modified standard ISO20. The company contacted for the construction is Shellbox from Ravenna. Shocks during transportation, heating cycle time, and at last an accurate insulation from vibrations have to be taken into account.

Dimensions of a ISO20 container are those indicated by UNI 7011-72, classified as type 1C with height, width and length respectively of 2591 x 2438 x 6056 mm, and a maximum gross capacity of 30480 tons (see table 3-12).

They are usually designed for a stack height of nine or more full containers.

Two main types of stresses may be distinguished: structural and climatic ones. Strucutral stresses include static and dynamic cases, while climatic include chemical and thermal. Storage and handling shocks are present too, rather difficult to measure, as they depend on a lot of parameters. Transport conditions are related to ship and transverse from Baia di Terranova to Dome C.

It is known that an ordinary ship, of medium tonnage, while pitching in a heavy sea, receives a loss of speed from 21 knots to 9.3 knots (1 knot= 1852 m) within 2 seconds, undergoing an average deceleration of 3 m/s². Estimated frequencies due to transportation are within the range of 3-10 Hz, and maximum shocks about 4g.

| | | |
|---|---|---|
| External size | Length [mm] | 6062 |
| | Width [mm] | 2438 |
| | Height [mm] | 2591 |
| Internal size | Length [mm] | 5750 |
| | Width [mm] | 2200 |
| | Height [mm] | 2480 |
| Doors dimensions | Aperture [mm] | 2340 |
| | Height [mm] | 2280 |
| Capacity | 22 $m^3$ | |
| Maximum Gross Weight [kg] | 30480 | |
| Maximum payload [kg] | 28310 | |
| Tara [kg] | 4600 | |

**Table 3-13 Container sizes, provided by Shellbox.**

### 3.9.1 Heat transfer analysis

It was thought to heat the container during transportation, since the mechanical electronics components, as parts of the controllers and drive boxes of both telescope and focal plane instrument, are designed to withstand at least a temperature of -20 °C and, even if insulation boxes are present, the temperature could go down far below this value. Therefore, an estimation of the steady-state thermal power, which is necessary to provide, to maintain a temperature of T≤ -20° C inside the container, has been done. The problem was ascribed to a 2D model of heat transfer by conduction through a metallic wall of known thickness, keeping in mind the contribution of convection by means of a coefficient of adduction (laminar flow).

Supposing the following values for the coefficients:

$\lambda_A$=60W/mK for steel ;
$\lambda_{ins}$ (provided by ISOBOX) = 0.26 W/mK for an insulation thickness of 80 mm (0.18 for 120 mm).

$h_{in} = 8\frac{W}{m^2 \cdot K}$ for internal adduction;

$h_{ex} = 20\frac{W}{m^2 \cdot K}$ for external adduction;

Walls thickness:
$s_1$=156;
$s_2$=86 mm;

Thermal flux is given by: $\dot{q} = \frac{T_{in} - T_{ex}}{\frac{2s_1}{\lambda_{acc}} + \frac{s_2}{\lambda_{ins}} + \frac{1}{h_{in}} + \frac{1}{h_{ex}}}$

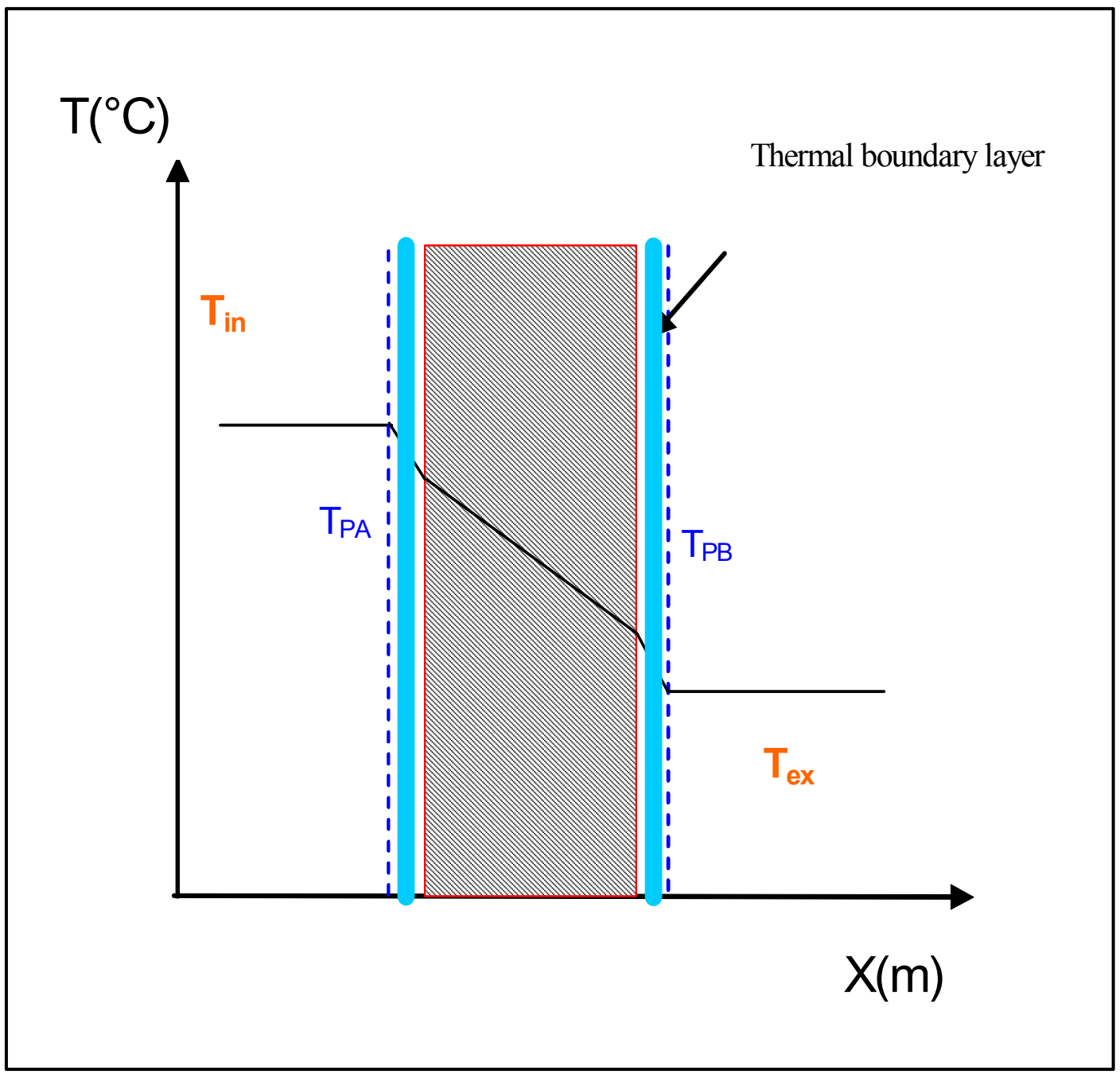

**Figure 3.49** Scheme of the conduction through container wall.

The heating power is given by:

$\dot{q}S = P_g$ [W] , being S the global internal area of the container.

For the total energy equilibrium: $Q_{in}=Q_{out}= m_{air}\, c_p \Delta T$, because there's no heat generation inside the wall. Assuming a minimum external temperature of -50 °C and internal fixed temperature of -20°C, we have:

$\dot{q} = 62.108 W/m^2$ , $P_g$ =3.685 kW. This value is compatible with an available heater power. The condition is valid supposing an instantaneous heating, and if, once reached the regime temperature, it keeps constant inside the container.

Letting a steady-state condition, heating time can be determined by the formula:

$$\frac{m_a C_{pa} \Delta T}{\dot{q}S} = \Delta\tau$$ , where *a* stands for air .

The calculated cycle time to warm up and cool down the inner room with a thermal gradient of 30 degrees is about 1100 s, with a heating power $P_g$ (see fig. 3.50).

Anyway, for a transient state, as it is actually, a different type of analysis must be carried out.

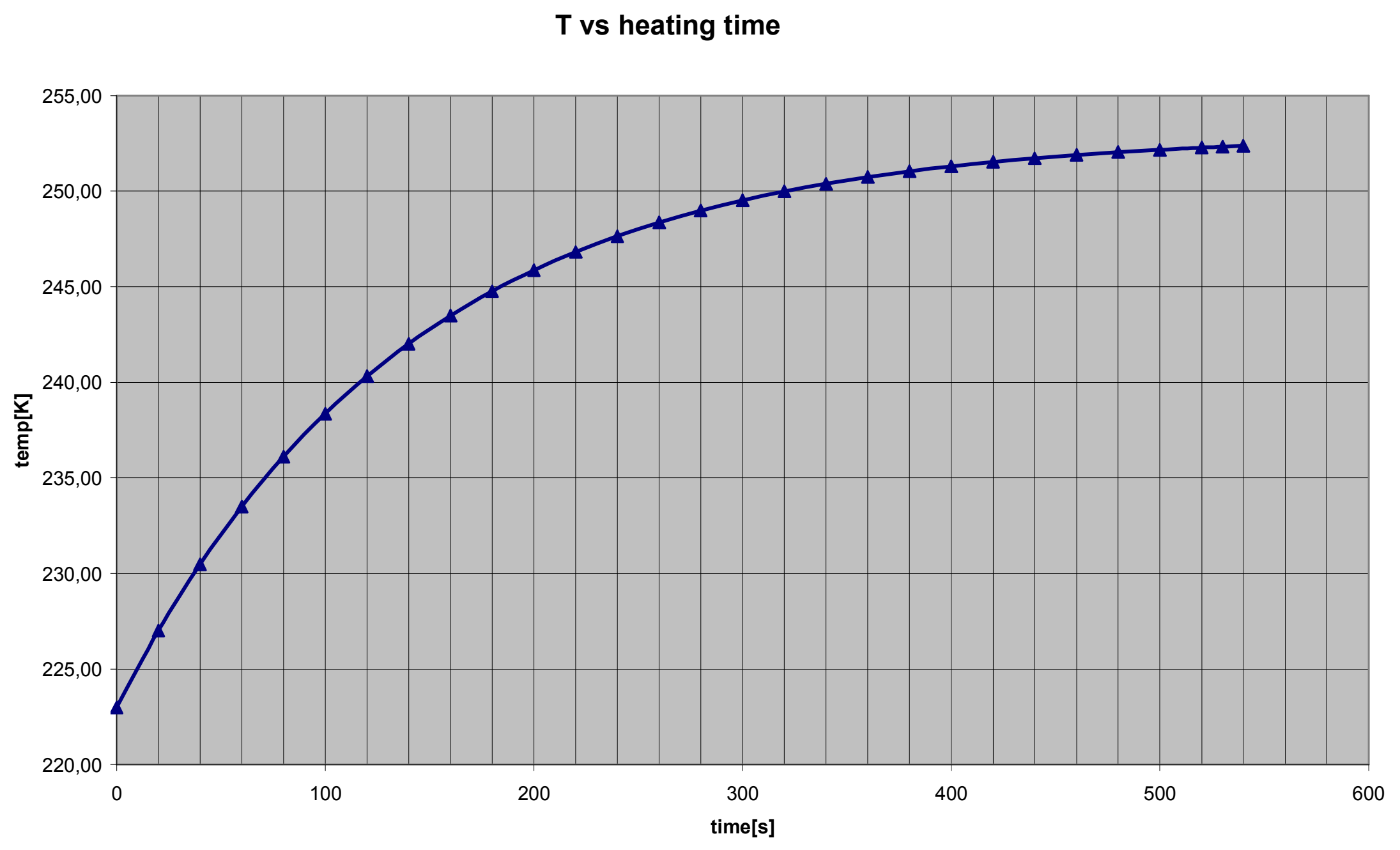

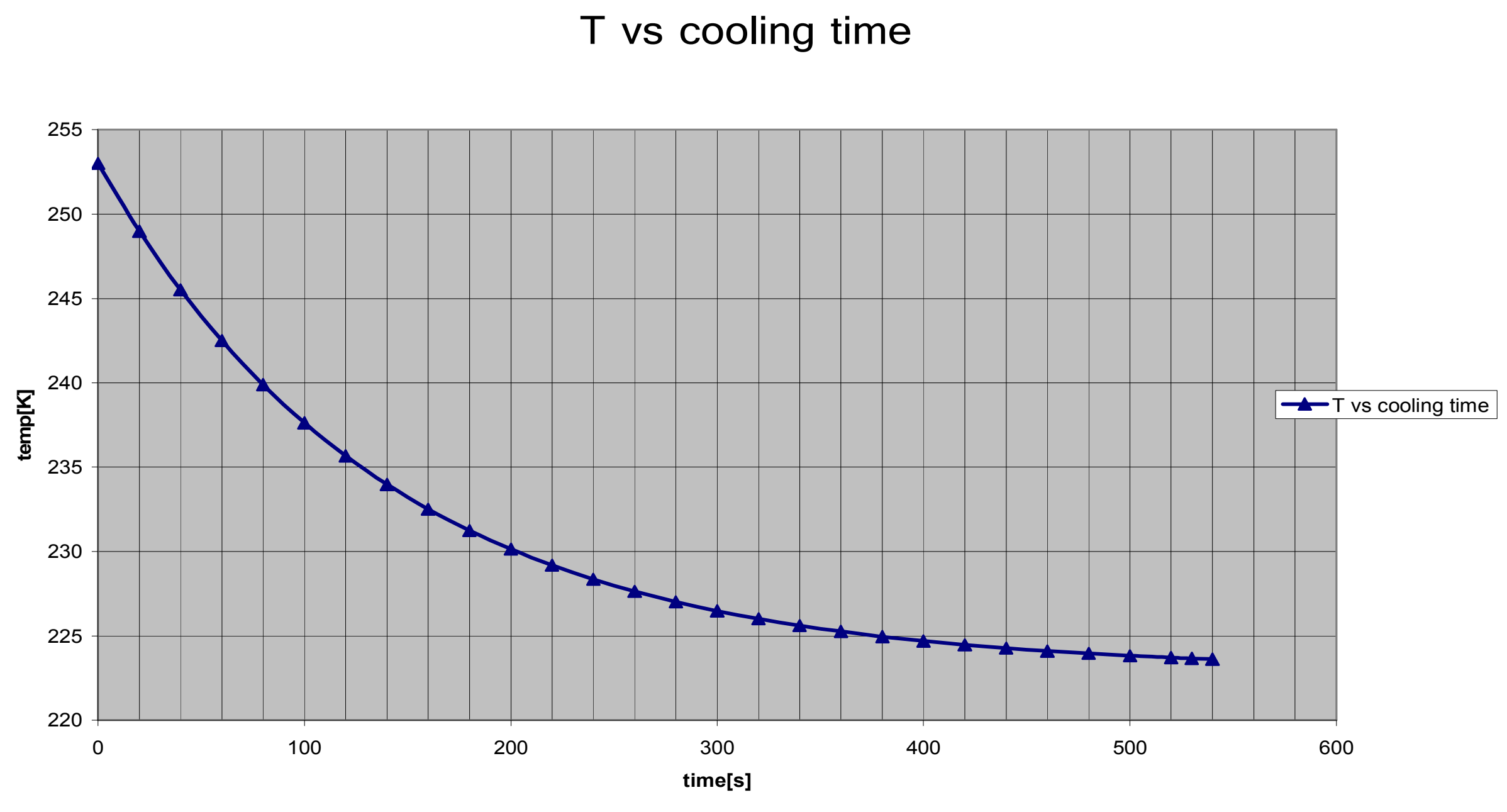

**Figure 3.50** Plots of respectively cooling and heating time.

# CHAPTER 4 Machine elements design

The present chapter contains the parameters that led to the design and sizing of the elements composing the drive train and the supports in azimuth and elevation. Looking back at the history of the project, the initial target, once determined velocity rates, was the selection of slewing azimuth bearing, as it is the component most prone to stresses and that has a greater influence on operating conditions of the telescope.

## 4.1 Bearings choice criteria

### 4.1.1 Azimuth slewing bearing

In the first phase of the project we thought to purchase a slewing bearing from the SKF catalogue for a prototype in order to test it in a climatic chamber. The 92115-0101 XD/X type, with the characteristics indicated in the table below, revealed to be the most suitable. Later on, because of a long realization time, in order to minimize supply costs, as a custom manufacturing was necessary for the extreme conditions of employment, we decided to purchase the azimuth bearing on the basis of some detailed lists. Cross roller bearings are largely used in excavators and cranes, therefore they must be adapted to rapid movements and resistant to shocks, designed for a charge capacity up to 170 kN. Anyway, this kind of bearings are not appropriate for movements where high precision is strongly required. In our case, for the very low rotation speed, backlash and misalignments must be absolutely reduced as far as possible, and the repeatability of movements must be guaranteed too. Almost static load conditions can be considered and fatigue conditions can be neglected in first approximation. An external set of teeth with external motors has been selected, so that there's enough space inside the inner ring for the passage of cables.

For a better stability and compactness of the system we decided to have a fixed external ring, bolted to a mechanical interface with the basement chassis, and an internal ring which rotates together with the entire upper structure. They are teeth of a spur gear mechanism so that axial forces, parallel to azimuth axis, are absent. A proper design of the pinion is necessary, so fundamental parameters like the module must be settled, and it must be checked if it satisfies the criteria of fatigue and wear resistance of the teeth. The catalogue contains also the values of the minimum and maximum backlash as a function of the module.

For bearings in the catalogue the values are respectively m=10 and J=300÷450μ, where J is the backlash deriving from the positioning of the pinion in the point of maximum radial run-out. With a module m =3.5 backlash is reduced to J=250÷300 μ.

| Type | External diameter [mm] | Internal diameter [mm] | Height [mm] | Weight [kg] | Teeth | Module *m* | Number of teeth *Z* |
|---|---|---|---|---|---|---|---|
| 92115-0101 XD/X | 403.5 | 233 | 55 | 24 | External | 4.5 | 88 |

**Table 4-1** Characteristics of a standard catalogue bearing.

If we choose the module and the number of the teeth z as constraints to develop the design, the pitch diameter is determined by the formula :

$$D_p = m*z = 1260 \text{ mm.}$$

All other parameters characterizing the gear derive from the module. They are:

- addendum: a = m ;
- dedendum: d= 1.25 m;
- tooth thickness : 10m;
- tooth height: h = a +d.

Maintaining as fixed parameters a total reduction ratio of 1/720, the variation of other dependent quantities is shown in the next table.

| Reduction ratio | Bearing Nominal Diameter[mm] | Z bearing | Z pinion | module m | Pinion Nominal Diameter[mm] | Total reduction |
|---|---|---|---|---|---|---|
| 1/100 | 1440 | 144 | 20 | 10 | 200 | 1/720 |
| 1/60 | 1440 | 240 | 20 | 6 | 120 | 1/720 |
| 1/80 | 1440 | 180 | 20 | 8 | 160 | 1/720 |
| 1/72 | 1440 | 180 | 18 | 8 | 144 | 1/720 |
| 1/40 | 1440 | 360 | 20 | 4 | 80 | 1/720 |
| 1/66 | 1260 | 360 | 33 | 3.5 | 115.5 | 1/1440 |
| 1/54 | 1440 | 240 | 18 | 6 | 108 | 1/720 |

| τ pinion-bearing | Addendum [mm] | Dedendum [mm] | Tooth height[mm] | Pitch p[mm] | Tooth thickness b[mm] |
|---|---|---|---|---|---|
| 5/36 | 10 | 11,67 | 21,67 | 31,42 | 60 |
| 1/12 | 6 | 7,5 | 13,5 | 18,85 | 80 |
| 1/9 | 8 | 10 | 18 | 25,13 | 80 |
| 1/10 | 8 | 10 | 18 | 25,13 | 40 |
| 11/120 | 3.5 | 4.37 | 7.87 | 11 | 35 |
| 7/120 | 4 | 5 | 9 | 12,57 | 60 |

**Table 4-2 Pinion - gear parameters.**

Reduction boxes of ALPHA joint-stock corporation were chosen (model TP 050 BUT, of coaxial type, 3 stages, using planetary gears), with a reduction ratio of 1/66, the only option compatible is the one evidenced in grey. The peak torque of the motor, with a tension of 560 V, is 8.5 Nm, so that the output torque is 462 Nm.

### 4.1.2 Estimation of the forces acting on the azimuth bearing

Two fundamental load conditions have been considered: the first, due to the transportation, which is the most critical, the second referred to work conditions in place, under the action of wind. Taking into account the inertial accelerations evaluated as 4g in all the directions, during transport there are two contributions, radial and axial, applied to the azimuth bearing. For the telescope initially estimated mass of 1500 kg, we have:

$F_r = m(4g) = 60$ kN;

$F_a = m\ (5g) = 75$ kN.

Considering a maximum offset of the centre of mass equal to 150 millimeters, in the horizontal position of the optical tube, the resulting tilting moment is:

$M_{tilt} = 0{,}15\ F_a + 1.5\ F_r = 101.125$ kNm.

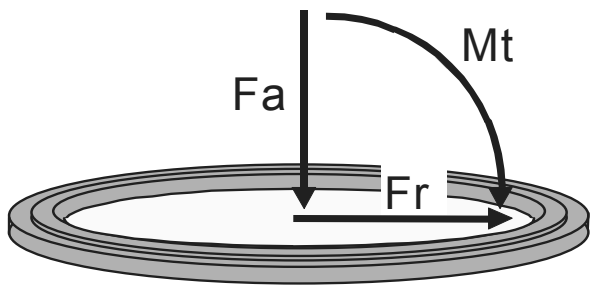

**Figure 4.1 Schematic of the loads acting on the azimuth bearing.**

According to the second load case, there are the contributions of wind and centrifugal force. Imposing a maximum speed of 1.5 rpm, we obtain:

$F_c = m\omega^2 r = 27.73$ N.

As far as the wind action is concerned, $F_V = \frac{1}{2}\rho_{air} v^2 c_R A$
where it is assumed that:
$\rho_{air}$=0.909 kg/m$^3$ at T = -20 °C;

$v$ = mean wind velocity=16 m/s ;
$C_R$ = resistance factor, depending on the geometric shape and on Reynolds number:
$Re = \frac{vD}{\nu_{air}}$ ,
$\nu_{air}$ is the cinematic viscosity, which varies with the altitude.

Knowing that the 3000 m upon sea level atmospheric conditions of Dome C are comparable with those of 5000 m at our latitudes, $\nu_{air}$ =1.86 * 10-5 m/s.
*A* is the frontal area of the rectangle, whose size is H and D. In the first approximation, since we were not interested in detail in the dynamic analysis of wind forces, the whole structure has been assumed as two overlapped cylinders, the first one constituted by the optical tube, the other by the mount fork and the basis, whose sizes are respectively: D x H = 830x1800 mm; d x h = 1500x1400 mm.

Therefore, the estimated values for load conditions, are as follows:

$Re = 1.29x10^{6;}$, $C_R = 0.14;$

$F_V = 58.5\ N;$

$F_r = F_C + F_V = 85.5*10^{-3}\ kN;$

$F_a = mg = 15\ kN;$

$M_{tilt} = M_{t\,v} + 0.15\ F_a = 2.3\ kNm.$

It is clear that loads in the second case are almost negligible. During the development of telescope design, different types of motor boxes, fork braces, and optical tube truss have been considered. Also the overall size of devices attached to the Nasmyth foci, as well as counterweights have changed, leading to an increase of total weight. Now, the mass moved by azimuth bearing is supposed to be 4500 kg, so that an increase factor of 3 for the loads must be considered. Anyway this underestimation luckily had no influence on the choice of azimuth bearings, because these components have always elevated safety factors, capable of assuring extreme loads applications with elevated shocks and tilts.

Table 4-4 illustrates the variation of torque in relation with two different preloads.

| Loads / **Load conditions** | Axial load [kN] | Radial load [kN] | Tilting moment [kN] |
|---|---|---|---|
| 1 Static-transport | 75 | 60 | 101.2 |
| 2 Static-operating | 30 | 0.18 | 4.6 |
| 3 Dynamic/operating | 30 | 0.18 | 4.6 |

**Table 4-3 Load conditions on azimuth bearing.**

| ***Precharge [mm]*** | ***Load conditions*** | ***Torque without any load applied[Nm]*** | ***Friction torque due to the sealings [Nm]*** | ***precharge torque [Nm]*** | ***load torque [Nm]*** | ***Total torque + 20% [Nm]*** |
|---|---|---|---|---|---|---|
| 0.080 | **2** | 39 | 117 | 273 | 104 | 640 |
| | **3** | 39 | 117 | 273 | 104 | 640 |
| 0.185 | **2** | 39 | 117 | 642 | 104 | 1080 |
| | **3** | 39 | 117 | 642 | 104 | 1080 |

**Table 4-4 Torque values provided by RKS.**

### 4.1.3 Pinion design

Supposing an overload of a factor 2 it is possible to determine the stresses and verify both the pinions fatigue and wearing in altitude and azimuth.

$M_p = M_{az} \cdot \tau = 2.16 kN \cdot 11/120$ ;

$M_p = F_t \cdot d_p/2$ .

Hence $F_t = 3.429$ kN.

#### 4.1.3.1 Fatigue verification

It has to be checked that the duty cycle stress condition $\sigma_f < \sigma_{adm} = \sigma_n * C_L * C_G * C_S * K_R K_T K_S$

$C_L$ = load factor =1

$C_G$ = stress gradient factor =1

$C_S$ =surface roughness factor= 0.8

$K_R$ = reliability factor=0.814

$K_T$ = temperature factor=1

$K_S$ = medium stress factor =1.4 .

And $\sigma_f = \frac{F_t K_0 K_m}{mbJ}$,

where:

$K_o$= overload factor =1.5;

$K_m$ = mounting factor;

b=tooth width=35;

J=Lewis factor=0.37;

m=module=3.5.

Substituting all the values it was found that the pinion satisfies the fatigue criterion, as it is:

$\sigma_f = 147.51 < \sigma_{adm} = 547.01$ .

#### 4.1.3.2 Hertz verification

It has to be verified that the pressure exchanged from a tooth to another is lower than the maximum one, i. e.:

$P_{max} < P_{amm}$

$P_{max} = \sigma_H = C_P \sqrt{\frac{F_t}{bd_P l} K_V K_O K_m}$ and $l = \frac{\sin\phi\cos\phi}{2}\frac{R}{R+1}$ , being φ=20° the pressure angle,

and $P_{amm} = \frac{2.5H_d}{n_H^{1/6}}$ , being $H_d$ surface hardness=2100N/mm$^2$

and $n_h$ =44000 h the design life .

The pressure check is fully satisfied as it is:

$P_{max}$=587.57< $P_{amm}$=883.59.

### 4.1.4 Verification of the duty cycle of azimuth bearing

If we suppose that the telescope works all the time for ten years, 6 months per year, it is equivalent to approximately 44000 h of operation. For the verification to duty cycle it is, therefore, necessary that duration L is lower than $L_{10h}$ , which is the duration expressed in hours of service, overrun by 90 % of a batch of bearings:

$$L_{10h} = \frac{1 \cdot 10^6}{60 \cdot 1.5}\left(\frac{C}{P}\right)^n$$

- *n* depends on the geometry of rollers, and for a cylindrical shape it is 10/3;
- 1.5 is the maximum velocity, expressed in *rpm*;
- *C = 169,2 kN* is the charge capacity of the model 42506-0101 XD/X;
- *P = xFr +yFa* with x,y as indicated on the catalogue.

Considering that the axial force acts mostly during motion (Fa=7.5 kN), it follows that:

$L_{10h} = 296.15 \cdot 10^6$ h

Thus, the condition is completely verified.

### 4.1.5 Altitude taper roller bearings

In order to give the hollow axles a certain freedom in axial movement, we have chosen a couple of taper roller bearings with O mounting: in fact they can compensate thermal variations and, at the same time, keep a precise alignment. For a single row taper roller bearing, internal clearance can be obtained only after mounting and adjustment against each other. In general, they have a high friction torque during the first hours of operation, which decays soon after the running-in period. As for azimuth bearings, they are provided of a preload too, indicated by SKF at our specific request.
They tested the components applying different levels of axial preload (10, 20, 30, 40, 50 micron), noticing where contact of all the rollers in the crown occur. For the two types selected they have specified preload on inner and outer rings. The highest preload moment taken into account for structural analysis relative to a penetration of 50 micron, is 133 Nm.
The distance between the centres of two bearings is 100 mm.
The choice of bearings was made taking axle diameter, D= 140 mm, as input parameter. To size the most suitable bearing the dynamic charge capacity C must be known. If this value matches with that provided by the catalogue, then the choice is good.
With the same relationship between duration $L_{10}$ and capacity seen previously, and converting it from millions of cycles into million kilometres, we obtain:

$$L_{10} = \left( \frac{C}{F} \right)^n$$

$$L_{mil.km} = \pi D \times 10^{-6} L_{10} \, ,$$

substituting reaction force, written in paragraph 3.6.5:
$C_{min}$ =4505.63×39.01=175.778 kN,
that is inferior to 429 kN, provided by SKF.
The chosen bearing is designated as 32032X. Same considerations can be made for the other bearing with inner diameter of 180mm, type 32936 with a dynamic load capacity of 352 kN.

## 4.2 Shaft dimensioning for azimuth transmission motion

### 4.2.1 Forces calculation

The two components of forces exchanged between pinion and azimuth bearing need to be defined.
We now consider limit conditions: supposing that maximum output torque delivered by the motor is 10 Nm and therefore on the pinion, without friction, the torque is 1280 Nm. Being the pitch diameter of the pinion 112 mm, the two force components acting on a tooth are:

$F_t = 2 \cdot C/D = 26785$ N;
$F_R = F_t \tan 20° = 9747$ N.

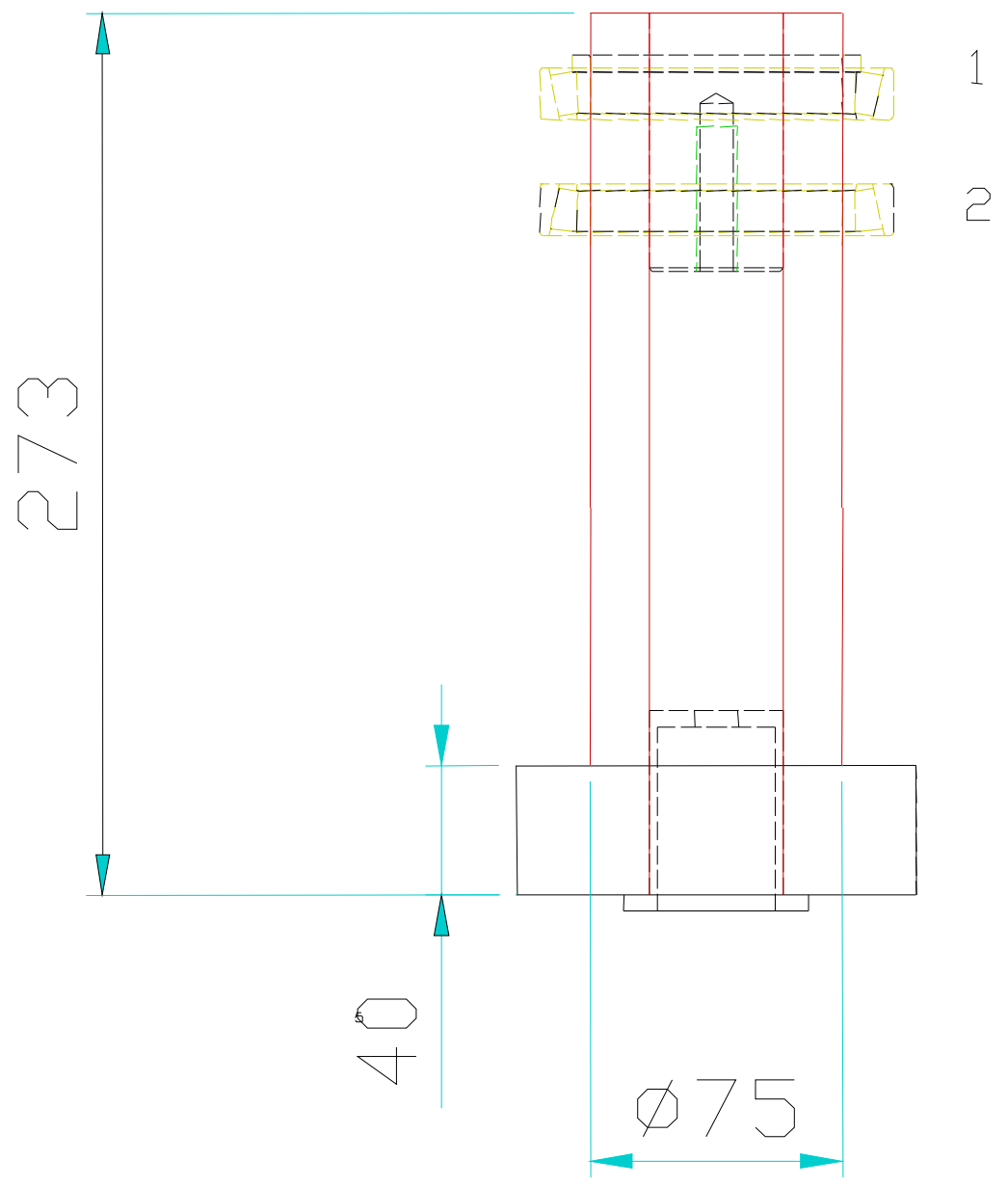

**Figure 4.2** Theoretical shaft dimensions.

A compression force of about 200N can also be considered, induced by shocks. To dimension a shaft means to check the resistance and stiffness of the component. It is necessary to retrieve the stresses, in order to find bending and twisting moments. The shaft can be seen as a hammer beam with two hinges, as indicated in fig. 20. Estimated reactions are respectively:

$X_1$=200N; $Y_1$=35.211 kN; $Y_2$= 44.992 kN and lay on XY plane.

On XZ plane they are respectively: $Z_1$=96.750kN $Z_2$=123.625kN.

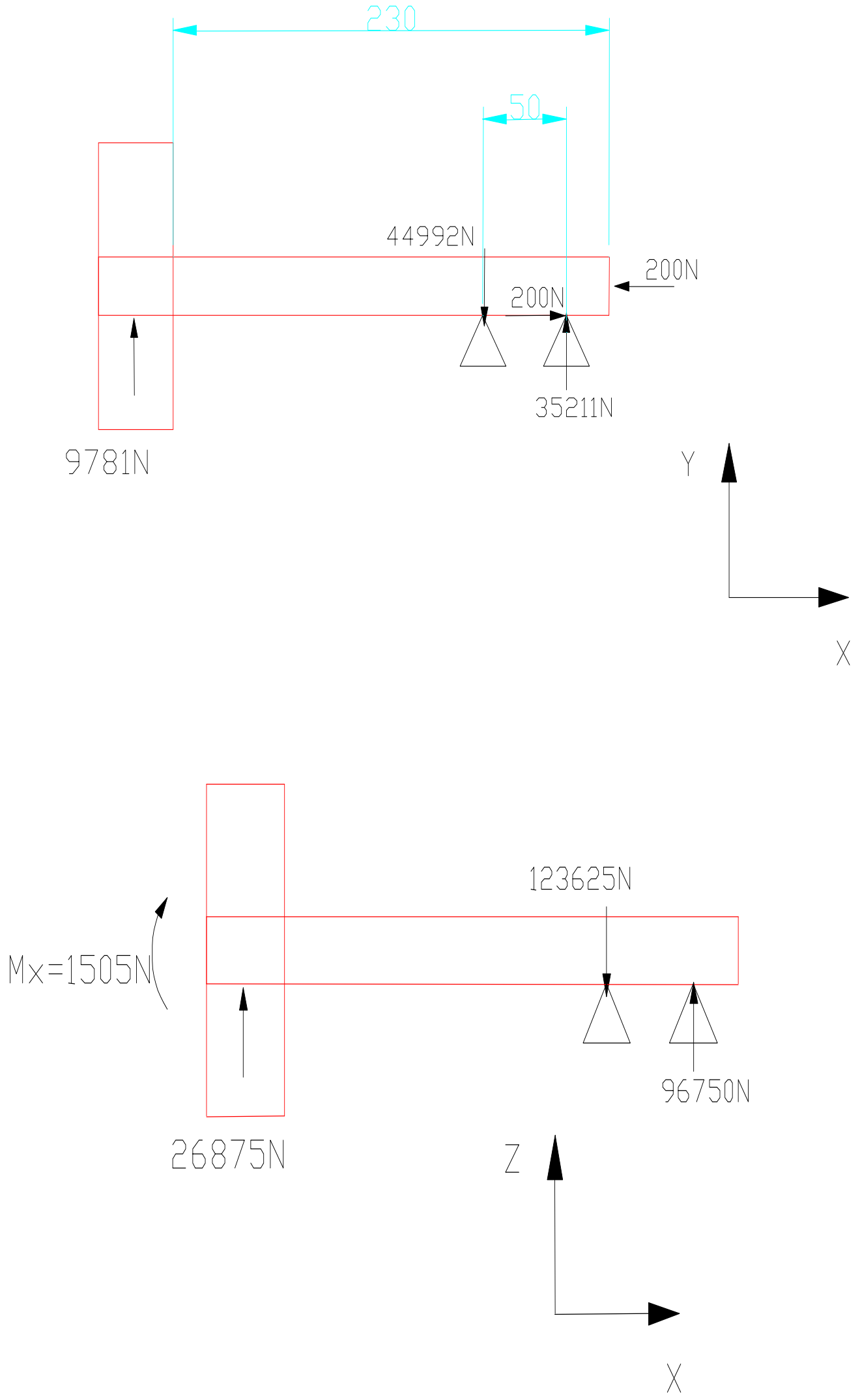

**Figure 4.3** Forces acting on shaft and external joints reactions.

$M_x$ represents torque moment acting on the shaft axis, due to the tangential component of the force exchanged between pinion and azimuth bearing, and it is constant on the whole shaft.

### 4.2.2 Stresses plot

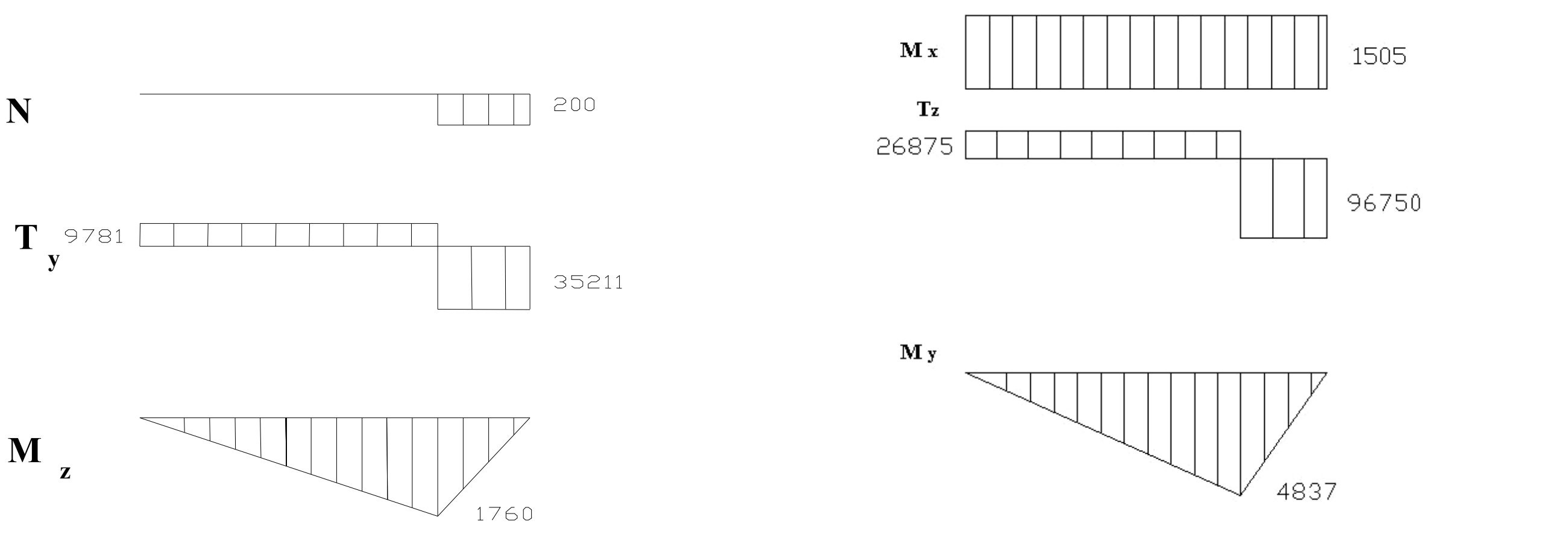

**Figure 4.4** Plots of : normal tension; shear in two perpendicular directions; bending moments in two perpendicular directions.

### 4.2.3 Maximum stresses determination

Goodman fatigue criterion was taken into account to determine the diameter of the shaft where the pinion is fastened: it consists in comparing the maximum stress acting on an element with break stress of material. In a static case , the limit value is given by the maximum material stress (σ yield). This, however, does not occur in dynamic case, as alternate stress, defined as $\sigma_a = \frac{\sigma_{\max} - \sigma_{\min}}{2}$ and average stress $\sigma_m = \frac{\sigma_{\max} + \sigma_{\min}}{2}$ vary in a different way according to active loads. Indicating a generic tension state with P on the Haigh diagram, an increasing external load causes a shift to the limit straight line ( passing through points (0;$\sigma_l$) and (0;$\sigma_R$) ). Safety factor is determined as the ratio between the segments of load line O'P e O'P', that is: $n = \frac{OP'}{OP}$, and

$$\frac{OP'}{OP} = \frac{\sigma_{a'}}{\sigma_a} = \frac{\sigma_{m'}}{\sigma_m} = 1 \Big/ \left( \frac{\sigma_m}{\sigma_r} + \frac{\sigma_a}{\sigma_f} \right).$$

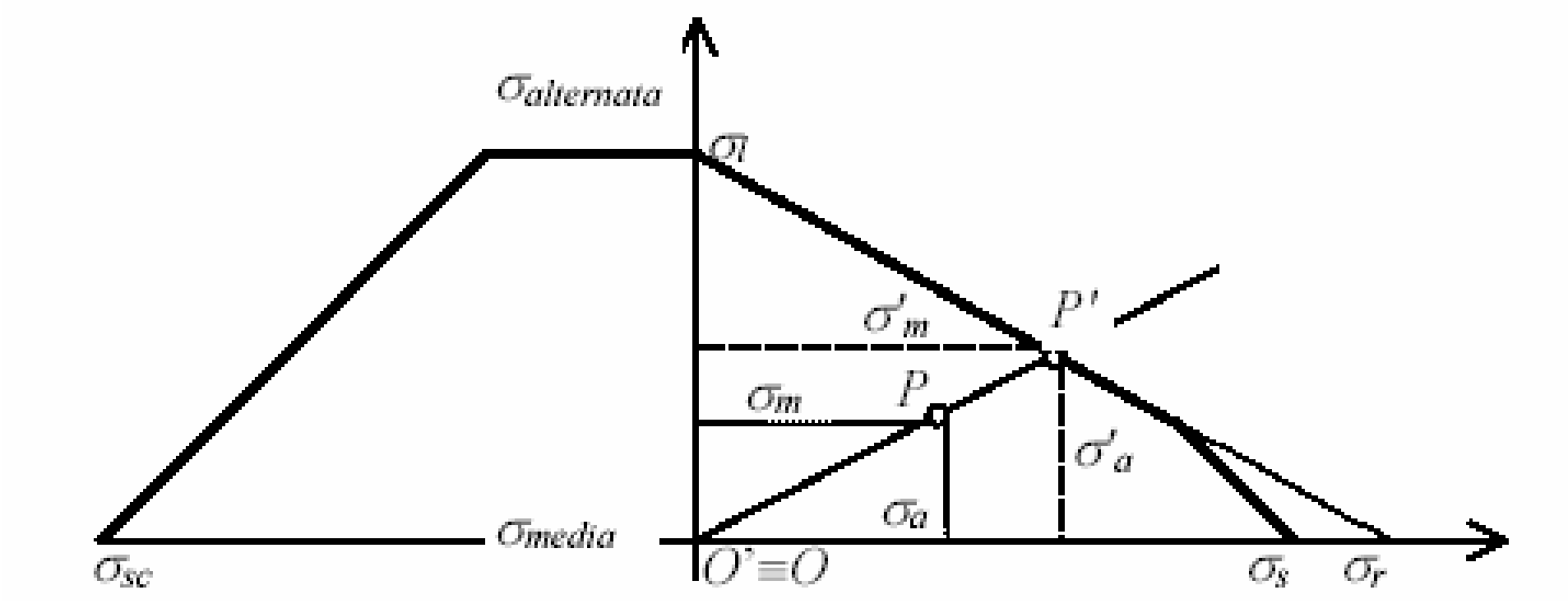

**Figure 4.5** Haigh diagram with constant ratio $\sigma_a / \sigma_m$.

Using a 16NiCr4 steel with limit tension $\sigma_R$= 1080 N/mm$^2$, and supposing a safety factor SF=0.5 ,it leads to: $\sigma_{amm} = 0.5\sigma_R = 540$ N/mm$^2$.

Limit fatigue stress must be modified and recalculated as follows:

$S_e = k_a k_b k_c k_d k_e \sigma_{amm}$ , where:

- $k_a$ is a surface factor, that we put 0.8;
- $k_b$ depends on the diameter, and assuming it to be 75 mm, then: $k_b = \left(\frac{D}{7.62}\right)^{-0.1133}$ = 0.77;
- $k_c$ is load factor, supposed to be equal to 1;
- $k_d$ =1 is a factor depending on temperature;
- $k_e$ is a compound safety factor, put equal to 0.9.

Thus, $S_e$ ,as a diameter function, is: $S_e$ =376.61$D^{-0.1133}$.

Alternate stress, due to bending, is given by: $\sigma_a = \frac{32\sqrt{M_y^2 + M_z^2}}{\pi d^3} = \frac{52231.544 * 10^3}{d^3}$ N/mm$^2$.

With a crack scale factor $k_t$=1.6 , divided by a coefficient:

$\delta_{st} = 1 + 0.83(k_t - 1)\sqrt[4]{\frac{300}{\sigma_{P0.2}}}$ = 1.33, with $\sigma_{P0.2}$=600 N/mm$^2$, and $K_{st}$ = 1.2.

We obtain:

$$\sigma_{a'} = \frac{62677.852 * 10^3}{d^3}$$

Tangential stress is equivalent to:

$\tau_m = \frac{16M_t}{\pi d^3} = \frac{7639.4 * 10^3}{d^3}$. Von Mises stress is $\sigma_m = \sqrt{3}\tau_m = \frac{13231.8 * 10^3}{d^3}$ .

Applying Goodman criterion with a coefficient taking into account reliability and safety equal to 3, we have:

$$\frac{\sigma_{a'}}{S_e}+\frac{\sigma_m}{\sigma_R}\leq\frac{1}{3}$$

This condition is fulfilled for d ≥ 83 mm. Therefore it is possible to select a shaft with an external diameter of 85 mm, rather than 75 mm considered previously.

### 4.2.4 Strain verification

A necessary but not sufficient condition to design a shaft is the strength verification; for more precise calculations in fact a strain verification must be lead. In particular, strain is due to flexure, so that misalignments and irregularity in mating of the spur gears need to be avoided.

In literature a maximum value of displacement for spur and helical wheels equal to 0.13 mm is reported. Moreover, the relative inclination between the axes must be inferior to 0.03°. Maximum beam deflection, relative to one end, where force is transmitted by the pinion, can be obtained by integrating the equation of deformation curve $y''=\frac{M\,(y)}{EI}$ twice. These values are usually already scheduled. Maximum deflection can be found through the following formula[4]:

$$\delta_{\max}=\frac{Pb^2L}{3EI}=0.048\text{ mm.}$$

Where E=206000 N/mm$^2$; $I=\frac{\pi(D_e^4-D_i^4)}{64}$; $P=F_r$;

*L=a+b,*

$D_e$=85 mm; $D_i$ = 40 mm.

$$\alpha=\delta_{\max}/b=0.0152°.$$

Both conditions are fulfilled.

[4] Juvinall, R.C., Marshek, K. M., *Fundamentals of machine component design,* John Wiley & Sons, New York 2003, pp. 851.

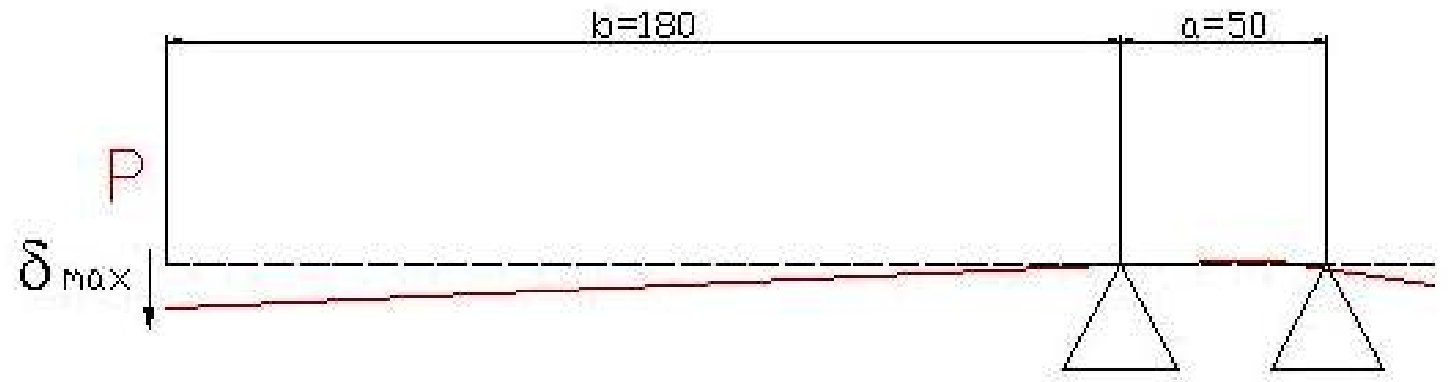

**Figure 4.6** Deformed shape of the beam subject to lumped load at one end.

## 4.3 Drive train

The drive train providing rotary motion is structured in the same way on both elevation and azimuth axis: double pinions rotating in opposite direction are coupled to gearboxes of 1/66 ratio. With a very small difference of torque transmitted by pinions, this system permit to compensate any back-lash between the teeth. There are two angular gearboxes of ½ ratio with two output shafts: one goes to the 1/66 planetary reduction box, the other is coupled to its “twin” through an elastic joint. In such a way the two subsystems are mechanically linked with a torsion bar.

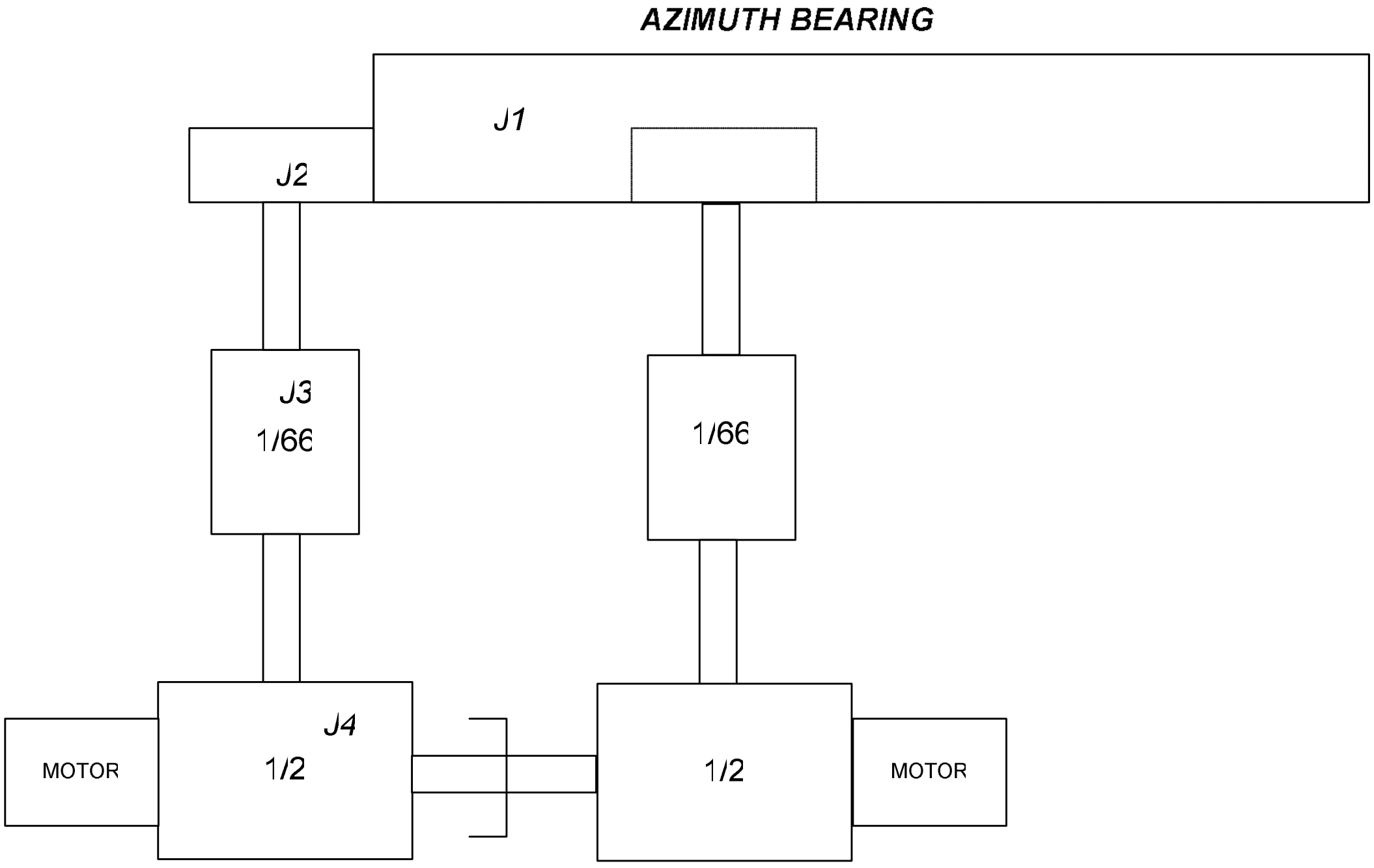

**Figure 4.7** Schematic of drive train.

By the free body diagram we can obtain following moment equations:

$$C_1 - M_p = J_1\dot{\omega}_1$$

$$C_2 - \frac{C_1}{10.9\eta_m} = J_2\dot{\omega}_2$$

$$C_3 - \frac{C_2}{66\,\eta_m} = J_3\dot{\omega}_3$$

$$C_4 - \frac{C_3}{2\eta_m} = J_4\dot{\omega}_4$$

$$C_m - C_4 = J_m\dot{\omega}_m$$

Being: $M_p$ the above mentioned preload in azimuth,
$\dot{\omega}_1$ =0.052 rad/s the maximum acceleration, $\dot{\omega}_2 = 10.9\dot{\omega}_1$ , $\dot{\omega}_3 = 64\,\dot{\omega}_2$ , $\dot{\omega}_4 = 2\dot{\omega}_3$ ,
$\dot{\omega}_m = 1440\;\dot{\omega}_1$ ,

$\eta_m$= 0.98 mechanical yield of gears. The highest moment of inertia occurs in azimuth, and the motor torque and inertia calculated refer to this case. Estimated value of azimuth moment is $J_1$=3000 kg·m$^2$; for the other components values are those provided by companies: $J_2$=4.233·10$^{-3}$ kg·m$^2$, $J_3$=9.52·10$^{-4}$, $J_4$=0.01747 kg·m$^2$.
Substituting all the known values, we have $C_4$=2.547. In order to calculate $C_m$, we have to know $J_m$, which can be determined by imposing the energy equilibrium equation, referred to motor shaft:

$$J_m = \left(\left(\frac{J_1}{10.9^2} + J_2\right)\frac{1}{66^2} + J_3\right)\frac{1}{4} + J_4 = 1.915\cdot 10^{-2}\text{kg·m}^2$$

### 4.3.1 Selection of stepper motors

We have decided to use the same DC stepper motors on both axes. The main advantages presented by this type of motor, among all, are: possibility of using it without a feedback closed loop, eliminating costs of encoders and their computer interfaces, and easy control by computer through a simple source of pulses (Tosti,G., Busso, M., et al., 2003).
The value of $C_m$ corresponding to calculated $J_m$ is 4 Nm. Therefore a suitable stepper motor for our application is Pythron ZSH 87/3 200.5 PLE80/1 3:1, with 200 steps, an accuracy of 1.8 °, high dynamics and high holding torques, up to 17 Nm (after the gearbox), and it's also compatible with ministep mode. It has an internal reduction box with 1:3 ratio, so that $J_m$ calculated, referred to the actual motor axis, is reduced of a factor 3, and the total transmission ratio is 1:4320. In such way the torque provided can be further amplified. With the selected controller SLS-4x PAB93-70 with four axes, we have a nominal global resolution pf 153600 points per revolution.

The maximum velocity of 1.5 °/s of the azimuth bearing or gear sector corresponds to 360 rpm of the motor. However, even if the $J_m$ indicated by the catalogue is lower ($1.6 \cdot 10^{-3}$ kg·m²), as speed and acceleration performances are quite moderate, the choice can be considered acceptable.

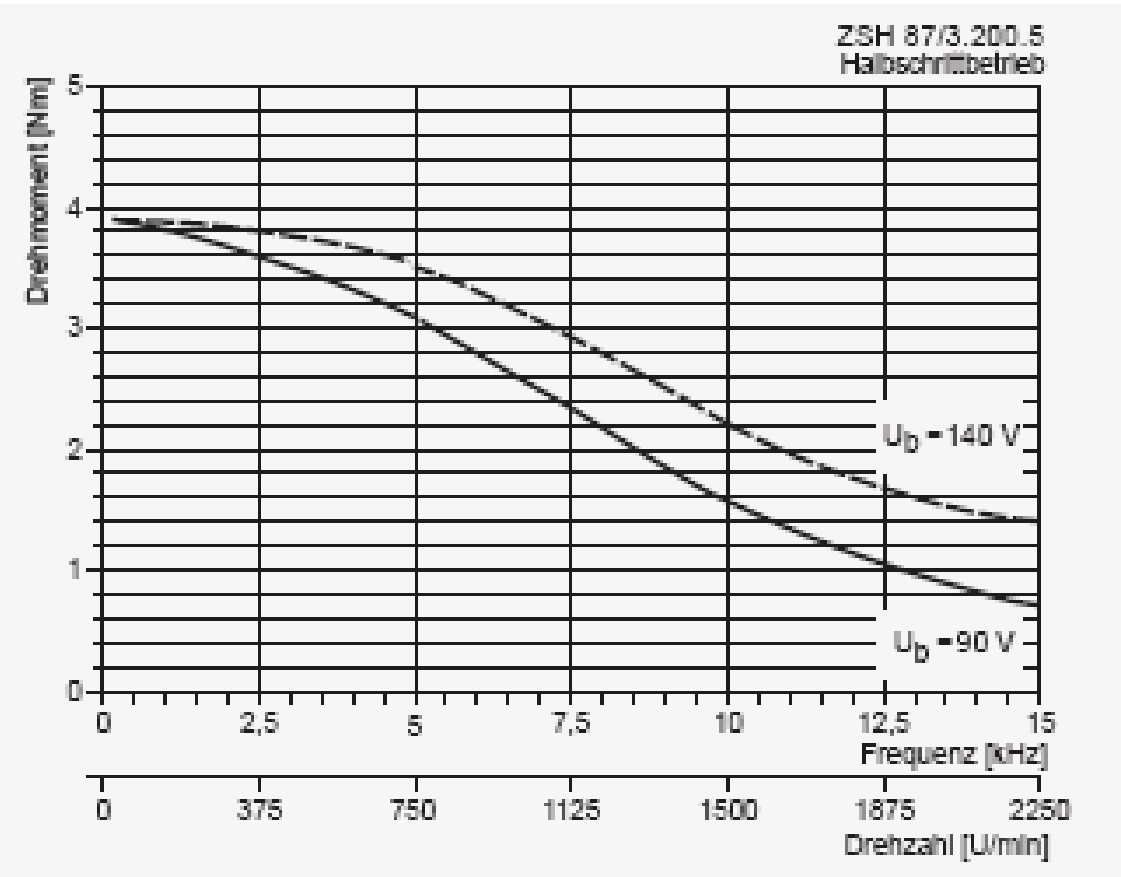

**Figure 4.8** Torque plot of selected motor vs velocity, given in rpm.

## 4.4 Joints

Bellow joints are interposed between the motor and gearbox: this kind of joints are able to compensate thermal variations, keeping the right coaxiality between the two elements. In fact, unlike rigid joints, they allow little axial displacements, angular plays, and, moreover, they can absorb shocks and sharp velocity or power changes. They are designed to resist, in particular, to shear stresses. Model PF- WK

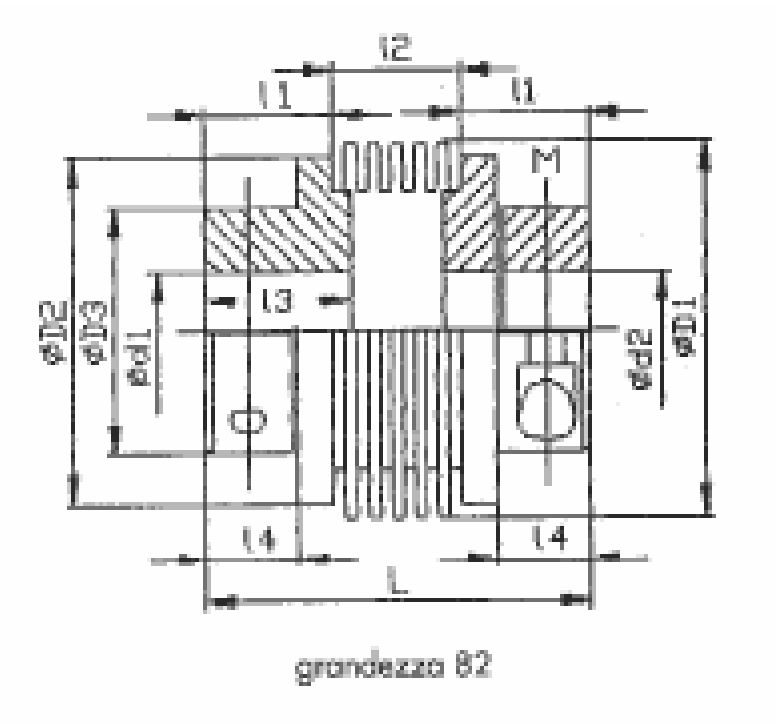

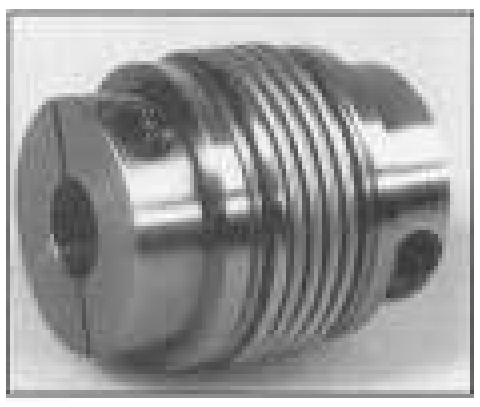

**Figure 4.9 A drawing and a picture of bellow joint by Favari.**

Indications provided by Favari company for a correct choice refer to three employment conditions with different service factors associated: normal, medium and exacting.

*TAN=9550·PAN/n*, where *TAN* is the required torque, in Nm;
*PAN* is the power (kW), *n* velocity ($min^{-1}$).

It must be checked that:
*TKN > TAN·SB·ST*,
Where *TKN* is the nominal torque transmitted by the joint, *ST* a thermal factor which is 1 for low temperatures, and *SB* the service factor, that for normal condition is comprised between 0.75 and 1.5. A proper joint for application was found to be WK 1050, with a length of 50 mm and a hole 16H7 for the shaft coupling. In fact TKN> 12.75 Nm is lower than that on the catalog (50 Nm, $TK_{max}$=100).

## 4.5 Welded joints

Another important aspect in structure design and during manufacturing process is the weld dimensioning. In fact, beyond bolts and rivets, another technique of linking together plates, profilate bars and pipes is through welded joints. In order to realize a permanent and continuous joint of metal sheets in a complex configuration as the case of fork arms, to confer enough stiffness and to enhance the eigenfrequencies of the system, we have to choose the right welding joint type.
The first Nasmyth focus fork arm has been taken into consideration, and, in particular, four welding joints:
the upper two by the housing bearing and the lower two by the base plate.
A metal inert gas welding process has been considered, which is of good quality and also very easy for almost all metals.
A fillet weld, T-type, has been selected, fixing a guess (starting) thickness *t* of 4 mm, on the basis of the available layout. Main dimensions of a welded joint are shown in fig. xxx.
For the resistance check of the joint I made reference to Standards UNI 100011-88 for steel structures, where yield and admissible stresses are indicated. Choosing a Fe430 for a t ≤ 40 mm $\sigma_{adm}$= 190 N/mm$^2$. We can consider that the proper weight of AMICA rack, assumed to be equal to 5 kN at the utmost, produces an effect of bending and twisting moments beyond a shear force.

The tangential stress due to twist can be written as: $\tau_{tors} = K_\tau = \frac{M_t}{J_P} r$ , where $J_p$ is the polar moment of inertia: $J_P = \int_A r^2 dA = \int_A (x^2 + y^2) dA = I_x + I_z$ .

For symmetry condition, analysis can be reduced to the solution of two joints with half the load.

As twisting moment acts along a direction perpendicular to the distance from the centre of mass of the welding sections, it can be divided in two components, as indicated in fig.4.7, which are given by the formulas:

$$\tau_{//} = \frac{FL}{J_P}\frac{h}{2}, \quad \tau_{\perp} = \frac{FL}{J_P}\frac{B}{2}$$

Moreover, it must be considered the shear tangential stress adding to the orthogonal twisting component, so that the sum is:

$$\tau_{\perp} = \tau_{\perp tw} + \tau_{\perp sh} = \frac{F}{2tl} + \frac{FL}{J_P}\frac{B}{2} .$$

Extending the polar moment of inertia, referred to welding section centre of mass, with all its terms, we obtain:

$$J_P = 2\left\{\left[\frac{1}{12}t^3 B + tB\left(\frac{h}{2}\right)^2\right] + \frac{1}{12}tB^3 + tB\left(\frac{B}{2}\right)^2\right\}$$

Substituting the known parameters B=200, h=966, and ignoring the first term we have:

$J_P$= 1.97297·10$^8$ N/mm$^2$, $\tau_{//}$ =3.12 N/mm$^2$ , and $\tau_{\perp}$= 2.20 N/mm$^2$.

The force projected on the plane containing the two welded joints produces also a tension effect due to bending laying on XZ (see fig.4.8) . It is given by:

$$\sigma_{b,max} = \frac{F \cdot x \cos 16}{I_z}\frac{H}{2}, \text{ where } I_z = 2\left[\frac{1}{12}Bt^3 + tB\left(\frac{H+t}{2}\right)^2\right].$$

Knowing that x = 437 mm we have: $\sigma_{max}$=1.34 N/mm$^2$.

The standards (CNR 10021, 1986) prescribe two conditions to check the resistance of welded joints:

$$\begin{cases} \sigma_{id} = \sqrt{\sigma_{max}^2 + \tau_{//}^2 + \tau_{\perp}^2} \le 0.85\sigma_{adm} \\ |\sigma_{max}| + |\tau_{//}| \le \sigma_{adm} \end{cases}$$

As the values determined are far below the limit, the conditions are completely fulfilled.

For this reason it is possible to use segments of joints along portion of the surface that must remain in contact, as indicated in fig.4.9.

**Figure 4.10** Typical dimensions of a welded fillet joint.

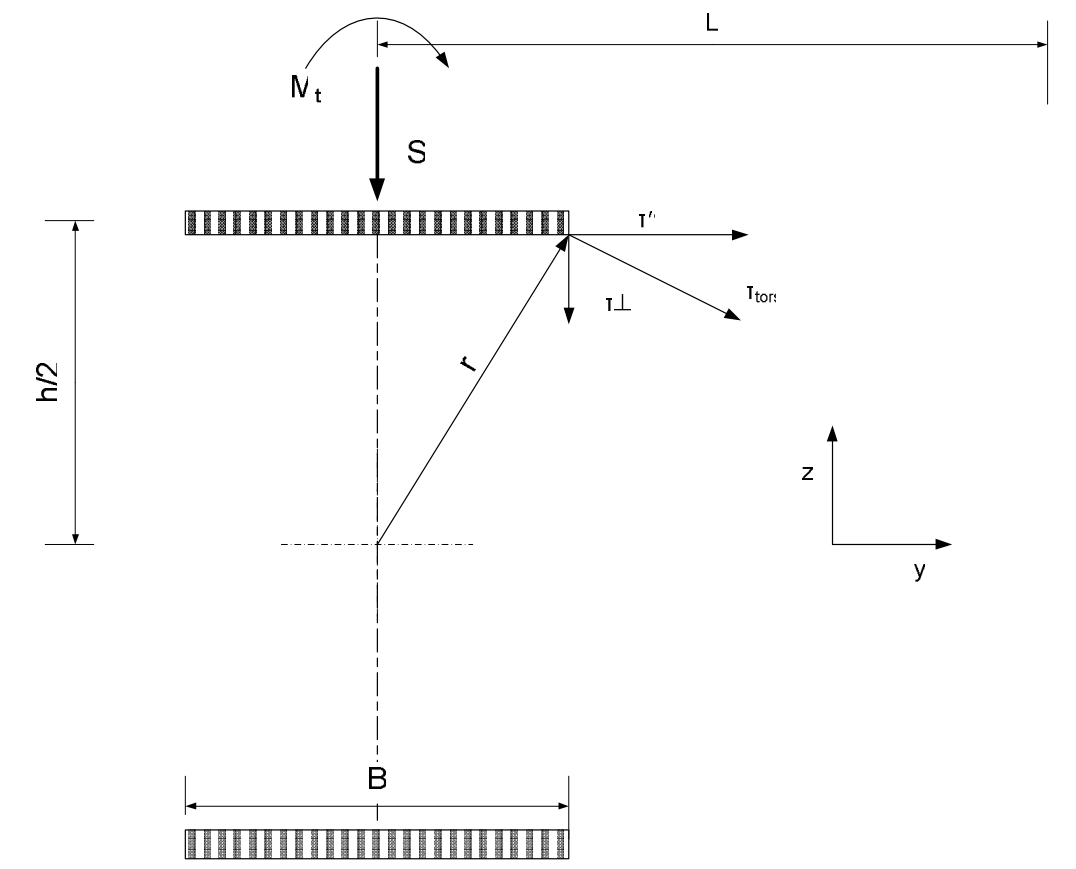

**Figure 4.11** A schematic of the swinging moment acting upon the welding sections.

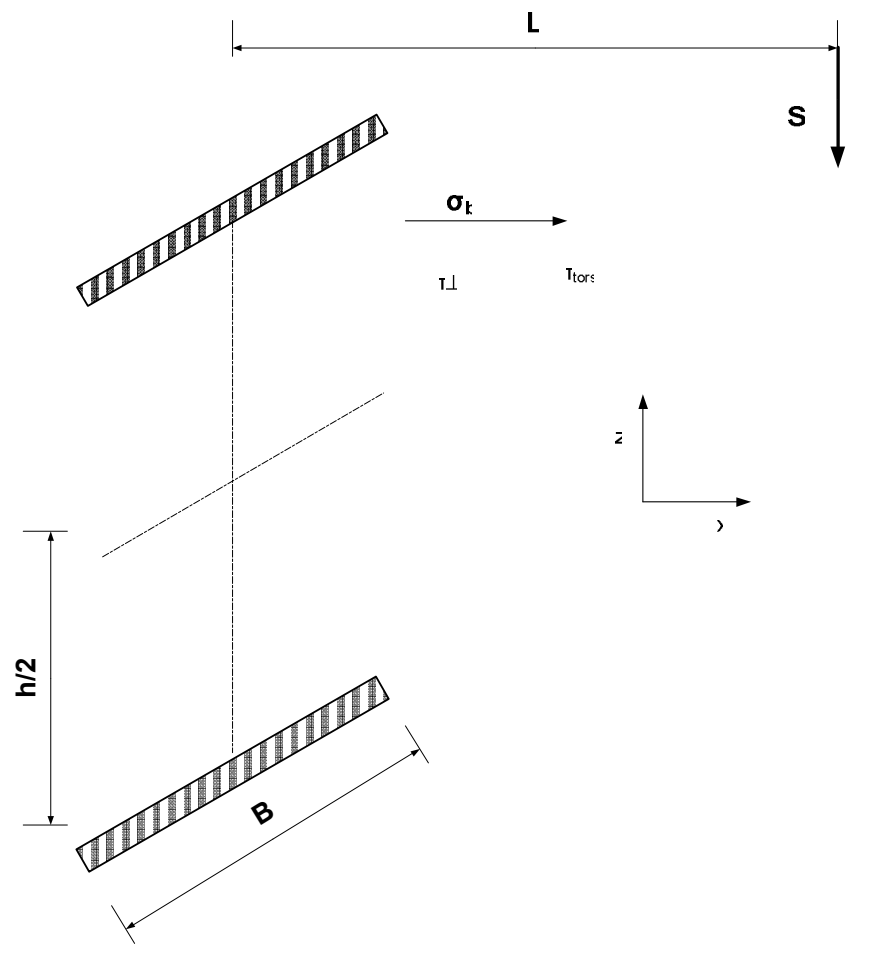

**Figure 4.12** A schematic of the bending moment acting upon the welding sections.

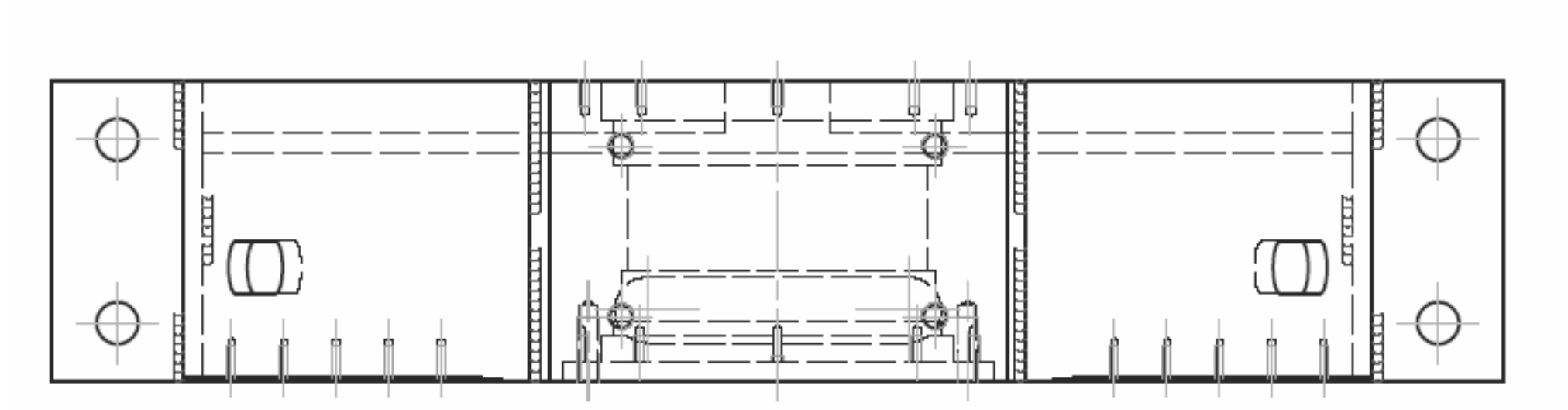

**Figure 4.13** A top view with the layout of welding segments: the six external joints are on the fork base, while the internal ones connect the side plates to the housing bearing.

# CHAPTER 5 Systematic errors

## 5.1 Astrometric errors

Once observing schedule is started, the first thing the telescope must do is pointing the object. The procedure can be divided in three main stages, which are necessary to convert star's catalog position into settings, namely to transform *mean* position into the actual observed position.

As a matter of fact, usually an object's right ascension and declination from the catalog are referred to standard epoch, which is the *mean* position, so it is not corrected for time and location effects. In the first stage mean coordinates must be converted to apparent: this involves such astronomical corrections as proper motion, precession and nutation. Then, in the second stage, we pass from apparent to topocentric position through correction for diurnal aberration and refraction, obtaining the observed position. Final conversion is based on pointing corrections and it involves systematic optical and mechanical errors (Wallace, on Internet).

Most of them can be described by correct physical models and easily implemented via software (see Sick, J., at http://homepage.mac.com/jonathansick/).

### 5.1.1 Annual aberration

The light which comes from a star takes a finite amount of time to reach the observer. During this time the Earth moves in its orbit along the sun and this appears as an effect of displacement of the star in the sky. Calling $\Delta\theta$ the displacement , $\theta$ the elevation angle, and v the velocity of the observer, the correction is given by:

$$\Delta\theta = \frac{v}{c} \sin\theta$$

Its maximum value is 20''.

### 5.1.2 Stellar Parallax

It is the apparent displacement of a nearby star relative to more distant stars, as seen by the different orbital positions of the Earth.It is measured in [arscec/parsec].It is less than 0.1″ so that this effect is too small and is undetectable without extremely precise measurements. It becomes significant only for large telescopes that use adaptive optics facilities.
Anyway, for targets such as characterization of solar system bodies and planets solar or planetary parallax, values increase up to 9 arcscec/UA.

### 5.1.3 Precession

Earth's axis rotates about the pole of the ecliptic due to the combination of forces exerted by the Sun, Moon, and, in a lesser degree, by planets. The ecliptic plane tilts so that the equinox precesses about 12" per century, and the obliquity of the ecliptic decreases at 47 " per century.
A method of computing precession consists in using rotating matrices, with three fundamental passages:

- transforming from Right Ascension and declination to geocentric equatorial coordinates;
- rotating the equatorial coordinates;
- transforming the precessed coordinates back to the precessed Right Ascension and declination.

It corresponds to a set of rotations: the first about z-axis, another about x-axis, and the third about z-axis. Assuming two coordinate systems so that the plane x′y′ is parallel to the ecliptic at time $T_0$ and x" y" to the ecliptic at time $T_0 + T$, $\Pi$ and $\Lambda$ can be defined respectively as the angles between the x′-x" and the vernal equinox $\gamma_0$ at epoch $T_0$ and the vernal equinox $\gamma$ at $T_0 + T$; $p$ is the precession in longitude and $\pi$ is the angle between ecliptic at $T_0$ and at $T_0 + T$.

$$\begin{Bmatrix} x \\ y \\ z \end{Bmatrix} = [R] \begin{Bmatrix} x_0 \\ y_0 \\ z_0 \end{Bmatrix}$$

where $[R] = R_z(-\Pi - p)\, R_x(\pi) R_z(\Pi)$

$$\pi = (47''.0029 - 0''.06603 \cdot T_0 + 0''.000598 \cdot T_0^2) \cdot T - (0''.03302 + 0''.000598 \cdot T_0) \cdot T^2 + 0.00006 \cdot T^3$$

$$\Pi = (174''.876383889 + 3289''.4789\mathrm{T}_0 + 0''.60622\mathrm{T}_0^2) - (869''.8089 - 0''.50491\mathrm{T}_0)\mathrm{T} + 0.03536\mathrm{T}^2$$

$$p = (5029.0966 + 2.22226\,T_0 - 0.000042\,T_0^2)T + (1.11113 - 0.000042\,T_0)T^2 - 0.000006\,T^3$$

$\Lambda = \Pi + p$

Returning again to the equatorial coordinates, it can be written:

$$[R] = R_z(-z)\,R_x(\theta) R_z(-\zeta),$$

where:

$$\zeta = (2306''.2181 + 1''.39656\ T_0 - 0''.000139\ T_0^2)T + \\ + (0''.30188 - 0''.000345\ T_0)T^2 + 0''.017998\ T^3$$

$$\theta = (2004''.3109 - 0''.85330\ T_0 - 0''.000217\ T_0^2)T + \\ - (0''.42665 - 0''.000217\ T_0)T^2 - 0''.041833\ T^3$$

$$z = \zeta + (0''.79280 + 0''.000411\ T_0)T^2 + 0''.000205\ T^3 .$$

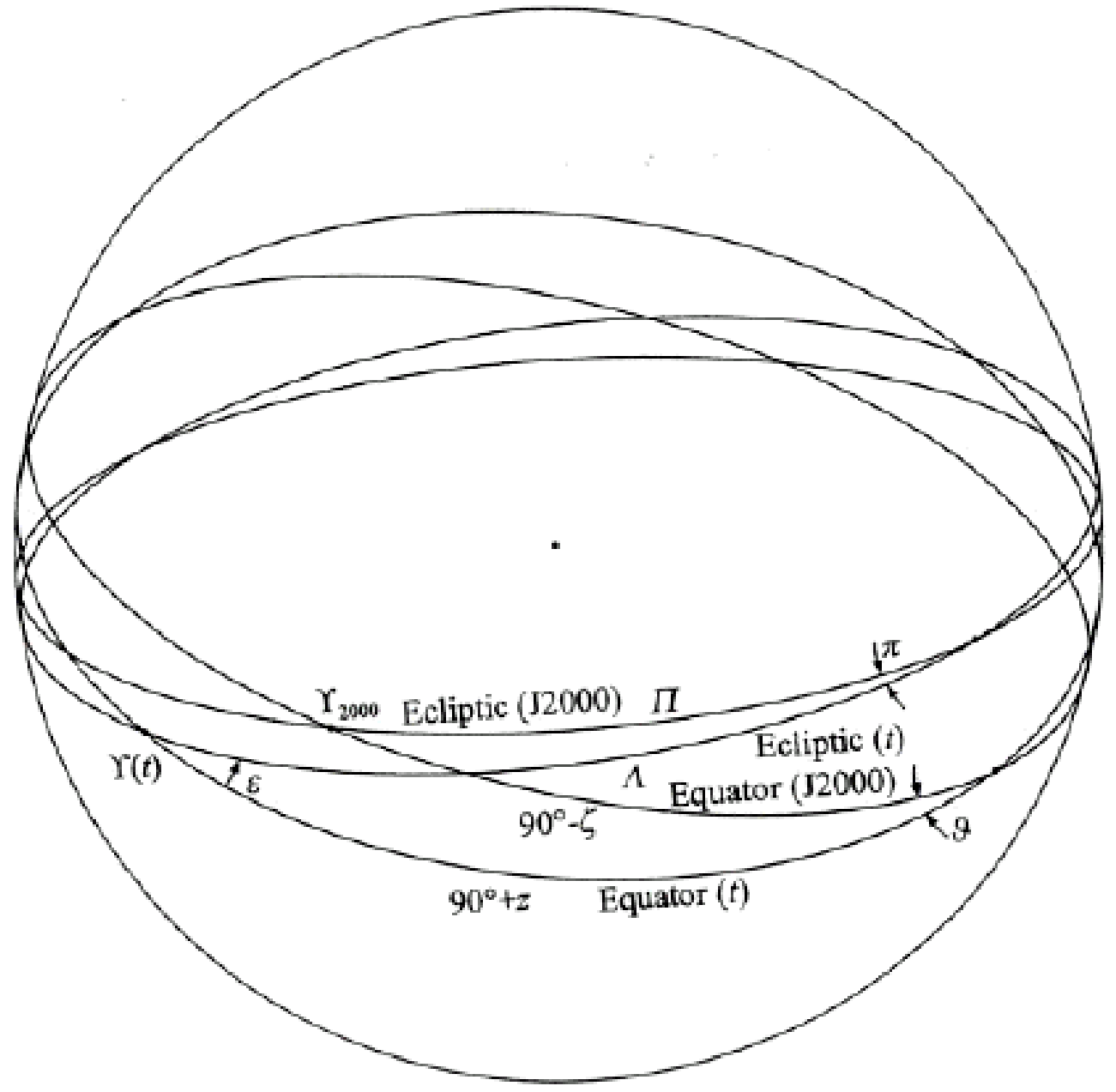

**Figure 5.1** A representation of the precession angles.

### 5.1.4 Nutation

While the precession indicates the variation of orientation of the polar axis along the centuries, there's another phenomenon to consider due to small, periodic variations, called nutation. As a consequence the Earth's true pole rotates every 18.6 years around the mean pole, whose motion is predicted by the precession. It can be described assuming three rotations: one around x of $-\varepsilon' = -(\varepsilon+\Delta\varepsilon)$ , another of an angle $\Delta\Psi$ in longitude about z axis, then once again about x of $+\ \varepsilon$. The measures amplitudes are $\Delta\Psi=17''$ , and $\Delta\ \varepsilon=9''$ ,

$$\begin{Bmatrix} x \\ y \\ z \end{Bmatrix} = [\ N\ ] \begin{Bmatrix} x_0 \\ y_0 \\ z_0 \end{Bmatrix}$$

Where $N= R_x(-\varepsilon')\ R_z\ (\Delta\Psi)R_x(\varepsilon),$ and

$$\Delta\psi = -17''.200 \cdot \sin(\Omega) + 0''.206 \cdot \sin(2\Omega) + 0''.143 \cdot \sin(l') - 1''.319 \cdot \sin(2(F - D + \Omega)) + \\ - 0''.227 \cdot \sin(2(F + \Omega))$$

$$\Delta\varepsilon = 9''.203 \cdot \cos(\Omega) - 0''.090 \cdot \cos(2\Omega) + 0''.574 \cdot \cos(2(F - D + \Omega)) + \\ + 0''.098 \cdot \cos(2(F + \Omega)).$$

### 5.1.5 Polar motion

Other perturbations that are not computed in precession an nutation concern the change in position of the pole, due to random realignments, which are due to tectonic plate drift and earthquakes. This usually amounts to no more than 50 meters per year which, referred to Earth radius, is less than 0.1″ in an year.

### 5.1.6 Proper stellar motion

The proper motion of a celestial body represents its motion with respect to the so called *fixed stars*, which don't change significantly their position over a long period time. Catalogs give corrected values for Δα and Δδ over a certain period. Correction formulae (by Green) for proper motion are:

$$\alpha' - \alpha = \left( \mu_\alpha + \frac{1}{2} t \frac{d\mu_\alpha}{dt} \right) t$$
$$\delta' - \delta = \left( \mu_\delta + \frac{1}{2} t \frac{d\mu_\delta}{dt} \right) t$$

where $\mu_\alpha$ is the annual proper in α in arcsec per year or century;
$\mu_\beta$ indicates the annual proper in β in arcsec per year or century.
A large amount of them is less than 1″, so its contribution compared to other errors is irrelevant.

### 5.1.7 Diurnal aberration

It is the same as annual, but instead of velocity around the sun , rotational motion is considered. So

$$\frac{v}{c} r \cos L = 0''.320\, r \cos L \, ,$$

where L is the geocentric latitude.

Passing to coordinates (α,δ), the correction is:

$$\Delta\alpha = 0''.0213 r \cos L \cosh \sec\delta$$

$$\Delta\delta = 0''.320 r \cos L \sinh \sin\delta$$

As its value is about 0.1'', it is almost negligible.

### 5.1.8 Atmospheric refraction

A light beam that penetrates the atmosphere passes from interstellar space of refraction index $n_0=1$ to a mean of $n > n_0$ .

Unlike precession, nutation, and aberration errors, which can be theoretically predicted, refraction depends on a lot of factors such as pressure and temperature inversions, therefore data tables are necessary to appreciate it.

If we consider pure refraction, supposing that the layers are plane-parallel and have no discontinuity, we can calculate the refraction angle R by the formula: R=( $n_f$-1) tanZ', being Z' the apparent angle seen by the observer, and $n_f$ the refraction index of the lowest layer. A more precise equation taking into account the variation of refraction with the zenith distance, is that given by Eisele and Shannon :

$$R = \frac{17P}{460 + T_F}(57.626039 \tan Z - 0.05813517 \tan^3 Z)$$

valid for 0 ≤ Z ≤ 85°, and

$$R = \frac{17\,P}{460\ + T_F} 871\ .94412\ \left[e^{-0.53520501\ (90^\circ - Z)} + e^{-0.107041\ (90^\circ - Z)}\right]$$

For 85° ≤ Z ≤ 90.6°, being:

Z = zenith distance

P = atmospheric pressure expressed in inches of Hg on the Earth's surface

$T_F$ = atmospheric temperature in °F on the Earth's surface.

The change of the refraction index is mainly due to temperature gradients, and less to pressure variations. Therefore all the structure that surrounds the telescope must be kept isothermal as much as possible.

Its typical value is 30'.

## 5.2 Mechanical errors

### 5.2.1 Azimuth axis misalignment

The zenith misalignment for an alt-az mount can be seen as an effect of two rotations: one around North axis, indicated as x, of an $\alpha$ angle, the other around azimuth axis z, of $\beta$. Rotations can be represented through the product of two matrices:

$$A = \begin{bmatrix} 1 & 0 & 0 \\ 0 & \cos\alpha & \sin\alpha \\ 0 & -\sin\alpha & \cos\alpha \end{bmatrix}, \qquad B = \begin{bmatrix} \cos\beta & \sin\beta & 0 \\ -\sin\beta & \cos\beta & 0 \\ 0 & 0 & 1 \end{bmatrix}$$

Resulting matrix sets the transformation of coordinates from (x,y,z,) to (x″,y″,z″) :

$[x'' \; y'' \; z'']^T = B \cdot A [x \; y \; z]^T$.

In general $\alpha$ is small, while $\beta$ can be even larger. Using approximation for small angles ($\sin\alpha = \alpha$, $\cos\alpha = 1$), the following system of equations is obtained:

$$\begin{cases} x'' = x\cos\beta + (y + z\alpha)\sin\beta \\ y'' = -x\sin\beta + (y + z\alpha)\cos\beta \\ z'' = -y\alpha + z \end{cases}$$

The angles $\alpha$ and $\beta$ can be determined once known the difference between (A, Z) computed from the catalogue, and (A″ ,Z″) retrieved by the encoder:

$$\alpha = \frac{z - z''}{y}; \quad \sin\beta = \frac{x''(y + z\alpha) - xy''}{x^2 + (y + z\alpha)^2}$$

Coordinates x,y,z can be calculated with a spherical to rectangular system transformation:

$$\begin{cases} x = \sin Z \cos A \\ y = \sin Z \sin A \\ z = \cos Z \end{cases} \qquad \begin{cases} x'' = \sin Z'' \cos A'' \\ y'' = \sin Z'' \sin A'' \\ z'' = \cos Z'' \end{cases}$$

### 5.2.2 Collimation error

Collimation errors between optical and mechanical axes are mainly due to two complimentary effects:

- a static one, related to the construction of the structure, when optics was mounted, causing a constant offset;
- a dynamic one, deriving from the flexure of the tube.

The correction given by Wallace (1988), for an alt-az telescope is:

$$\Delta A = C_{ew} \cos Z$$

$$\Delta Z = C_{ns}$$

where $C_{ew}$ is the collimation angle measured in the direction East-West, $C_{ns}$ collimation angle in North-South. We can assume a maximum error of 15 arcsec for both coefficients, and, in any case, these values can be verified only after a complete assembling of the structure and a set of tests, in agreement with the constructor.

### 5.2.3 Driving rates

See paragraph 3.2.2.

### 5.2.4 Field rotation corrections

See paragraph 3.2.1.

### 5.2.5 Mount misalignments (intrinsic errors)

#### 5.2.5.1 Misalignment of altitude axis due to fork arm flexure

Another problem is that of non-perpendicularity of the two axes, due to mounting errors of the two fork arms or relative to their different dilatation which is due to sun exposure. Another source of mechanical mount error, besides the flexure, is given by different elongations of the two fork arms, due to different load conditions and, in a larger measure, to thermal gradient. The effect of such displacement is the non perpendicularity of the two axes.

Considering the maximum distance between the two arms of 1610 mm, the difference of elongation is given by:

$$p = \frac{0.3035 + 2.91 * 10^{-2}}{1610} = 42.66 \quad arc \ sec \ .$$

Using Napier's rule for an alt-az mount, expressing with (90°+ $p$) the real angle between azimuth and altitude axis due to the maximum deformation of an arm with respect to the other, the error in alt-az coordinates is:

$$\Delta A = p \sin(alt) = p \cos Z$$
$$\Delta Z = p \Delta A$$

Substituting the value of $p$ =42.66 arcsec , the two error curves are obtained for alt-az coordinates.

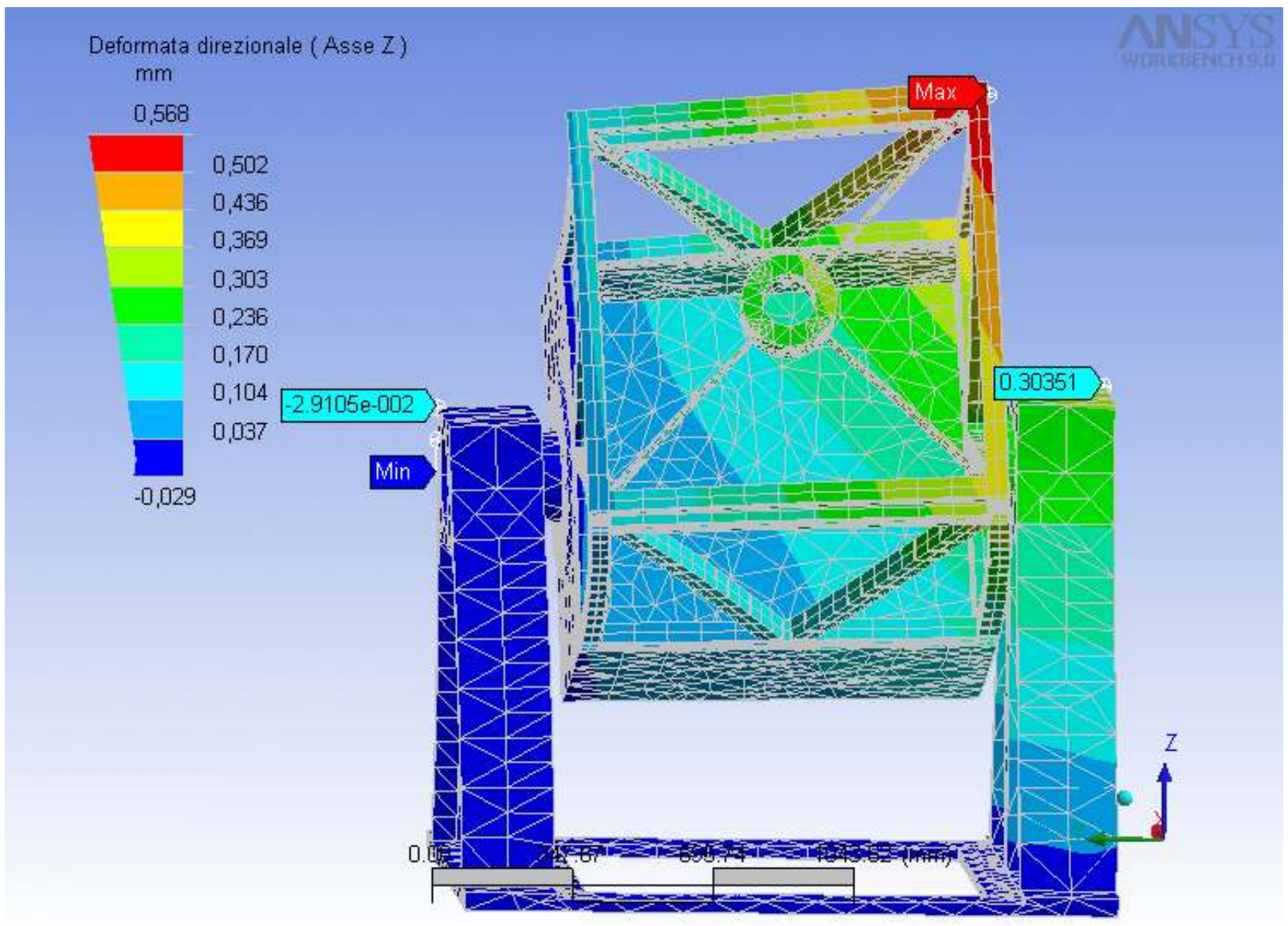

**Figure 5.1 I**n this figure displacements along Z, concerning two extreme points of the arms, are highlighted.

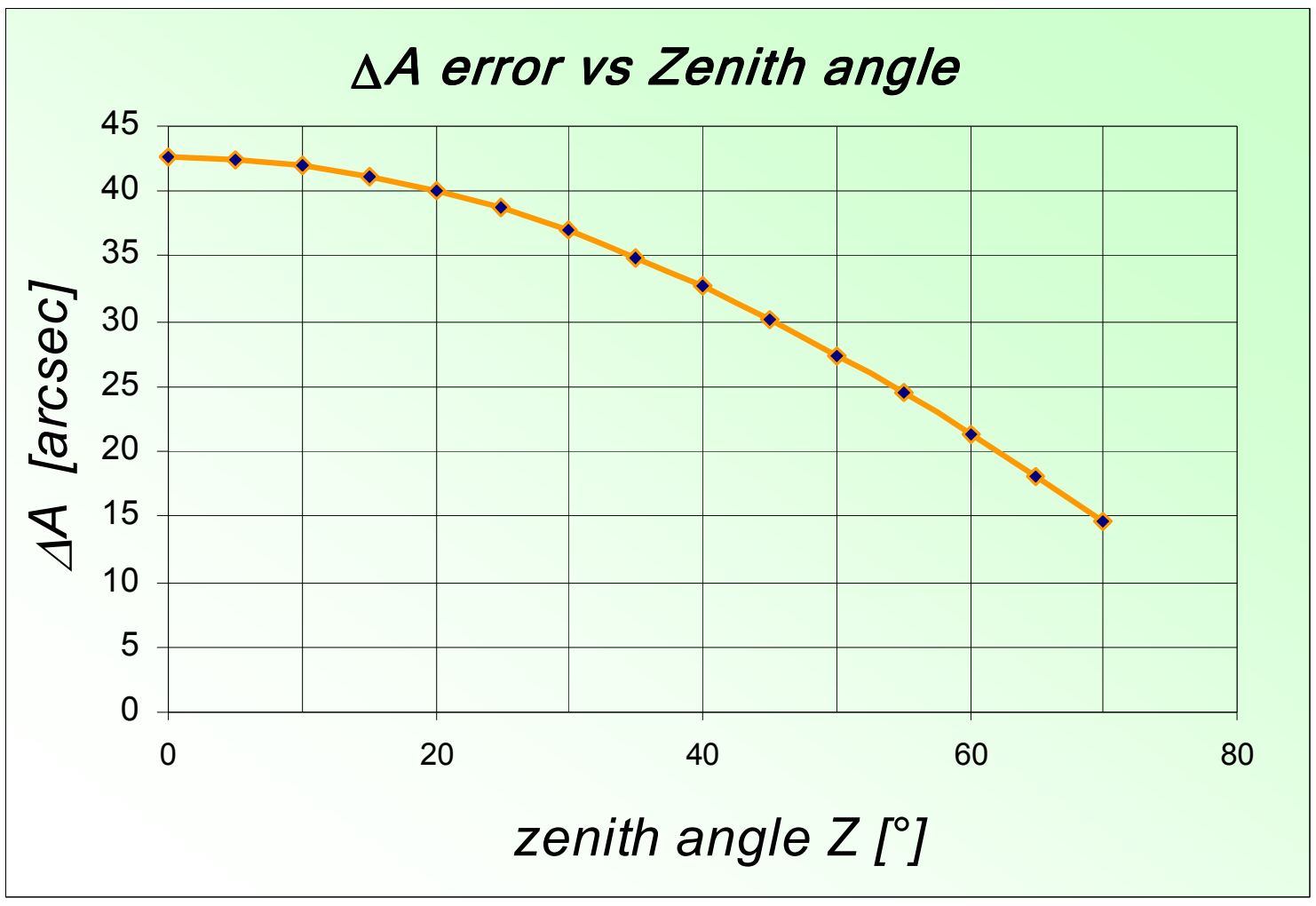

**Figure 5.2** Relationship between azimuth error and zenith angle due to non-perpendicularity of the altitude axis.

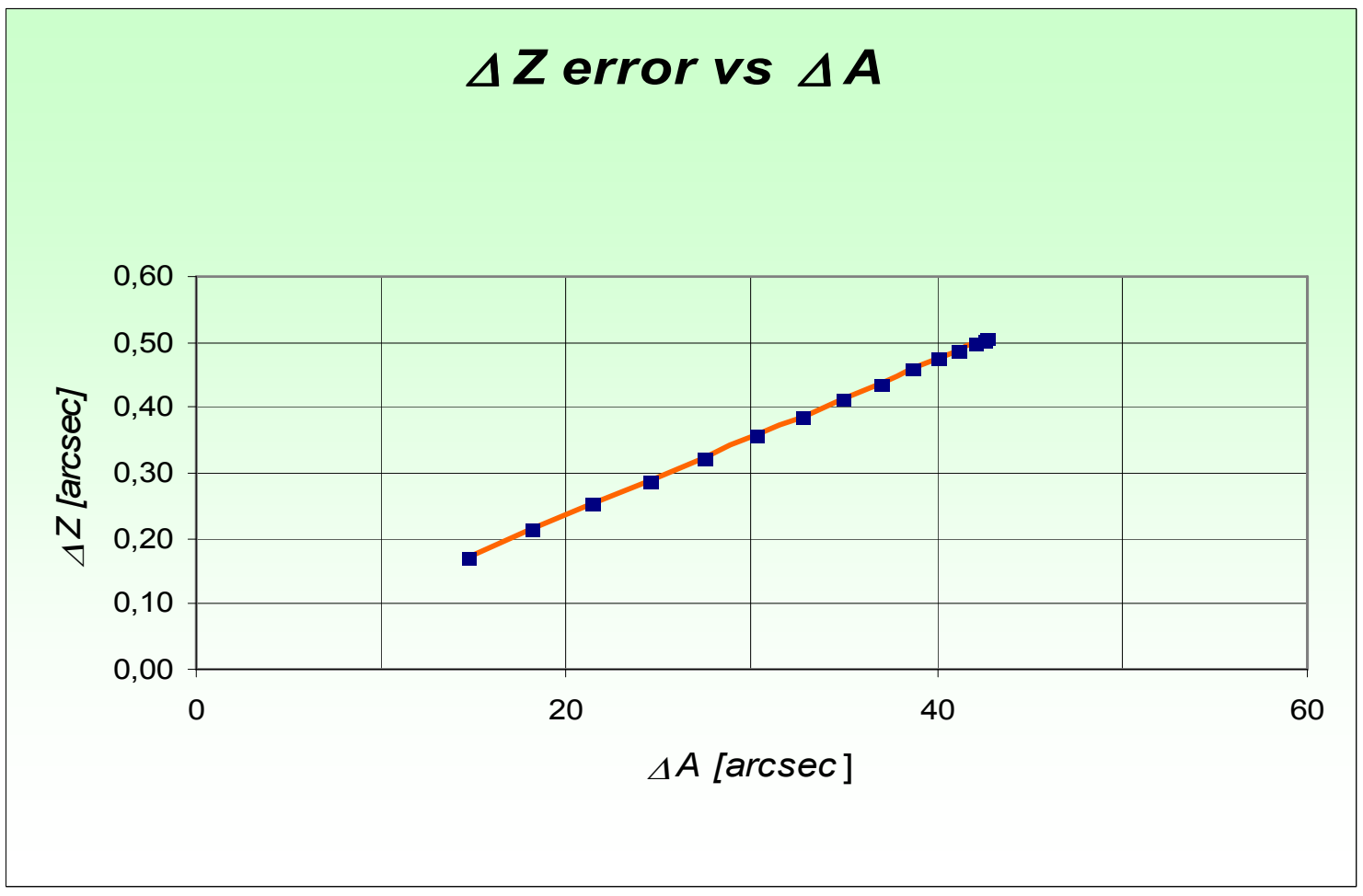

**Figure 5.3** Relationship between zenith and azimuth error due to non-perpendicularity of the altitude axis.

### 5.2.6 Periodic gearing errors

There are many types of gearing errors, depending on the precision of machine tool and linked to manufacturing process. The first to be mentioned is the centre distance variation. It is the variation in distance from the centre to the outer diameter of the gear. Even if it doesn't affect indexing (relative distance between centres), it may cause backlash problems.

Another source of error is the tooth thickness variation, which represents the variation in thickness of a tooth compared to its neighbors. Its value in arcmin is given by:

$e_{th} = 3428\left(\frac{\Delta B_0 - \Delta B_1}{D}\right)$, where $\Delta B_0$ , $\Delta B_1$ is the width of two adjacent teeth, D is the pitch diameter.

Pitch error is the difference in spacing along the pitch line, and is given, in arcmin, by:

$e_p = 6875\left(\frac{\Delta p_0 - \Delta p_1}{D}\right)$ where $\Delta p_0$, $\Delta p_1$ are the pitch errors of adjacent gear teeth.

For a spur gear with a grade JIS=1, the empirical formula for the pitch error is:

$P = \Delta p_0 - \Delta p_1 = 0.71W+2.0$, $W$ being the tolerance unit, defined as: $W = \sqrt[3]{d} + 0.65\,m$ . Substituting the known values: $e_p$=0.06 arcmin.

The involute error represents the deviation of tooth from true involute profile, and it is equal to:

$e_{ip} = 6875\left(\frac{\Delta a_1 - \Delta a_2}{D\cos\varphi}\right)$ being $\Delta a_1$ and $\Delta a_2$ the profile errors of adjacent gear teeth, φ the pressure angle (usually 20°).

The tooth profile error is also given by relationship: $\Delta a_1 - \Delta a_2 = 1.0m + 3.15$ µm, that converted to arcmin is: $e_{ip}$= 0.038 for the azimuth bearing.

Pitch diameter eccentricity is the eccentricity of the pitch line of the gear, and it is:

$e_{pd} = 3438\left(\frac{e\sin\theta_e}{R}\right)$ where e is eccentricity, R the radius, and $\Theta_e$ the rotation angle about the axis.

Then there's the lead angle error, which consist of a tilt in the face of the tooth away from perpendicularity:

$e_{la} = 6875\left(\frac{F\tan\lambda_e}{D}\right)$ where F is the face width and $\lambda_e$ the lead angle error.

Another experimental relationship to determine lead error, on the basis of parameters characterizing gears, is:

$la = 0.71\,(0.1b + 10)$ , where b is the tooth width. In our case its value is: $la=9.585$, and $e_{la}$=0.052arcmin.

At last, there is the lateral runout, produced by misalignment of the gear to rotation axis, given by:

$e_{lr} = 6875\left(\frac{F\sin\lambda}{D}\right)$, where F is the face width of the gear, λ the tilt angle of tooth. The empirical formulas (JIS B 1704) to determine lateral runout are:

$j = 1.1\sqrt[3]{d_a} + 5.5$ ;

$q = \frac{6d}{b+50} + 3$ ,where $d_a$ is outside diameter, b the tooth width and d the pitch diameter (all in mm). For a precision grade 1 the lateral runout is: $l_r=1.0q$ μm. For the larger gear in azimuth the calculated value is $l_r$=91.94, and $e_{lr}$=0.5 arcmin. It has a large impact on the gear tooth accuracy and depends mostly on gear size.

### 5.2.7 Bearing errors

Elevation bearings provided by SKF have a very little runout, that is typically held to be 5 μ at maximum. Considering a clearance between the two pairs being of 1438.5 mm, the runout generates a non-perpendicularity between the two axes of about 0.716 arcsec: the result is an increase of pointing error at small zenith distance. Indeed, the largest contribution of non perpendicularity is caused by the inclination of the upper structure hanging on azimuth bearing. In fact, examining the effective loads acting on the subassembly base chassis, it has been found out a vertical displacement of the outer ring of $7.98\cdot10^{-2}$ mm, so that deviation angle from vertical direction is: $\Delta Z = 7.98\cdot10^{-2}/1260 = 13$ arcsec. The loads and reaction forces are illustrated in table 5.1. The plots of stresses and strains are in fig. 5.4. Four restraint conditions are applied symmetrically on bolting holes close to each other to reproduce actual tilting moments. Torque moment due to azimuth inertia, earlier discussed, was spanned on the lower surface, which interfaces to the bearing.

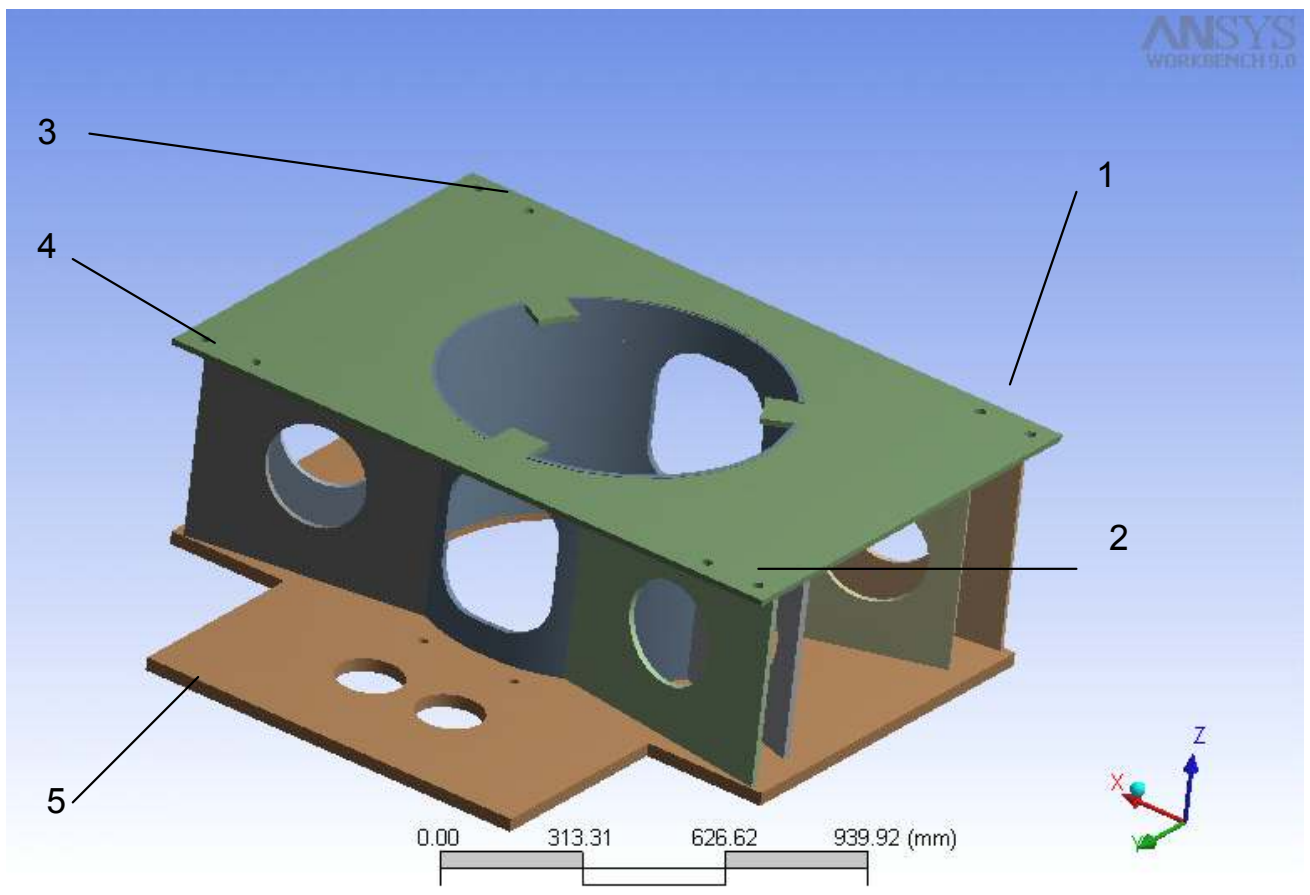

| Load influence Area | $F_X$ [N] | $F_Y$ [N] | $F_Z$ [N] | $M_X$ [Nmm] | $M_Y$ [Nmm] | $M_Z$ [Nmm] |
|---|---|---|---|---|---|---|
| Side Hole 1.1 | 42.516 | 35.944 | 1286 | 40503 | -48164 | 2561.2 |
| Side Hole 1.2 | 8.4066 | -864.12 | -3983.7 | $-1.5607*10^5$ | $-1.2683*10^5$ | 8080 |
| Side Hole 2.1 | 50.032 | 1109.7 | -3437.9 | $1.2959*10^5$ | $-1.1124*10^5$ | -9913.9 |
| Side Hole 2.2 | -100.7 | 237.9 | 2996.5 | -97517 | 95797 | -8725.6 |
| Side Hole 3.1 | 210.98 | 3023 | -12968 | $-4.7062*10^5$ | $1.1372*10^5$ | -23141 |
| Side Hole 3.2 | 85.516 | -5352.4 | 10226 | $3.67480*10^5$ | $1.26090*10^5$ | -29000 |
| Side Hole 4.1 | 124 | -4674 | 8350.2 | $-2.9892*10^5$ | $1.09730*10^5$ | 23990 |
| Side Hole 4.2 | 169.81 | 2364 | -11128 | $4.034*10^5$ | 92696 | 25121 |
| Moment on plate 5 due to motor torque | - | - | - | - | - | $1.158*10^6$ |

**Table 5-1** Layout of active loads.

| #node ANSYS | Position | | | Displacement Z |
|---|---|---|---|---|
| | X | Y | Z | |
| 14331 | 485.19 | 23.17 | 498 | $-4.8832*10^{-2}$ |
| 20425 | 448.14 | 315.77 | 514 | $-4.1762*10^{-2}$ |
| 15446 | -444.48 | 150.6 | 530 | $3.10*10^{-2}$ |

**Table 5-2** Most significant displacements at the mechanical interface with azimuth bearing, with positions in global co-ordinate system.

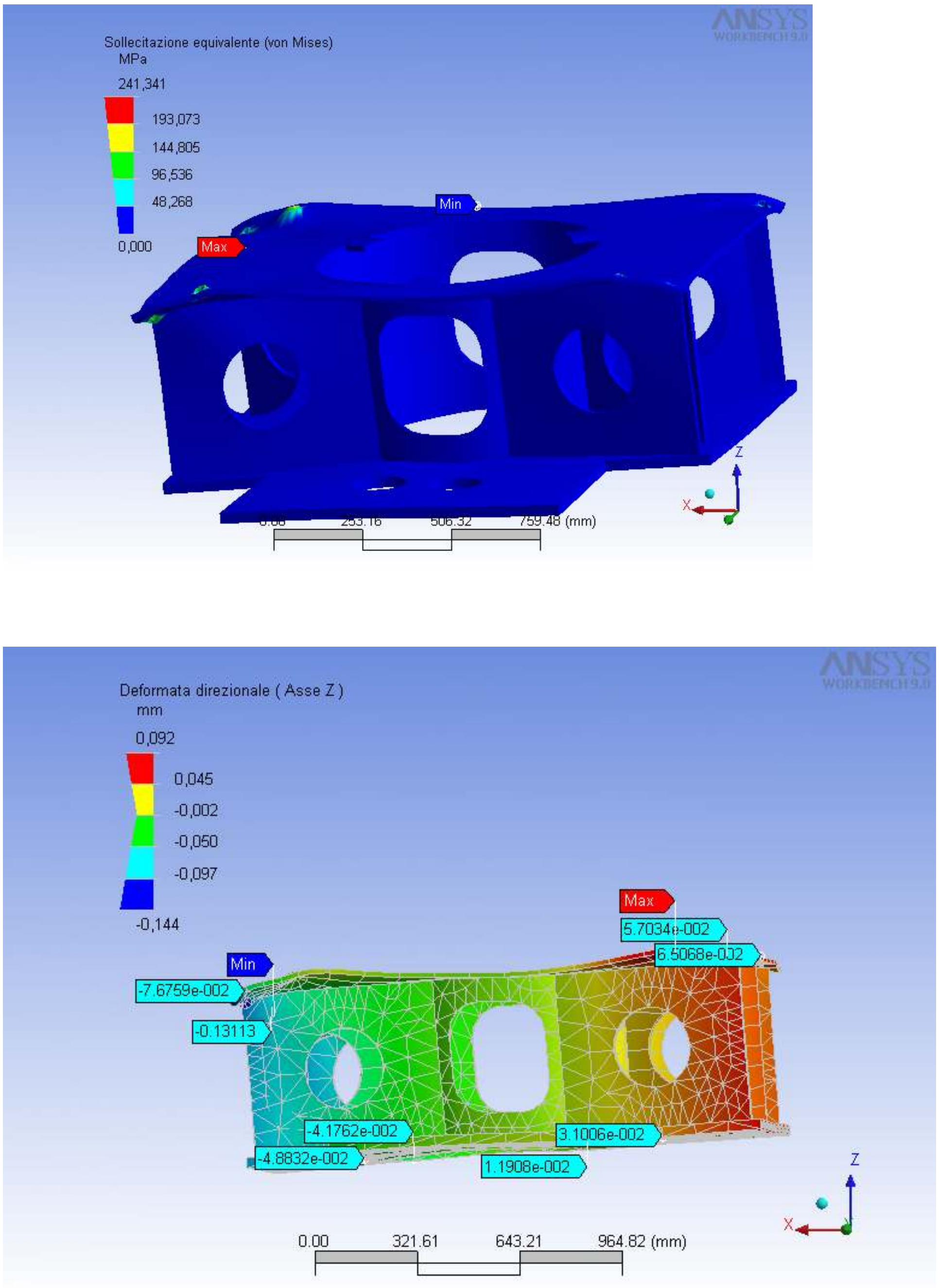

**Figure 5.4** Distribution of Von Mises stress and deformation in Z with highlighted the maximum displacements in Z.

### 5.2.8 Tube flexure

Same type of static analysis, with the same load conditions as those seen in § 3.6.4 , has been extended to the telescope structure including the fork and the base plate, with five different inclinations of the optical tube, to plot the flexure versus the altitude angle.

A restraint to all degrees of freedom has been set at the whole base surface, and four concentrated forces in addition, each of 104 N, have been applied at the corners of the top ring, to simulate lump forces. Also three eigenfrequencies for each position have been determined, to simulate the dynamic behavior of the telescope structure.

The overall reaction moment applied in the constraints increases with the variation of inclination, reaching the highest value of 2929 Nm when alt=20°.

The next plots collect data retrieved from analysis about maximum displacements between a node corresponding to M1 vertex and a node of the top ring corresponding to M2. We can notice that at the minimum altitude displacement it is significant and comparable to the dimension of NIR array.

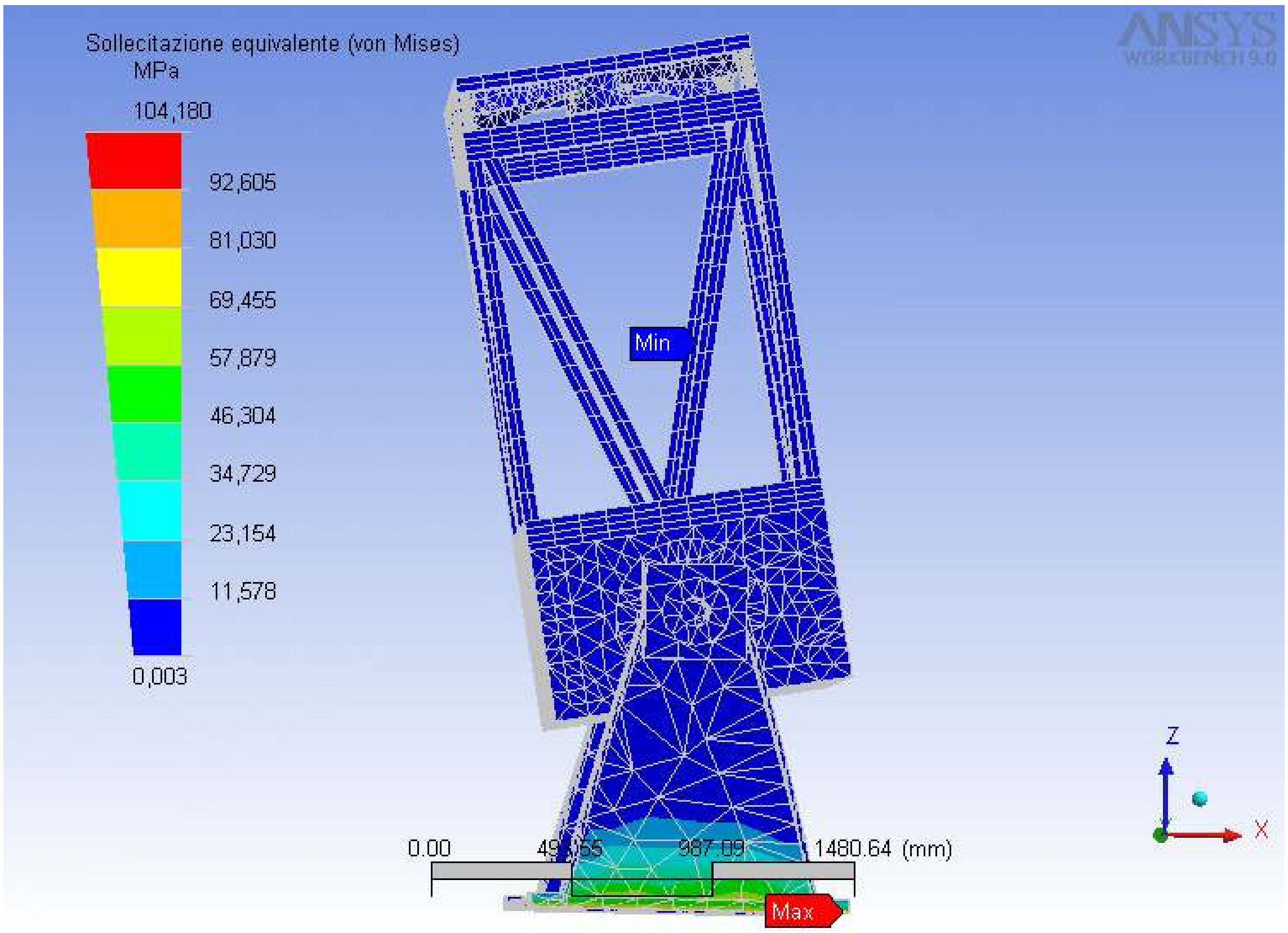

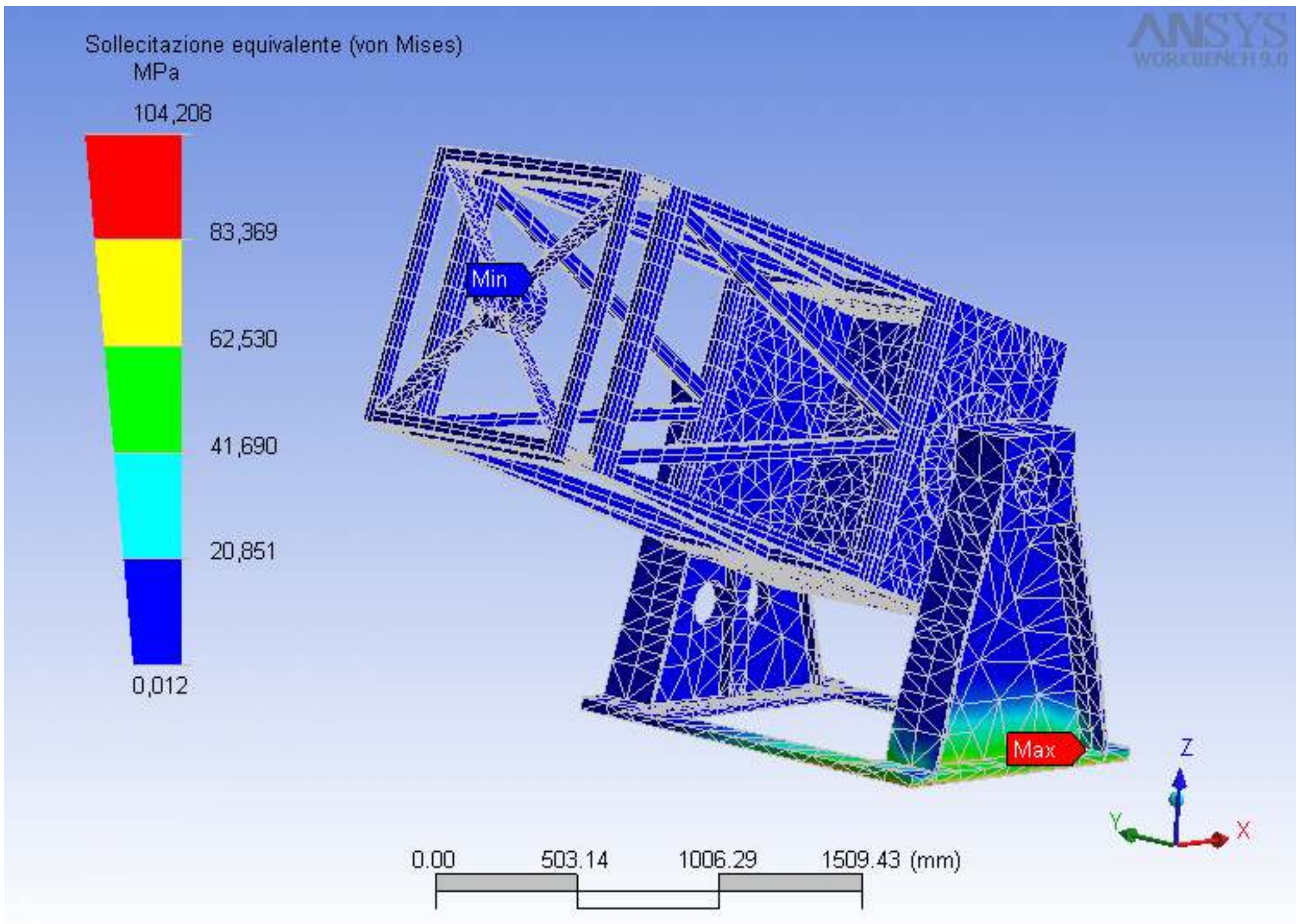

**Figure 5.5 and Figure 5.6** : Plots of Von Mises stresses relative to inclinations of 80, and 20 degrees.

| BASE PLATE | $F_X$ [N] | $F_Y$ [N] | $F_Z$ [N] | $M_X$ [N mm] | $M_Y$ [N mm] | $M_Z$ [N mm] |
|---|---|---|---|---|---|---|
| 80° | $-255,14 \cdot 10^{-2}$ | $3,8247 \cdot 10^{-3}$ | 16223 | $-2,221 \cdot 10^{5}$ | $-1,3318 \cdot 10^{6}$ | -54150 |
| 70° | $-2,2809 \cdot 10^{-2}$ | $-5,0099 \cdot 10^{-3}$ | 16140 | $-2,0183 \cdot 10^{5}$ | $-1,5624 \cdot 10^{6}$ | -88725 |
| 60° | $1,9065 \cdot 10^{-2}$ | $-8,2461 \cdot 10^{-2}$ | 16140 | $-2,2062 \cdot 10^{5}$ | $-1,9585 \cdot 10^{6}$ | -57516 |
| 45° | $-2,8408 \cdot 10^{-2}$ | $-1,4516 \cdot 10^{-2}$ | 16140 | $-1,3973 \cdot 10^{5}$ | $-2,426 \cdot 10^{6}$ | $-1,4067 \cdot 10^{5}$ |
| 20° | $1,708 \cdot 10^{-2}$ | $-6,7358 \cdot 10^{-3}$ | 16140 | -91381 | $-2,9255 \cdot 10^{6}$ | $-1,2234 \cdot 10^{5}$ |

**Table 5-3 Joint reaction forces.**

| ANGLE[°] | FLEXURE ERROR [MM] | FLEXURE ERROR [ARCMIN] |
|---|---|---|
| 90 | -0,012 | -0,020 |
| 80 | 0,261 | 0,44 |
| 70 | 0,570 | 0,97 |
| 60 | 0,843 | 1,44 |
| 45 | 1,140 | 1,95 |
| 20 | 1,517 | 2,60 |

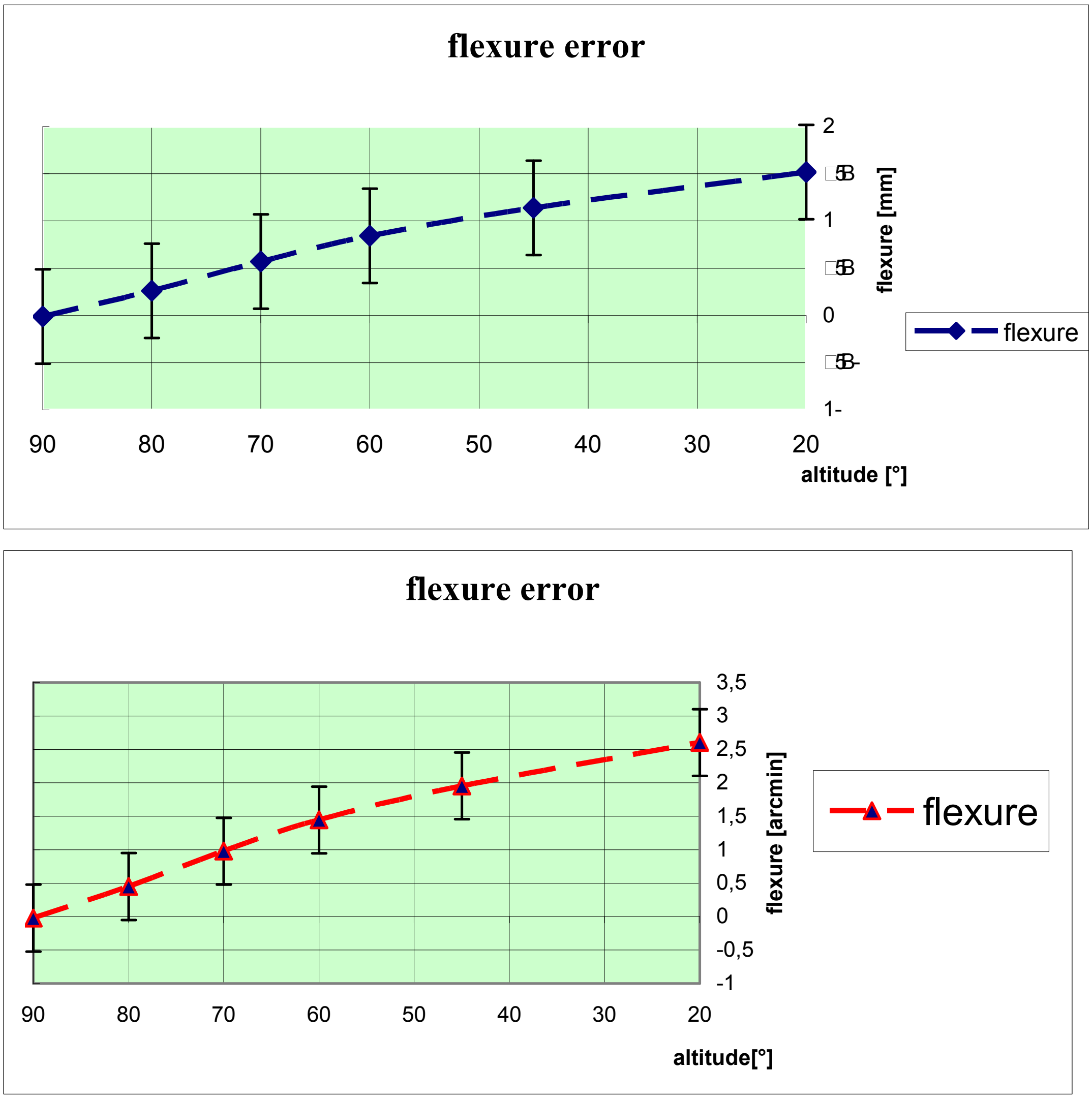

**Figure 5.7** Plot of flexure global errors vs inclination in case of local thermal gradient of 20 °C.

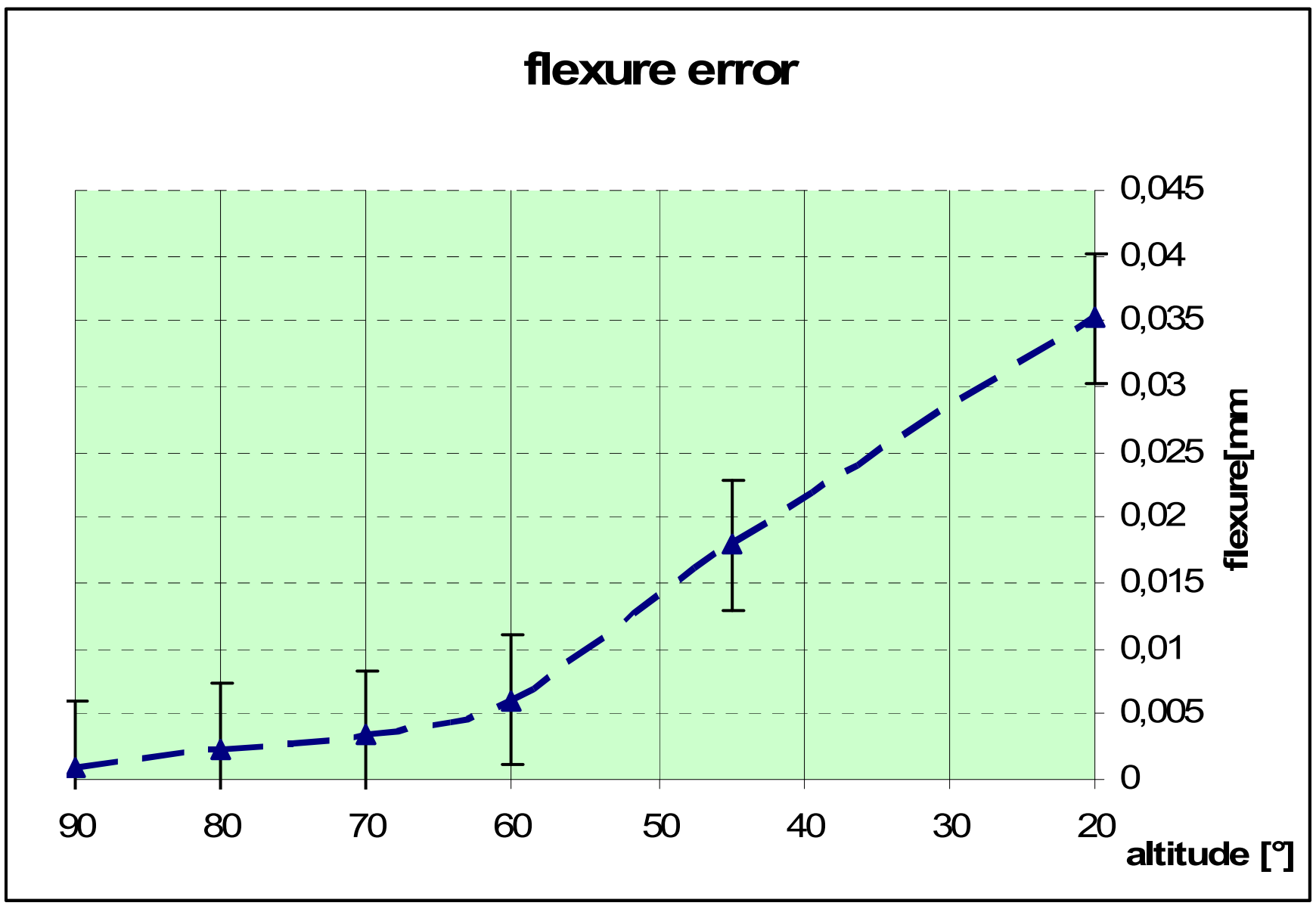

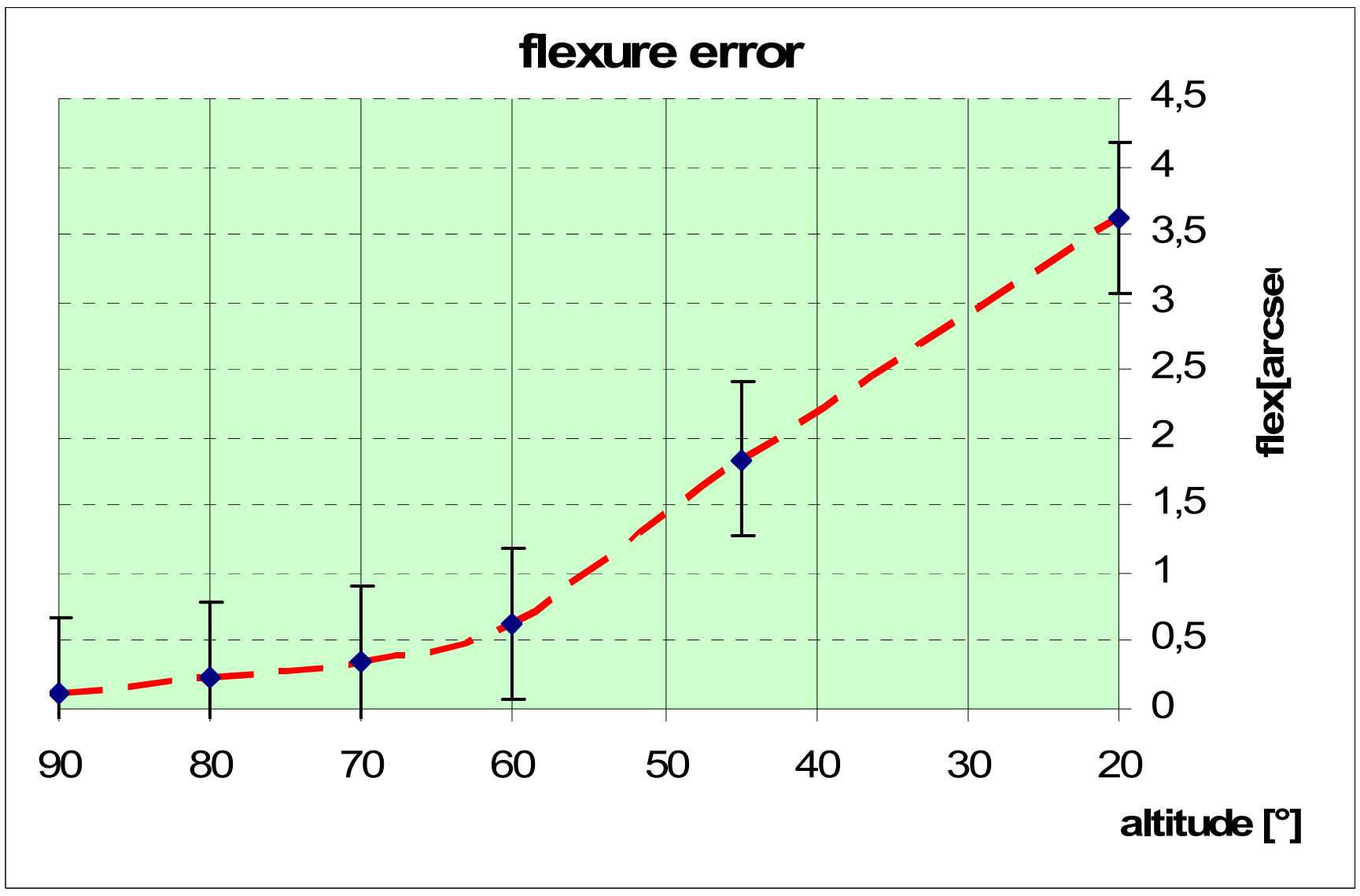

**Figure 5.8** Plots of flexure error in mm and arcsec vs inclination in winter, when the whole structure is subject to the same temperature condition (T=-80 ° C).

### 5.2.9 Mount flexure

It can be ignored in alt-az mounts, as the axial and radial projections of the weight load do not change with hour angle, like in equatorial and alt-alt mounts.

### 5.2.10 Tilt errors

Tilt errors are detected via an inclinometer[5] with two working axes, $\theta_1$ and $\theta_2$. Its task is to measure the inclination angles of the horizontal plane XY. It is planned to be attached possibly to the primary cell, or anyway along the azimuth axis, mounted on the fork, in order to reveal actual misalignments of the upper rotating structure (including the fork, optical tube and focal plane instruments).

Apart from the repeatable errors, like those mentioned above, due to axis misalignments, encoders scale factors, periodic gear errors, elastic flexure, which can be correctly modeled, there are non-repeatable errors, that are hard to estimate: they regard, for example, hysteresis, unbalancing of instruments and thermal deformations (Kibrick, R.,et al., 1995).

[5] See Appendix, section C.2, for technical features of the tiltmeter.

Our issue is to find the transfer function that converts the position retrieved by the sensor into telescope coordinate system, in order to correct encoder readings to the actual position.
An inclinometer provided by Geomechanics, type 701-2B(4X) in stainless steel, with mounting bracket, PC board, connectors and switches has been chosen.

#### 5.2.10.1 Coordinate correction by data retrieved from inclinometers

Every transformation from a coordinate system to another can be seen as the sum of rotation and translation effects:

$$\begin{Bmatrix} x' \\ y' \\ z' \end{Bmatrix} = \begin{bmatrix} \cos(x'x) & \cos(y'x) & \cos(z'x) \\ \cos(x'y) & \cos(y'y) & \cos(z'y) \\ \cos(x'z) & \cos(y'z) & \cos(z'z) \end{bmatrix} \begin{Bmatrix} x \\ y \\ z \end{Bmatrix} + \begin{Bmatrix} a \\ b \\ c \end{Bmatrix}$$

and it can be rewritten, in accordance with the Denavit-Hartenberg convention, in a compact manner:

$$\begin{Bmatrix} x' \\ y' \\ z' \\ 1 \end{Bmatrix} = \begin{bmatrix} \cos(x'x) & \cos(y'x) & \cos(z'x) & a \\ \cos(x'y) & \cos(y'y) & \cos(z'y) & b \\ \cos(x'z) & \cos(y'z) & \cos(z'z) & c \\ 0 & 0 & 0 & 1 \end{bmatrix} \begin{Bmatrix} x \\ y \\ z \\ 1 \end{Bmatrix}$$

The threesome *xyz* is deduced on the basis of (A,Z) angles converting polar to cartesian coordinates:

$$\begin{cases} x = sin\ Z\ cos\ A \\ y = sin\ Z\ sin\ A \\ z = cos\ Z \end{cases}$$

This relation widely applied in robot kinematics is suitable for problem like this, with the only difference that instead of actuators, here there are sensors. In fact the instrument behaves as a spherical joint detecting the two Euler angles and the transformation is the product of two rotation matrices. The signs are chosen so that a positive angle corresponds to an anticlockwise rotation.

$[X'] = R_x(\theta_1)\ R_y\ (\theta_2)[X]$, or in a better way:

$$\begin{Bmatrix} x' \\ y' \\ z' \\ 1 \end{Bmatrix} = \begin{bmatrix} 1 & 0 & 0 & 0 \\ 0 & \cos\theta_1 & \sin\theta_1 & 0 \\ 0 & -\sin\theta_1 & \cos\theta_1 & 0 \\ 0 & 0 & 0 & 1 \end{bmatrix} \cdot \begin{bmatrix} \cos\theta_2 & 0 & -\sin\theta_2 & 0 \\ 0 & 1 & 0 & 0 \\ \sin\theta_2 & 0 & \cos\theta_2 & d \\ 0 & 0 & 0 & 1 \end{bmatrix} \begin{Bmatrix} x \\ y \\ z \\ 1 \end{Bmatrix} \Rightarrow$$

$$\begin{cases} x' = x\cos\theta_2 - y\sin\theta_2 \\ y' = x\sin\theta_1\sin\theta_2 + y\cos\theta_1 + z\sin\theta_1\cos\theta_2 + d\sin\theta_1 \\ z' = x\cos\theta_1\sin\theta_2 - y\sin\theta_1 + z\cos\theta_1\cos\theta_2 + d\cos\theta_1 \end{cases}$$

$d$ represents the distance between the origins of the 2 systems as in the figure below.

The study of the Jacobi matrix allows cinematic calibration and an accurate control of the stability for the shift of parameters from ideal values. Indicating the deviation with $\boldsymbol{\Delta x = x - x_m}$, and $\boldsymbol{x_m}$ as a function of the sensor parameters $\boldsymbol{x_m = k(\theta_1, \theta_2)}$, it can be written:

$$\mathbf{\Delta x} = \frac{\partial \mathbf{k}}{\partial \theta_1}\Delta\theta_1 + \frac{\partial \mathbf{k}}{\partial \theta_2}\Delta\theta_2 ,$$

$$J = \begin{bmatrix} \partial x / \partial \theta_1 & \partial x / \partial \theta_2 \\ \partial y / \partial \theta_1 & \partial y / \partial \theta_2 \\ \partial z / \partial \theta_1 & \partial z / \partial \theta_2 \end{bmatrix}.$$

Such sensor parameters will be achievable in the context of testing all the structure when it's assembled. An evaluation of correction factors was made under MATLAB (Appendix section B.2), selecting a given star when it transits the meridian. For example , choosing Canopus (RA=6 h 24.092 m, DEC=-52°41.812'), with measures of tilt angles $\theta_1$= 30 µrad, $\theta_2$= - 30 µrad, at different position nearby the meridian, correction errors have been estimated. We noticed that a large influence is given by the offset *d* along z, namely more precisely the ratio *d/r*, being *r* the polar radius in spherical coordinates. Actually as this ratio is very close to 0, *d* parameter was neglected. Correction of position for tilt errors are in table 5-4.

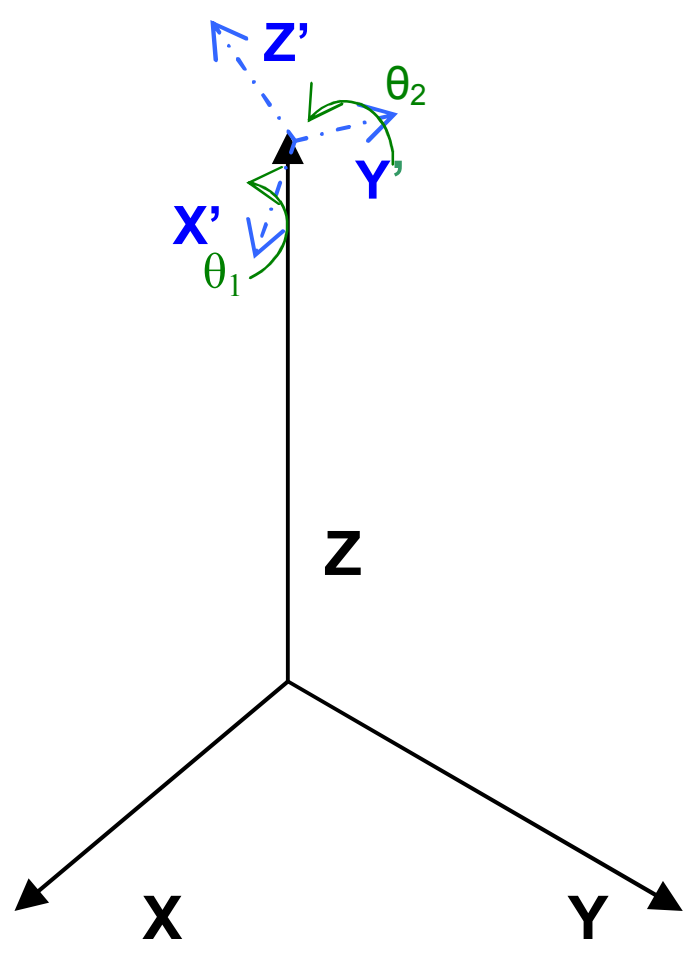

**Figure 5.9** Orientation of the system integral with the inclinometer.

| Hour angle [°] | A [°] | Z [°] | A' [°] | Z' [°] | Correction in arcsec A | Cor-rection in arcsec Z | Relative error po-sition A | Relative error po-sition Z |
|---|---|---|---|---|---|---|---|---|
| -0.4 | 0.75767 | 22.4036 | 0.76179 | 22.4053 | 14.80934 | 6.2694 | 0.00540 | $7.772*10^{-5}$ |
| -0.3 | 0.56827 | 22.4033 | 0.57240 | 22.4051 | 14.85972 | 6.2492 | 0.00721 | $7.747*10^{-5}$ |
| -0.2 | 0.37886 | 22.4031 | 0.38300 | 22.4049 | 14.90988 | 6.2289 | 0.01081 | $7.722*10^{-5}$ |
| -0.1 | 0.18943 | 22.4030 | 0.19359 | 22.4048 | 14.95983 | 6.2086 | 0.02147 | $7.697*10^{-5}$ |
| 0 | 0 | 22.4030 | 0.00417 | 22.4047 | 15.00957 | 6.1882 | - | $7.672*10^{-5}$ |
| 0.1 | -0.1894 | 22.4030 | -0.1852 | 22.4047 | 15.05909 | 6.1677 | 0.02258 | $7.646*10^{-5}$ |
| 0.2 | -0.37886 | 22.4031 | -0.37466 | 22.4049 | 15.10839 | 6.1471 | 0.01120 | $7.621*10^{-5}$ |
| 0.4 | -0.75767 | 22.4036 | -0.75345 | 22.4053 | 15.20632 | 6.1058 | 0.00561 | $7.569*10^{-5}$ |

**Table 5-4** Correction of position in alt-azimuth coordinates for a set of hour angles nearby the meridian for Canopus.

# CHAPTER 6 Thermal analysis

## 6.1 Heat mass transfer analysis for electronic boxes

The present paragraph illustrates the results retrieved from a heat transfer analysis inside the electronic boxes, which preside at the system control and supply.

We employ electronic boxes with blind panels, provided by ABB (SR series). Most of the components are located on the bottom plate, mounted on 35 mm DIN rails (BS 5584:1978 standard). Cable connections are realized through stiff cable trays, made of self-extinguishing PVC (UL94V-0) with open slots, according to DIN 43659.

Firstly we thought to cover thin external panels with polyurethane as insulator, of 100 mm thickness but, after the last campaign at Dome C in December 2005, having encountered a general overheating of sensors and other elements, we have decided to reduce thickness to 60 mm.

Another important issue is to equip the box with a fan system for a better heat dissipation, as convection is highly reduced with respect to temperate sites. Each box will have at least one 100 W heater, properly installed for a uniform distribution as much as possible.

An array of Pt 100 temperature sensors will be arranged inside the boxes, at different heights, to monitor internal thermal flow. According to CEI 17-43 Standards, the power dissipation of a system in steady state, with a thermal gradient $\Delta T$= 50 °C (average internal temperature is supposed of 10 °C, the outer one on Summer is -40°C) is given by the formula:

$P = KA_e\Delta T = 440$ W

where $K=5.5$ W/m$^2$ K, is the average thermal exchange coefficient, assumed for a galvanized metal sheet;

$A_e = 1.6$ m$^2$ is the equivalent heat transfer area. This is true within the assumption of thin walls and supposing that insulating contribution is negligible.

The boxes will be located in three different zones of the telescope: one on the fixed platform, another one on the fork base, over the rotating platform, the third on the optical tube. This layout is optimal for the connection of cables, coming from the slip-ring. Hereafter the outcomes provided by the analytical solution are showed: they concern transient cooling time. Beside that numerical data about thermal fluxes and temperature distribution on the walls, calculated by CFD (computational fluodynamics) software GAMBIT/FLUENT, are presented.

## 6.2 Optical tube box

The components layout is that of fig. 3.1. It includes the input/output modules of PLC, that switches on the heaters and retrieve data from Pt100 sensors. This box provides power supply and network links to chopping and focusing of M2.

It has been studied, at first, the transient problem with analytical approach, to determine time steps as input to obtain a convergent solution in FLUENT. Then, a model has been implemented under FLUENT 6.0, keeping the time steps to reach the convergence of the solution and the additional boundaries conditions required by the program itself.

For this purpose a model constituted by external fluid, internal walls, internal fluid, has been considered, and the time spent for cooling down to the thermal equilibrium, at a minimum allowable temperature (T=-20°C), has been examined.

The history of fluid temperature, called T1 (see fig.2), is obtained through the system of differential equations for both fluid and box, whose values are indicated in next table 1.

$$-(\rho\ c\ V\ )_1 \frac{dT_1}{d\tau} = \overline{h_1} A_1 (T_1 - T_2\ ) \qquad air \qquad (3.1)$$

$$-(\rho\ c\ V\ )_2 \frac{dT_2}{d\tau} = \overline{h_2} A_2 (T_2 - T_\infty\ ) - \overline{h_1} A_1 (T_1 - T_2\ ) \qquad insulator \qquad (3.2)$$

Boundary conditions are: $T_1=T_2=T_0$, $\frac{dT_1}{d\tau} = 0$ .

The system of differential equation can also be written in this way:

$$(D+K_1)T_1-K_1T_2=0 \qquad (3.3)$$

$$-K_2T_1+(D+K_2+K_3)T_2=K_3T_\infty \qquad (3.4)$$

where D is the differential with respect to time and:

$$K_1 = \frac{\overline{h_1} A_1}{\rho_1 c_1 V_1} \quad , K_2 = \frac{\overline{h_1} A_1}{\rho_2 c_2 V_2} \quad , K_3 = \frac{\overline{h_2} A_2}{\rho_2 c_2 V_2}$$

Solving the system we obtain the differential equation in the variable $T_1$:

$$[D_2+(K_1+K_2+K_3)D+K_1K_3]\,T_1=K_1K_3T_\infty$$

whose primitive is: $T = T_\infty + Me^{m_1\theta} + Ne^{m_2\theta}$

and:

$$m_1 = -\frac{K_1 + K_2 + K_3 + [(K_1 + K_2 + K_3)^2 - 4K_1K_3]^{1/2}}{2}$$

$$m_2 = -\frac{K_1 + K_2 + K_3 - [(K_1 + K_2 + K_3)^2 - 4K_1K_3]^{1/2}}{2}$$

*m, n* are determined by boundary conditions introduced above .

Therefore the final solution is:

$$\frac{T_1 - T_\infty}{T_0 - T_\infty} = \frac{m_2}{m_2 - m_1} e^{m_1\theta} - \frac{m_1}{m_2 - m_1} e^{m_2\theta} \qquad (3.5)$$

Solution relating to $T_2$ is obtained substituting (3.5) in (3.1).

Above all, we are interested in determining the time, knowing $T_1$. It increases noticeably with the variation of thickness, because with 60 mm the average conductivity reduces from 5 to 0.44 $W/m^2K$, being: $\overline{h_2} = \frac{1}{\frac{1}{h_2} + \frac{s_i}{K_i}}$ .

The plot in fig.3 illustrates the trend of cooling time, one with the initial temperature of 303 K, the other with a 273 K condition, both considering a 60 mm insulation, the third neglecting its presence. The magenta curve in particular shows that a gradient of 20 °C is reached in almost 6000 s, whereas more than 100 minutes are necessary to cool the system from 300 K down to the minimum admissible temperature of 253 K.

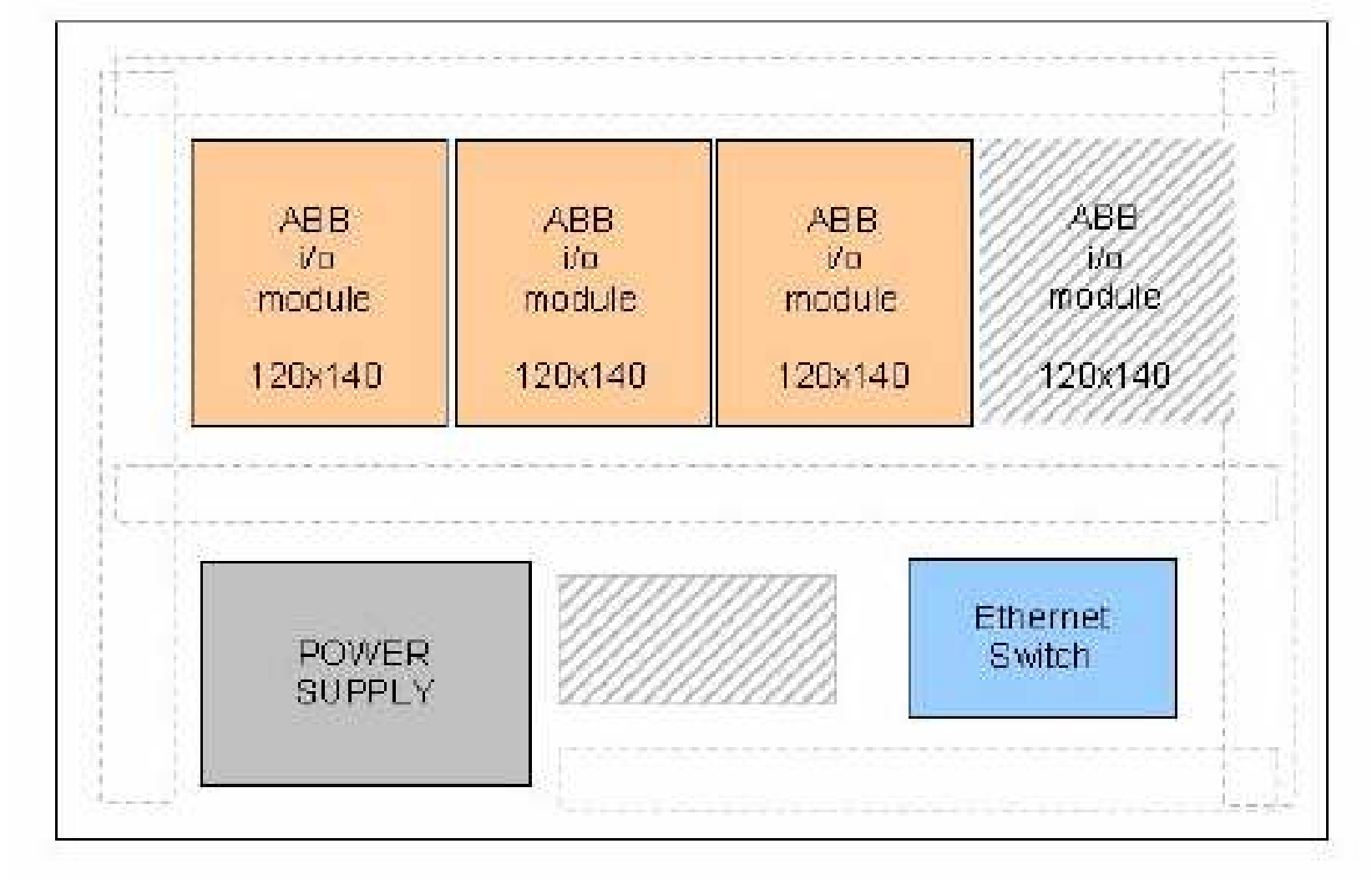

**Figure 6.1** A schematic layout of internal modules of optical tube box.

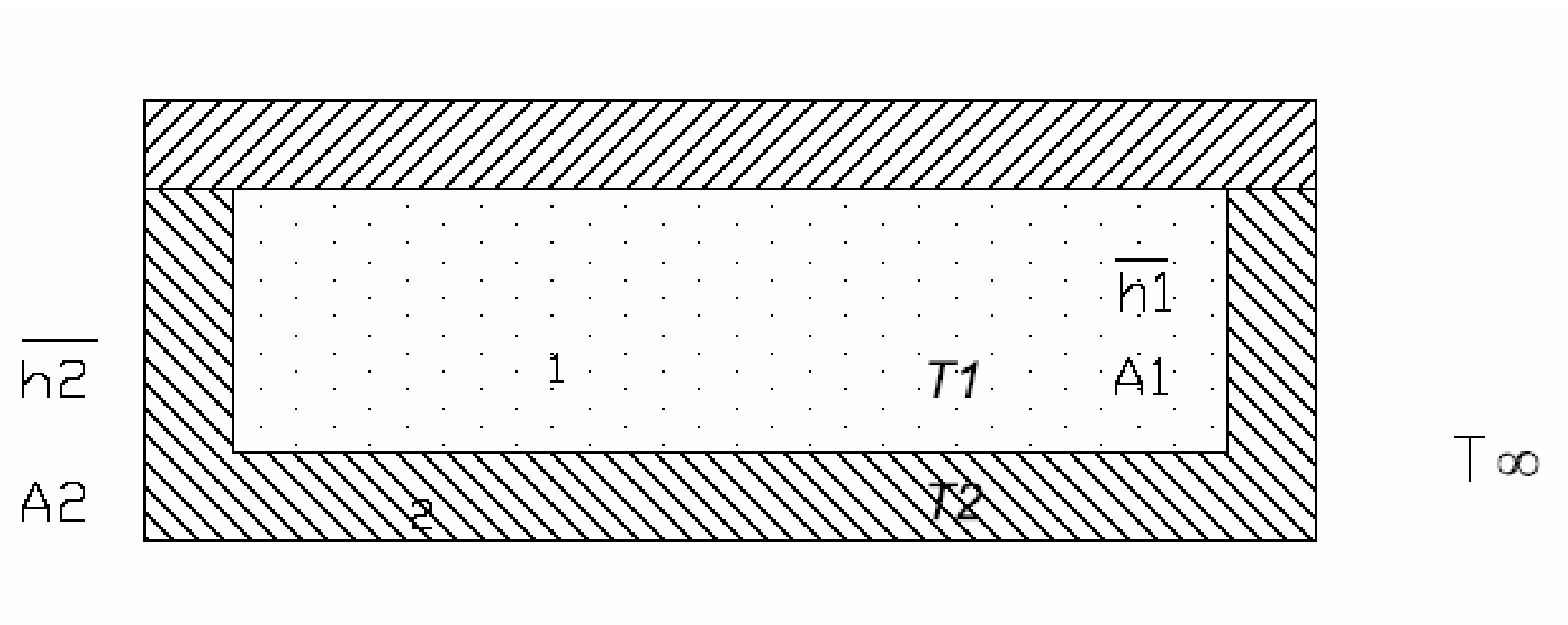

**Figure 6.2** Model of the analyzed system.

| | Average conductivity | Density [kg/m³] | Area [m²] | Heat capacity [J/kgK] | Volume [m³] |
|---|---|---|---|---|---|
| Inner Air | $h_1$=8 | $\rho_1$=1.225 | $A_1$=0.64 | $C_1$=1006 | $V_1$=0.144 |
| Air/Polyurethane | $h_2$=5 ignoring insulation $h_2$=0.44 with insulation | $\rho_2$=30 | $A_2$=1 | $C_2$=1600 | $V_2$=0.108 |
| Boundary conditions | $T_0$=300 , $T_\infty$=223K | | | | |

**Table 6-1.** Input parameters for analytical solution.

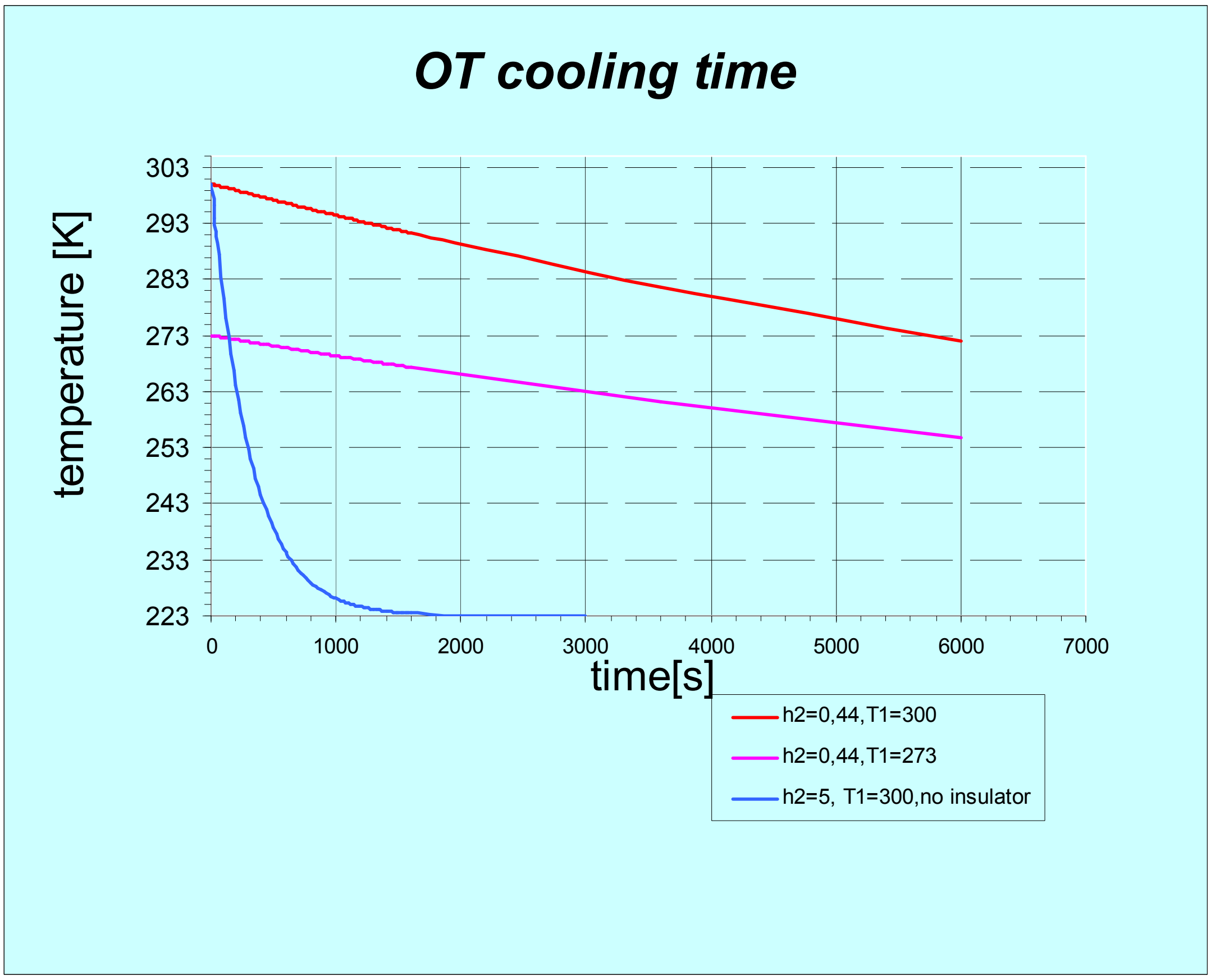

**Figure 6.3** Plots of thermal gradient through the walls, evaluated in presence of insulator and without it, with two different initial conditions (T=273 K, T=303 K).

The mesh of the problem for the numerical analysis has been generated by GAMBIT preprocessor: a 3D model of the box was created, as the layout of the inside components has no evident symmetry plane, even if geometry is rather simple.

As boundary conditions an inlet velocity $v_y$=0.01 m/s has been considered, simulating the internal convection. Atmospheric pressure is 64400 Pa, with gravity lying along –Y direction.

A constant external temperature $T_l$=223 K on the five walls exposed to environment was taken into account, and the base plate is supposed to be fixed on the M1 unit box. Four modules, each radiating 22500 W/m$^3$ and a power unit with an energy of 142000 W/m$^3$ per unit volume, were set (defined) as heat sources. A polyurethane layer of 60 mm thickness was considered, with a heat transfer coefficient of 2.8 W/mK. With 0.1s time step and 20 maximum iterations for each time step, the solution reached convergence after 160 iterations.

A 3D segregated solver, with implicit formulation for unsteady problems, has been used. The figure below shows the isothermal field, which anneals at the edges, and gradually becomes uniform in the colder part where there are no heat sources. Temperature range inside the box is comprised between 263 and 301 K. Then the flux curves, normal to isotherms are presented in the next plots. Highest dissipated power has a value of 20.384 $W/m^2$, corresponding to the centre and top plate of the box.

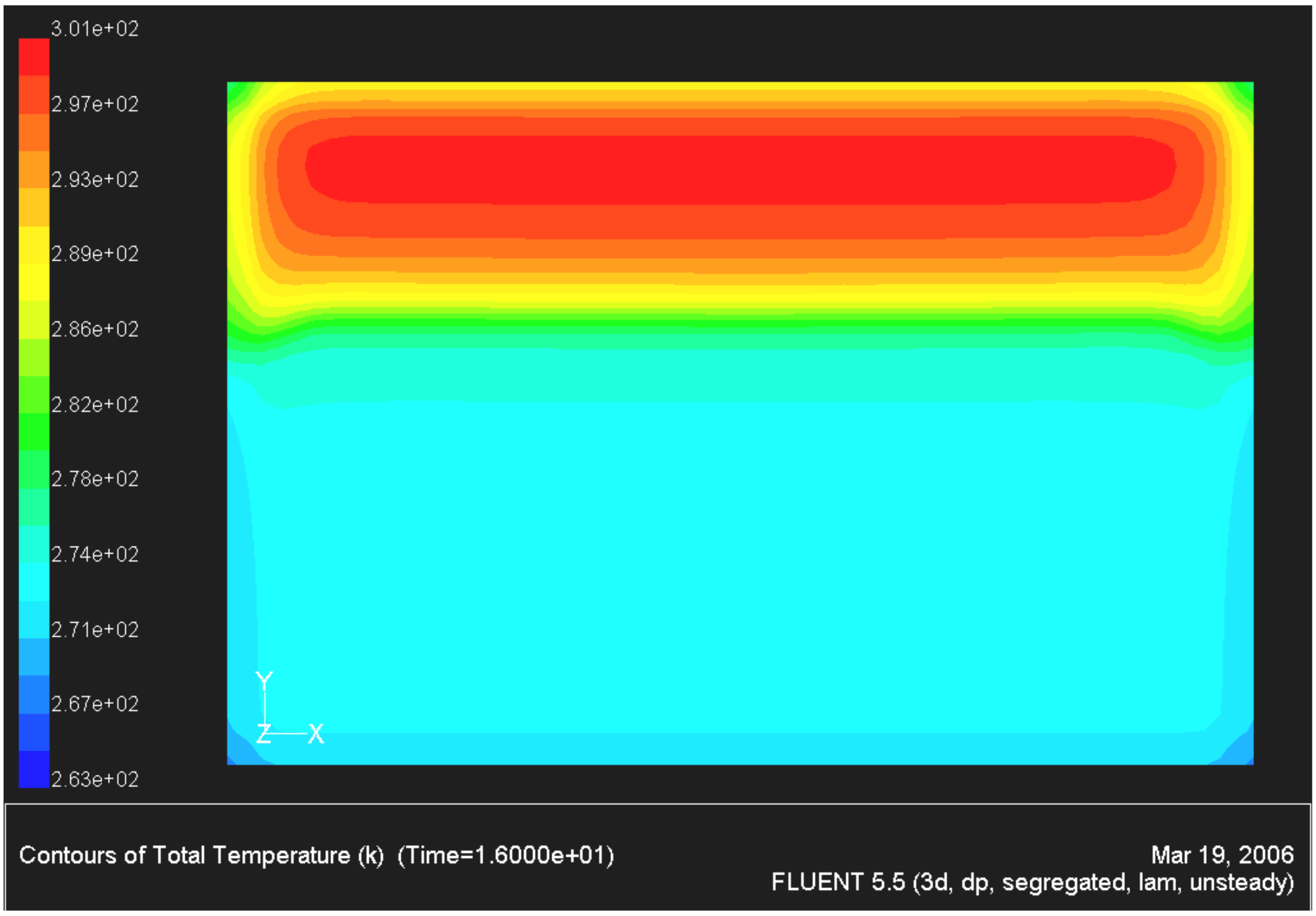

**Figure 6.4** A top view of temperature distribution on the base plate surface.

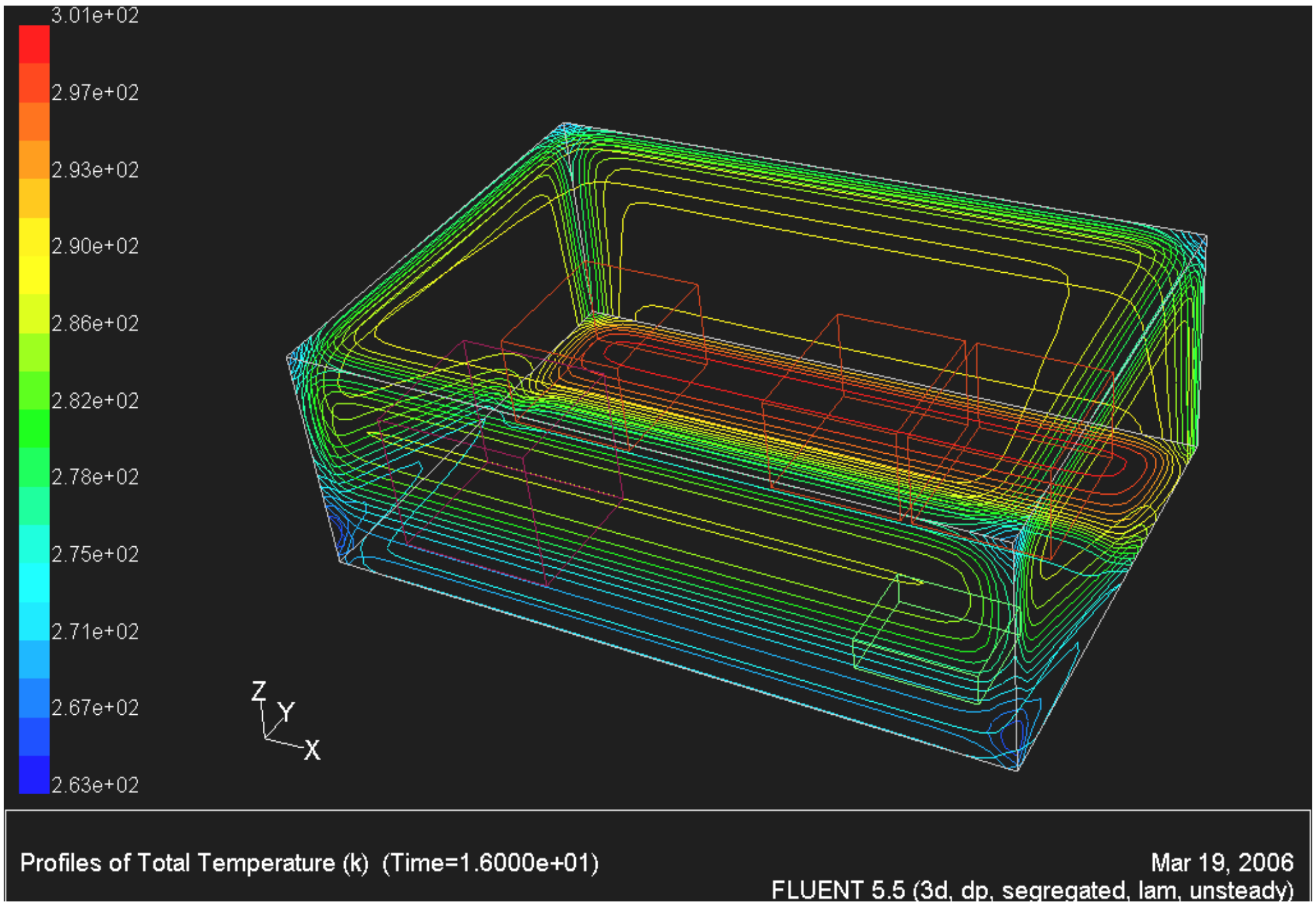

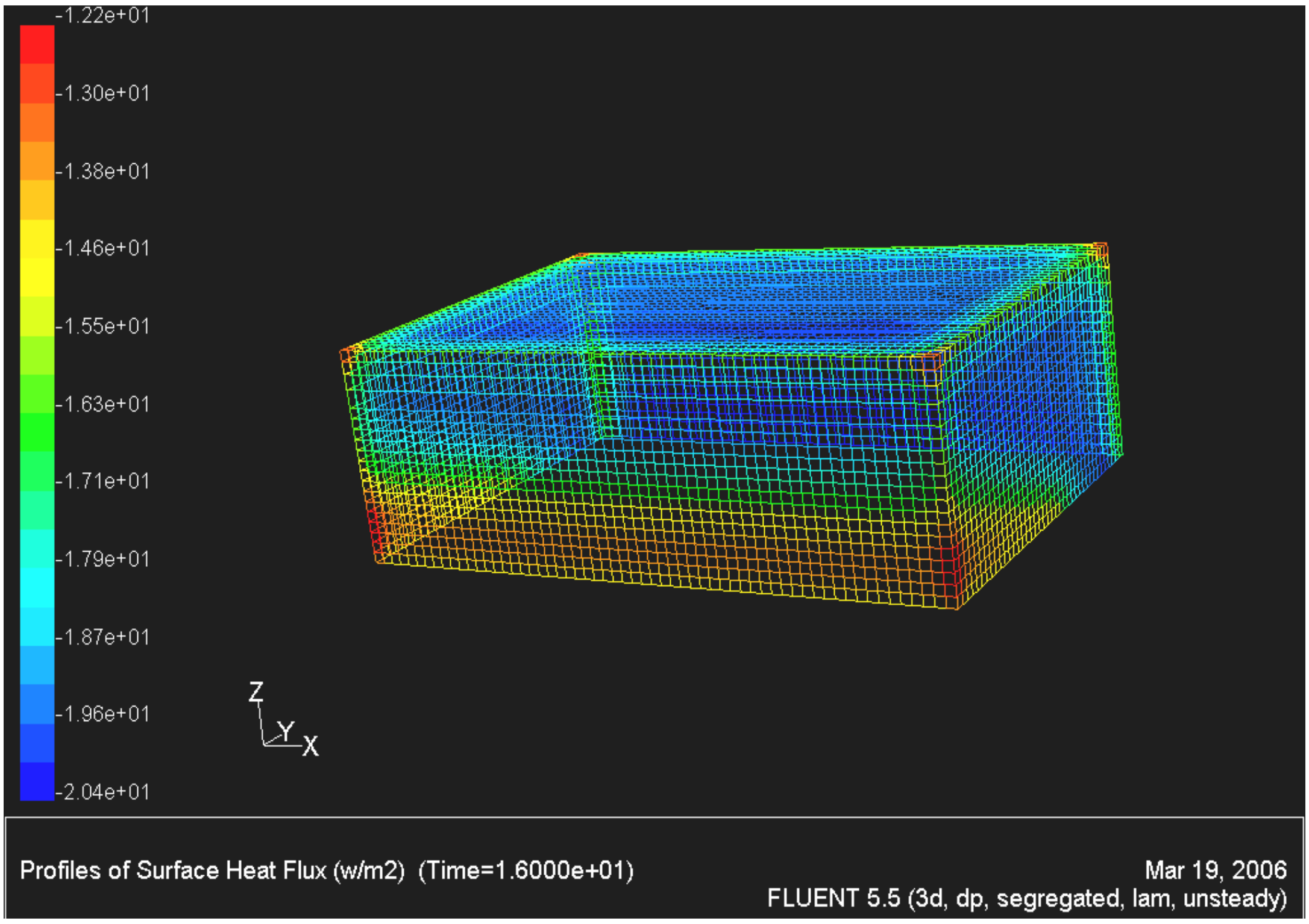

**Figure 6.5** Contour plots of total temperature and surface heat flux along the walls of the optical tube box.

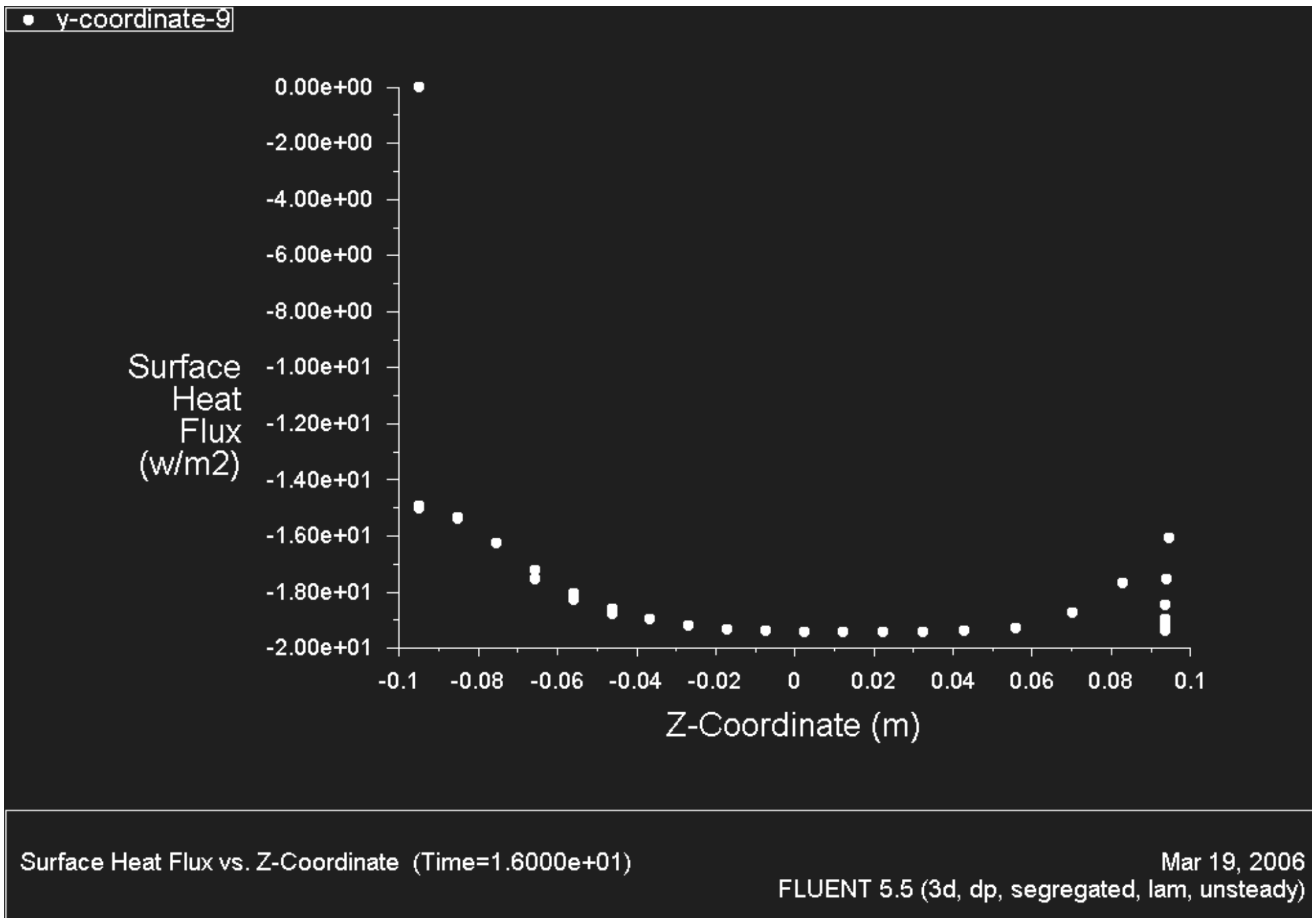

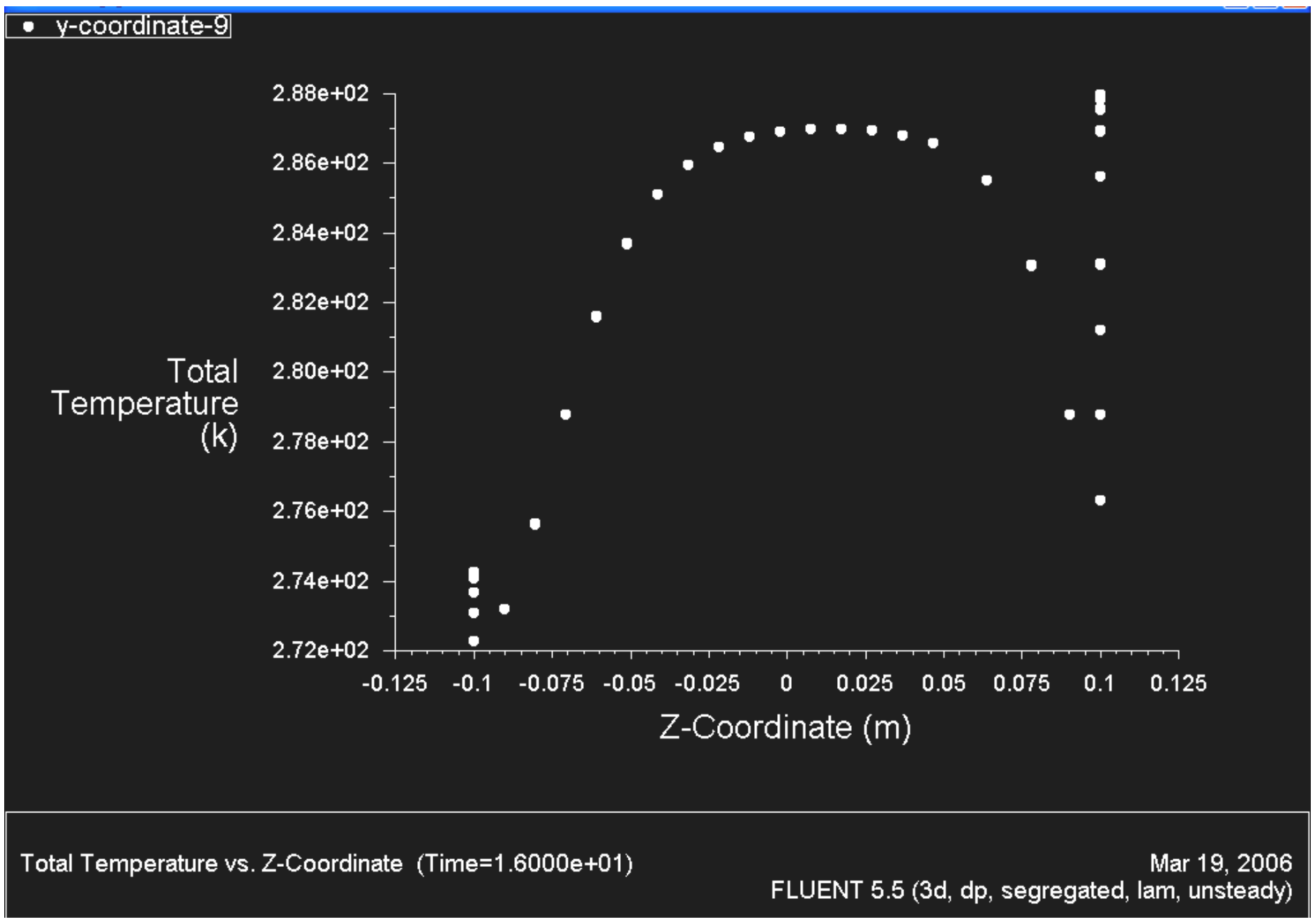

**Figure 6.6** Distributions respectively of surface heat flux and temperature on a central section plane (X=0).

## 6.3 Fork mount box 1

The same kind of analytical solution has been carried out for one of the three fork mount boxes. Only the elements layout and volume are different. $A_1$=1.6 and $A_2$=2.39 m$^2$ are the effective heat transfer areas. In fact its dimensions are 600x800x 300 (720x920x420 including insulator). In this case, with a starting temperature of 302 K, so that a decrease of 20 ° is reached in about 8000 s.

The main elements generating heat are:

- a power supply with energy per unit of volume 95200 W/m$^3$ ;
- a PLC with 10500 W/m$^3$ energy;
- GALIL controllers with 20000 W/m$^3$;
- Four Pythron modules PAB 93-70, Micro 256, with an energy contribution of 28000 W/m$^3$;
- One rack SLS 4 PAB 70 V.

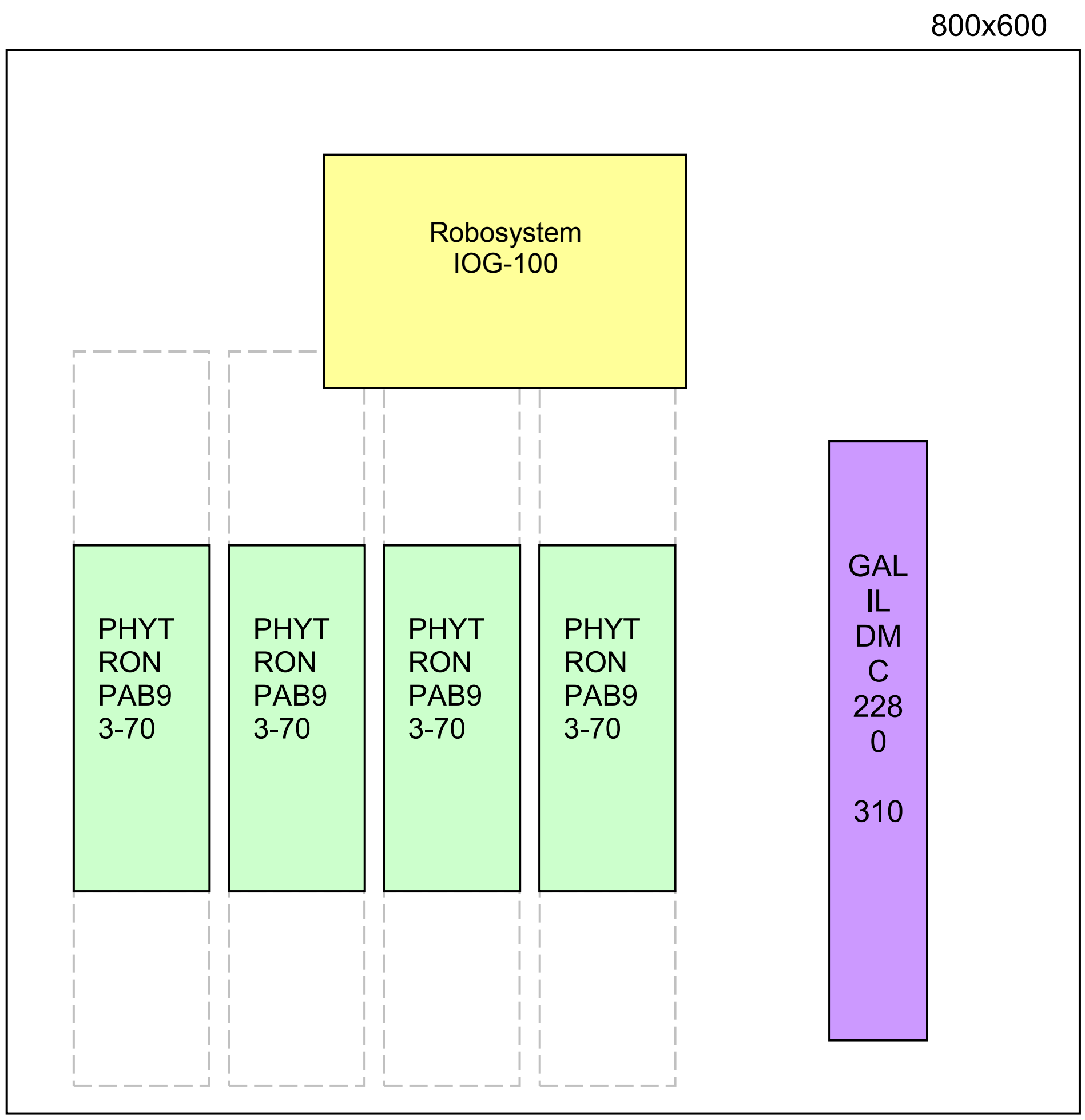

**Figure 6.7** Layout of the elements inside a fork box.

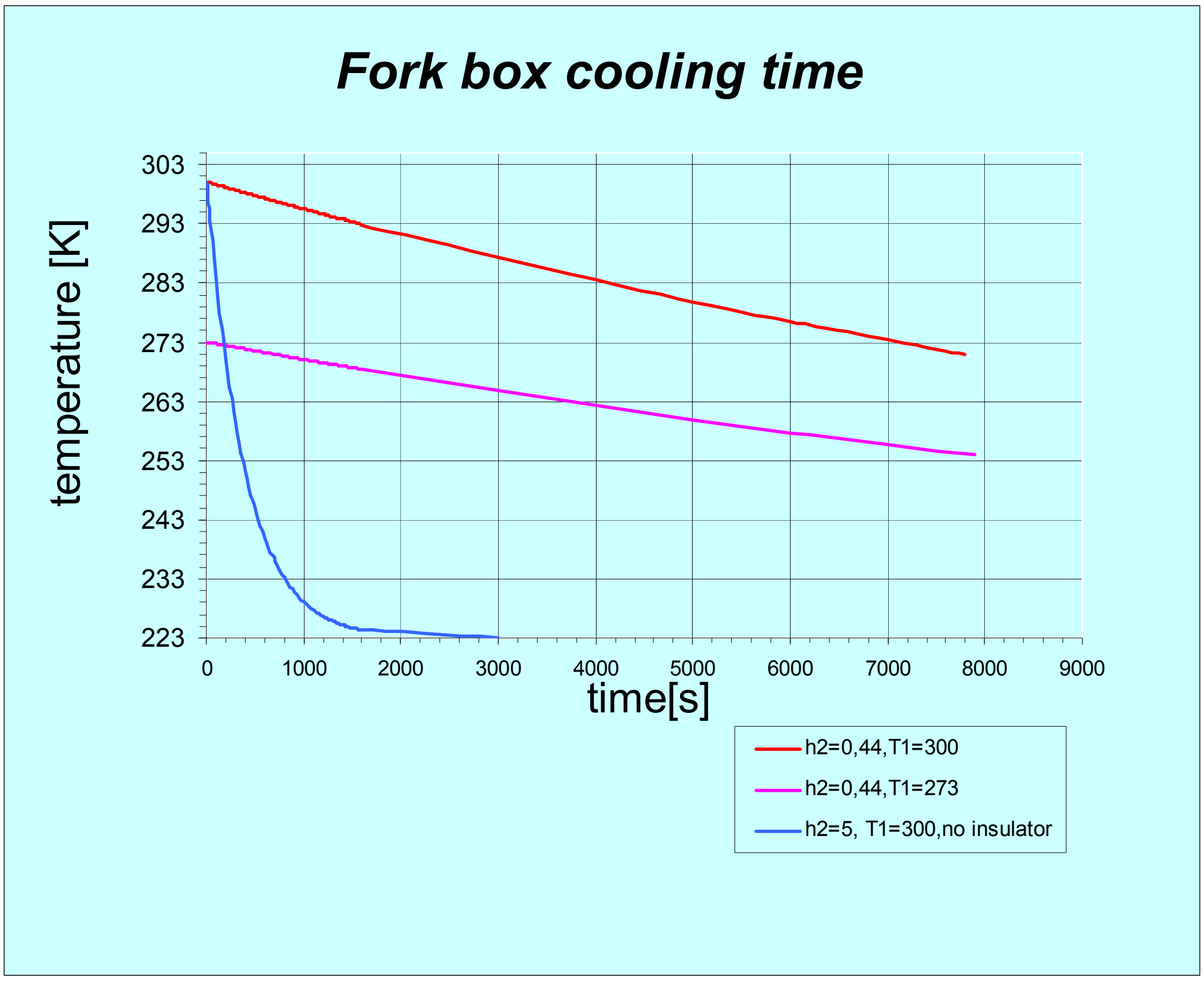

**Figure 6.8** Plots of thermal gradient through the walls, with and without insulator, considering two different initial conditions (T=273 K, T=303 K).

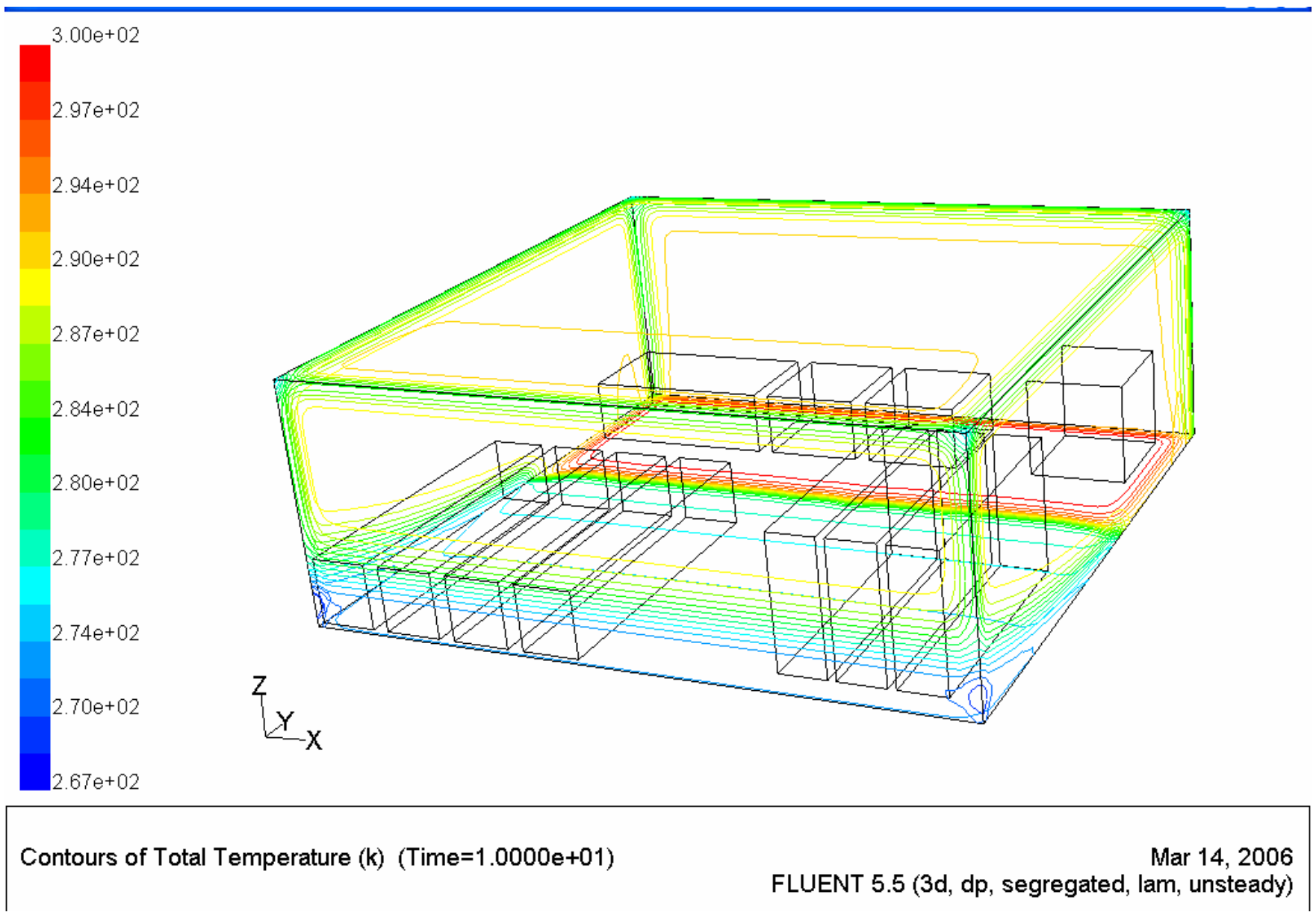

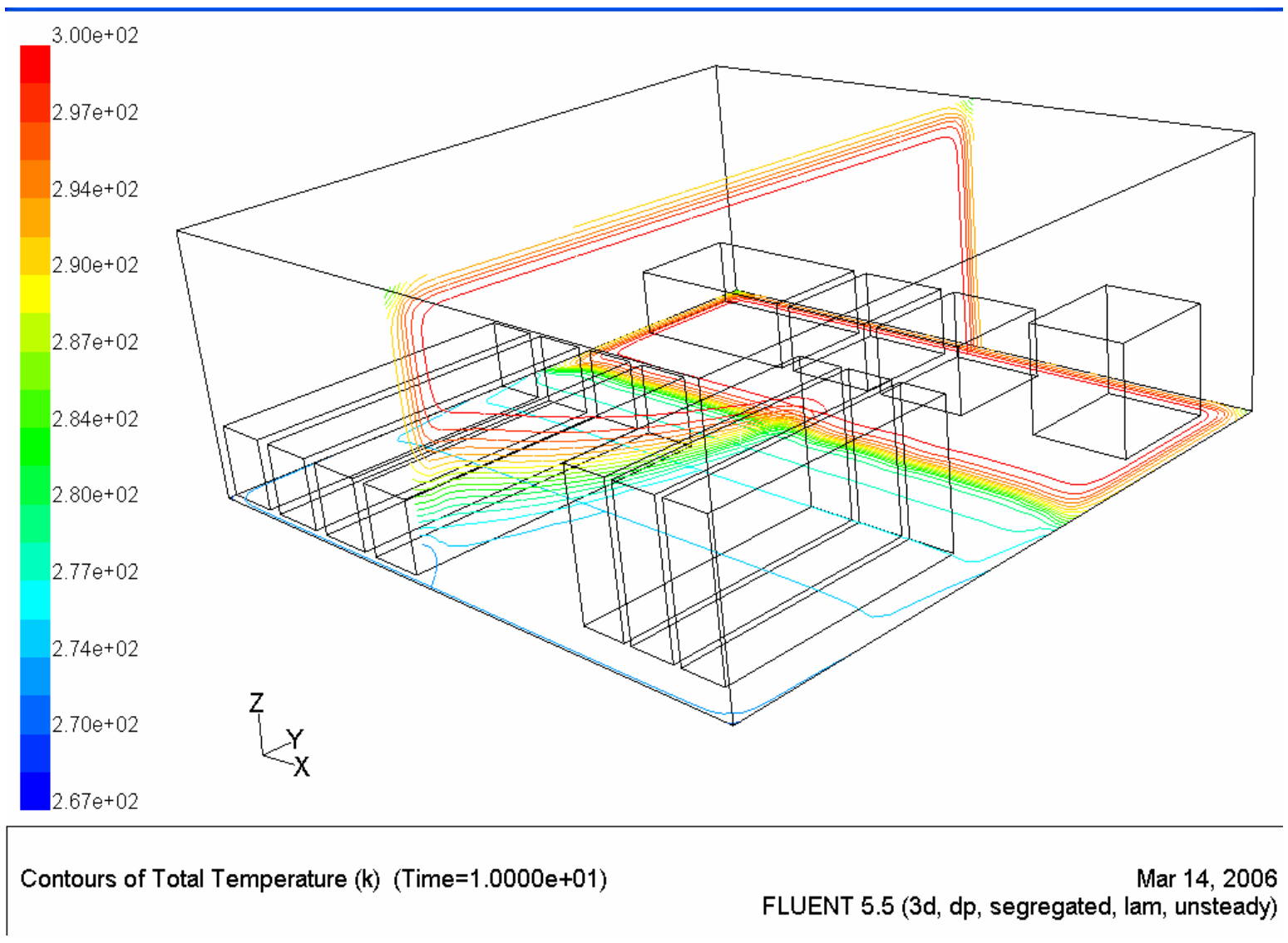

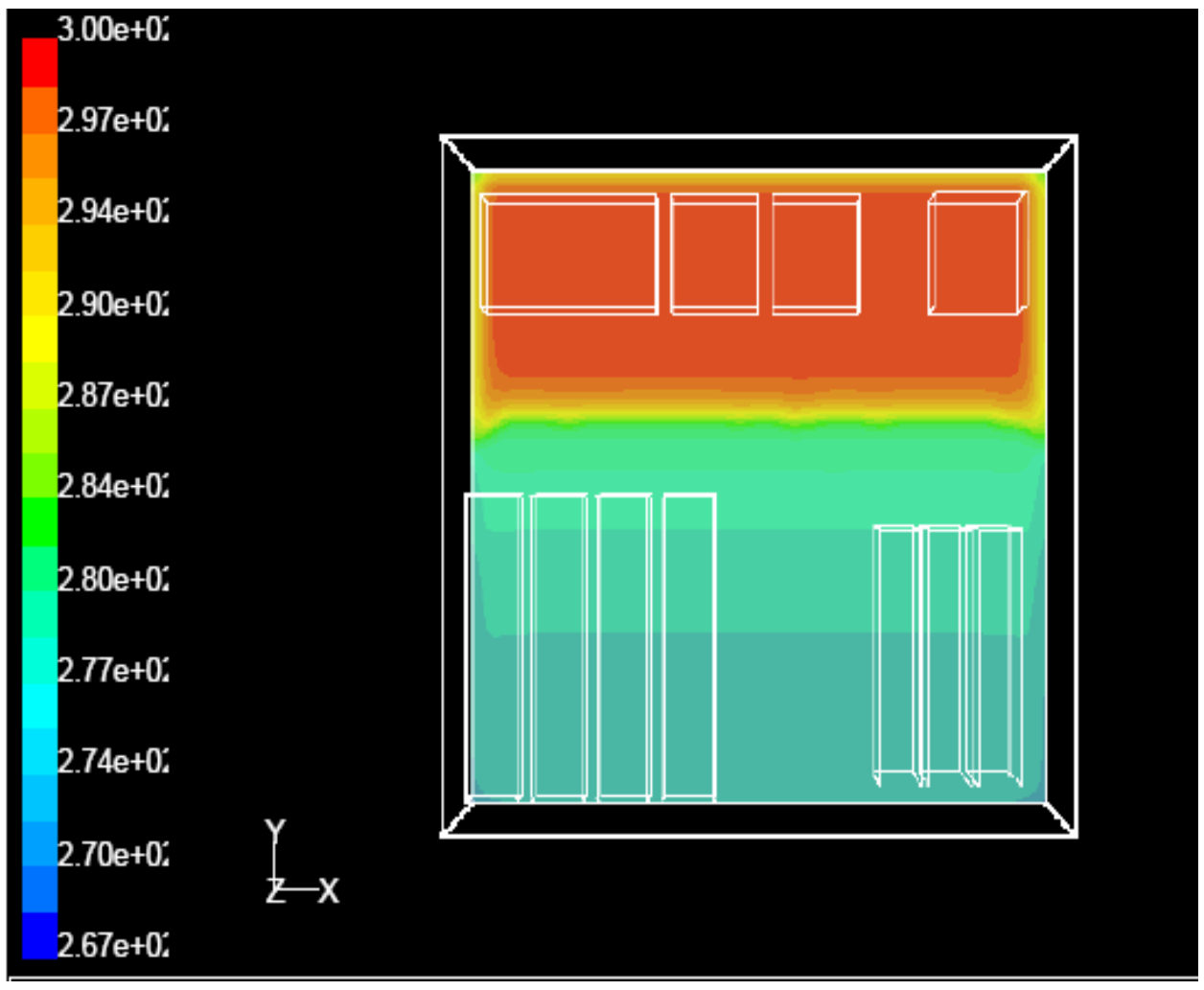

**Figure 6.9** Contour plots of temperature on the outside walls and on a central cross plane parallel to YZ.

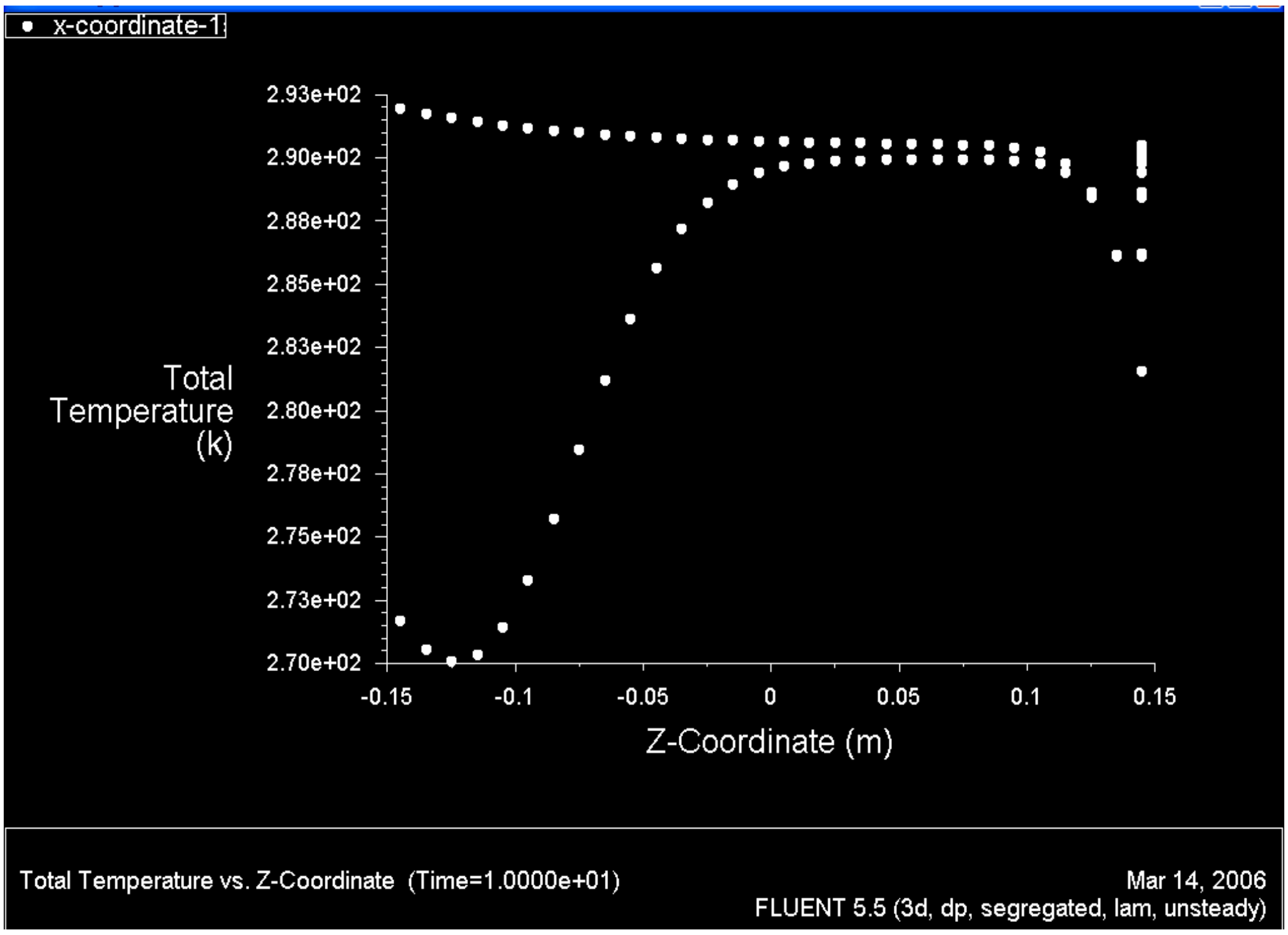

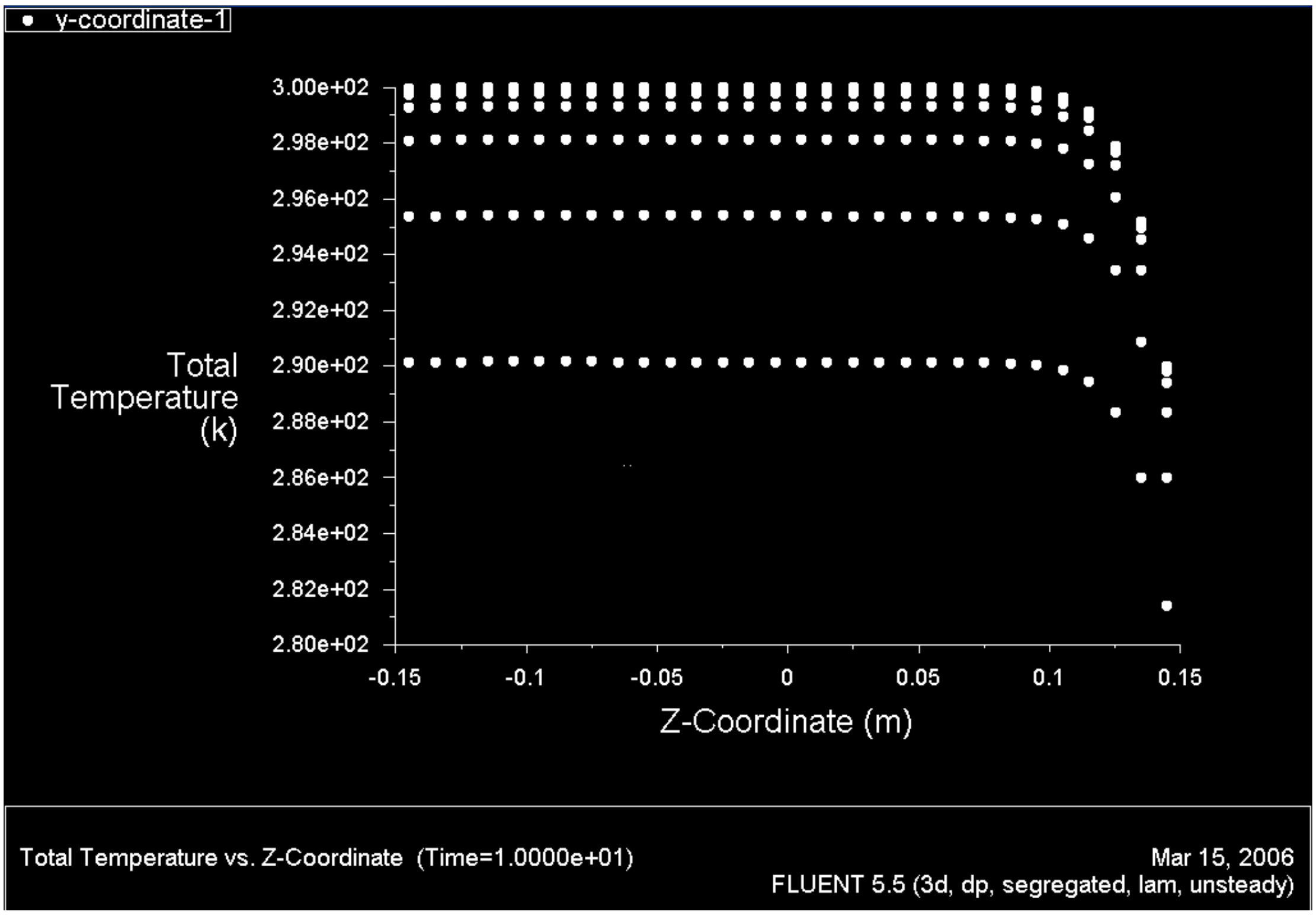

**Figure 6.10** Temperature distribution on two orthogonal central planes.

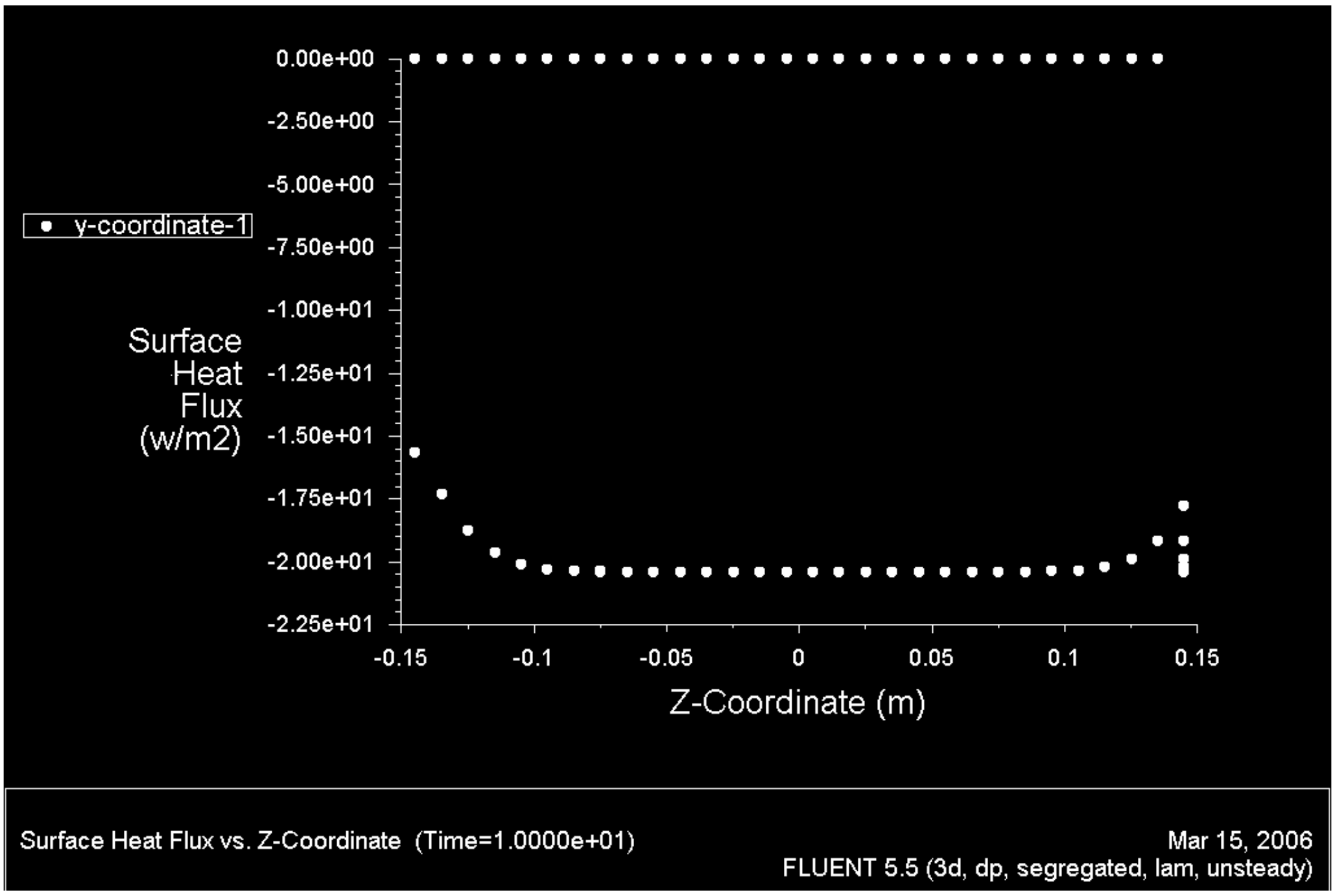

**Figure 6.11** A plot of heat flux on plane y=0.

## 6.4 Container box

For the analytical solution, the plot of temperature versus time is almost the same as the previous case since internal and external size is the same. Anyway a small difference is given by the actual internal volume, taking into account the different dimensions of the components. In fact the main parts included are:

- a General Power Supply
- a box power supply
- 2 ethernet switches
- a Point-to-Point Wireless LAN Connection to Dome C Base
- a power supply to heating sensors
- Analog signal from Pt 100 sensors
- a Container Open/Close System

The recalculated values of areas and volumes are respectively: $A_1$= 1.6 m$^2$, $A_2$= 2.39 m$^2$, $V_1$=0.184 m$^3$ and $V_2$=0.355 m$^3$. The cooling curve, considering a 100 mm insulation layer is also plotted: the two curves in red and blue show that time due to a 20 ° C gradient increases of a factor about 1.8. The following pictures illustrate data obtained by the numerical solution under FLUENT. The boundary conditions are the same as the other analyses. Specific energies supplied by the component included, are:

- GPS with 90000 W/m$^3$
- Power Supply with 75000 W/m$^3$
- Ethernet Switch: 41000 W/m$^3$

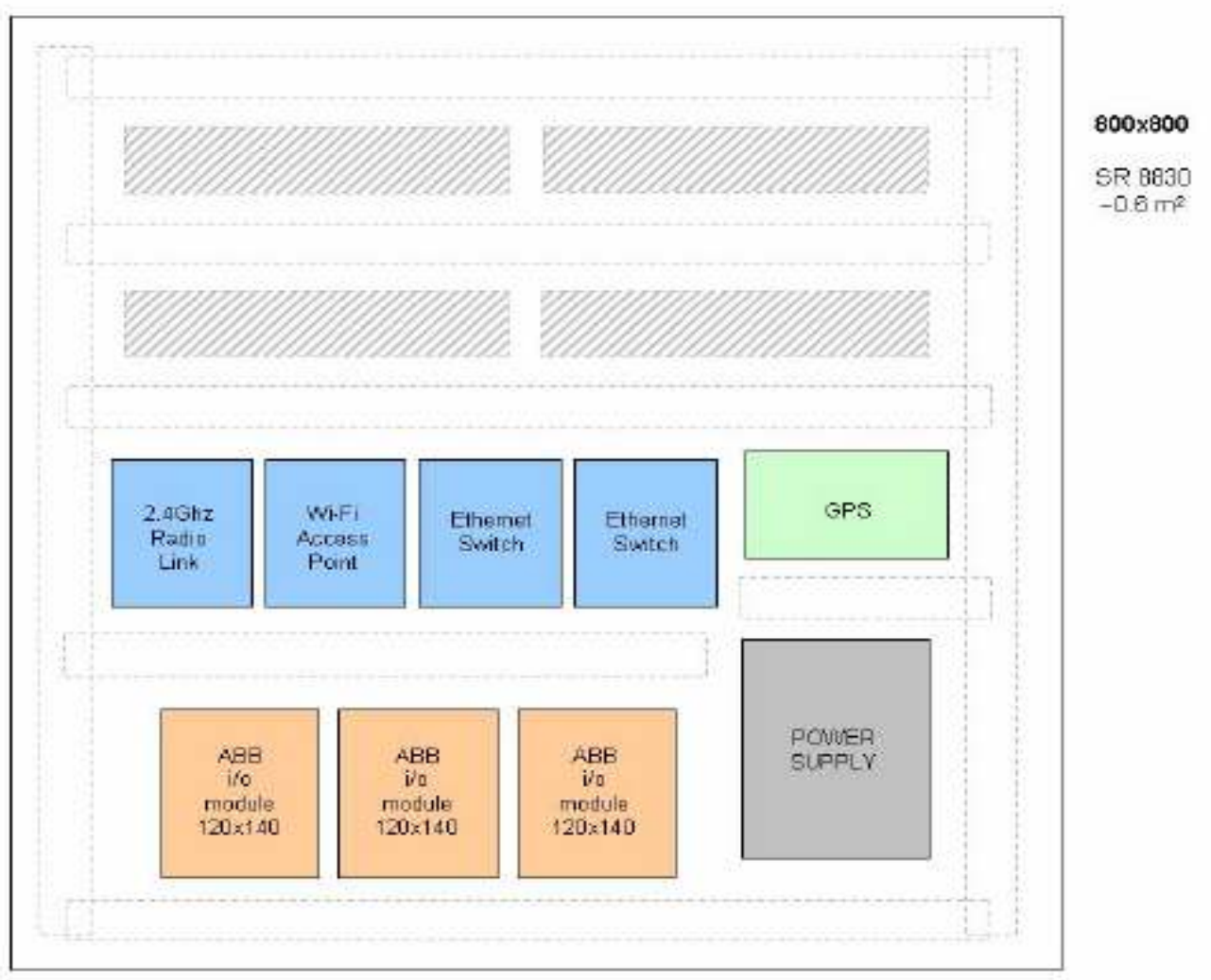

**Figure 6.12** Layout of the container box elements.

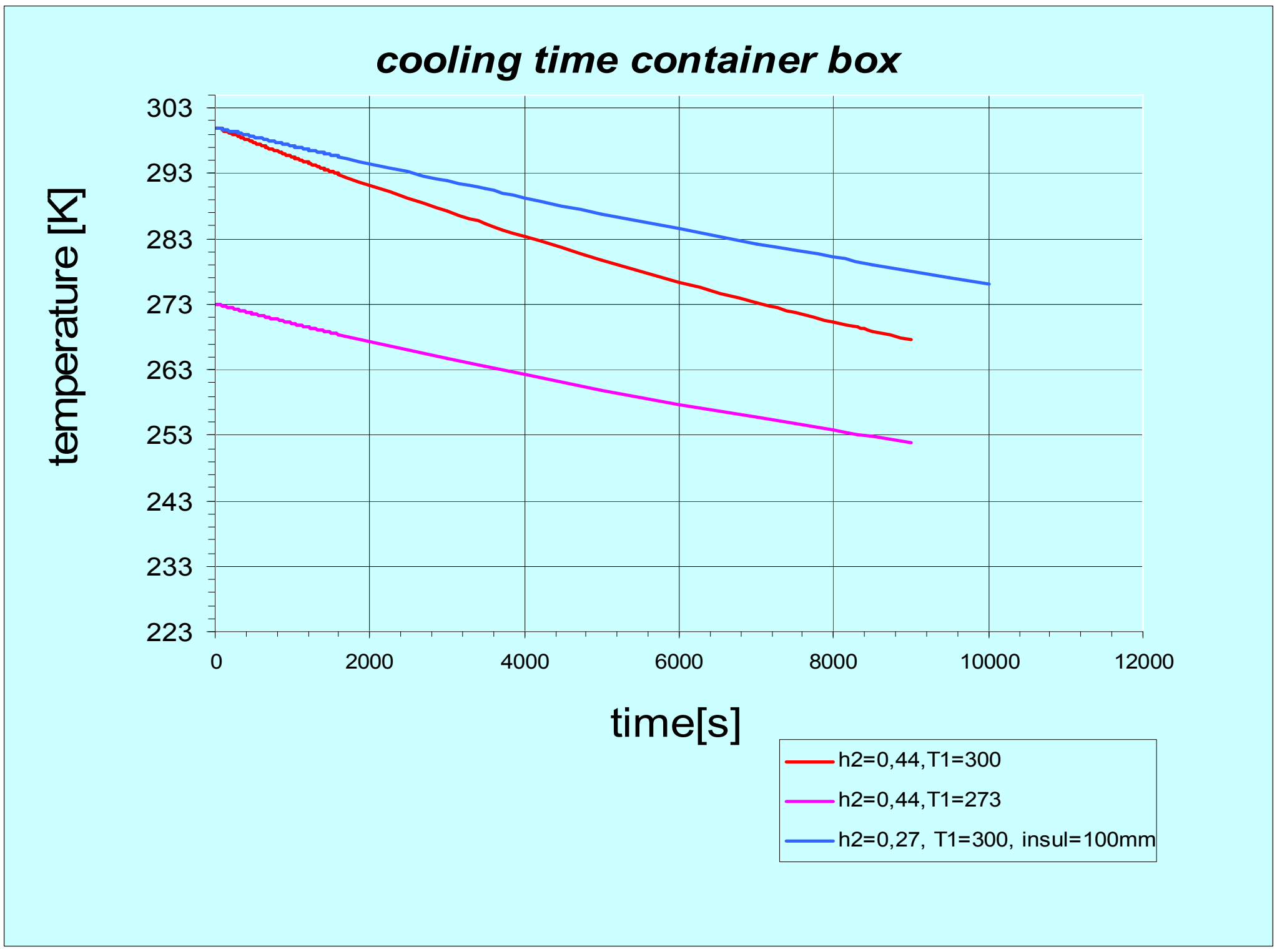

**Figure 6.13** Plots of temperature vs time, with two different initial conditions (T=273 K, T=303 K), varying the insulation thickness.

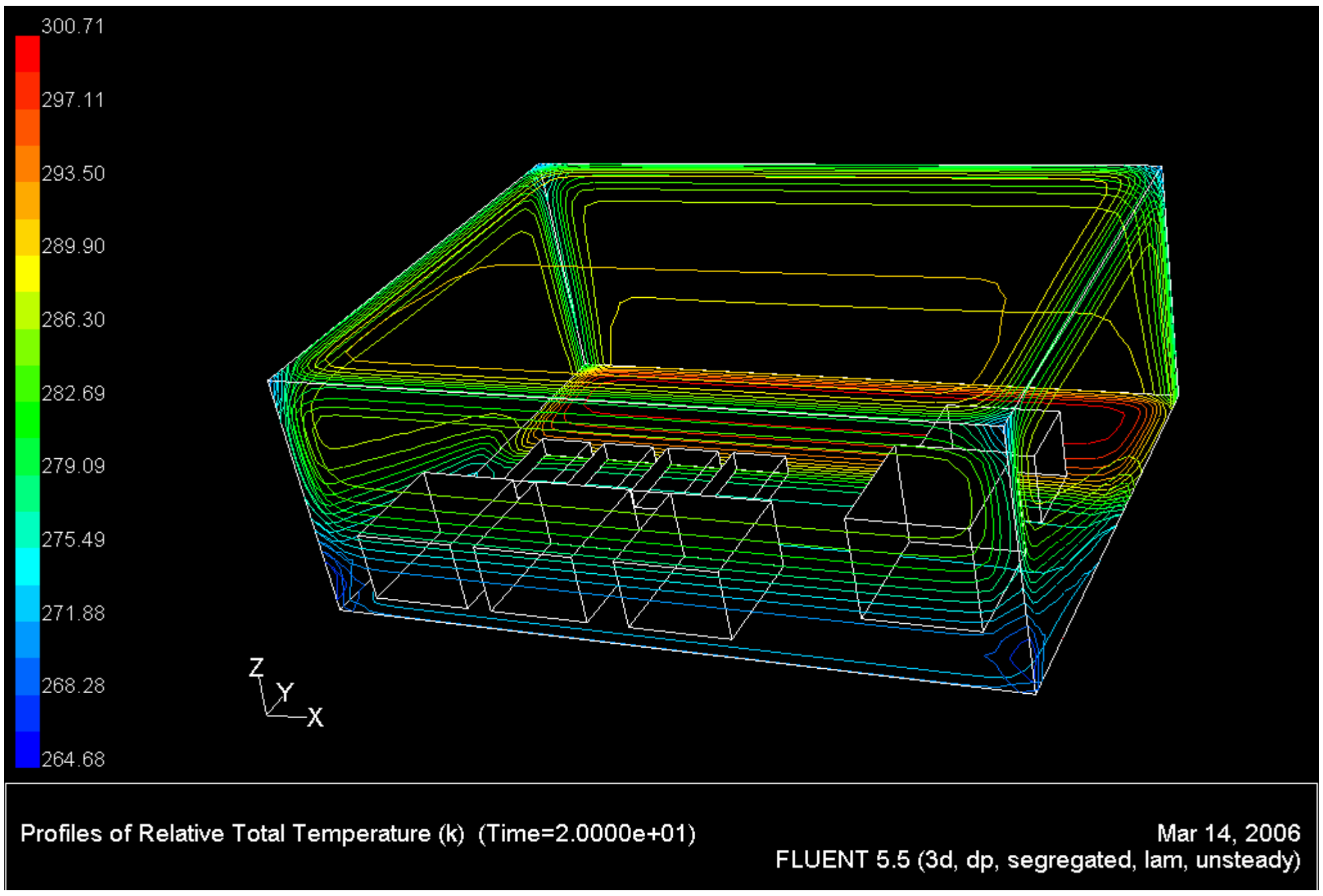

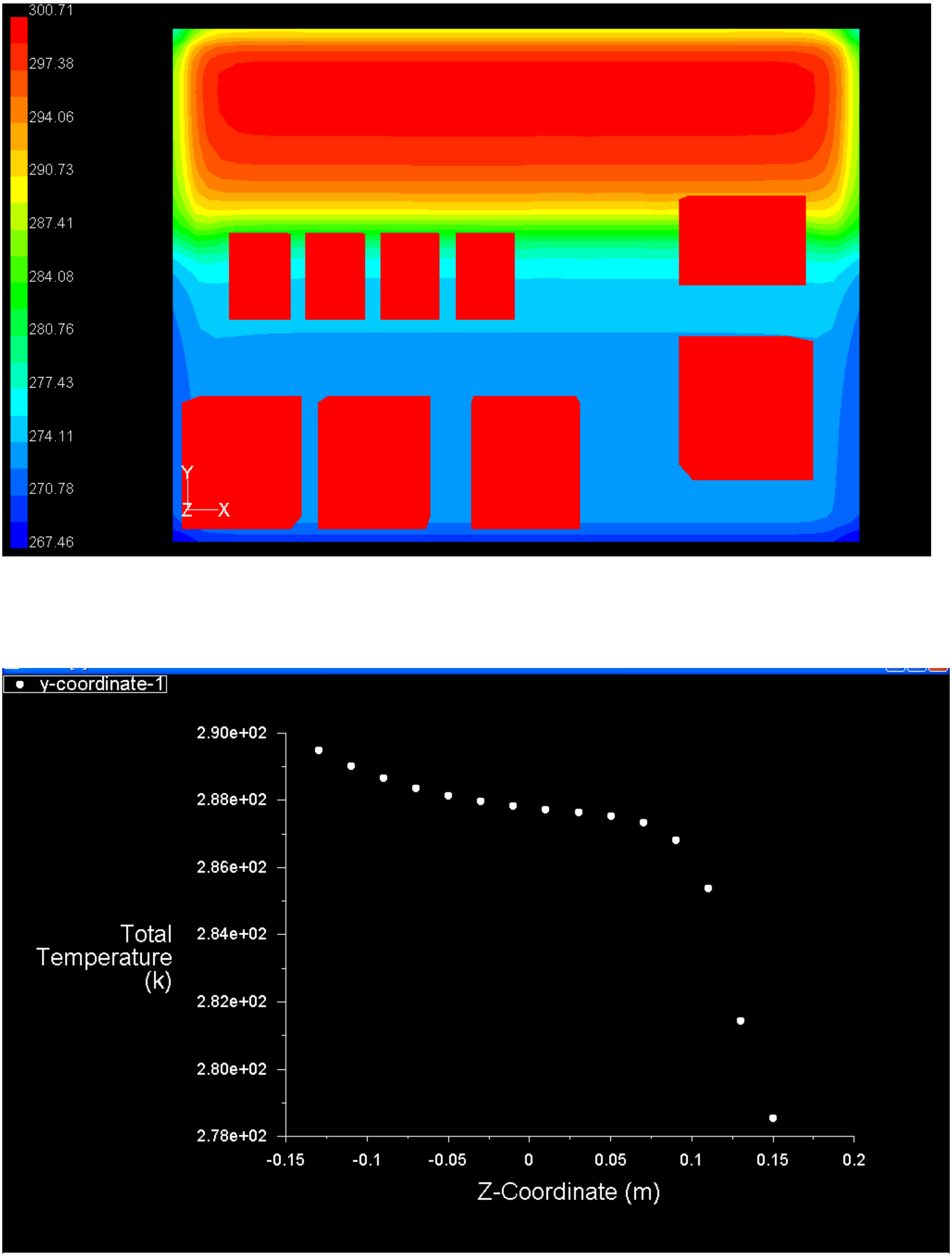

**Figure 6.14** Temperature trend along Z on a central plane (Y=0) .

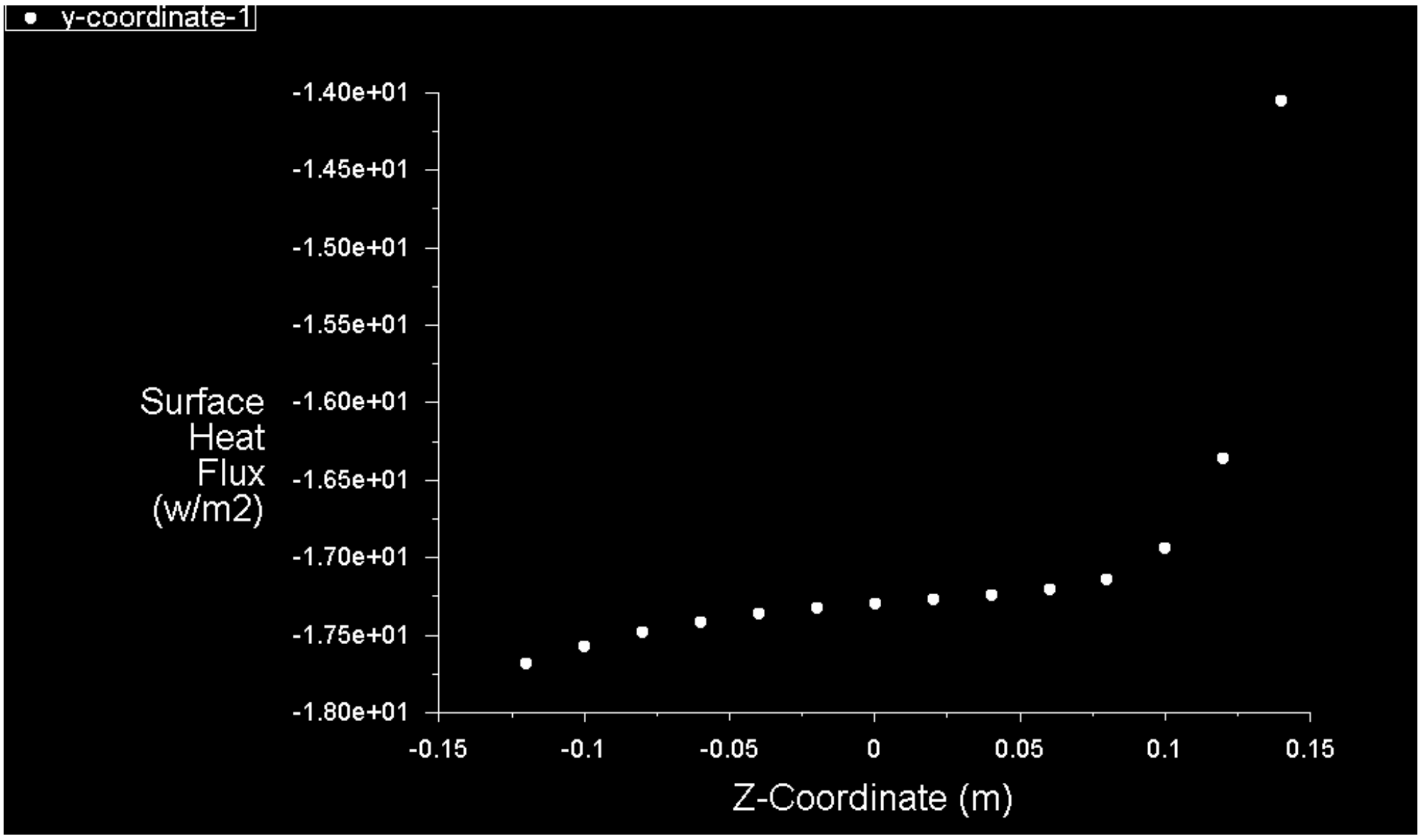

**Figure 6.15** Heat flux plot vs Z on the same plane.

# CHAPTER 7 The AMICA rack

## 7.1 Introduction

In this chapter an outline of the state of art of the design of the mechanical interface between telescope and AMICA camera is given. Considering that almost all electronic devices cannot stand the extremely hostile environmental conditions, we have decided to host the different camera subsystems inside a thermally controlled rack. In particular, following a modular logic settlement, the various parts have been arranged in two cabinets, up and down the co-rotating platform which rotates together with the telescope (Dolci et al.,2006).

## 7.2 Upper cabinet

In order to establish upper box dimensions we have taken into account the tent inner size and that of the co-rotating platform; therefore, the whole rack together with the telescope has to move inside a circle whose maximum diameter is 3.5 m and clearance between upper box and platform must be not lower than 10 mm, in order to prevent that maximum cabinet deflection can cause interference between the two parts (see fig.7.1). Another restraint is given by the distance between the centre of the telescope, assumed on primary mirror vertex, and interface flange, whose initially fixed value was 833.6 mm. Supposing that the insulating layer with external panels reaches a thickness of 70 mm, it has been found a rack provided by ABB, suitable for our use. The standard series IS 2000x1200x600 (height x width x depth) for automation is typically equipped with side, rear, bottom panels, door and a stiffening socket at the base. On the other end we are not interested in all the parts, but mainly in the bearing structure, given by the special bars designed by ABB with 7 folded profile. Such a profile may guarantee resistance to bending moments and more lightness at the same time. Also a solid bottom plate where to fix vacuum pump, piping and related accessories, is important. It must have holes to allow cable passage and linkage to the compressor contained in the lower cabinet.

The cryostat provided by IRLAB, located around the central hole with a box shape (25x46x30 $cm^3$), with a cylindrical cryocooler coldhead (14 cm diameter, 29 cm high), will be surrounded by an insulating layer too. Around the entrance window, made of CdTe, the rack will have a larger hole, in order to consent bolting and calibration to the interface flange: camera mounting must be as far as possible independent from that of the cabinet. Furthermore the camera equipment must be removable for ordinary maintenance operations: for this reason we have thought to use two telescopic rails fastened to the vertical struts. Control and readout electronics with cable carriers is distributed along three rows in agreement with handling clearance around the cryostat: they are attached on DIN 25 bar. The upper row, at a distance of 850 mm over the entrance window, hosts two expansions I/O , a multimeter , a PLC and a power supplier. All components are provided by ABB. The medium row, at a distance of 300 mm below the window, carries twenty relays controlled by PLC. The lower row, lastly, includes seven switches and motor switches.

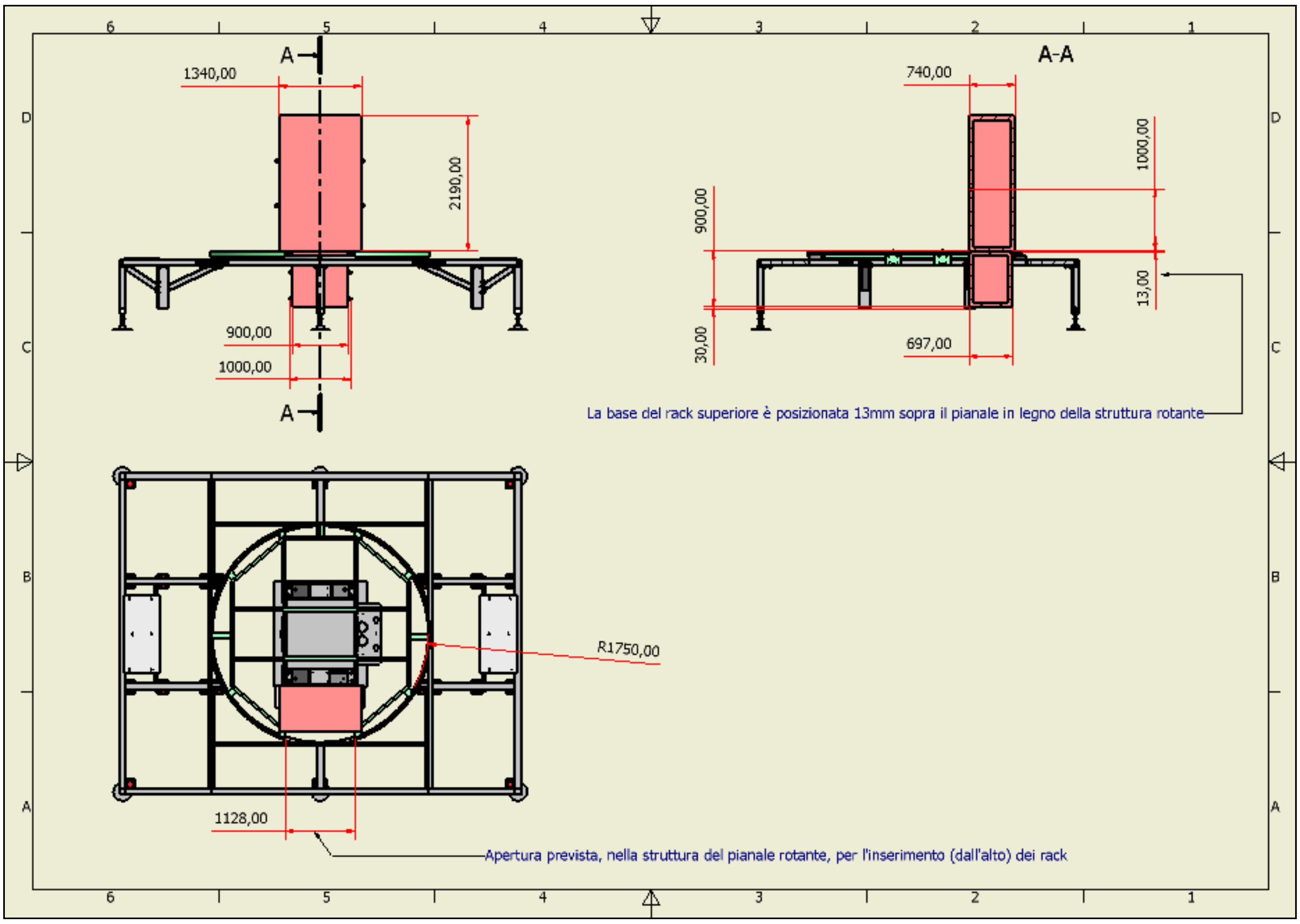

**Figure 7.1 Mechanical layout of overall dimensions and allowable clearance.**

For the insulating enclosure we are now in contact with the firm Baulificio Perugino, the same that has built boxes for the last campaign at Dome C. Since the bearing structure is only provided by the internal struts, it is necessary that the joints for external connection to the telescope start from here. Now the design of the mechanical connection between upper rack and IRAIT system is under development. Two beams running along the rear struts are expected, with a length of 1100 mm. A standard angle steel shape 80x40x6 (DIN 1029) has been chosen, because it fits the ABB feature better.

At a distance of 175 mm from the entrance window, brackets of angle section (75x75x8) are welded to the vertical beams, coming out the rack for 80mm; two sleeves with a 20 mm diameter, for M16 screw insertion, with an overall length of 120 mm, are in charge of bolting the rack to an appropriate beam, located behind the interface flange.

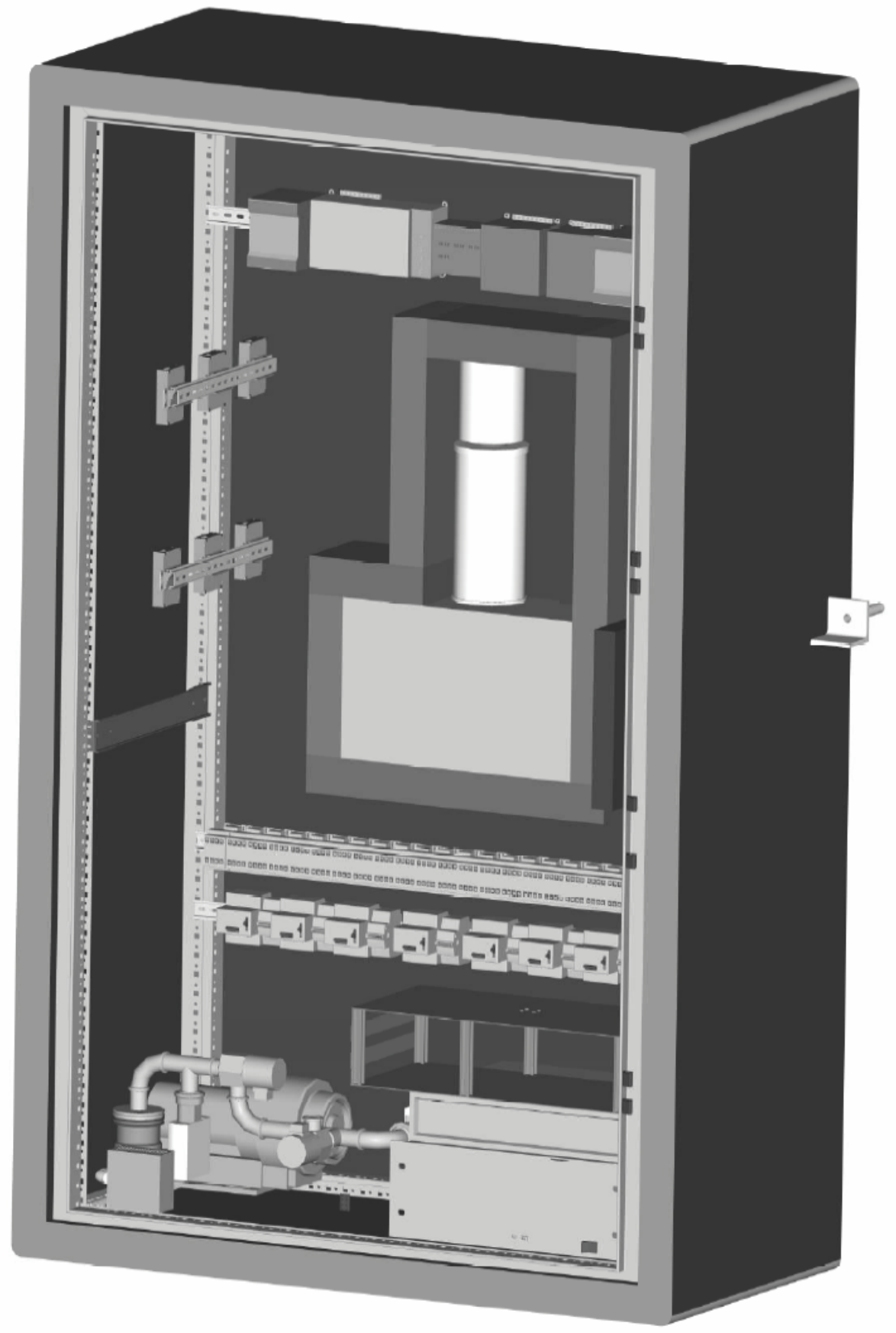

**Figure 7.2** A graphical layout of upper cabinet: cryostat in the central part with the cryocooler coldhead can be distinguished; in the lower part vacuum pump and turbo pump with piping system are visible. Computer and electronic rack are on the lower right side.

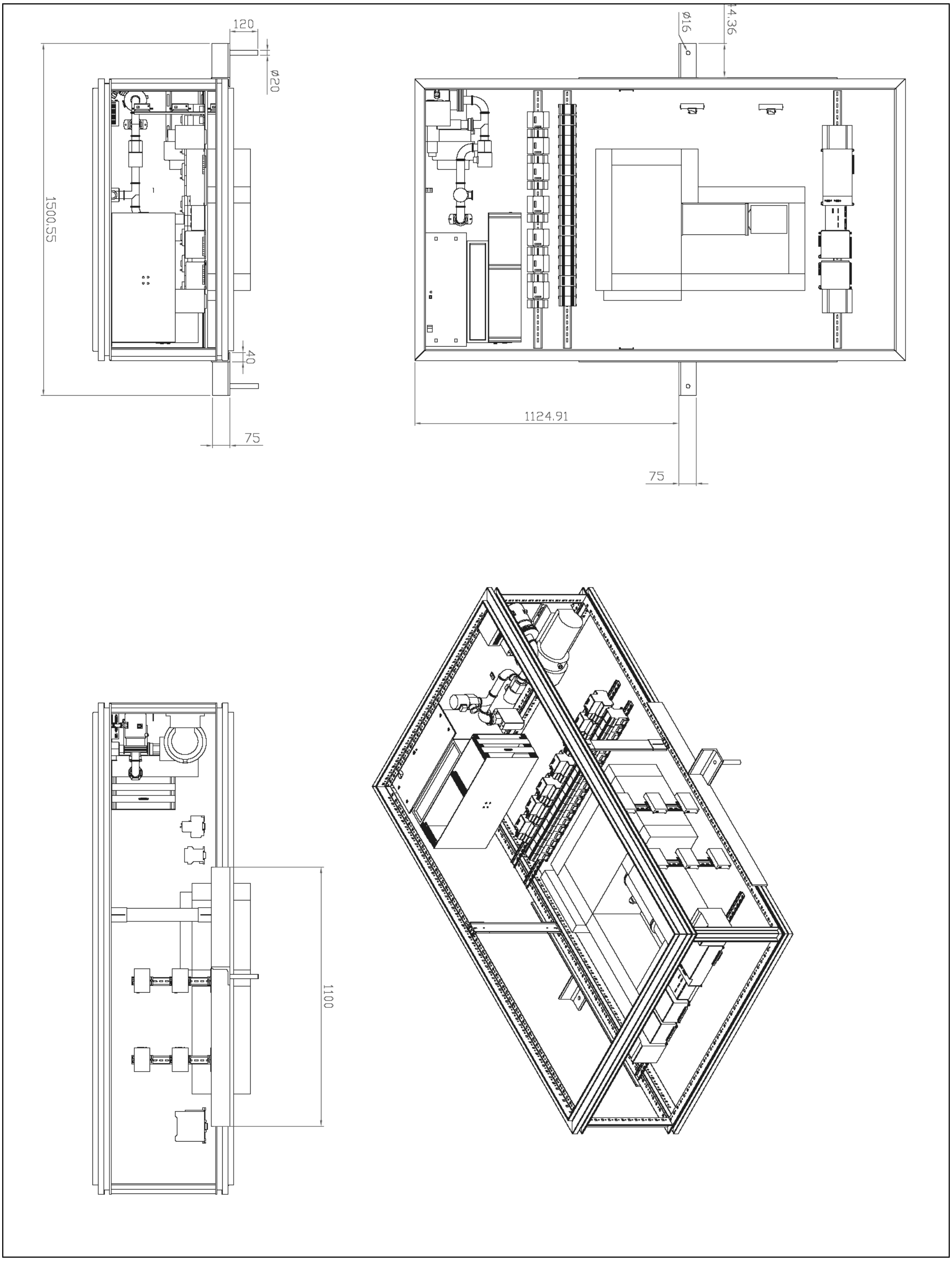

**Figure 7.3** Technical drawing of the upper cabinet assembly with dimensions of the joint elements for the mounting on IRAIT structure.

## 7.3 Lower cabinet

The lower box is 900 mm high, 900 wide, and has a depth of 700 mm. It is internally fixed to the upper cabinet. It includes cryo-compressor with Helium liquid line and refrigerating fluid pipeline. We are currently discussing about what type of compressor to choose: initial dimensions of 627 mm for height, 465 for depth, and 455 for width, were set up in order to determine the outer useful size. A fundamental issue is the dampening system, which must limit large vibrations. A custom system with dimensions compatible with inner cabinet volume is necessary, and a possibility is given by wire rope isolators, on the pattern of those initially conceived for telescope transportation. As in upper cabinet case, two angle joints, mounted on the inner struts, are expected. They have external M24 holes for insertion in the studs located on the basis console, belonging to the mechanical interface (see next paragraph).

## 7.4 Mechanical interface with the telescope

Mechanical interface is devised as a subsystem already installed on the telescope, on which the whole rack has to be mounted. It is composed by two parts: a channel section beam (UPN 45x80x8 mm), bolted on the fork arm behind the flange, with an overall length of 1500 mm and holes of 18 mm diameter to fasten the upper cabinet. This will prevent the rack from tilting.
The other part is a small chassis, shown in the fig. 7.4 ,which is fixed on prearranged side bars below the wood platform by means of 6 M14 screws.
L shaped beams have been selected as upper beams: they have the function of supporting part of the dead weight together with the basis console, so that the chassis is designed to bear the upper and lower cabinet. The beams are linked to the rear plate by square welded joint having dimensions: 100x100x8. The plate has an average thickness of 10 mm. Two ribs at both sides give more stiffness to the assembly (see fig. 7.4 ).
In brief, the sequence of operations planned to be done can be distinguished in the preliminary and *in situ*:

1. preliminary: mounting upper U section bar on the fork arm behind the interface flange (part number 44);

2. preliminary: bolting small chassis of the mechanical interface on the bars below the platform;
3. *in situ*: the cabinets are lifted by a crane and inserted from above by means of four ringbolts at the edges of the top panel until the holes on the brackets of the lower cabinets match the M24 studs;
4. fixing the upper cabinet with interposed sleeves;
5. clamping the studs by special nuts on the basis console, with am articulated wrench (chiave a snodo), as the compartment is accessible only from above the co-rotating platform;
6. adjusting the camera. The rack is opened and the cryostat is pushed on the rails until it is in contact with the flange. Through a mounting system now under development, including flange and counter-flange with a truncated cone centering step, the camera is aligned along the optical axis. Tilt and centering allowed errors are in any case within the estimated mechanical tolerances of a tenth of millimeter.

In fig. 7.5 there is a perspective view of the rack mounted on the telescope.

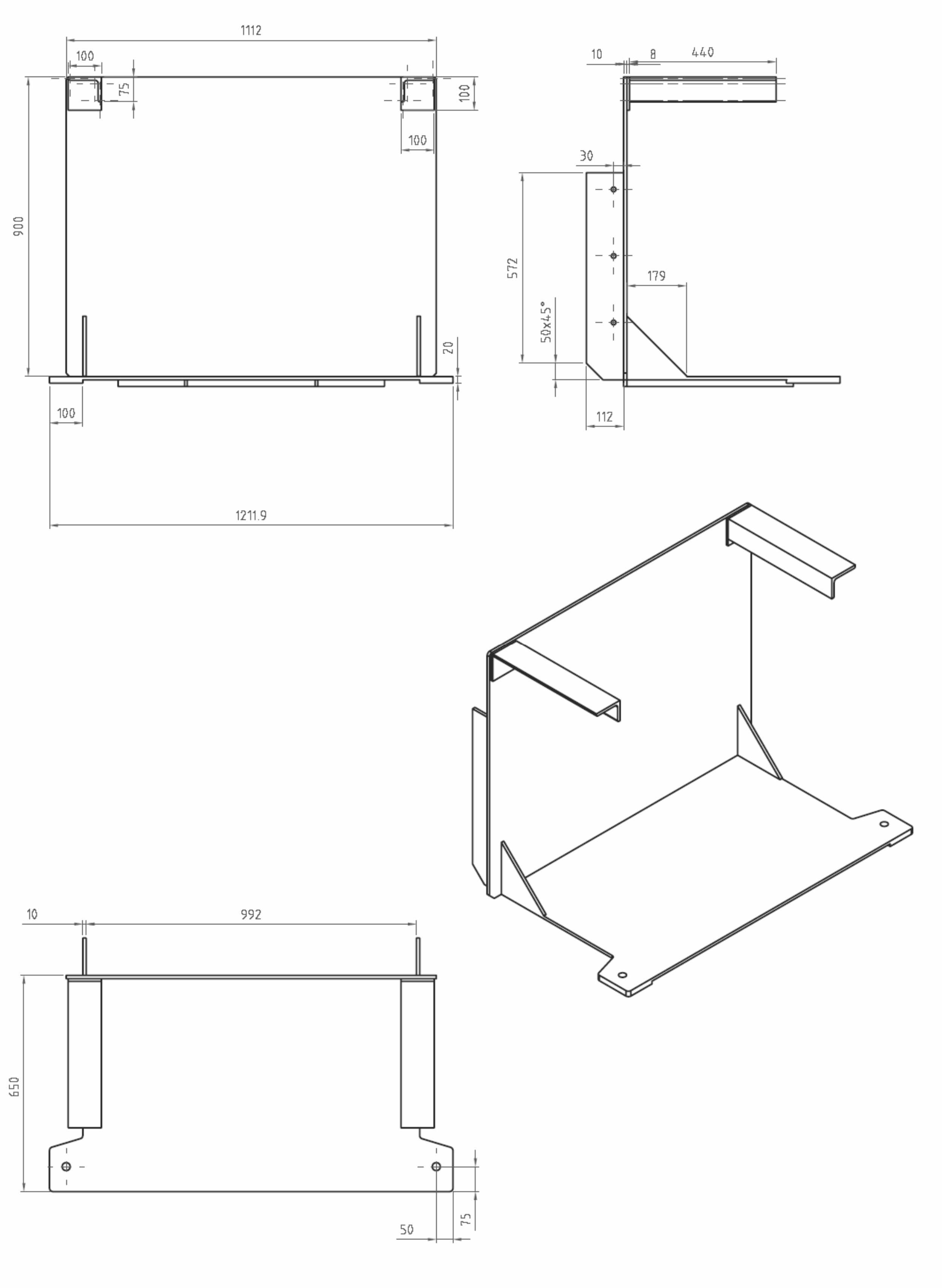

**Figure 7.4** A technical drawing of the mechanical interface bolted to the telescope.

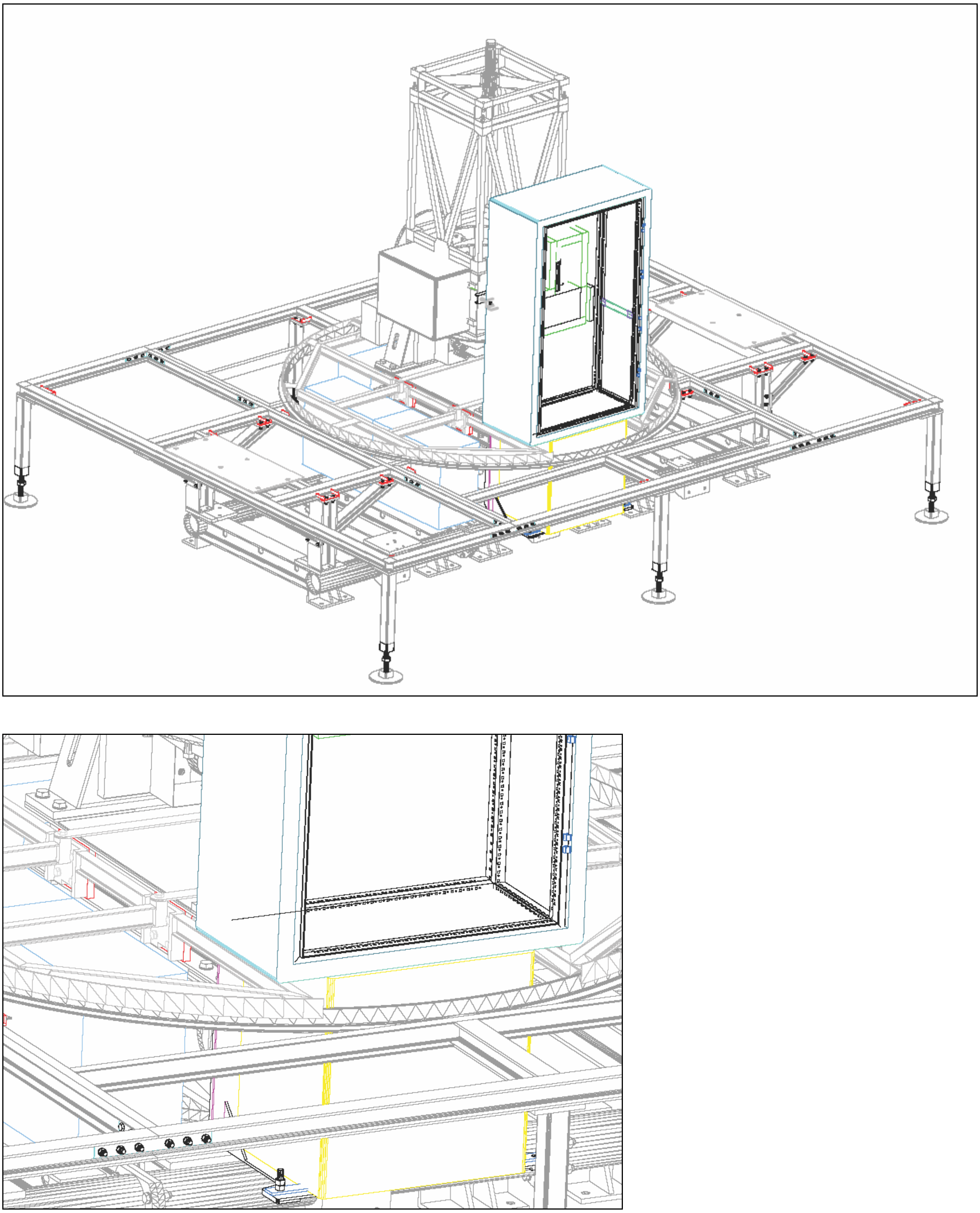

**Figure 7.5** A picture with a zoom view of how the rack is planned to be assembled to the telescope structure.

# Conclusions and remarks

The purpose of the present work has been that of examining the steps followed in designing the IRAIT telescope, pointing out, on the basis of structural analysis (stresses and displacements) through FEM software, the significant contribution of thermal loads, and the motivations of the choice we opted for, as far as transmission gears, drives and machine elements are concerned. Moreover, a study of systematic mechanical errors has been carried out, which together with astrometric errors affects every observation: this correction needs to be considered and put into telescope control software.

Now some integrated tests of both telescope and assembled camera are being planned: they will permit to compare simulated errors with those detected by experiments, for producing a model of periodic disturbs of the different subsystems (drive trains and bearing), and for substantially reduce their influence. In the same time some test on lubricants and motors are under development in a climatic chamber in Perugia.

Simulations under structural analysis revealed to be meaningful for test planning and useful to foresee the operative mode of mechanics, once IRAIT will be at work at Dome C, and it will start to observe the first objects.

# APPENDIX

# A. FEM method analysis

## A.1. Introduction

Finite elements method is the discretization of a continuous structure in elements, which are in the simplest case one-dimensional, or tetrahedral to simulate three-dimensional analysis, and it corresponds in mathematic terms to a passage from differential analytical equations to algebraic equations. The unknowns, which are displacements, must be determined at the nodes and along the beams by interpolating form functions (linear or splines): on the base of applied loads reaction on the external restraints are determined at first. Then the elements are reassembled and a system of equations, in which the unknowns are the nodes displacements, is obtained. Once the displacements are calculated as primary variables, the secondary ones, tensions and moments, are determined.
The amount of operations rapidly increases passing from a beam to a plate or a shell: indeed, stiffness matrix goes from 2x2 to 12x12 dimension in the spatial case.

As a simplifying hypothesis, in general, volumes are equivalently considered as areas with constant thickness. This involves an approximation for the fact that it's not always guaranteed to resolve a complex domain. The choice of elements vary with the complexity of the problem and is not at all arbitrary, but rather is left to the skill and experience of the engineer. For an element *e* defined by n nodes, the followings relations can be written:

$$\{u\}^e = [\Phi]\{a\}^e \qquad (1)$$

Where $\{u\}$ is a compact way to indicate horizontal and vertical displacements internal to the element, i.e. $\begin{Bmatrix} u(x,y) \\ v(x,y) \end{Bmatrix}$, $\Phi(x,y)_{2x2n}$ is the matrix of the form functions, while $\{a\} = \begin{Bmatrix} u_i(x,y) \\ v_i(x,y) \end{Bmatrix}$ is the *2n* vector of displacements in the nodes. For example for a triangular element three vector functions $\Phi_i$ $\Phi_j$ $\Phi_k$ are defined, so that in the nodes the following conditions must be satisfied:

$$[\Phi_i(x_i,y_i)] = \underline{1}, \quad [\Phi_j(x_i,y_i)] = \underline{0}, \quad [\Phi_k(x_i,y_i)] = \underline{0}$$

Then from the displacement vector we pass to the strain vector using compatibility equations:

$$\{\varepsilon\} = \begin{Bmatrix} \frac{\partial u}{\partial x} \\ \frac{\partial v}{\partial y} \\ \frac{\partial u}{\partial y} + \frac{\partial v}{\partial x} \end{Bmatrix} = \begin{bmatrix} \frac{\partial}{\partial x} & 0 \\ 0 & \frac{\partial}{\partial y} \\ \frac{\partial}{\partial y} & \frac{\partial}{\partial x} \end{bmatrix} \begin{Bmatrix} u \\ v \end{Bmatrix} = \left[B(x,y)\right]_{3x2n} \{a\}_{2nx1} \qquad (2)$$

where [B] is the matrix of the form functions derivatives.

By means of Hooke's law for an elastic isotropic (and isothermal) body, we determine the stress components:

$$\{\sigma\} = \begin{Bmatrix} \sigma_x \\ \sigma_y \\ \tau_{xy} \end{Bmatrix} = \frac{E}{1-\nu^2} \begin{bmatrix} 1 & \nu & 0 \\ \nu & 1 & 0 \\ 0 & 0 & (1-\nu)/2 \end{bmatrix} \begin{Bmatrix} \varepsilon_x \\ \varepsilon_y \\ \gamma_{xy} \end{Bmatrix} = [D]\{\varepsilon\} \qquad (3)$$

This is valid in the hypothesis of plane stress. Assuming the presence of residual stresses, as it actually happens, indicating them as $\sigma_0$, and the presence of strains due to temperature or to history of the elastic material by $\varepsilon_0$, we can write:

$$\{\sigma\} = [D]\{\varepsilon - \varepsilon_0\} + \{\sigma_0\} \qquad (4)$$

The problem is defined on a closed domain *Ω,* with the hypothesis that it is restrained on part of the boundary $\Gamma_u$ and is subject to a field of lumped and surface forces on the rest of the contour $\Gamma_f$, then to volume forces applied in V. The boundary of the domain is $\partial\Omega = \Gamma_f \cup \Gamma_u$.

To solve the problem also equilibrium equation must be used, given by:

$$div\boldsymbol{\sigma} + \boldsymbol{b} = 0 \qquad (5)$$

where {b} is the vector of the volume forces.

Contour conditions must be added to equation (4), which are respectively static and cinematic, defined in $\Gamma_u$:

$$\begin{cases} \sigma_{ij} n_j = \hat{t}_j \\ u_i = \hat{r}_i \end{cases} \qquad (6)$$

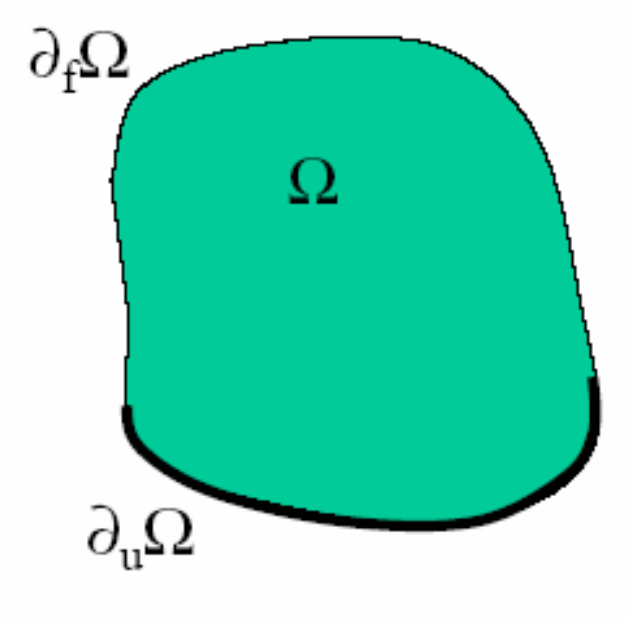

### A.2. Virtual works principle

Virtual works principle consists of applying virtual displacements to the homologous components of node forces, indicated as {q} . For the energy conservation, we know that work due to external forces must be equivalent to that of internal forces, or in other terms:

$$\{\delta a^e\}^T \{q^e\} = \{\delta\varepsilon\}^T \{\sigma\} - \{\delta u\}^T \{b\} \qquad (7)$$

Substituting in the second member the relationship $\{\delta a^e\}^T \left([B]^T \{\sigma\} - [\Phi]^T \{b\}\right)$, then we obtain:

$$\{\delta a^e\}^T \{q^e\} = \{\delta a^e\}^T \left(\int_V [B]^T \{\sigma\} dV - \int_V [\Phi]^T \{b\} dV\right) \qquad (8)$$

As the displacements are virtual, putting them equal to 1 we have:

$$\{q^e\} = \int_V [B]^T \{\sigma\} dV - \int_V [\Phi]^T \{b\} dV \qquad (9)$$

Substituting equation (4) in (8), it leads to:

$$\{q^e\} = \int_V [B]^T [D]\{\varepsilon\} dV - \int_V [B]^T [D]\{\varepsilon_0\} dV + \int_V [B]^T \{\sigma_0\} dV - \int_V [\Phi]^T \{b\} dV \qquad (10)$$

Substituting relation (2) inside (10), we obtain at last:

$$\{q^e\} = \int_V [B]^T [D][B]\{a\} dV - \int_V [B]^T [D]\{\varepsilon_0\} dV + \int_V [B]^T \{\sigma_0\} dV - \int_V [\Phi]^T \{b\} dV \qquad (11)$$

Definitely, setting the quantities as following:

$$\begin{aligned} [K^e] &= \int_V [B]^T [D][B] d(V) \\ \{f^e\} &= \int_V [B]^T [D]\{\varepsilon_0\} d(V) + \int_V [B]^T \{\sigma_0\} d(V) - \int_V [N]^T \{b\} d(V) \end{aligned} \qquad (12)$$

we can write, in a more compact way:

$$\{q^e\} = [K^e]\{a\}^e + \{f^e\} \tag{13}$$

If external distributed loads are present, per unit surface, they generate a virtual work, that must be taken into account. Indicating by $\{\bar{t}\}$ the vector of these forces, we can write the relationship:

$$\delta\overline{L} = -\{\delta u\}^T \{\bar{t}\} \tag{14}$$

Then we obtain the following expression:

$$\{f^e\} = -\int_A [N]^T \{t\} d(A) \tag{15}$$

Such loads are applied only at the nodes, not along the element.
Equation $\{q^e\} = [K^e]\{a\}^e$ can be expressed in terms of displacements of the whole structure; then all the equations written for a single element must be assembled in a single structural matrix.
Referring to equilibrium condition of a single node, all forces converging to it must equal joint reactions and applied loads. Thus, we can write:

$$\sum_{i=1}^{n} \{q^i\} = \{q\} = \left( \sum_{i=1}^{m} [K^i] \right) \{a\} = [K]\{a\} \tag{16}$$

where the summation at the second member indicates the assembling stiffness matrix.
Displacements can be easily converted from local coordinate system to the global one by directional cosines matrix:

$$\{a_l\} = [L]\{a_g\}, \tag{17}$$

indicating with $[L]$ the directional cosines matrix.

As the work does not depend on the coordinate system, it can be written:

$$\{q_g\}^T \{a_g\} = \{q_l\}^T \{a_l\}.$$

And substituting the (17):

$$\{q_g\}^T \{a_g\} = \{q_l\}^T [L]\{a_g\} \tag{18}$$

Then, after some steps:

$$\{q_g\} = [L]^T [K_l]\{a_l\} \Rightarrow \{q_g\} = [L]^T [K_l][L]\{a_g\} = [K_g]\{a_g\} \qquad (19)$$

where the global stiffness matrix id given by:

$$[K_g] = [L]^T [K_l][L] \qquad (20)$$

### A.3. FEM Theory of Plates

Ansys, like any other structural software, based on FEM philosophy, make use of shell elements, among all, descending from Kirchhoff theory of thin plates[6]. Considering a quadrangular element as in fig. A.1, it can be theoretically demonstrated that membrane components are uncoupled from the bending ones. We can treat the two problems separately.

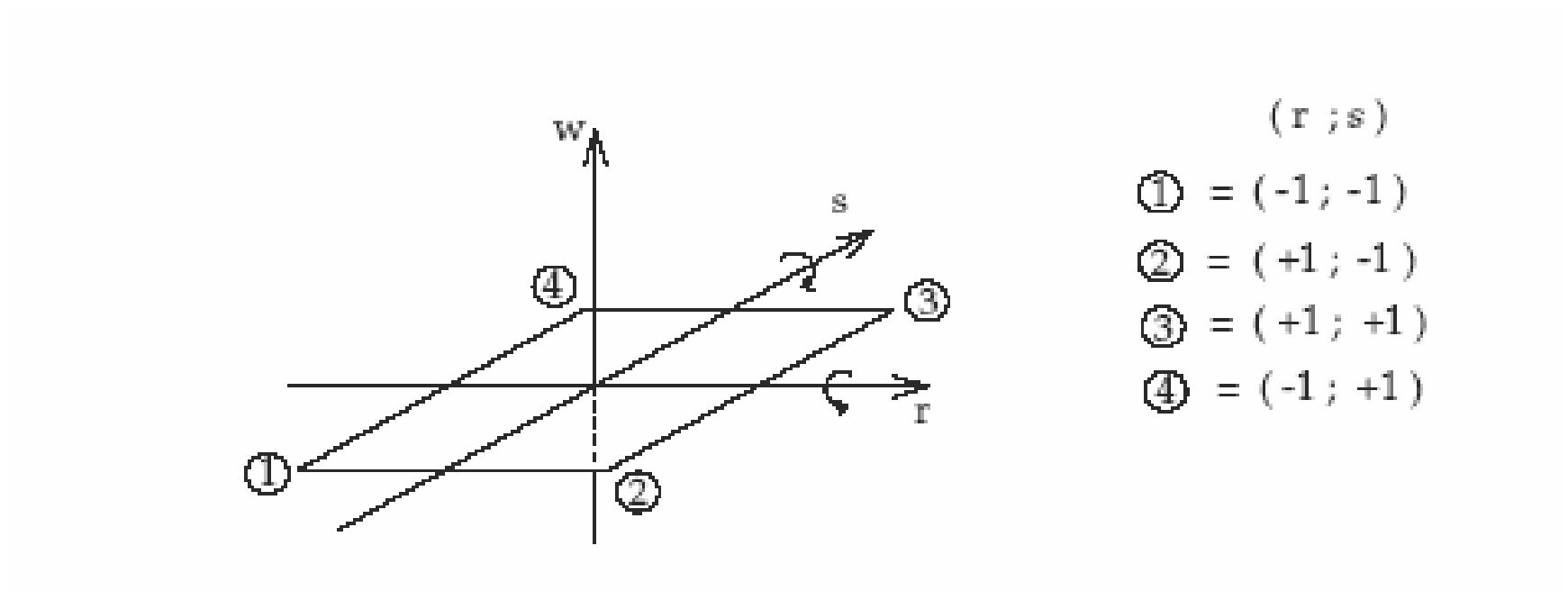

**Fig. A.1** Planar shell element with four nodes.

Membrane displacements: In order to describe the displacement field in plane *(r,s)*, it is necessary to assign two degrees of freedom to each node, along two directions parallel to the axes. Thus, the shape functions, assumed to recompose X displacement components (the same thing is for Y), can be identified by 4 parameters, and the displacements of the four node along *X,* are indicated as $u_{xi}$, with $i=1,2,3,4$. A suitable shape function is the following:

$$\Phi_s(r,s)=(\alpha_1+\alpha_2 r)(\alpha_3+\alpha_4 s) \tag{21}$$

where the four parameters are calculated under condition that the function is equal to 1 at the node, and 0 elsewhere. Therefore, we obtain 4 shape functions:

$$\begin{aligned}
\Phi^1(r,s)&=\frac{(r-1)(s-1)}{4}\\
\Phi^2(r,s)&=\frac{(r+1)(1-s)}{4}\\
\Phi^3(r,s)&=\frac{(r+1)(s+1)}{4}\\
\Phi^4(r,s)&=\frac{(1-r)(s+1)}{4}
\end{aligned} \tag{22}$$

[6] It is supposed that the plate is not subject to any tension in a direction parallel to its plane.

In this way, we can rebuild the displacements field:

$$\begin{aligned}
u_r(r,s)&=\phi^1 u_r^1+\phi^2 u_r^2+\phi^3 u_r^3+\phi^4 u_r^4=\phi^j u_r^j\\
u_s(r,s)&=\phi^1 u_s^1+\phi^2 u_s^2+\phi^3 u_s^3+\phi^4 u_s^4=\phi^j u_s^j
\end{aligned} \tag{23}$$

Bending displacements:

Bending displacements are determined by a component of displacement normal to the plane, and by a rotation around *y* (see fig. A.2): in general it can be ascribed to two rotations around *r* and *s.* We can assign to each node three degrees of freedom: a displacement in the direction normal to the plane, $u_w$, and rotations $u_{\theta r}$ and $u_{\theta s}$. There is a total of twelve bending degrees of freedom, so that 12 parameters must be introduced. The following shape functions can be assumed:

$$\begin{aligned}
f(r,s)&=\alpha_1+\alpha_2 r+\alpha_3 s+\alpha_4 r^2+\alpha_5 rs+\alpha_6 s^2+\alpha_7 r^3+\alpha_8 r^2 s+\alpha_9 rs^2+\\
&+\alpha_{10}s^3+\alpha_{11}r^3 s+\alpha_{12}rs^3
\end{aligned} \tag{24}$$

As for membrane displacements, imposing that the general degree of freedom is equal to 1 and all the others are zero, it is possible to obtain the 12 parameters. For a generic node *i* the shape functions are assumed to be the following:

$$\begin{aligned}
\phi_w^i&=\frac{1}{8}(rr_i+1)(ss_i+1)(2+rr_i+ss_i-r^2-s^2)\\
\phi_w^i&=\frac{1}{8}r_i(rr_i+1)^2(rr_i-1)(ss_i+1)\\
\phi_w^i&=\frac{1}{8}s_i(rr_i+1)(ss_i+1)^2(ss_i-1)
\end{aligned} \tag{25}$$

When passing from the local reference *(r,s)* to the global one *(x,y,z),* element planarity must be preserved. For this reason, an isoparametric transformation needs to be adopted, which can be written as:

$$u_w(r,s) = \phi^1 u_w^1 + \phi^2 u_w^2 + \phi^3 u_w^3 + \phi^4 u_w^4 = \phi^j u_w^j \tag{26}$$

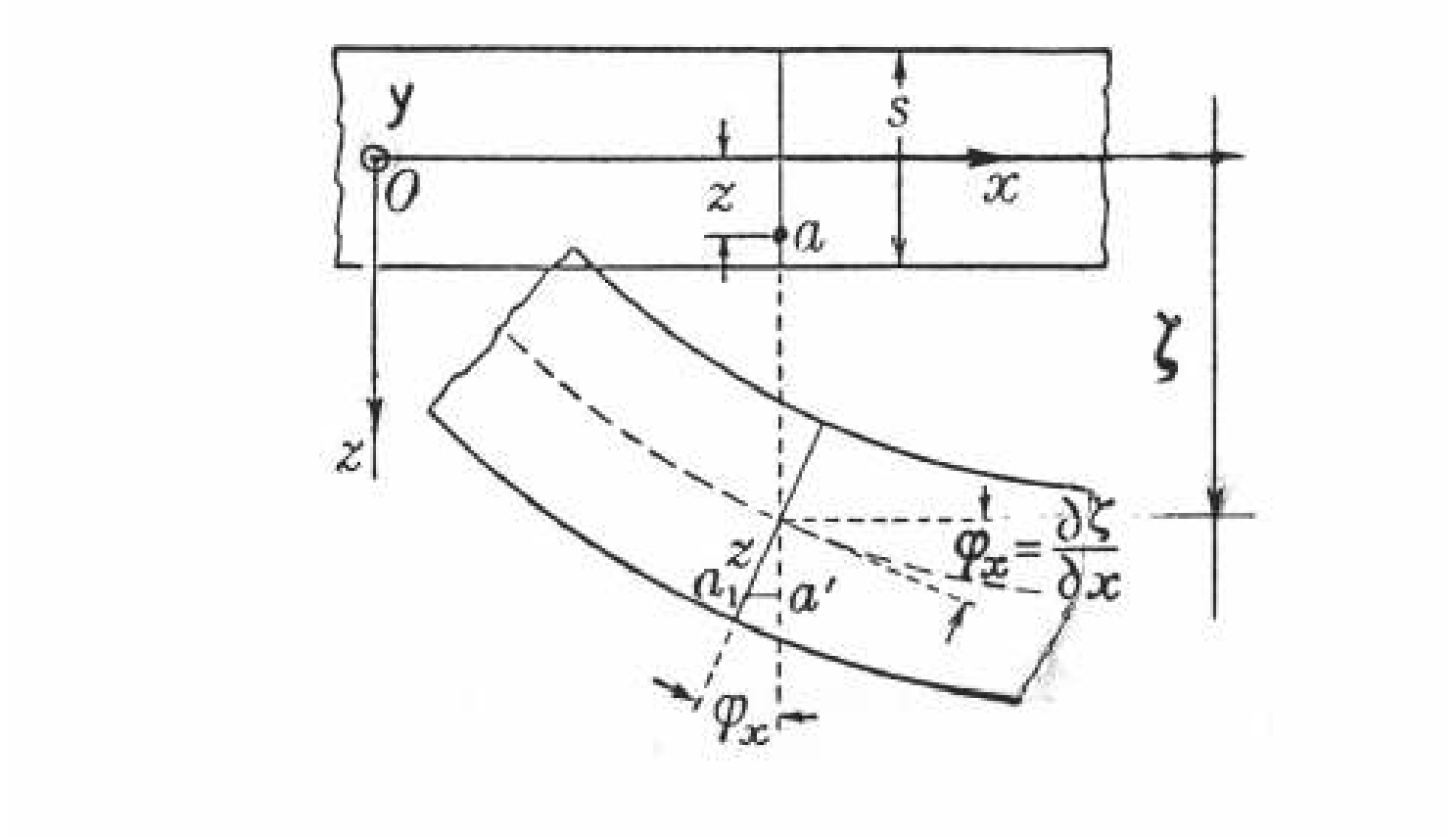

**Fig. A.2** Strain parameters for an inflected plate.

# B. Programs codes

## B.1. A preprocessing ANSYS program to set up a Serrurier strut (geometry)

Here is a routine in order to set up the geometry of a typical Serrurier strut for a small aperture telescope. The user can specify the following parameters: whether top ring is square or circular (as default), the number of spiders (3 at least), diameter of M2 unit, diameter or side of top ring, height of the strut, dimension of beams (pipe or box). The beams are supposed to be made of isotropic, elastic steel. As output the program shows the three dimensional geometry of the shape of the strut. The program is written in APDL, ANSYS Parametric Design Language, a scripting language used to automate common tasks or even build a parametrical model.

M2 unit is assumed as a solid element, namely a hollow cylinder.

```
!*******PREPROCESSING SERRURIER STRUT  PROGRAM****************
!SYSTEM OF UNIT IS M K S
/UNITS,SI
/FILNAME, SERRURIER
!SELECT GEOMETRY OF TOP RING
/PREP7
*SET,PI,3.14159265359
*ASK,N,"NUMBER OF SPIDERS? (DEFAULT VALUE=3)",3
*ASK,D2,"MEDIUM DIAMETER OF THE M2 UNIT?",0.2
*ASK,D3,"MEDIUM APERTURE OF TOP RING?",1
*ASK,D1,"BASE DIAMETER OF M1 UNIT? (MINIMUM=0.40 m)", 1.2
*ASK,Z2,"HEIGHT OF THE M2 MOUNT? (DEFAULT=1.5 m)",1.5
!SPECIFY THE HEIGTH OF TRUSS RESPECT TO  M1
*ASK,Z3,"HEIGHT OF THE STRUT? (DEFAULT=1.5 m)",1.5
*ASK,S,"WIDTH OF M2 FLANGE? (DEFAULT=0.02 m)",0.02
*ASK,Z_C,"LENGTH OF M2 FLANGE? (DEFAULT=0.25 m)",0.25
*ASK,G1,"SQUARE OR CIRCULAR TOP RING? (TYPE S OR C)",'C'
!geometria circolare simulata da poligoni regolari di n lati
!DEFINE CONSTANTS
*SET,R1,0.5*D1
*SET,R3,0.5*D3
*SET,R2,0.5*D2
*SET,W,R2-S
*IF, G1,EQ,'C', THEN
!trovo le coordinate dei 4 vertici del top ring su cui cotruisco il quadrato
*DIM,X2,,N
*DIM,Y2,,N
*DIM,X3,,2*N
*DIM,Y3,,2*N
*DIM,X_C,,N
*DIM,Y_C,,N
!generation of keypoints of M2 mount
*DO,I,1,N,1
        X2(I)=R2*COS(PI*(4*I-1)/(2*N))
        Y2(I)=R2*SIN(PI*(4*I-1)/(2*N))
        K,I,X2(I),Y2(I),Z2
*ENDDO
KEYP=N
K_IN=N+1
*DO,II,1,2*N,1
        X3(II)=R3*COS(PI*II/N)
        Y3(II)=R3*SIN(PI*II/N)
        KEYP=KEYP+1
        K,KEYP,X3(II),Y3(II),Z3
        L,KEYP-1,KEYP
        *IF,KEYP,EQ,(3*N),THEN
        L,K_IN,KEYP
        *ENDIF
*ENDDO
LDELE, 1
KEYP2=3*N
*DO,J,1,2*N,1
        X3(J)=R1*COS(PI*J/N)
        Y3(J)=R1*SIN(PI*J/N)
        KEYP2=KEYP2+1
        K,KEYP2,X3(J),Y3(J),0
        L,KEYP2,KEYP2-1
        *IF,KEYP2,EQ,(5*N),THEN
```

```
            L,KEYP2,(3*N+1)
            *ENDIF
*ENDDO
!PLOTLINES
LDELE,1
LDIV,ALL,2
*DO,JJ,(3*N+1),(5*N),2
            L,JJ,(JJ+2*N)
*ENDDO
*DO,KK,(3*N+2),(5*N),2
            L,KK,(KK+2*N-1)
*ENDDO
I=0
*DO,I,1,N,1
            L,I,(5*N+2*I-1)
*ENDDO
!BUILD THE M2 MOUNT LOWER NODES
            *DO,K_C,1,N,1
            X_C(K_C)=R2*COS(PI*(4*K_C-1)/(2*N))
            Y_C(K_C)=R2*SIN(PI*(4*K_C-1)/(2*N))
            K,(K_C+9*N),X_C(K_C),Y_C(K_C),(Z2-Z_C)
            *ENDDO
J=0
!BUILD THE BEAMS OF TOP RING
*DO,J,1,N,1
            L,(5*N-1+2*J),(9*N+J)
*ENDDO
!draw areas between the spider nodes
            *DO,J2,1,N,1
            A,J2,(5*N+2*J2-1),(9*N+J2)
            *ENDDO
*ELSEIF, G1,EQ,'S',THEN
!SQUARE TOP RING(4 SPIDERS)
*SET,N,4
*DIM,X1,,N
*DIM,Y1,,N
*DIM,X2,,N
*DIM,Y2,,N
*DIM,X3,,N
*DIM,Y3,,N
*DIM,X_C,,N
*DIM,Y_C,,N
*DO,I,1,N,1
            X2(I)=R2*COS(2*PI*I/N+PI/4)
            Y2(I)=R2*SIN(2*PI*I/N+PI/4)
            X3(I)=R3*COS(2*PI*I/N+PI/4)
            Y3(I)=R3*SIN(2*PI*I/N+PI/4)
            K,I,X2(I),Y2(I),Z2
            K,(I+4),X3(I),Y3(I),Z3
            X1(I)=R1*COS(2*PI*I/N+PI/4)
            Y1(I)=R1*SIN(2*PI*I/N+PI/4)
            K,(I+8),X1(I),Y1(I),0
            L,I,I+4
            L,(I+N),(I+2*N)
*ENDDO
L,5,6
L,6,7
L,7,8
L,5,8
L,9,10
L,10,11
L,11,12
L,9,12
LDIV,13,2
```

```
LDIV,14,2
LDIV,15,2
LDIV,16,2
J=0
*DO,J,1,N,1
        L,(J+4),(J+12)
*ENDDO
L,5,16
L,6,13
L,7,14
L,8,15
        *DO,K_C,1,N,1
        X_C(K_C)=R2*COS(2*PI*K_C/N+PI/4)
        Y_C(K_C)=R2*SIN(2*PI*K_C/N+PI/4)
        K,(K_C+4*N),X_C(K_C),Y_C(K_C),(Z2-Z_C)
        *ENDDO
L,1,17
L,2,18
L,3,19
L,5,17
L,4,20
L,6,18
L,7,19
L,8,20
!draw areas between the spider nodes
A,1,5,17
A,2,6,18
A,3,7,19
A,4,8,20
*ENDIF
CYLIND,R2,W,Z2-Z_C,Z2,,
/PNUM,KP,1
!CHOOSE ELEMENT TYPE-> BEAM4, BEAM3, OR ELSE
!SUPPOSE THAT ALL ELEMENTS ARE OF THE SAME SHAPE AND TYPE
!STEEL ELEMENT
!CHOOSE SECTION TYPE: HOLLOW BOX OR TUBE
!INSERT SECTION DIMENSIONS
*ASK,HSHAPE,"HOLLOW BOX OR TUBE?",'HB'
        *IF,HSHAPE,EQ,'HB',THEN
                *ASK,B,"INSERT BOX DIMENSIONS:B ",0.05
                *ASK,H,"INSERT BOX DIMENSIONS:H ",0.05
                *ASK,W1,"INSERT BOX THICKNESS:W1 ",0.005
                *ASK,W2,"INSERT BOX THICKNESS:W2 ",0.005
                ET,1,BEAM4
                MP,EX,1,210000
                MP,PRXY,1,0.3
                R,1,(B*H-(B-2*W1)*(H-2*W2)),(B*H**3-(B-2*W1)*(H-2*W2)**3)/12,(H*B**3-(H-2*W2)*(B-
2*W1)**3)/12,B,H,,
        !RMORE,W1,W2,
        *ELSEIF,HSHAPE,EQ,'HT',THEN
        *ASK,DIA,"What's outside diameter?",0.06
        *ASK,TH,"What's wall thickness?",0.003
        !circular element
        ET,1,PIPE16
        R,1,DIA,TH,
        MP,EX,1,210000
        MP,PRXY,1,0.3
        *ENDIF
                *IF,G1,EQ,'S',THEN
                LSEL,S,LINE,,9,28
                LSEL,A,LINE,,4
                LSEL,A,LINE,,6
                LSEL,A,LINE,,8
                LSEL,A,LINE,,2
```

```
                        ET,2,SHELL131
                        R,2,0.003,0.003,
                        ASEL,S,AREA,,1,9
                        ESIZE,,6
                        AMESH,1,9,1
                        LMESH,ALL
                        *ENDIF
FINISH
/ESHAPE,1,1
```

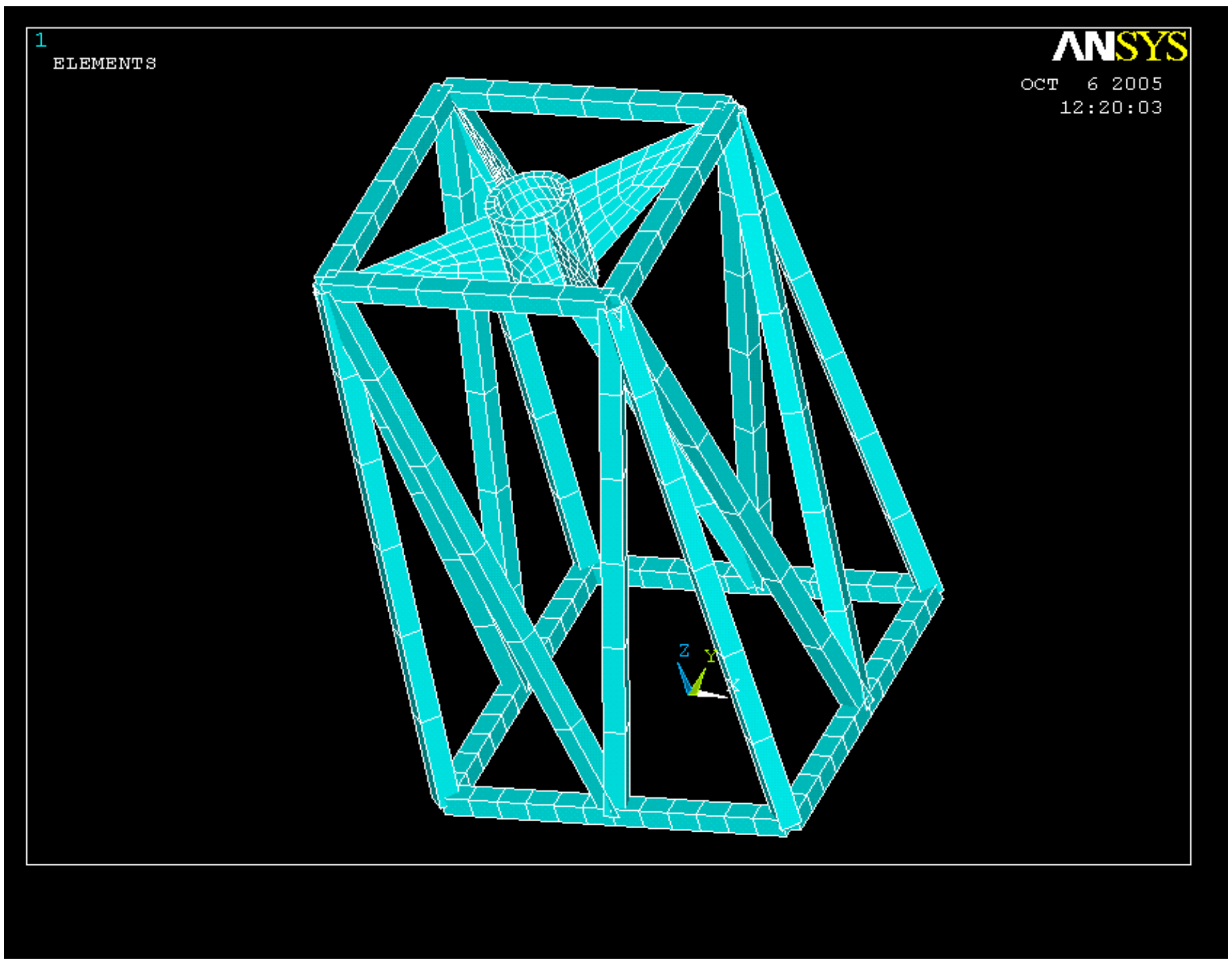

**Fig. B.1** A schematic of the layout of truss elements with default input sizes.

## B.2. MATLAB program for calculating tilt errors

The *tiltmeter.m* program, written in MATLAB has the function of calculating correction in altazimuth coordinates, elaborating data retrieved from a tiltmeter.

```
%tiltmeter program
clear all; clc;
format  compact;
%input AR e DEC of a star at transit: Canopus
%RA=6 h 24.092 m, dec=-52°41.812', fi=-75°6'25"
%theta1=15 microrad,theta2=-15 microrad ..  tiltmeter resolution
%d= offset along azimuth axis Z=0.01
theta1=30E-06;theta2=-30E-06;
d=0.00;
% supposing the star is nearby the meridian: 1 min offset
%h=deg2rad(0);
h=deg2rad(0.4);
%latitude
fi=-75.10;
```

```
z=acos(sin(deg2rad(fi))*sin(deg2rad(-52.697))+cos(deg2rad(fi))*cos(deg2rad(-
52.697))*cos(h));
a=atan(sin(h)/(sin(fi)*cos(h)-cos(fi)*tan(-52.697)));
az=rad2deg(a);
Z=rad2deg(z);
x0=sin(z)*cos(a);
y0=sin(z)*sin(a);
z0=cos(z);
%rotation matrices
Rx=[1 0 0 0; 0 cos(theta1) sin(theta1) 0; 0 -sin(theta1) cos(theta1) 0; 0 0 0
1];
Ry=[cos(theta2) 0 -sin(theta2) 0; 0 1 0 0; sin(theta2) 0 cos(theta2) -d; 0 0 0
1];
A=[x0 y0 z0]';
A_1=Rx*Ry*[x0 y0 z0 1]';
xx1=A_1(1,:);
yy1=A_1(2,:);
zz1=A_1(3,:);
a1=atan(yy1/xx1);
z1=acos(zz1);
%value of A Z in degrees
A1=rad2deg(a1);
Z1=rad2deg(z1);
disp('compare the original values of (A,Z) of a star with the actual
ones(A1,Z1)');
fprintf('A =   %8.5f  Z=   %6.4f \n',az,Z);
fprintf('A1=   %8.5f  Z1=   %6.4f \n',A1,Z1);
disp('correction in arcsec');
fprintf('errA=  %8.5f  errZ1=   %6.4f',3600*(A1-az),3600*(Z1-Z));
fprintf('\n');
disp('estimate relative error');
fprintf('rel_err A=  %8.5f  rel_err Z=   %6.4e\n',abs((A1-az)/A1),abs((Z1-
Z)/Z1));
fprintf('h =  % 5.2f° \n', rad2deg(h));
```

## B.3. Program to determine blind spot size

This is a program called *blinspot.m*, written in MATLAB, used to determine the blind spot size in the neighbourhood of zenith position, starting from a declination point very close to declination:

```
%PROGRAM- DETERMINATION OF BLIND SPOT
format compact;
p=(75.1-75.059)/500;%step
dec=75.059:p:75.1;%declination angle
fi=75.1; %latitude
p1=1.5/500;
h=-0.5:p1:1; %hour angle (in degrees)
% zenith angle
Z=acos(sin(deg2rad(fi)).*sin(deg2rad(dec))+cos(deg2rad(fi)).*cos(deg2rad(dec)).*
cos(deg2rad(h)));
A_h= atan(sin(deg2rad(h))./(sin(deg2rad(fi)).*cos(deg2rad(h))-
cos(deg2rad(fi)).*tan(deg2rad(dec))));
h_0=-acos((sin(deg2rad(dec)).*(1/720-
sin(deg2rad(fi)))+(1+(sin(deg2rad(dec))/720).^2-
sin(deg2rad(fi))/360).^(1/2))./...
    (cos(deg2rad(dec)).*cos(deg2rad(fi))));
```

```
A_0= atan(sin(h_0)./(sin(deg2rad(fi)).*cos(h_0)-
cos(deg2rad(fi)).*tan(deg2rad(dec))))+pi.*(1-sign(fi-dec));
h_1=h_0-pi*sin(A_0)/360;
%guess point
A_1=atan(sin(h_1)./sin(deg2rad(fi)).*cos(h_1)-
cos(deg2rad(fi)).*tan(deg2rad(dec)))+pi.*(1-sign(fi-dec));
A_primo=(sin(deg2rad(fi))-sin(deg2rad(dec)).*cos(Z))./(sin(Z).*sin(Z))*0.25;
h_plus=(h_1-(A_1-A_0-(h_1-h_0)*360)/(A_primo-360))*57.29*4;% hour angle in min
of time
plot(h_plus,(fi-dec),'r.');
xlabel('hour angle [min]','Fontsize',12);
ylabel(' \delta - \phi [deg] ','Fontsize',12);
```

## B.4. Base chassis

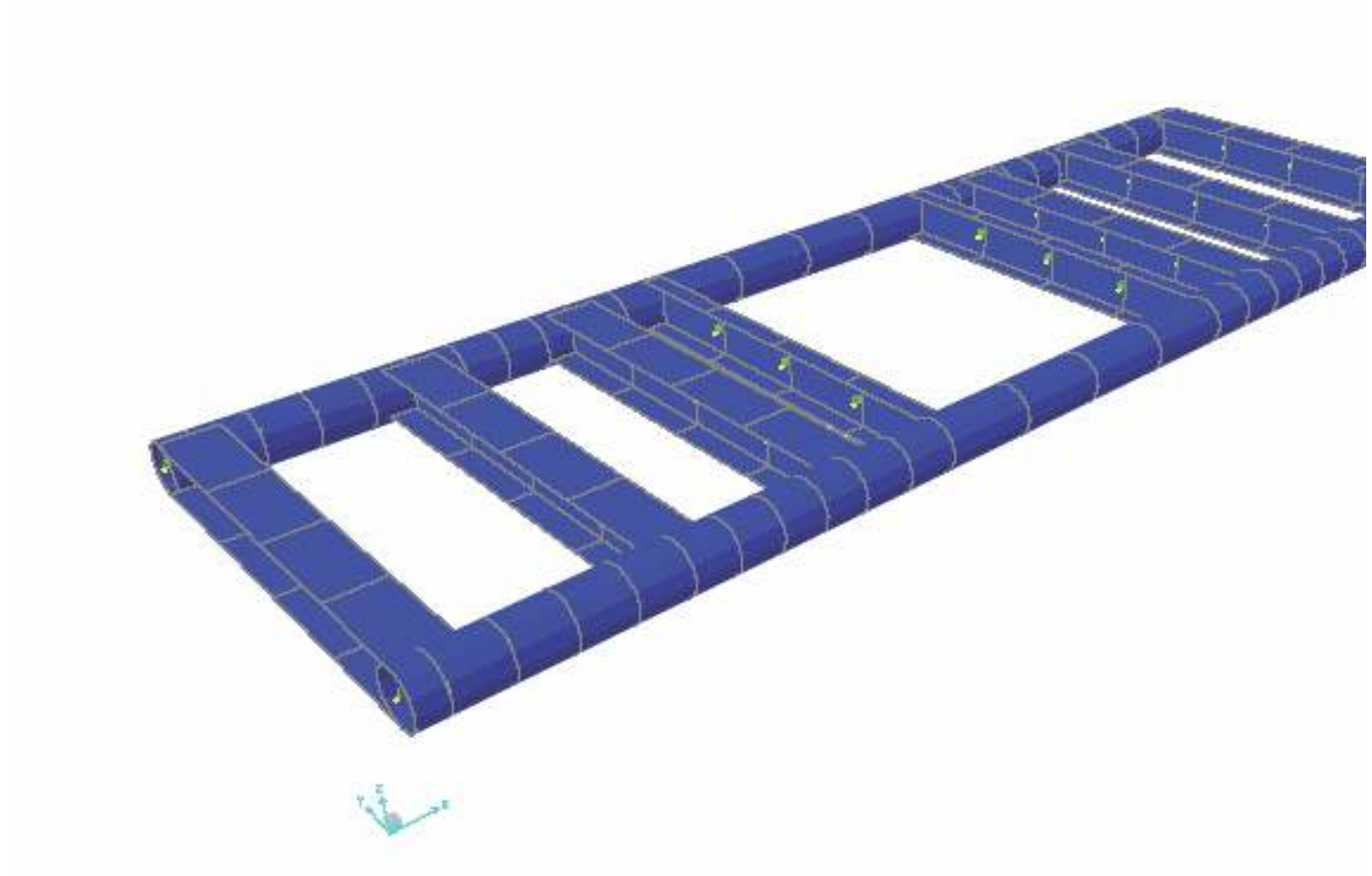

**Fig. B.2** View of the beams composing basis chassis.

Characteristics of chosen beams are illustrated in the next table.

| Section type | Dimensions [mm] | Area [$mm^2$] | Ix [cm4] | Iy [cm4] | Torsional mom J[cm4] | Radius of gyration X [mm] | Radius of gyration Y[mm] |
|---|---|---|---|---|---|---|---|
| 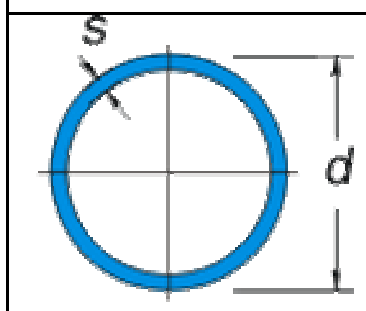  PIPE230 | s=15<br>d=230 | 9189 | 4393.5664 | 4393.5664 | 8787.1328 | 69.146 | 69.146 |
| 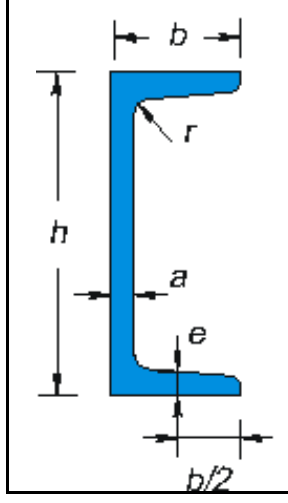  | b=80<br>h=220<br>a=9<br>b=12.5 | 3740 | 2691 | 196 | | 84.8 | 22.9 |
| UPN220 | | | | | | | |
| H200x200<br>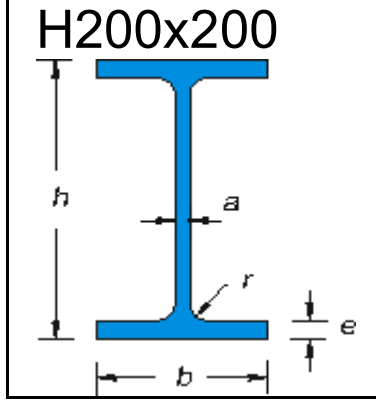<br> | b=200<br>h=200<br>e=15<br>a=9 | 7530 | 5513.4750 | 2001.0328 | 46.8669 | 85.56 | 51.55 |

**Table B-1 Features of the beams used for base chassis.**

## B.5. Wire rope isolators project

Wire rope isolators are special devices that absorb energy through the deformation of stainless steel stranded cable. It exhibits excellent damping characteristics, can provide support for dead weight loads, and has a high cycle fatigue life. They can be mounted in different configuration , depending on how loads are oriented. For our kind of application we thought of compression layout. The company provides an application worksheet with formulas included to dimension the proper isolators: they concern vibration sizing and shock sizing.
First of all we start calculating the weight supported by each of them, supposing that active load is equally distributed along the chassis. With an overall weight of 10500 and the selection of 12 isolators, supported weight is W= 8.75kN. The reaction of the cables to tension and compression is non linear and the dissipated energy is given by the difference of the areas subtended by the two curves. IN order to reach 80% of isolation, the natural frequency has to be at least 1/3 of the active force, so the requested value is 5 Hz. Shock resistance can be calculated by half-sine acceleration formula: $V = \frac{2g}{\pi} A_0 t_0$ where $A_0$ stands for peak acceleration, $t_0$ the duration time. Then, other parameters characterizing dynamic response can be also calculated:

Min response deflection: $D_{min} = \frac{V^2}{g(G_T - 1)1000}$, where $G_T$ is the maximum allowed transmitted load.

Average force: $F_{avg} = \frac{WV^2}{2gD_{min}1000}$

Average deflection: $d_{avg} = \frac{D_{min}}{2}$

As the weight is very high, compared to the average load that a single isolator can bear, ENIDINE provided us a custom solution.

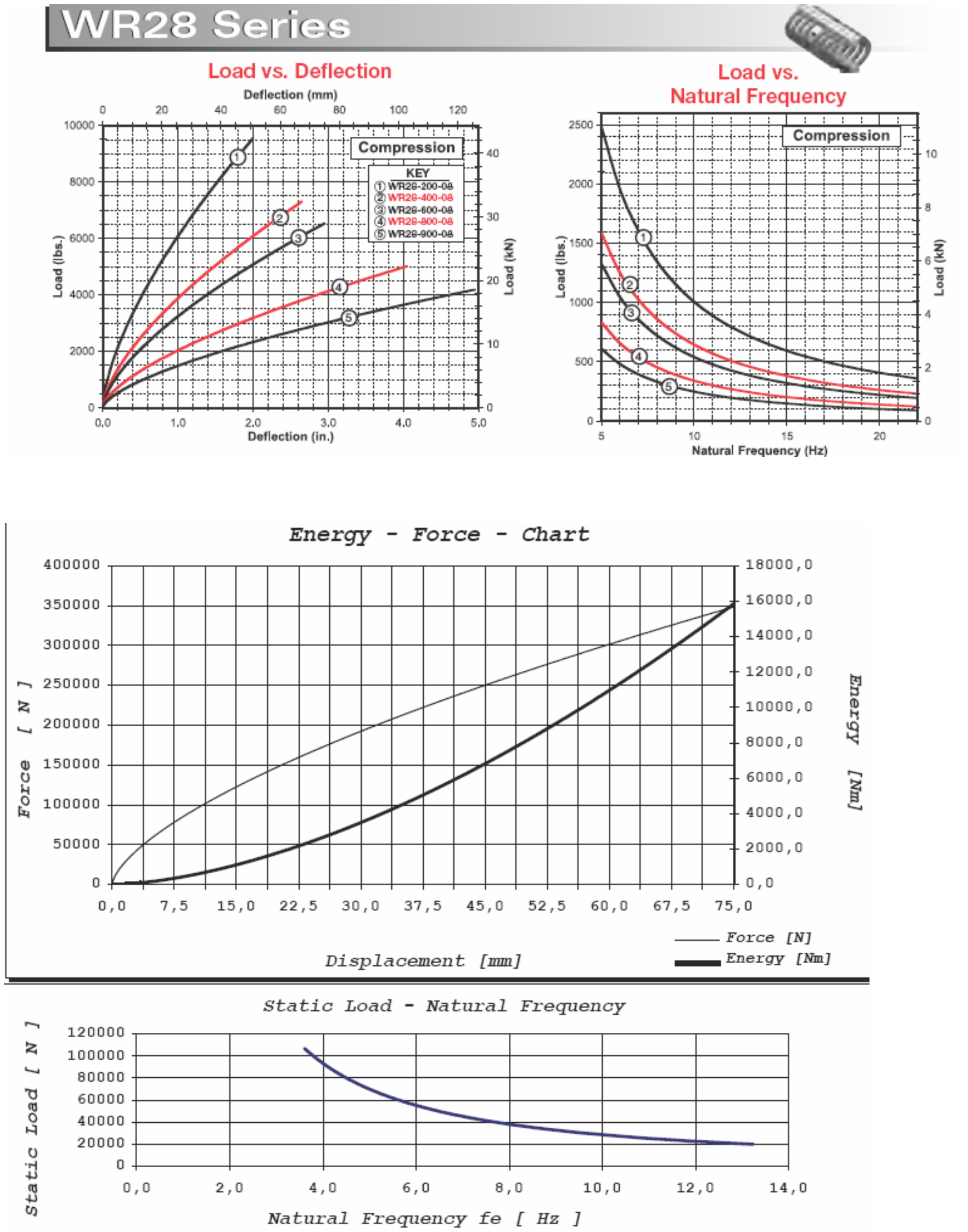

**Fig. B.3** Deflection and natural frequency curves of a wire rope isolator.

| | |
|---|---|
| Mass to damp | 10500 kg |
| acceleration | 3g |
| Maximum allowable deflection | 75 mm |
| dynamic deflection | 16.3 mm |
| Static deflection | 11.4 mm |
| Static stiffness | 5841.27 N/mm |
| Damped natural frequency | 3.7 Hz |
| Acceleration response | 1.3g |

**Table B-2** Characteristics of a wire rope isolator for our application.

# C. Catalogues and products technical features

## C.1. Gearbox

| | | |
|---|---|---|
| Grandezza riduttore | | TP050 |
| Versione riduttore | | MA – High Torque |
| Guarnizione | | FPM |
| Rapporto di riduzione | i | 66 |
| Flangia di trasmissione in uscita | | standard |
| Diametro del calettatore in ingresso | D10 | Max 38 mm |
| Posizione di montaggio | | **2 pz - B5 orizzontali 2 pz - V1 verticale verso il basso** |
| Lubrificazione | | PG220 |
| ****)** Coppia di accelerazione max con 1000 cicli/ora | $T_{2B}$ | 950 Nm |
| ****)** Coppia di emergenza | $T_{2Not}$ | 2375 Nm |
| ****)** Coppia nominale continuativa | $T_{2N}$ | 675 Nm |
| Velocità max in ingresso | $n_{1max}$ | 5000 rpm |
| Velocità nominale in ingresso | $n_{1nom}$ | 2600 rpm |
| Gioco angolare | j | < 1 arcmin (standard) |

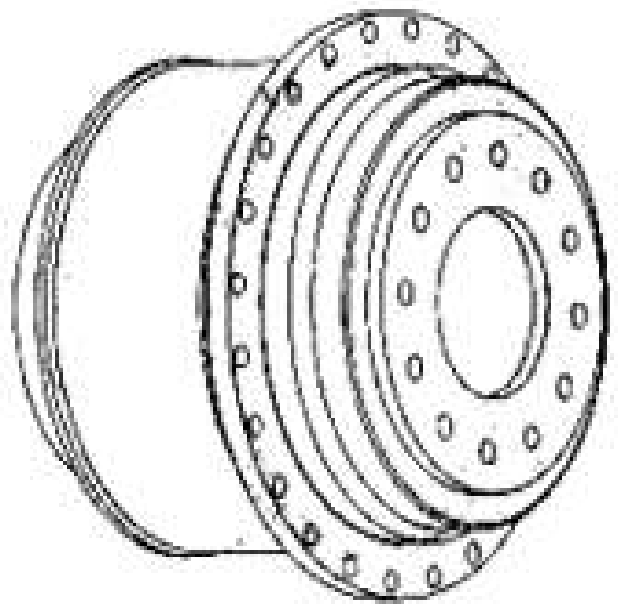

**Table C-1 Technical data of Alpha Riduttori.**

## C.2. Tiltmeter provided by GEOMECHANICS

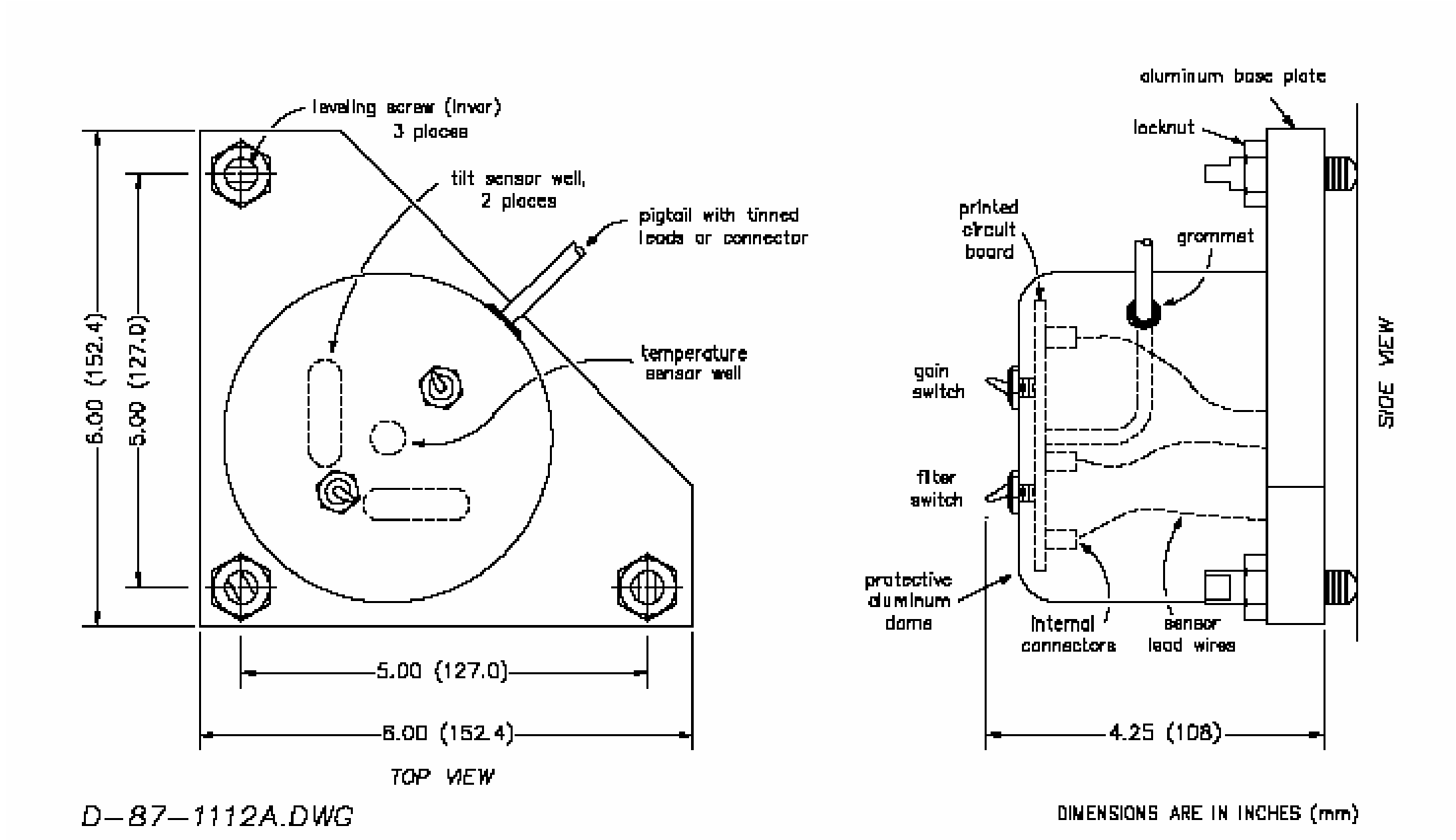

**Fig. C.1** Platform tiltmeter,Model 701-2.

## C.3. Joints (by FAVARI)

TSCHAN

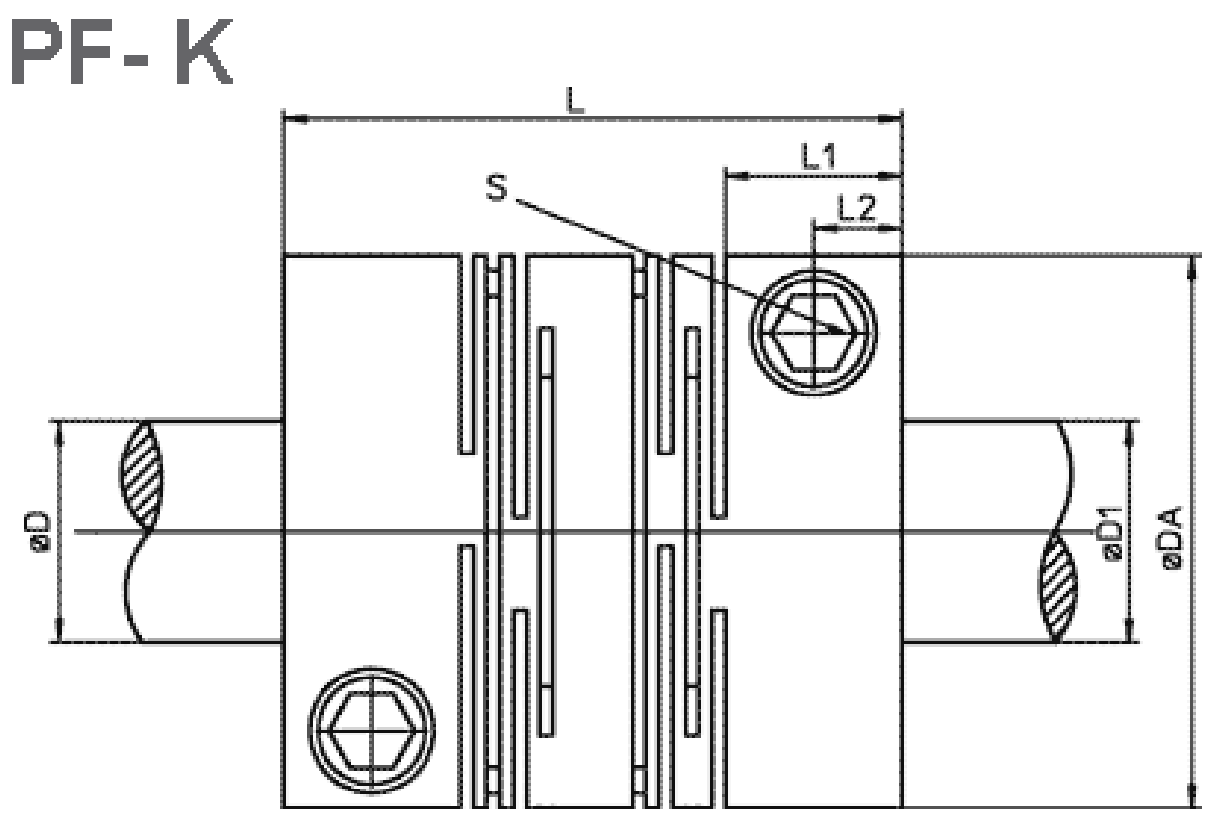

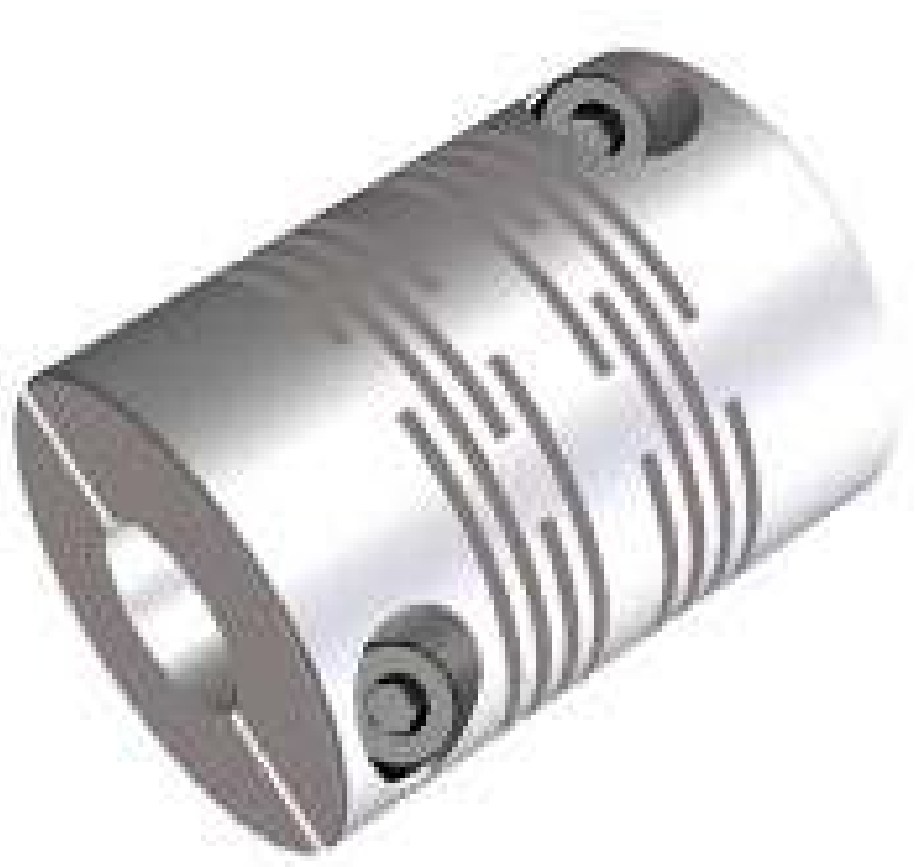

| Abmessungen / Dimensions / Cotes | | | | | | | | |
|---|---|---|---|---|---|---|---|---|
| Ident Nr<br>Id. No.<br>No. de code | Größe<br>Size<br>Tailles | L | L1 | L2 | D | D1 | DA | S<br>(DIN 912) |
| WK1016 | 16 | 23 | 7 | 3,5 | 3-6 | 3-6 | 16 | M2,5x6 |
| WK1018 | 18 | 16,6 | 5,5 | 2,75 | 3-6 | 3-6 | 18 | M2,5x8 |
| WK1020 | 20 | 28 | 8 | 4 | 3-8 | 3-8 | 20 | M2,5x8 |
| WK1022 | 22 | 20 | 5,5 | 2,75 | 3-10 | 3-10 | 22 | M2,5x8 |
| WK1025 | 25 | 28 | 8 | 4 | 6-12 | 6-12 | 25 | M3x10 |
| WK1030 | 30 | 40 | 11 | 5,5 | 6-14 | 6-14 | 30 | M4x10 |
| WK1040 | 40 | 48 | 11 | 5,5 | 6-19 | 6-19 | 40 | M5x14 |
| WK1050 | 50 | 65 | 19 | 9,5 | 10-26 | 10-26 | 50 | M6x16 |
| WK1060 | 60 | 80 | 25 | 12,5 | 10-30 | 10-30 | 60 | M8x18 |
| WK1070 | 70 | 95 | 25 | 12,5 | 15-35 | 15-35 | 70 | M8x25 |
| WL1080 | 80 | 100 | 25 | 12,5 | 20-40 | 20-40 | 80 | M8x25 |

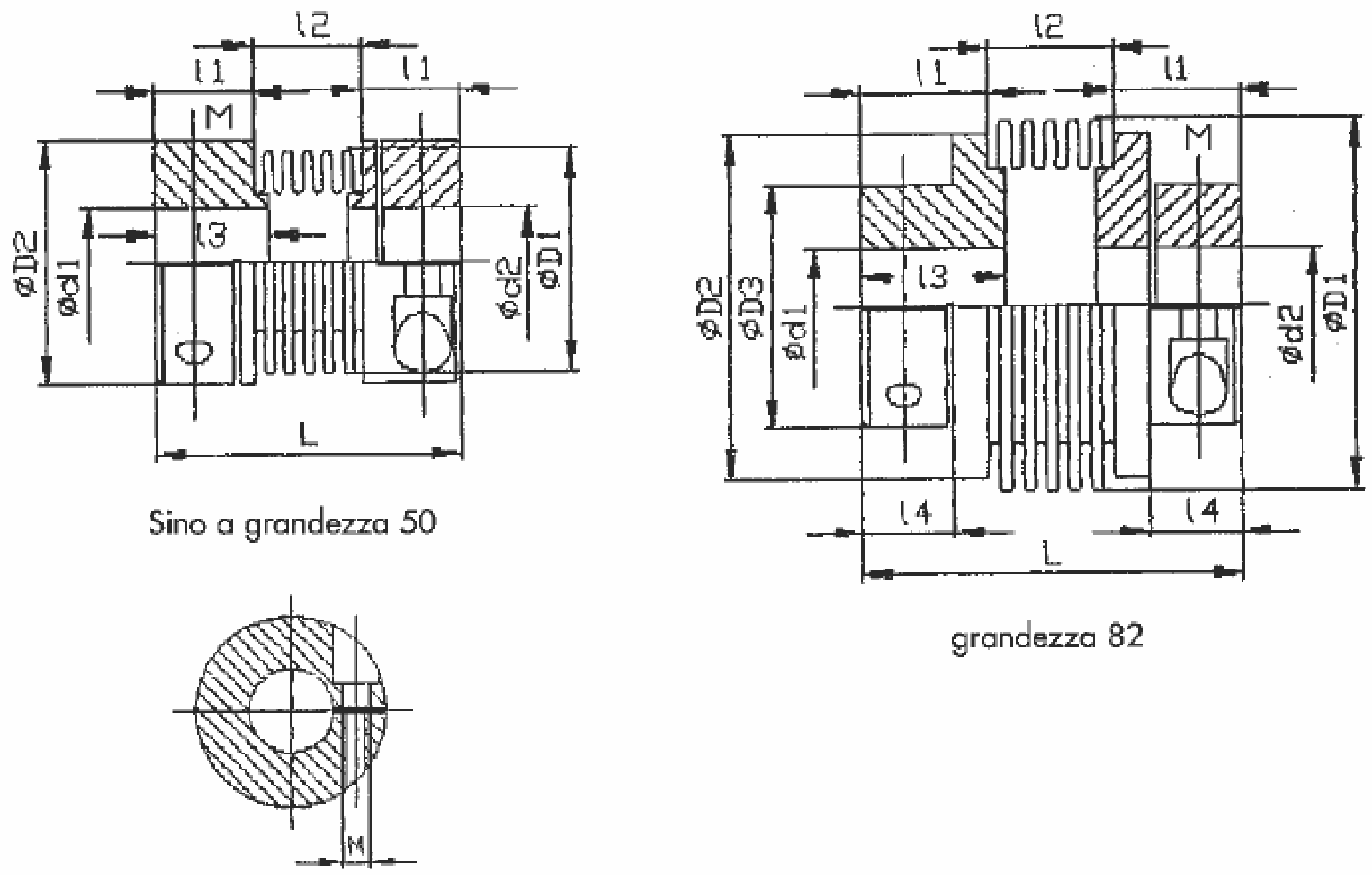

Serie FVK-K

| grandezza | $T_{KN}$ Nm | $T_{Kmax.}$ Nm | $d_1/d_2$ max mm | $D_1$ mm | $D_2$ mm | $D_3$ mm | $l_1$ mm | $l_2$ mm | $l_3$ mm | $l_4$ mm | L mm | M mm | ax +/- mm | rad mm | ang +/- Gradi | Momento d'inerzia in $Kgm^2 \times 10^{-6}$ calcolati per giunti non forati. |
|---|---|---|---|---|---|---|---|---|---|---|---|---|---|---|---|---|
| 15 | 4 | 8 | 7 | 15 | 18 | - | 10,5 | 11 | 12 | - | 32 | M3 | 0,08 | 0,08 | 0,56 | 2,30 |
| 20 | 7,5 | 15 | 8 | 19 | 22 | - | 12 | 12 | 14 | - | 36 | M4 | 0,07 | 0,06 | 0,45 | 3,812 |
| 28 | 35 | 70 | 12 | 28 | 30 | - | 14 | 17 | 16 | - | 45 | M5 | 0,11 | 0,1 | 0,4 | 17,44 |
| 37 | 50 | 100 | 18 | 36,5 | 38 | - | 17,5 | 20 | 20,5 | - | 55 | M6 | 0,14 | 0,17 | 0,5 | 58,73 |
| 50 | 85 | 170 | 25 | 50 | 52 | - | 21 | 28 | 24 | - | 70 | M8 | 0,2 | 0,18 | 0,5 | 248,20 |
| 66 | 120 | 240 | 28 | 66 | 58 | - | 23,5 | 33 | 27,5 | 18 | 80 | M8 | 0,3 | 0,22 | 0,45 | 382,8 |
| 82 | 160 | 320 | 30 | 82 | 78 | 65 | 31 | 33 | 35 | 21 | 95 | M10 | 0,32 | 0,24 | 0,44 | 1417 |
| 100 | 220 | 440 | 35 | 101 | 95 | 70 | 35 | 35 | 40 | 25 | 105 | M10 | 0,36 | 0,29 | 0,44 | 2844 |

A richiesta i fori possono venire forniti con cava per chiavetta secondo **DIN**

**N.B.** Il catalogo può subire variazioni senza preavviso

A richiesta, per minimo 10 pezzi, possiamo fornire giunti con TKN inferiore o superiore.

**Fig. C.2** Main features of joint by Favari.

## C.4. Cross roller bearing

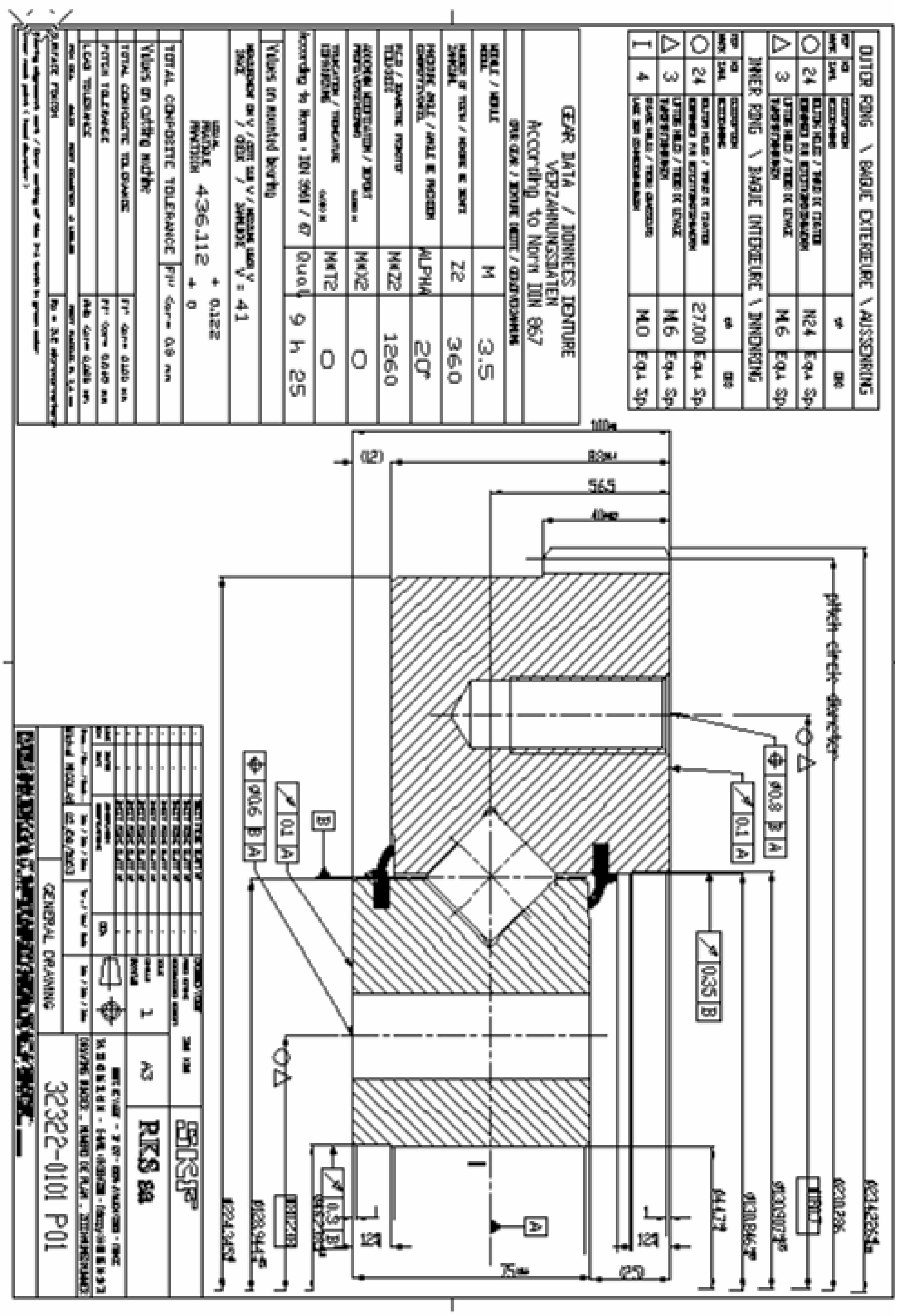

**Fig. C.3** Technical drawing of the cross roller bearing with geometric tolerances and gear datasheet.

## C.5. Taped roller bearings

| Principal dimensions | | | Basic load ratings | | Fatigue load limit | Speed ratings | | Mass | Designation |
|---|---|---|---|---|---|---|---|---|---|
| | | | dynamic | static | | Lubrication | | | |
| | | | | | | grease | oil | | |
| d | D | T | C | $C_0$ | $P_u$ | | | | |
| mm | | | N | | N | r/min | | kg | - |
| **160** | 240 | 51 | 429000 | 780000 | 78000 | 1300 | 1800 | 7,75 | **32032 X** |

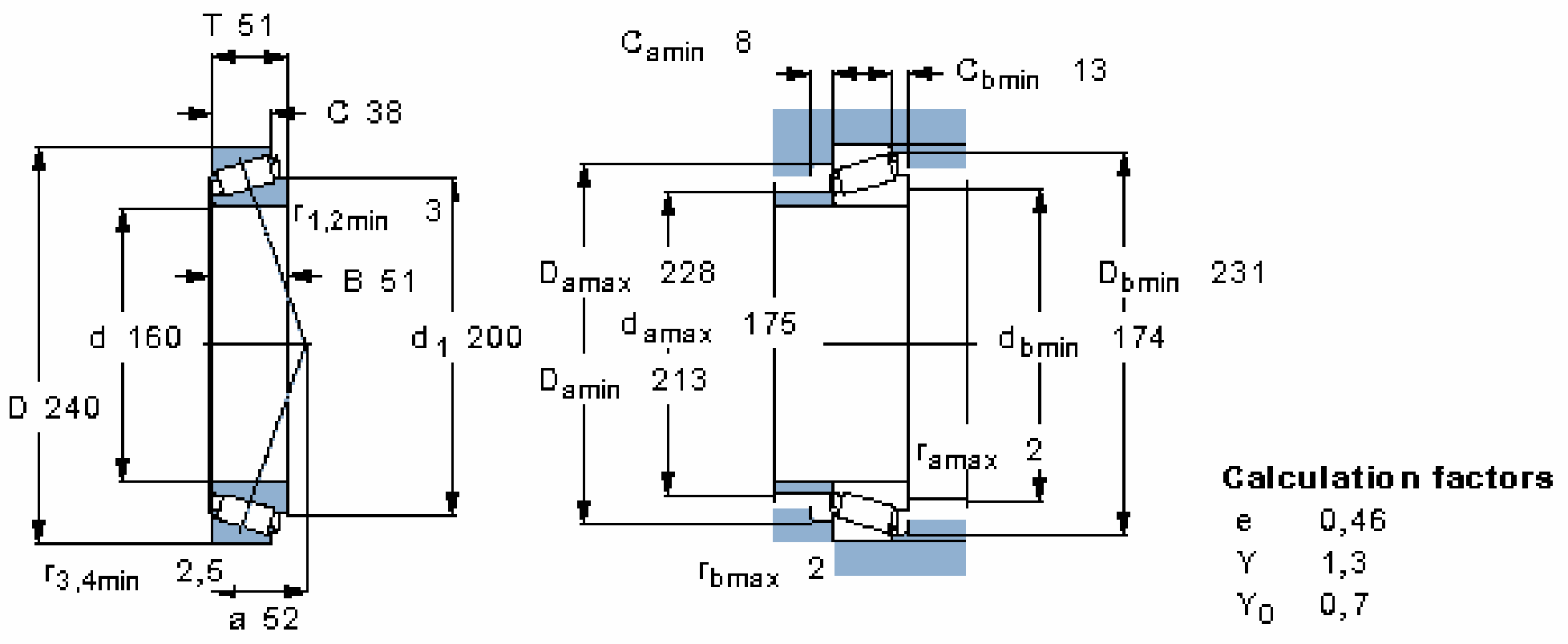

| Principal dimensions | | | Basic load ratings | | Fatigue load limit | Speed ratings | | Mass | Designation |
|---|---|---|---|---|---|---|---|---|---|
| | | | dynamic | static | | Lubrication | | | |
| | | | | | | grease | oil | | |
| d | D | T | C | $C_0$ | $P_u$ | | | | |
| mm | | | N | | N | r/min | | kg | - |
| **180** | 250 | 45 | 352000 | 735000 | 68000 | 1200 | 1700 | 6,65 | **32936** |

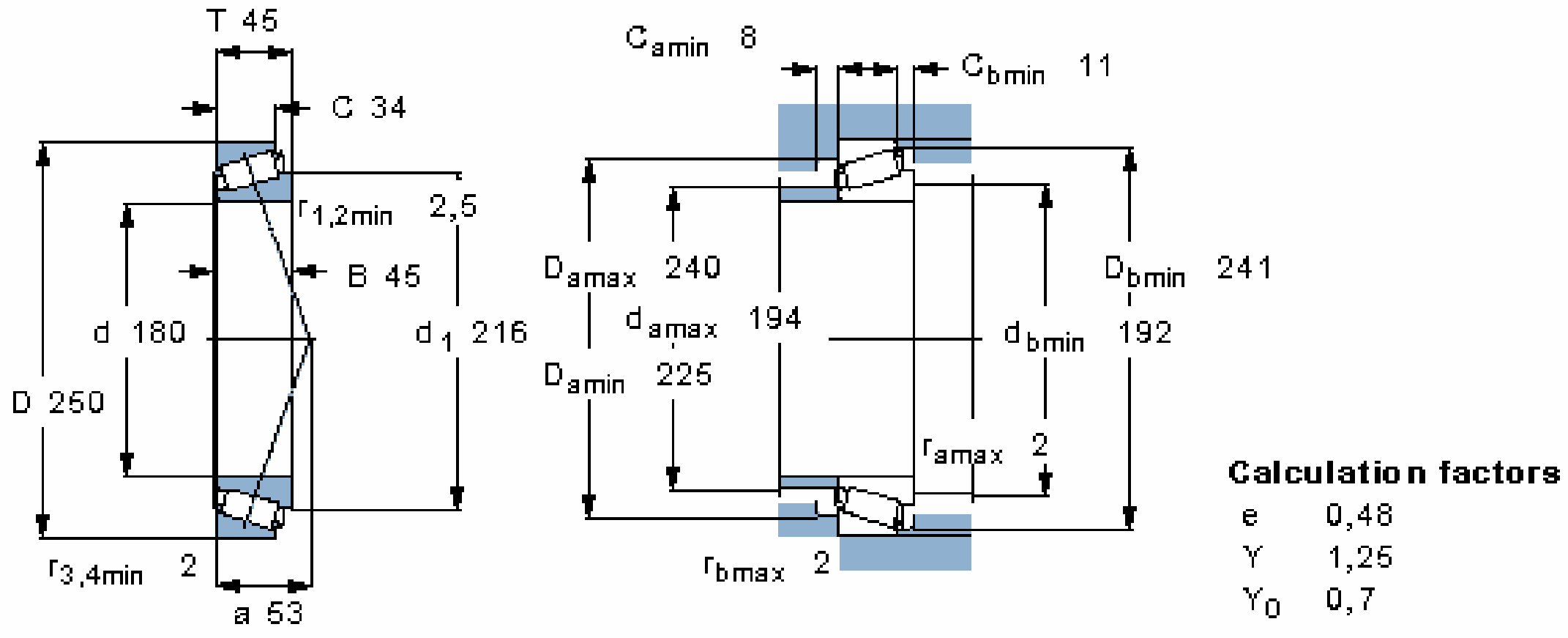

**Fig. C.4** Taper roller bearings main features.

| *Taper roller bearing* | *preload [mm]* | *Portion of loaded crown* | *Load on rollers (more loaded/less loaded) [N]* | *Specific pressure on rollers (max/min) [N/mm$^2$]* | |
|---|---|---|---|---|---|
| | | | | *anello int.* | *anello est.* |
| 32032 X | 0.0 | 0.484 | 755/0 | 493/0 | 479/0 |
| 32936 | | 0.487 | 580/0 | 506/0 | 503/0 |
| 32032 X | 0.010 | 0.742 | 672/0 | 474/0 | 461/0 |
| 32936 | | 0.744 | 515/0 | 486/0 | 483/0 |
| 32032 X | 0.020 | 1 | 882/197 | 519/313 | 505/305 |
| 32936 | | 1 | 673/150 | 532/320 | 529/319 |
| 32032 X | 0.030 | 1 | 1294/607 | 591/458 | 575/446 |
| 32936 | | 1 | 988/460 | 605/468 | 601/465 |
| 32032 X | 0.040 | 1 | 1796/1107 | 659/561 | 642/545 |
| 32936 | | 1 | 1372/845 | 675/574 | 671/571 |
| 32032 X | 0.050 | 1 | 2383/1701 | 725/647 | 705/630 |
| 32936 | | 1 | 1820/1294 | 742/662 | 738/658 |

**Table C-2** Variation of load with variation of preload.

## C.6. Pinion

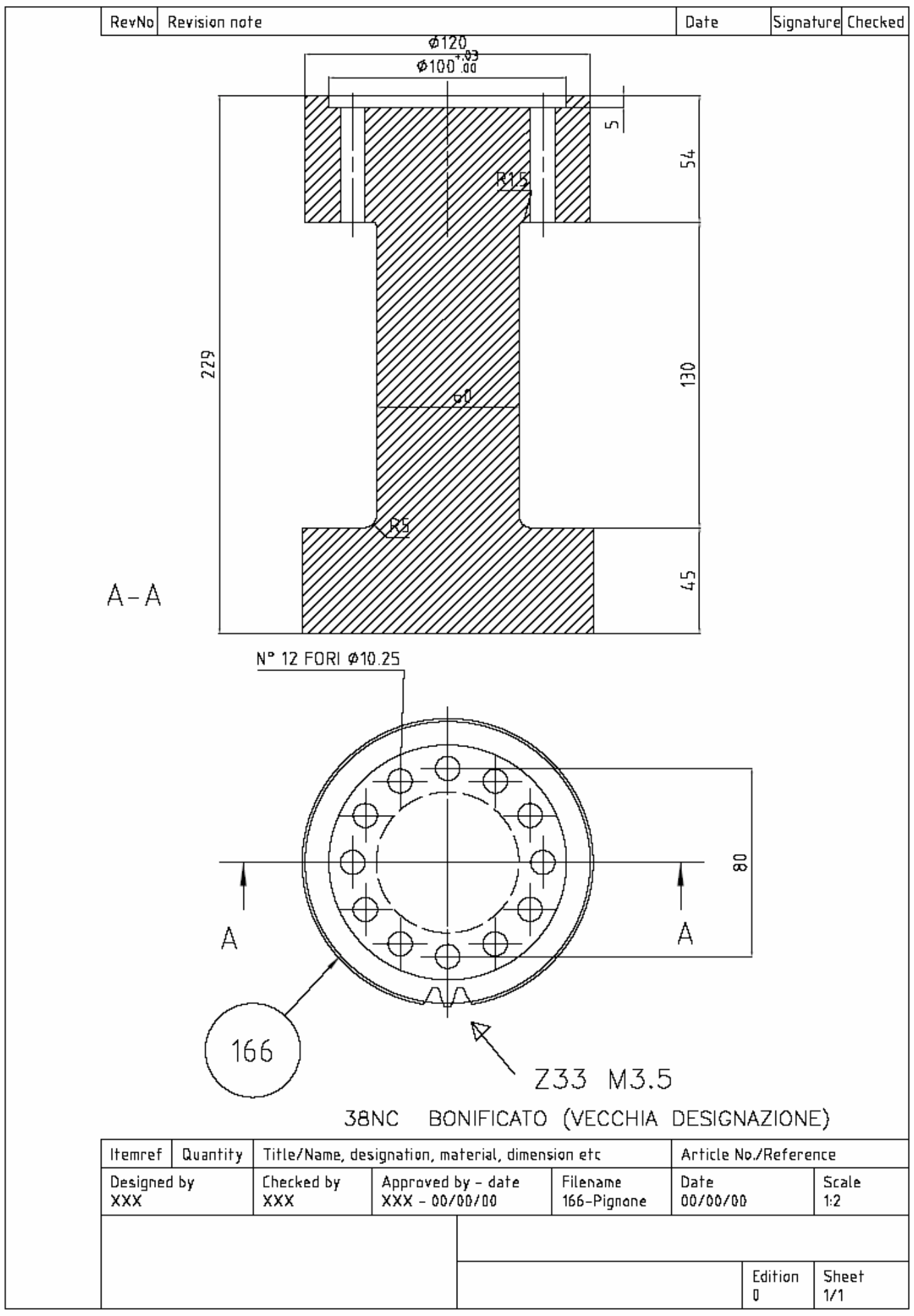

## C.7. Part number 44 AMICA interface flange

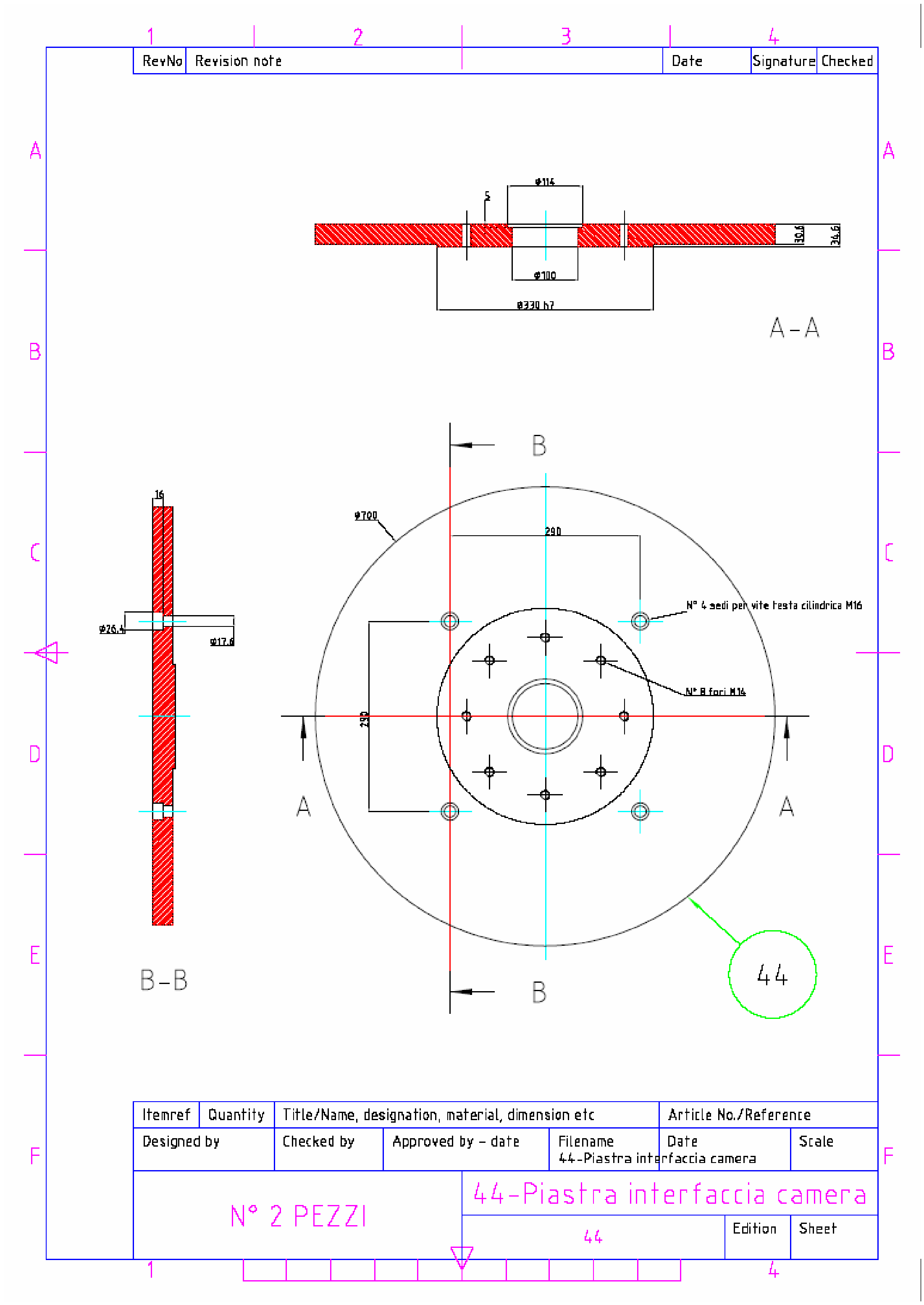

# Bibliography

Azouit, M.,Vernin, J., 2005, “Optical turbulence Profiling Balloons Relevant to Astronomy and Atmospheric Physics”, *PASP,* v.117, pp. 536-543.

Balser, D.S., & Prestage, R.M., “Systematic Elevation-Dependent Pointing Errors” , *PTCS/PN,* v. 24, 2003.

Bathe, K.J., *Finite Element Procedures in Engineering Analysis*, Prentice-Hall, Englewood Cliffs, N.J., 1982.

Borkowski, K.M., “Near zenith tracking limits for altitude-azimuth telescopes”, *Acta Astronomica*, pp.79-88, 1987.

Bru R., Catalan, A. et al., “IRAIT’s M2&M3 driver subsystems specification ”, Technical Report, NTE –SA, June 2005.

Bru, R., “Summary of activities until 12/2005 for the IRAIT M2&M3 driver subsystems”, Technical Report, NTE –SA 14, Dec 05.

Buckman, A., “Telescope pointing errors and correction”, AWR Technology, copy available from http://www.awr.tech.dial.pipex.com.

Busso, M., Tosti, G. et al., “The IRAIT project”, *Publ. Astron. Soc. Aust.*, v. 19, pp. 306–312, 2002.

Candidi, M., Lori, A., “Status of the Antarctic Base at Dome C and perspectives for Astrophysics”*, Mem. SAIt & IEPI* ,2001.

Catalan, A., “IRAIT’s M2 & M3 driver subsystem technical offer”*,* Technical Report, NTE –SA, 2004.

Dall’Oglio, G., De Bernardis P., “The OASI Project: Far Infrared Astronomy from Antarctica”, *Mem. S. A. It.*, v. 58, N°2-3, 1987.

Di Rico,G., Di Varano, I., “ The Antarctic Mid-Ir Camera (AMICA) for the IRAIT telescope”, *A-stronomische Nachrichten*, v. 325, Issue 6, p.664, 2004.

Di Varano, I., “Estimation of frequencies of the optical tube of IRAIT telescope”, Technical Report, INAF-OACT, 2004.

Dolci, M., “Stima del segnale di background sui rivelatori di AMICA al telescopio IRAIT, installato a Dome C”*,* Technical Report, INAF-OACT.

Eisele, J.A., Shannon, P.E.V., “Atmospheric Refraction Corrections for Optical Sightings of Astronomical Objects”, NRL Memorandum Report 3058, Naval Research Lab. (Washington 1975).

EN 1990:2002, *Eurocode 0: Basis of Structural Design*.

EN 1991:2001, *Eurocode1: Actions on structures.*

Gasperoni, F., Chomicz, R., Zordan, M., Dabalà, M., "IRAIT telescope and enclosure: engineering aspects for Antarctic operation", *Memorie SAIt*, **74**, p. 45, 2003.

Glass, I. S., *Handbook of Infrared Astronomy*, Cambridge Observing Handbooks for Research Series, UK 1999.

Juvinall, R.C., Marshek, K. M., *Fundamentals of machine component design,* John Wiley & Sons, pp. 851, New York 2003.

Kibrick R., Robinson L. et al., "An evaluation of precision tilt sensors for measuring telescope position", *Proc. SPIE,* v.2479, pp. 341-352,1995.

Lake, M., Hachkowski, M.R., "Mechanism Design Principle for Optical-Precision, Deployable Instruments", presented at the 41st AIAA/ASME/ASCE/AHS/ASC Structures, Structural Dynamics, and Materials Conference, AIAA Paper No. 2000-1409,2000.

Lawrence, J. S., Ashley, M., Tokovinin, A., Travouillon, T., "Exceptional astronomical seeing conditions above Dome C in Antarctica", *Nature*, v. 431, pp. 278-281, 2004.

Lawrence, J.S., "Infrared and Submillimetric Atmospheric Characteristics of High Antarctic Plateau Sites"*, PASP*, v. 116, pp. 482–492, 2004.

Lawrence, J.S., Ashley, M.C.B., "The AASTINO: Automated Astrophysical Site Testing International Observatory", in *Proceedings of the Conference on Towards Other Earths: DARWIN/TPF and the Search for Extrasolar Terrestrial Planets*, 22-25 April 2003, Heidelberg, Germany.

Marks, R.D., "Astronomical seeing from the summits of the Antarctic plateau"*, Astronomy and Astrophysics*, v. 385, pp. 328-336, 2002.

Marks, R.D., et al., "Antarctic site testing - microthermal measurements of surface-layer seeing at the South Pole", *Astron. Astrophys. Suppl.* Ser. 118, pp. 385-390, 1996.

Montenbruck, O., Pfleger,T., *Astronomy on the Personal Computer*, Springer, Heidelberg 2002.
Norme Tecniche C.N.R. 10021, *Prospetto 4-Ib –Costruzioni metalliche*, n.3, 1986.

PNRA, "Rapporto periodico di attività 4-24 dicembre 2004", in *XX Spedizione 2004-2005*, copy available from http://www.pnra.it/biblioteca/docs/.

Przemieniecki, J. S, *Theory of matrix structural analysis*, Mc Graw Hill Book Company, New York 1968.

RKS S.A., "Stock'n' roll Slewing bearings", Technical Documentation, 2003.

Roncella, F., "Minute IRAIT meeting 08/04/2004"*,* Technical Internal Report, 2004.

Schmidt,G., “Pointing/Derotator Coalignment for Alt-Az Telescopes”, MMTO Technical Memorandum 92-1, 2004.

Sick, J., “Introduction to Telescope control and errors”, copy available from http://homepage.mac.com/jonathansick/.

Sidgwick, J.B., *Amateur Astronomer’s Handbook*, Enslow Pub Inc, New Jersey 1980.

SKF, “Single row taper roller bearings” in *General Catalogue*, pp.628-630, June 2003.

Timoshenko, S.P., Woinoswski-Krieger, S., *Theory of Plates and Shells*, McGraw-Hill, New York 1987.

Tosti, G., Busso, M., Ciprini, S.,et al., “IRAIT: a Telescope for Infrared Astronomy from Antartica”*, SAIt*, v.74, p.37, 2003.

Tosti, G., Busso, M., Straniero, O. and Abia, C., “The Status of IRAIT project”, *Mem. S.A.It. Suppl.*, v. 5, 385, 2004.

Tosti, G., et al., *SPIE Proc.*,v. 5489, p. 742, 2004.

Travouillon, T., Ashley, M., Burton, M.G., Lawrence, J., Storey, J.W.V., “Low atmosphere turbulence at Dome C”*, Mem. S.A.It. Suppl.* v. 2, p.150, 2005.

Trueblood, M., Merle Genet, R., *Telescope control*, Willmann-Bell, Inc., 1997.

UNI 7011-72, *Container della serie 1*, Feb. 1972.
Walker, Christopher et al., “Forecast for HEAT on Dome A, Antarctica: the High Elevation Antarctic Terahertz Telescope”*, SPIE*, v. 5489, pp. 470-480, 2004.

Wallace, P., “Pointing and tracking algorithms for the Keck 10-Meter Telescope” , in L.B. Robinson, Ed. *Instrumentation for Ground-Based Optical Astronomy*, Springer-Verlag, New York 1988.

Zienkiewicz, O.C., Taylor, R., *The Finite Element Method*, v. 1, McGraw-Hill, London 1991.